\documentclass[usenatbib]{mn2e} \usepackage{epsfig}
\newcommand{\lsim}{\lower 2pt \hbox{$\, \buildrel {\scriptstyle
<}\over {\scriptstyle \sim}\,$}}  \newcommand{\gsim}{\lower 2pt
\hbox{$\, \buildrel {\scriptstyle >}\over {\scriptstyle \sim}\,$}}
\usepackage{appendix}
\usepackage{lscape}
\defcitealias{Robinson00}{Robinson et al., 2000}
\defcitealias{Veilleux99b}{Veilleux et al., 1999b}
\defcitealias{Veilleux99a}{Veilleux et al., 1999a}

\bibpunct{(}{)}{;}{a}{,}{,} \title[The properties of the stellar populations in ULIRGs I: sample, data and spectral synthesis modelling.]{The properties of the stellar populations in ULIRGs I: sample, data and spectral synthesis modelling.}
\author[J.Rodr\'iguez Zaur\'in, C.N.Tadhunter and R.M.Gonz\'alez
Delgado$^{2}$]{J.Rodr\'iguez
Zaur\'in$^{1,2}$\thanks{E-mail:jrz@damir.iem.csic.es}, C.N
Tadhunter$^{2}$ and R.M.Gonz\'alez Delgado$^{3}$\\ $^{1}$Department of
molecular and infrared astronomy, IEM (CSIC), 28006 Madrid, Spain\\
$^{2}$ Department of physics and Astronomy, University of Sheffield, Sheffield S3 7RH\\
$^{3}$Instituto de Astrofisica de Andalucia(CSIC), P.O.Box 3004, 18080
Granada, Spain}
\begin{document}

\pagerange{\pageref{firstpage}--\pageref{lastpage}} \pubyear{2002}

\maketitle

\label{firstpage}

\begin{abstract}

  We present deep long-slit optical spectra for a sample of 36
  Ultraluminous Infrared Galaxies (ULIRGs), taken with the William
  Herschel Telescope (WHT) on La Palma with the aim of investigating
  the star formation histories and testing evolutionary scenarios for
  such objects. Here we present the sample, the analysis techniques
  and a general overview of the properties of the stellar populations;
  a more detailed discussion will be presented in a forthcoming
  paper. Spectral synthesis modelling has been used in order to
  estimate the ages of the stellar populations found in the diffuse
  light sampled by the spectra in both the nuclear and extended
  regions of the target galaxies. We find that adequate fits can be
  obtained using combinations of {\it young stellar populations}
  (YSPs, t$_{YSP} \leq$ 2 Gyr), with ages divided into two groups:
  {\it very young stellar populations} (VYSPs, t$_{VYSP}$ $\leq$ 100
  Myr) and {\it intermediate-young stellar populations} (IYSPs, 0.1
  $<$ t$_{IYSP}$ $\leq$ 2 Gyr). Our results show that YSPs are present
  at {\it all} locations of the galaxies covered by our slit
  positions, with the exception of the northern nuclear region of the
  ULIRG IRAS 23327+2913. Furthermore, VYSPs are presents in at least
  85\% of the 133 extraction apertures used for this study, being more
  significant in the nuclear regions of the galaxies. {\it Old stellar
  populations} (OSPs, t$_{OSP} >$ 2 Gyr) do not make a major
  contribution to the optical light in the majority of the apertures
  extracted. In fact they are essential for fitting the spectra in
  only 5\% (7) of the extracted apertures. The estimated total masses
  for the YSPs (VYSPs + IYSPs) are in the range 0.18$\times10^{10}
  \leq M_{\rm YSP}\leq$ 50 $\times10^{10}$ M$_{\odot}$. We have also
  estimated the bolometric luminosities associated with the stellar
  populations detected at optical wavelengths, finding that they fall
  in the range 0.07 $\times$ 10$^{12}$ $<$ L$_{\rm bol} <$ 2.2
  $\times$ 10$^{12}$ L$_{\odot}$. In addition, we find that reddening
  is significant at all locations in the galaxies. This result
  emphasizes the importance of accounting for reddening effects when
  modelling the stellar populations of star-forming galaxies.

\end{abstract}


\section{Introduction}

The launch of the {\it Infrared Astronomical Satellite (IRAS)} in
1983 triggered a development of infrared astronomy that is still
ongoing. Although the majority of the sources discovered by {\it IRAS}
were modest infrared emitters, some objects had mid- to far-infrared
(MFIR) luminosities comparable to the bolometric luminosities of
optically-selected QSOs
\citep{Houck84,Houck85,Soifer84a,Soifer84b,Soifer87}. Such galaxies
are classified as Luminous (L$_{ir} > 10^{11}$ L$_{\odot}$) or
Ultraluminous ($L_{ir} > 10^{12}$ L$_{\odot}$) infrared galaxies
(LIRGs/ULIRGs) depending on their MFIR luminosities.

The prodigious infrared emission of ULIRGs is generally attributed to
the optical/UV light of luminous central sources reprocessed by
dust. Several infrared spectroscopic studies have provided evidence
that the power source of ULIRGs has a composite nature, with a mixture
of AGN and starburst activity
\citep{Genzel98,Lutz99,Armus07,Imanishi07,Farrah07}. Similar
conclusions are reached in the relatively few studies carried out at
UV/optical \citep{Veilleux95,Veilleux99a,Farrah03} and X-ray
\citep{Franceschini01,Franceschini03} wavelengths. Therefore these
objects provide us with an excellent oportunity to study the links
between AGN and starburst phenomena. In addition, ULIRGs are almost
unvariably associated with galaxy mergers and interactions
\citep[e.g.][]{Kim02,Veilleux02}. Thus, they also represent ideal
objects to test models of galaxy evolution via major galaxy mergers
(e.g. Mihos \& Hernquist, 1996; Barnes \& Hernquist, 1996; Springel et
al., 2005).

Stellar population studies have the potential to provide key
information about the histories of the merger events asscoaited with
ULIRGs. To date, two complementary approaches have been used to
investigate the stellar populations in such objects: one based on
photometric analysis of images of the galaxies taken at different
wavelengths, focussed mainly on studying the young stellar populations
associated with the bright knots
\citep{Surace98b,Surace99,Surace00a,Surace00b,Wilson06}; the other
comprising spectrocopic analysis of UV/optical spectra
\citep{Canalizo00a,Canalizo00b,Canalizo01,Farrah05,
Rodriguez-Zaurin07,Rodriguez-Zaurin08}. Overall, there have been
relatively few studies of the stellar populations in ULIRGs. The
existing studies mentioned above either concern relatively small
samples of ULIRGs, or concentrate on high surface brightness regions
which may give a misleading impression of the stellar populations of
the systems as a whole.

To remedy this situation we have obtained deep, wide spectral
coverage, intermediate-resolution long-slit spectroscopic observations
for a sample of 36 ULIRGs with redshifts z $<$ 0.18 using the ISIS
spectrograph on the WHT. With the aim of studing in detail the
properties of the stellar populations in the ULIRGs in our sample, and
thereby testing evolutionary scenarios, we model the spectra extracted
from a series of apertures using a large number of different
combinations of stellar populations synthesis model templates
\citep{Bruzual03}. In this paper we present the sample and the
spectral synthesis modelling results, while a more detailed analysis
of the results is presented in the forthcoming paper (Rodr\'iguez
Zaur\'in et al., in prep). Detailed analyses of two of the objects in
the sample, IRAS 13451+1232 (PKS1345+12) and IRAS 15327+2340 (Arp 220)
were presented in Rodr\'iguez Zaur\'in et al. (2007, hereafter RZ07)
and Rodr\'iguez Zaur\'in et al. (2008, hereafter RZ08) respectively.

Throughout this paper, we assume a cosmology with H$_{0}$ = 71 km
s$^{-1}$ Mpc$^{-1}$, $\Omega_{0} = 0.27$, $\Omega_{\Lambda} = 0.73$.
All the values presented in this paper have been adapted to this
cosmology.

\section{Observations and data Reduction}
\subsection{The Sample}

In the first place, we observed a complete RA- and
declination-limited sub-sample of the \cite{Kim98a} 1 Jy sample of
ULIRGs \footnote{This is in itself a complete flux-limited sample of
118 ULIRGs identified from the IRAS Faint Source Catalogue (FSC) and
selected to have a 60$\mu$m flux densities greater than 1 Jy in a
region of the sky with $\delta$ $>$ -40$^{\circ}$ and $|b|$ $>$
30$^{\circ}$.}, with RAs in the range 12 $<$ RA $<$ 1:30 h,
declinations $\delta$ $>$ -23 degrees, and redshifts z~$<$ 0.13. The
redshift limit was chosen to ensure that the objects are sufficiently
bright for spectroscopic study, and also to keep the size of the
sample tractable for deep observations on 4m class
telescopes. Throughout the paper, we will refer to this sample of 26
objects as the complete sample (CS). Table \ref{comp_sample}
summarizes some of the properties of the ULIRGs in the CS.

As shown in the table, only 8 ULIRGs with warm mid- to far-IR colours
(f$_{25}$/f$_{60}$ $>$ 0.2)\footnote{The quantities {\it f\/}$_{25}$
and {\it f}$_{60}$ represent the {\it IRAS} flux densities in Janskys
at 25$\mu$m and 60$\mu$m.} are included in the CS. However, one of the
aims of the project was to study the presence of an evolutionary link
between cool and warm ULIRGs. Therefore, we extended the sample to
include 6 higher-redshift objects classified as warm ULIRGs: IRAS
13305-1739, IRAS 14252-1739, IRAS 16156+0146, IRAS 17044+6720, IRAS
23060+0505 and IRAS 23389+0303. In addition, since we also aim to test
the evolutionary scenario in which ULIRGs evolve into QSOs or a radio
galaxies \cite[e.g.][]{Surace98b,Surace99,Canalizo01,Tadhunter05}, we
decided to include IRAS 17179+5444, which is classified as a cool
ULIRG with a Sy2-like optical spectrum. Finally, three ULIRGs were
included in the sample (IRAS 08572+3915, IRAS 10190+1322 and IRAS
10494+4424) which have redshifts z $<$ 0.13, but fall outside the RA
range of the CS. One of them, IRAS 08572+3915, is classified as a warm
ULIRG. Thus, the total sample discussed in this paper comprises 36
objects: 21 cool ULIRGs, and 15 warm ULIRGs. We will refer to this
sample as the extended sample (ES). With an upper limiting redshift of
z $<$ 0.18, the ES includes all the warm ULIRGs within the RA and
declination range of the CS with redshifts z $<$~0.15. Table
\ref{ext_sample} summarizes some of the properties of the 10 ULIRGs
which, together with the CS, comprise the ES. Optical and near-IR
images for all the objects in the ES are presented in \cite{Kim02}.

\begin{table*}
\centering
\begin{tabular}{lllrllllll}
\hline\hline
Object Name & z &RA & DEC & log & f$_{25}$/f$_{60}$ &Nuclear & NS & IC
& ST \\ IRAS & & (J2000.0) & (J2000.0) & L$_{ir}$ & &structure &
(kpc)& & \\ & & & & (L$_\odot$)& & & & & \\ (1) & (2) & (3) & (4) &
(5) & (6) & (7) & (8) & (9) & (10)\\
\hline
00091--0738 & 0.118& 00 11 43.4 & -07 22 06 & 12.23 & 0.08 & Double & 2.2 & IIIb & H\\ 
00188--0856 & 0.128& 00 21 26.0 & -08 39 29 & 12.37 & 0.14 & Single & -- & V & L\\  
01004--2237 & 0.129& 01 02 51.2 & -22 21 51 & 12.28 & 0.28 & Single & -- & V & H\\ 
12072--0444$^a$ & 0.129& 12 09 45.4 & -05 01 14 & 12.39 & 0.21 & Double & 2.8 & IVb & S2\\ 
12112+0305 & 0.073 & 12 13 47.3 & 02 48 34  & 12.32 & 0.06 & Double & 4.0 & IIIb & L\\ 
12540+5708 & 0.042 & 12 56 15.0 & 56 52 17  & 12.54 & 0.27 & Single & -- & IVb & S1\\ 
13428+5608$^a$ & 0.037 & 13 44 41.8 & 55 53 14& 12.14 & 0.10 & Double & 0.7 & IVb & S2\\
13451+1232 & 0.122 & 13 47 33.3 & 12 17 24  & 12.32 & 0.35 & Double & 4.2 & IIIb & S2\\ 
13539+2920 & 0.108 & 13 56 10.9 & 29 05 29  & 12.04 & 0.06 & Double & 7.4 & IIIb & H\\ 
14060+2919 & 0.117 & 14 .08 17.5 & 29 04 57 & 12.07 & 0.08 & Single & -- & IVa & H\\ 
14348--1447 & 0.083& 14 37 37.3 & -15 00 20 & 12.32 & 0.07 & Double & 5.1 & IIIb & L\\ 
14394+5332$^b$ & 0.105 & 14 41 04.3 & 53 20 08  & 12.08 & 0.17 & Multiple & 54.0 & Tpl & S2\\
15130--1958 & 0.109& 15 15 55.6 & -20 09 18 & 12.09 & 0.20 & Single & -- & IVb & S2\\ 
15206+3342 & 0.125 & 15 22 38.0 & 33 31 36  & 12.22 & 0.20 & Single & -- & IVb & H\\ 
15327+2340 & 0.018 & 15 34 57.1 & 23 30 10  & 12.21 & 0.07 & Double & 0.4 & IIIb & L\\ 
15462-0450 & 0.100 & 15 48 56.6 & -04 59 36 & 12.21 & 0.15 & Single & -- & IVb & S1\\ 
16474+3430 & 0.111 & 16 49 14.7 & 34 25 13  & 12.15 & 0.09 & Double & 6.9 & IIIb & H\\ 
16487+5447 & 0.104 & 16 49 47.8 & 54 42 34  & 12.16 & 0.07 & Double & 5.7 & IIIb & L\\
17028+5817 & 0.106 & 17 03 41.8 & 58 13 48  & 12.14 & 0.04 & Double & 24.6 & IIIa & L\\ 
20414-1651 & 0.086 & 20 44 17.4 & -16 40 14 & 12.18 & 0.08 & Single & -- & IVb & H\\ 
21208--0519 & 0.130& 21 23 28.7 & -05 06 59 & 12.05 & 0.13 & Double & 14.9 & IIIa & H\\ 
21219--1757 & 0.112& 21 24 42.5 & -17 44 40 & 12.10 & 0.42 & Single & -- & IVa & S1\\ 
22491--1808 & 0.076 & 22 51 49.0 & -17 52 27& 12.13 & 0.10 & Double & 2.4 & IIIb & H \\
23233+2817 & 0.114 & 23 25 48.7 & 28 34 19  & 12.04 & 0.22 & Single & -- & Iso. &S2\\ 
23234+0946 & 0.128 & 23 25 56.0 & 10 02 52  & 12.09 & 0.05 & Double & 7.8 & IIIb & L\\ 
23327+2913 & 0.107 & 23 35 12.5 & 29 30 05  & 12.10 & 0.10 & Double & 24.0 & IIIa &L\\
\end{tabular}
\caption [Properties of the 26 ULIRGs in the Complete sample (CS) with
z $<$ 0.13 and $\delta$ $>$ 0.23]{The Complete sample of 26 ULIRGs
with redshifts z $<$ 0.13 and declinations $\delta$ $>$ -23
degrees. Col (1): object designation in the IRAS Bright Galaxy survey
\citep{Soifer86,Soifer87}. Col (2): optical redshifts from
\cite{Kim98a}. Cols (3) and (4): right ascension (hours, minutes and
seconds) and declination (degrees, arcminutes and arcseconds) of the
IRAS source position as listed in the Faint Source Database
(FSDB). Col (5): IR luminosity from \cite{Kim98a}. Col (6):
f$_{25}$/f$_{60}$ mid- to far-infrared colour ratio, where f$_{25}$
and f$_{60}$ represent the {\it IRAS \/} flux densities (non-colour
corrected) in units of Jy at 25 and 60~$\mu$m respectively. Cols (5)
and (6): both from \cite{Kim98a}. Col (7): nuclear structure
\cite[from][]{Kim02}.  Col (8): nuclear separation (NS) from
\cite{Veilleux02}. Col (9): interaction class. Class I: first
approach; Class II: first contact; Class III: pre-merger; Class IV:
merger; Class V: old merger; Tpl: Triple; Iso:
isolated. \cite[see][for details]{Veilleux02}.  Col (10): optical
nuclear spectral type \cite[from][]{Veilleux95,Kim98a,Veilleux99a}. H:
HII; L: LINER; S1: Seyfert 1; S2: Seyfert 2.
  \newline $^{a}$ \cite{Kim02} classified these objects as single
  nucleus systems. However, a double nucleus structure is revealed in
  the studies of \cite{Dasyra06a} and \cite{Scoville00} for IRAS
  12072-0444 and IRAS 13428+5608 respectively. These objects will be
  classified as double nucleus systems for the work presented here.
  \newline $^{b}$ This source is a multiple ($>$ 2 nucleus) system.  
  The two main components are separated 54 kpc, but the eastern component 
  itself comprises two close nuclei.}
\label{comp_sample}
\end{table*}
\begin{table*}
%\vspace{-3.0cm}
\centering
\begin{tabular}{lllrllllll}
\hline\hline
Object Name & z &RA & DEC & log & f$_{25}$/f$_{60}$ & Nuclear & NS & IC & ST \\ 
IRAS & & (J2000.0) & (J2000.0) & L$_{ir}$ & &structure & (kpc) & & \\  
& & & & (L$_\odot$)& & & & & \\ 
(1) & (2) & (3) & (4) & (5) & (6) & (7) & (8) & (9) & (10) \\
\hline
08572+3915 & 0.058 & 09 00 25.0 & 39 03 56 & 12.15 & 0.22 & Double & 6.0 & IIIb &L\\ 
10190+1322 & 0.077 & 10 21 41.9 & 13 07 01 & 12.04 & 0.11 & Double & 5.9 & IIIb &H\\ 
10494+4424 & 0.092 & 10 52 22.2 & 44 08 59 & 12.17 & 0.04 & Single & -- & IVb &L\\ 
13305--1739 & 0.148& 13 33 15.2 &-17 55 01 & 12.25 & 0.33 & Single & -- & V & S2\\ 
14252--1550 & 0.149& 14 28 01.4 & -16 03 43 & 12.19 & 0.20 & Double & 8.8 & IIIb &L\\ 
16156+0146 & 0.132 & 16 18 08.2 & 01 39 21 & 12.08 & 0.24 & Double & 8.0 & IIIb &S2\\ 
17044+6720 & 0.135 & 17 04 28.5 & 67 16 34 & 12.17 & 0.28 & Single & -- & IVb &L\\
17179+5444 & 0.147 & 17 18 55.1 & 54 41 50 & 12.24 & 0.14 & Single & -- & IVb &S2\\ 
23060+0505 & 0.173 & 23 08 34.2 & 05 21 29 & 12.48 & 0.37 & Single & -- & IVb &S2\\ 
23389+0303 & 0.145 & 23 41 31.1 & 03 17 31 & 12.13 & 0.28 & Double & 5.2 & IIIb &S2\\ 
\end{tabular} 
\caption[Properties of the 10 additional objects included in the 
Extended sample(ES)] {Same as Table 1, but for the 10 additional objects
included in the extended sample (ES) discussed in this paper. }
\label{ext_sample}
\end{table*}

\begin{figure*}
\begin{tabular}{ccc}
\hspace{-1.0 cm}\psfig{file=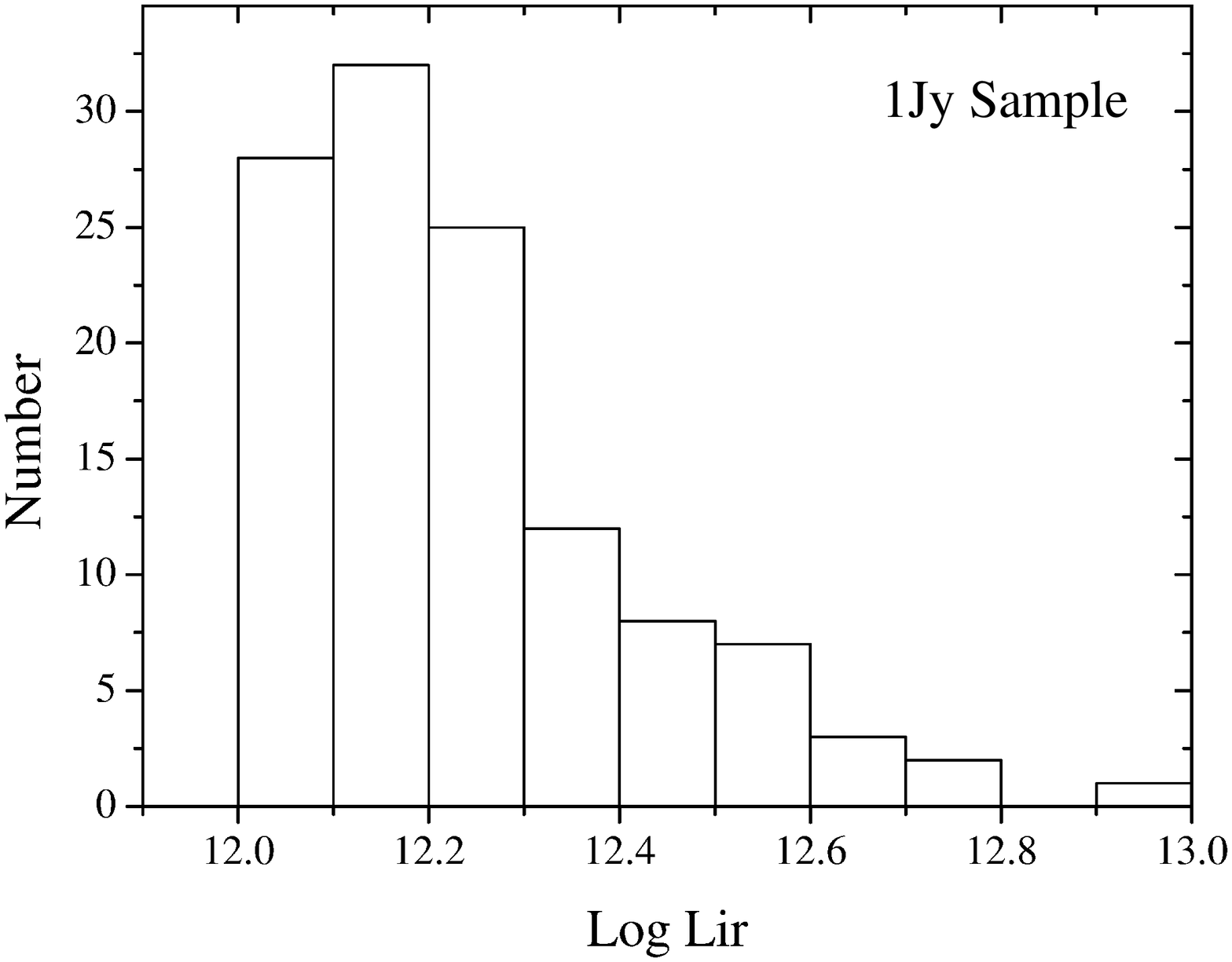,width=6.0cm,angle=0.}&
\psfig{file=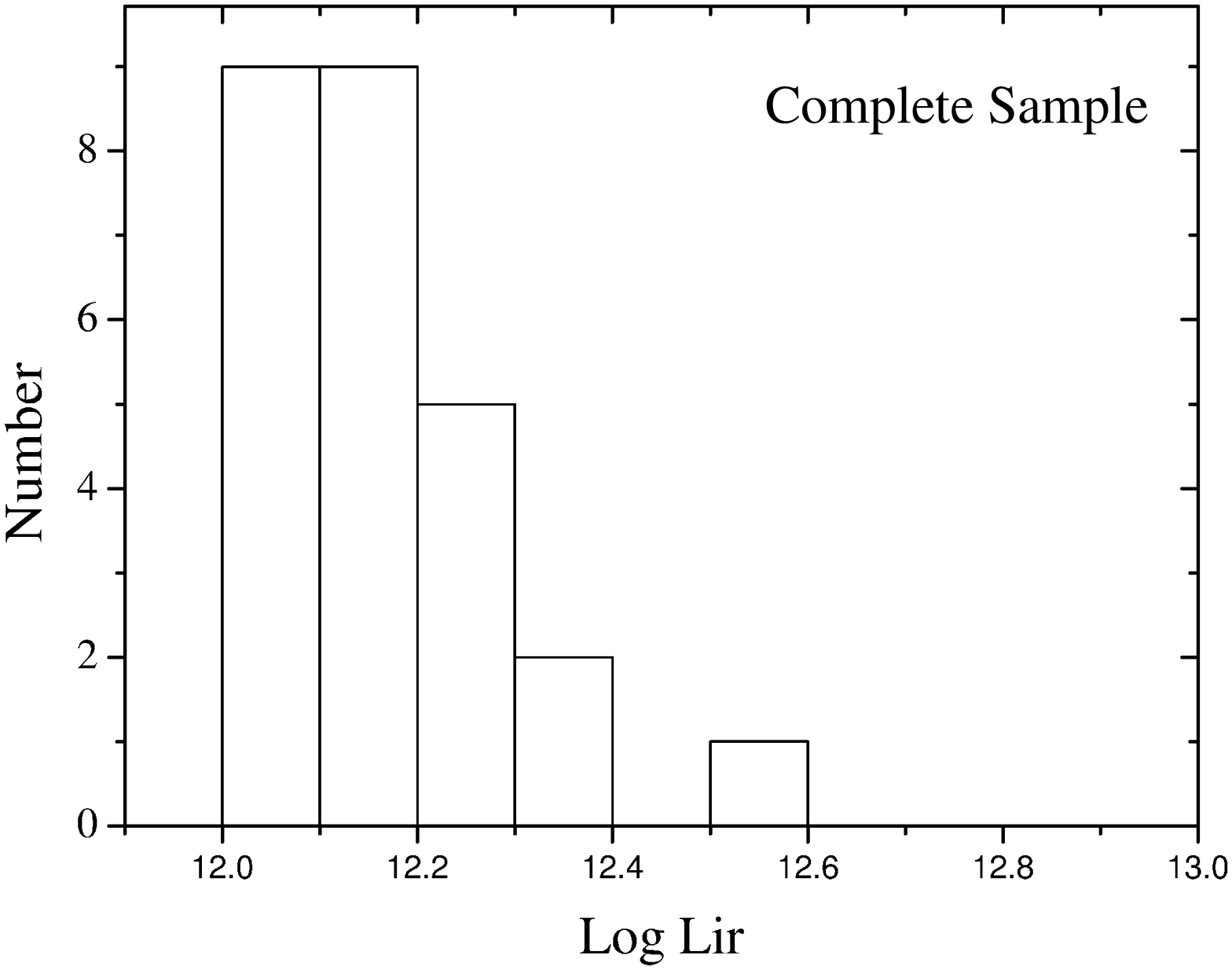,width=6.0cm,angle=0.}&
\psfig{file=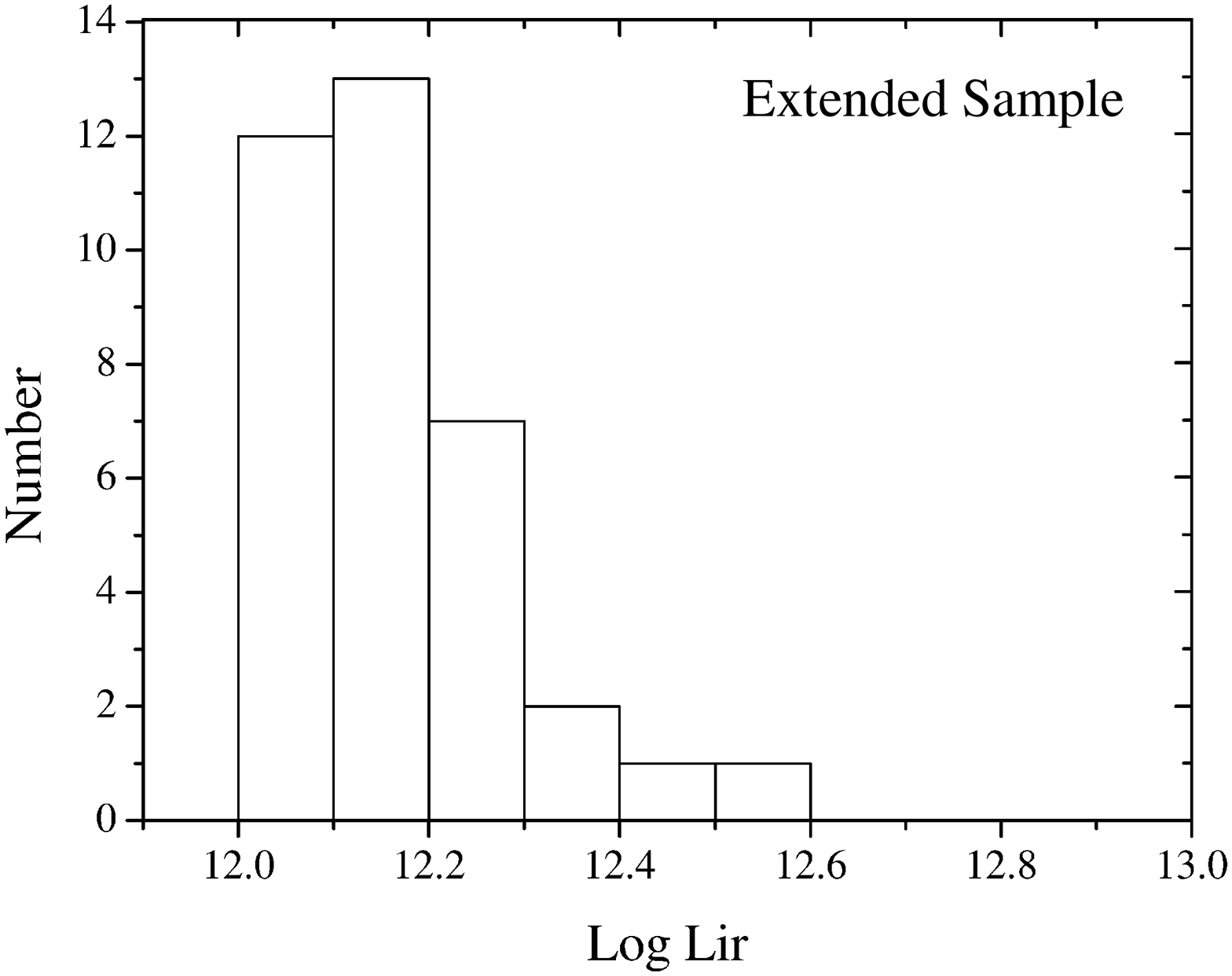,width=6.0cm,angle=0.}\\
\hspace{-1.0 cm}\psfig{file=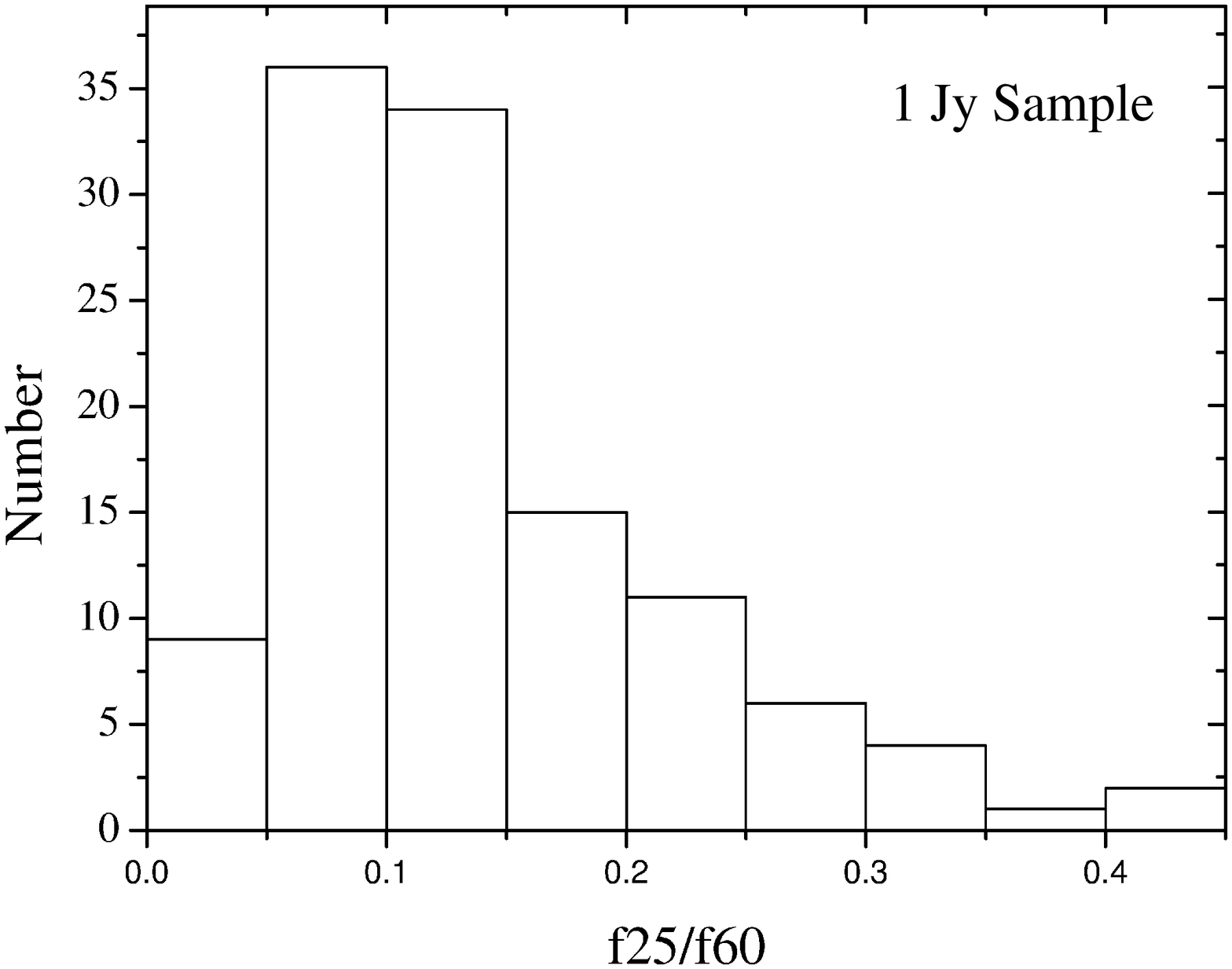,width=5.75cm,angle=0.}&
\psfig{file=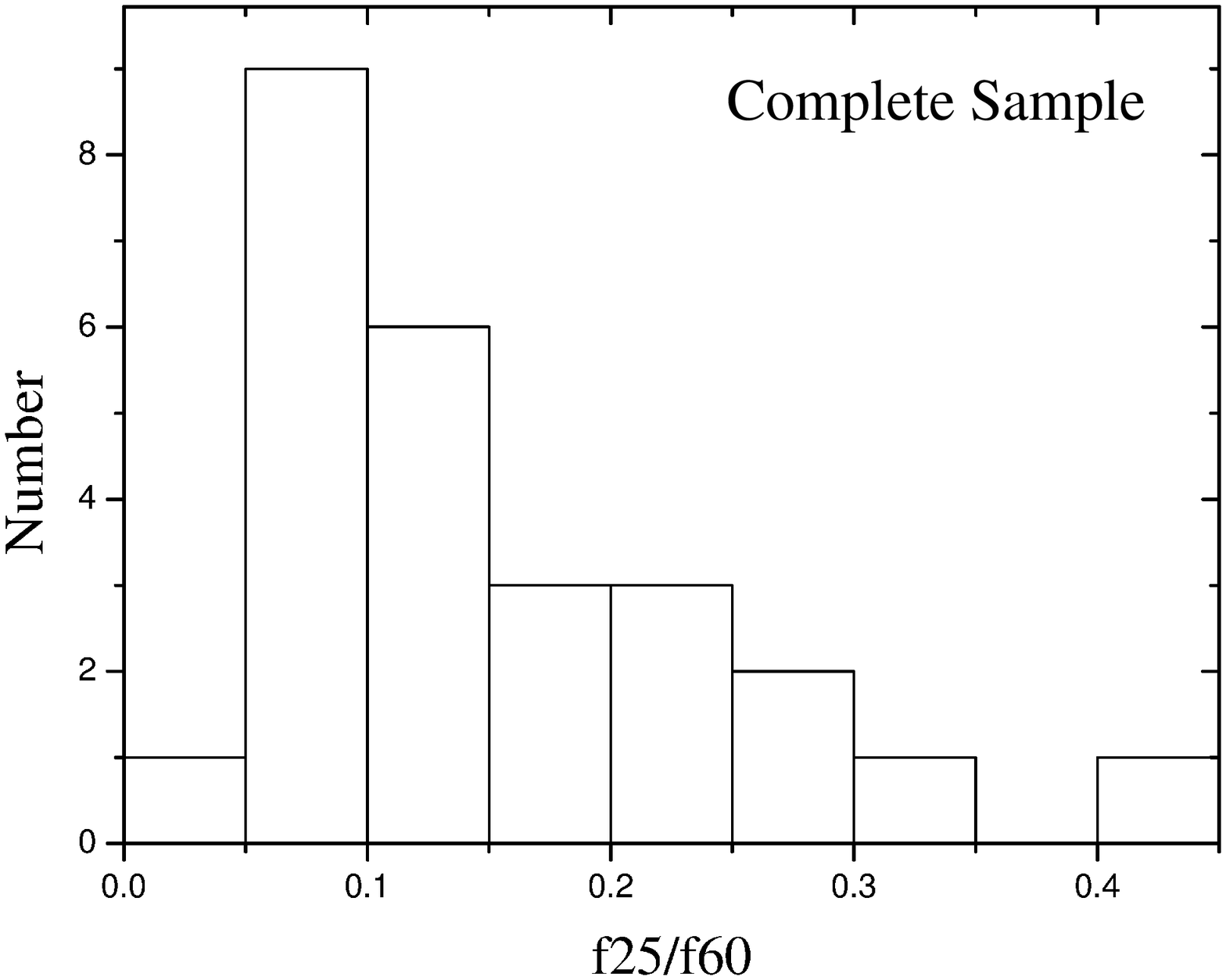,width=5.75cm,angle=0.}&
\psfig{file=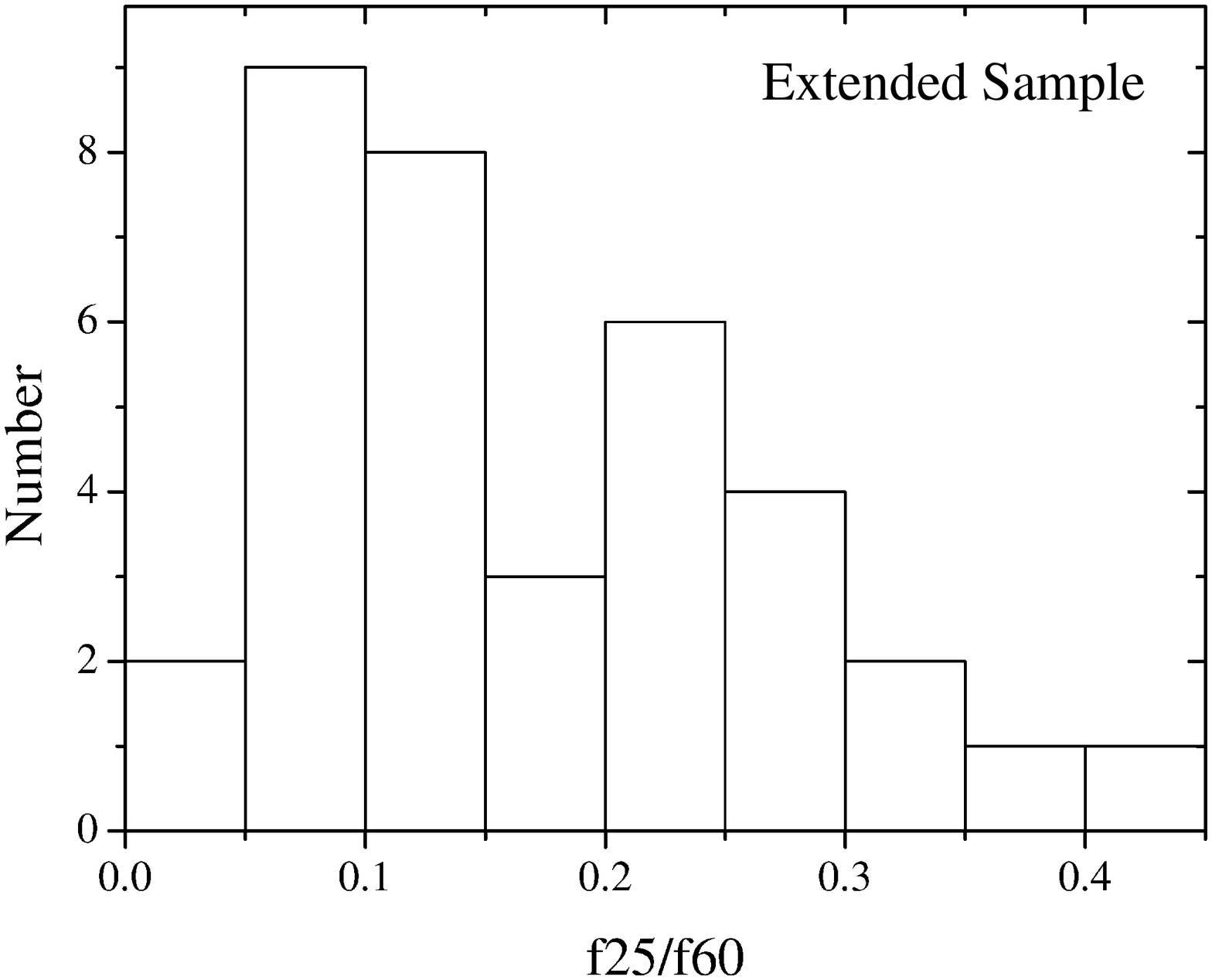,width=5.75cm,angle=0.}\\
\hspace{-1.0 cm}\psfig{file=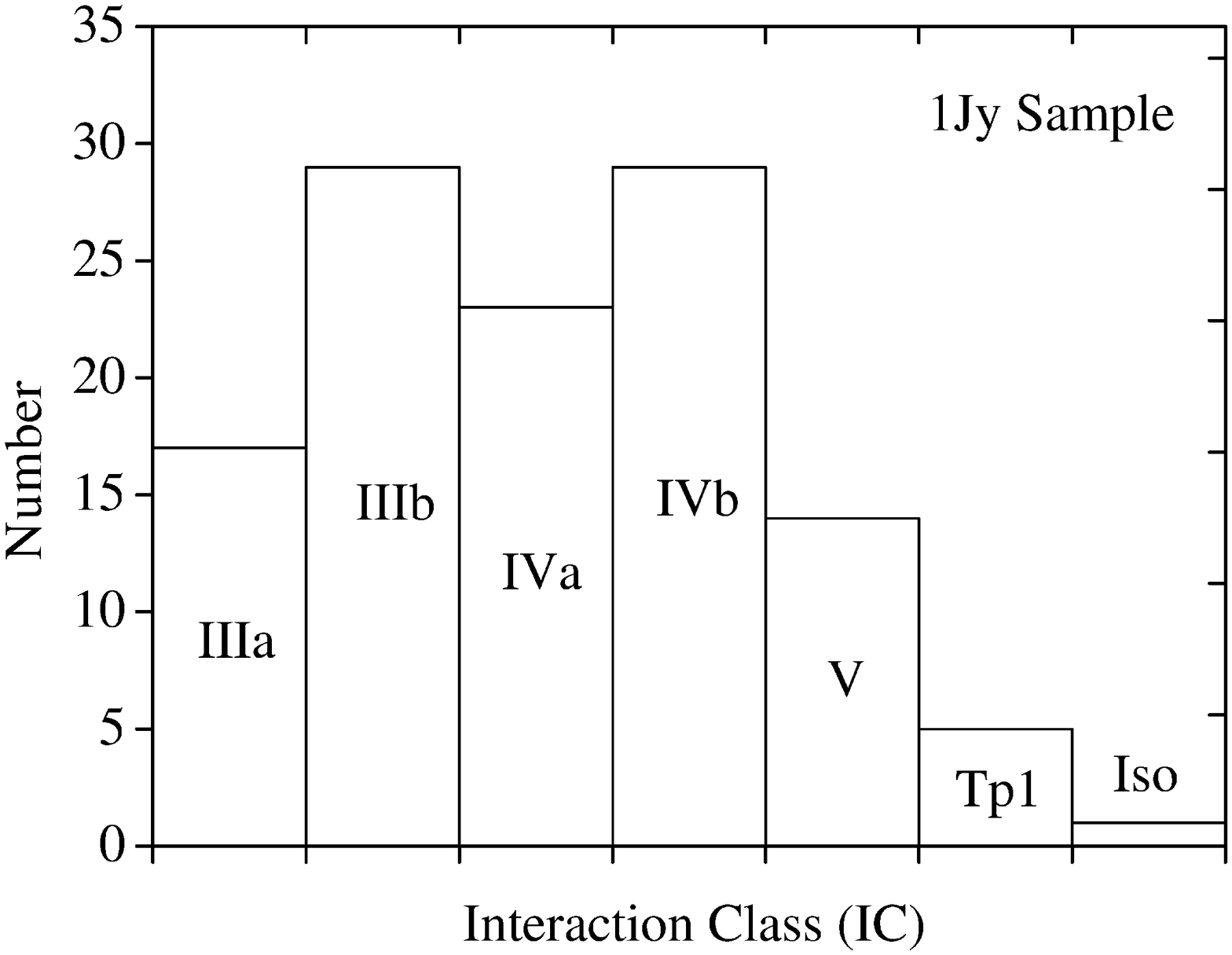,width=6.0cm,angle=0.}&
\psfig{file=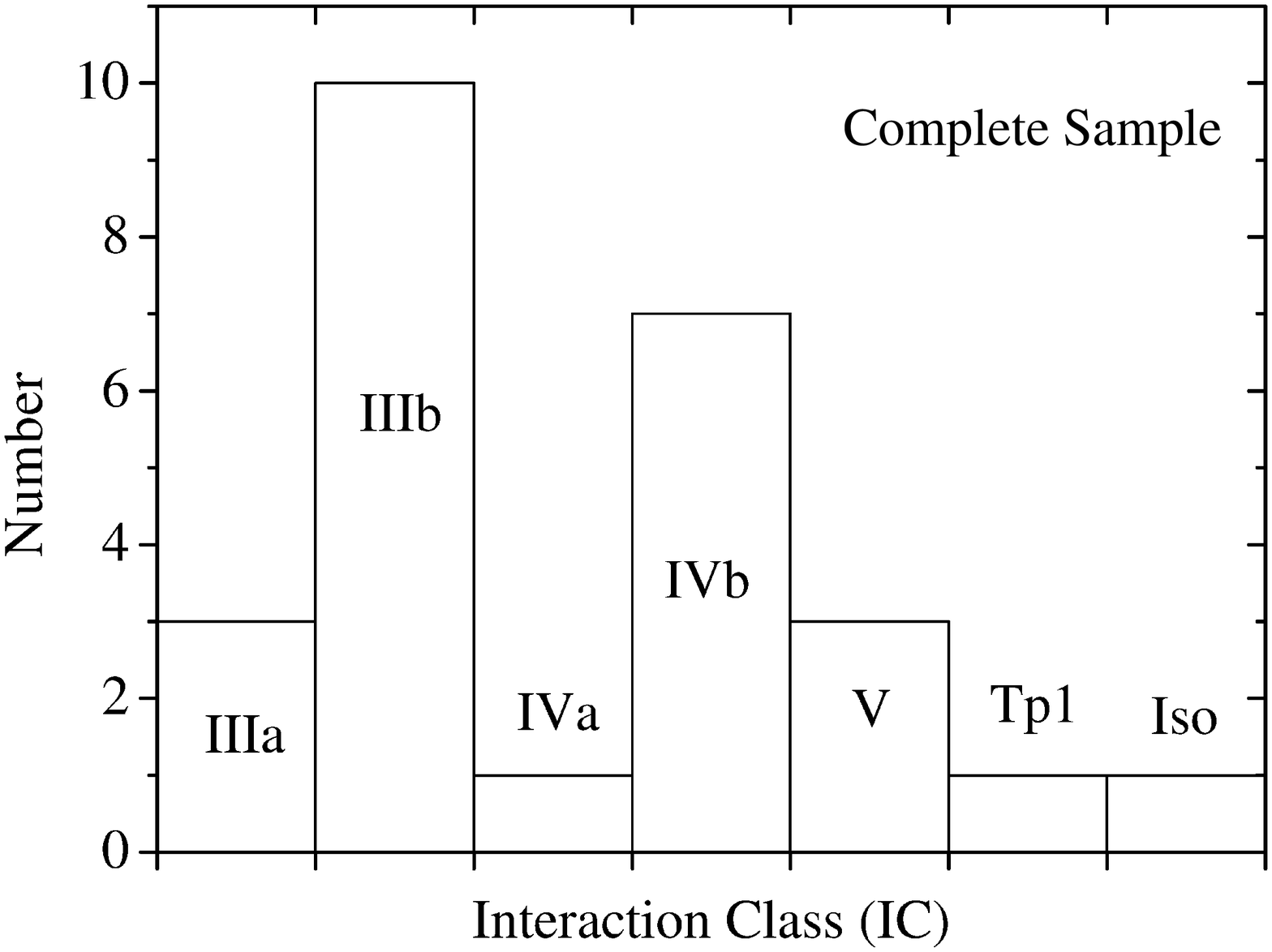,width=6.0cm,angle=0.}&
\psfig{file=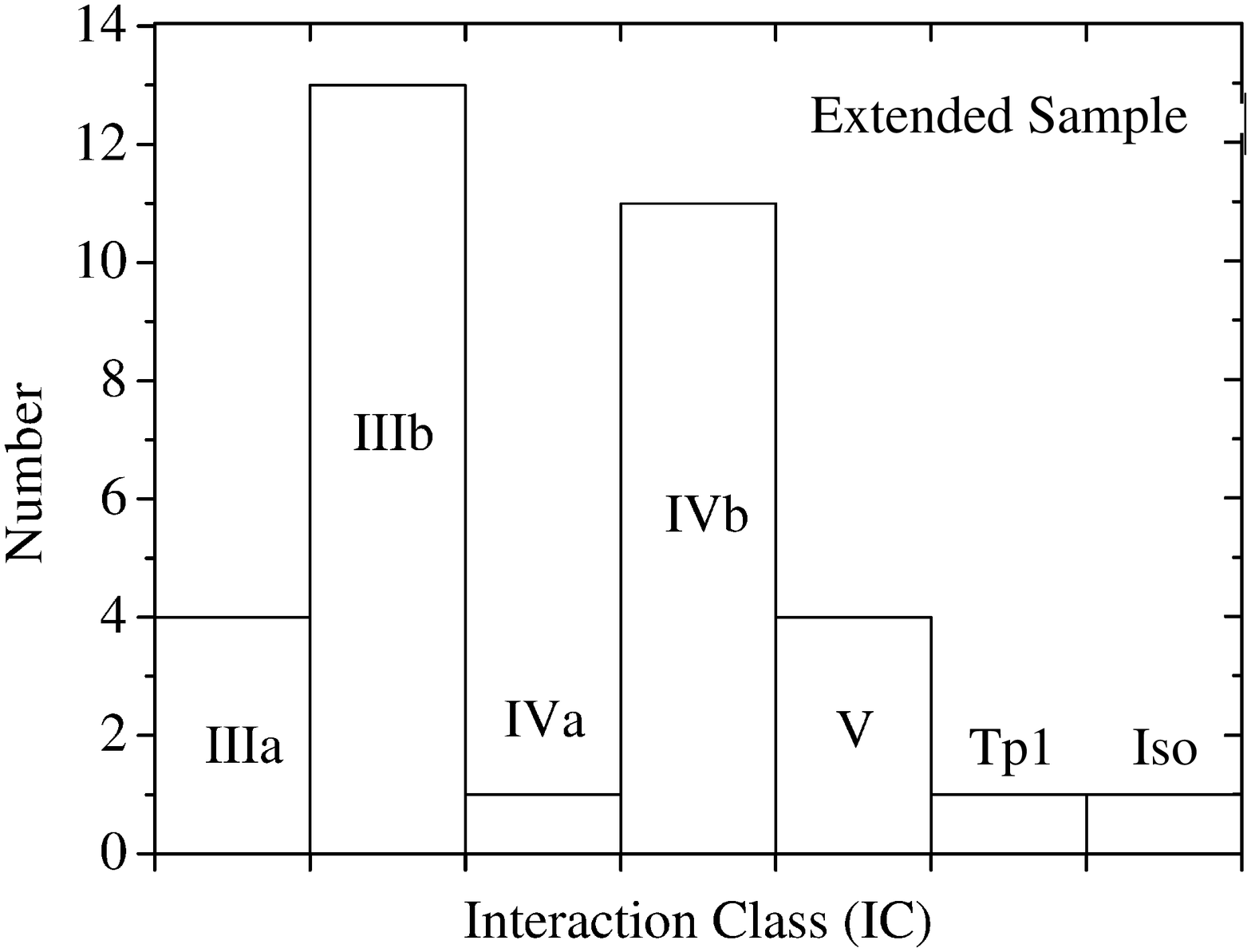,width=6.0cm,angle=0.}\\
\end{tabular}
\caption[Histograms comparing the 1 Jy, the complete and the extended
sampes]{Upper panel: Histograms showing the IR luminosity
  distributions for the full 1 Jy sample of \cite{Kim98a}, the complete
  sample (CS) and the extended sample (ES). The figure shows that both
  the CS and the ES are representative of the 1Jy sample in terms of
  IR luminosities. Middle panel: same as the upper panel, but for the
  f$_{25}$/f$_{60}$$\mu$m ratio, used to classifiy the ULIRGs as cold
  (f$_{25}$/f$_{60}$ $<$ 0.2) or warm (f$_{25}$/f$_{60}$ $\geq$ 0.2)
  objects. The figure shows that the CS is representative of the 1Jy
  sample in terms of the f$_{25}$/f$_{60}$-ratio distribution. On the
  other hand, as expected, the ES is biased towards warm
  objects. Lower panel: same as the upper panel, but for the
  interaction class distribution, using the morphological
  classification scheme described in Veilleux et al. (2002). The
  figure shows that the CS and the ES both have similar distributions
  of objects in interaction class to the 1Jy sample, with the
  exception of the IVa class (see the text for an explanation).}
\label{fig:Samples}
\end{figure*}

Figure \ref{fig:Samples} presents histograms comparing the full 1Jy
sample, the CS and the ES in terms of IR-luminosity, f$_{25}$/f$_{60}$
flux ratio and morphological type distributions, respectively. The
upper panel shows that both the CS and the ES are representative of
the 1Jy sample in terms of IR-luminosities, while the middle panel of
the figure shows that, as expected, the ES is biased towards objects
classified as warm ULIRGs (f$_{25}$/f$_{60} >$ 0.2). Using the
morphological classification scheme described in \cite{Veilleux02},
the lower panel in Figure \ref{fig:Samples} shows the distribution of
the different morphological types of the objects comprising each
sample. To summarize \cite[see][for further details]{Veilleux02}, the
different classes are: class I (first approach), class II (first
contact), class III (pre-merger), class IV (merger), class V (old
merger), Tpl (triple) and Iso (Isolated). As is clear in the figure,
both the CS and the ES samples account for most of the
morphological types of objects found in the full 1 Jy sample, with the
exception of those classified as IVa\footnote{In the \cite{Veilleux02}
morphological classification, objects classified as IVa are those with
prominent tidal features and diffuse extended central regions, but
only one nucleus.}. This is due to the fact that all but 4 of the IVa
sources in the 1 Jy sample are at redshifts z $>$ 0.13 (one of these 4
sources, IRAS 14060+2919, is included in the CS), and the objects in
the CS are selected at low (z $<$ 0.13) redshifts. In addition, among
the 23 objects classified as IVa in the 1Jy sample, only one is
classified as a warm ULIRG, and therefore, these objects were not
selected, a priori, to be observed and included in the ES. To
summarize, the CS and the ES samples are representative of ULIRGs in
the local universe, although the ES contains a higher proportion of
warm objects.

\subsection{Optical spectroscopy}

Long-slit spectra of all the objects in the ES were taken in July
2005, December 2005 and May/July 2006, with the ISIS dual-beam
spectrograph on the 4.2-m William Herschel Telescope (WHT), on La
Palma, Spain. With the exception of IRAS 22491-1808, described below,
we used the R300B grating with the EEV12 CCD, and the R316R grating
with the MARCONI2 CCD on the blue and the red arm respectively, along
with a dichroic cutting at $\sim$ 5300 \AA. A 2$\times$2 binning mode
was used during the observations, leading to a final spatial scale of
0.4 arcsec/pix for both arms, and a dispersion of 1.72 \AA/pix in the
blue and 1.65 \AA/pix in the red arm. The observed useful wavelength
range is $\sim$ 3300 -- 7800~\AA. A slit of width 1.5 arcsec was used
for all the objects in the ES. In the case of IRAS 22491-1808, the
settings in the blue arm were the same as described above. However, on
the red arm, the R600R grating with the MARCONI2 CCD was used, with
2$\times$2 binning mode. The final spatial scale was 0.4 arcsec/pixel
and the dispersion was 0.89~\AA/pixel. The observed useful wavelength
range is $\sim$ 3300 -- 7100~\AA. Details of the observations for
all the objects included in the ES are presented in Table
\ref{tab:log_observ}.

The position angles (PA) of the slit were carefully selected to cover
the main features of the objects (nuclei,tidal tail, bridges,
knots). In order to minimize the effects of differential atmospheric
refraction, the objects were observed either along parallactic angle
or at small airmass (AM $<$ 1.1). For those objects with a double
nucleus structure, the slits covered both nuclei, with the exceptions
of IRAS 08572+3915 and PKS 1345+12, for which it was not possible to
observe the object at the parallactic angle and we did not have
sufficient time to observe both nuclei during the night of the
observations. In the cases of the double nucleus objects IRAS
16487+5447 and IRAS 17028+5817, because of the parallatic angle
constraints, the two nuclei in each system were observed with separate
slit PAs.

\begin{table*}
\centering
\begin{tabular}{llccll}
\hline\hline
Object name & PA    & Arm & Exposure & Airmass &Seeing FWHM\\ 
IRAS        &$\circ$&     & (s)      &         &(arcsec)\\ 
(1)         & (2)   & (3) & (4)      & (5)     & (6)\\
\hline
00091-0738 & 0 & B & 2700   &  1.24 & 1.2\\
           &   & R & 2700   &  1.24 & 1.2\\
00188-0856 & 332 & B & 2100 &  1.33-1.37 & 1.3\\
           &   & R & 2100   &  1.33-1.37 & 1.3\\
01004-2237 & 352 & B & 2100 &  1.60 & 2.8\\
           &   & R & 2100   &  1.60 & 2.8 \\
08572+3915 & 265 & B & 3600 &  1.14-1.32 & 1.3\\
           &   & R & 3600   &  1.14-1.32 & 1.3\\
10190+1322 & 64 & B & 3600  &  1.04 & 1.2\\
           &   & R & 3600   &  1.04 & 1.2\\
10494+4424 & 167 & B & 3600 &  1.03-1.07 & 1.2\\
           &   & R & 3600   &  1.03-1.07 & 1.2\\
12072-0444 & 45 & B & 2700  &  1.30-1.42 & 1.1\\
           &   & R & 2700   &  1.30-1.42 & 1.1\\
12112-0305 & 15 & B & 2700  &  1.11 & 1.2\\
           &   & R & 2700   &  1.11 & 1.2\\
12540+5708 & 180 & B & 2700 &  1.13 & 0.7 - 1.6\\
           &   & R & 2700   &  1.13 & 0.7 - 1.6\\
13305-1739 & 342 & B & 2700 &  1.53-1.62 & 1.7\\
           &   & R & 2700   &  1.53-1.62 & 1.7\\
13428+5608 & 180 & B & 2700 &  1.12 & 1.7\\
           &   & R & 2700   &  1.12 & 1.7\\
13451+1232 & 160 & B & 3600 &  1.05 & 1.3\\
           &   & R & 3600   &  1.05 & 1.3\\
           & 230 & B & 3600 &  1.1 & 1.7\\
           &   & R & 3600   &  1.1 & 1.7\\
13539+2920 & 285 & B & 2700 &  1.01 & 1.1 - 1.6\\
           &   & R & 2700   &  1.01 & 1.1 - 1.6\\
14060+2919 & 83 & B & 2700  &  1.06-1.11 & 0.9 - 1.3\\
           &   & R & 2700   &  1.06-1.11 & 0.9 - 1.3\\
14252-1550 & 40 & B & 2700  &  1.63-1.82 & 0.8\\
           &   & R & 2700   &  1.63-1.82 & 0.8\\
14348-1447 & 35 & B & 2700  &  1.61-1.80 & 1.4 - 2.1\\
           &   & R & 2700   &  1.61-1.80 & 1.4 - 2.1\\
14394+5332 & 92 & B & 3600  &  1.47-1.70 & 1.6\\
           &   & R & 3600   &  1.47-1.70 & 1.6\\
15130-1958 & 358 & B & 2700 &  1.53 & 0.8 - 1.1\\
           &   & R & 2700   &  1.53 & 0.8 - 1.1\\
15206+3342 & 90 & B & 2700  &  1.23-1.28 & 0.9 - 1.8\\
           &   & R & 2700   &  1.23-1.28 & 0.9 - 1.8\\
15327+2340 & 160 & B & 2700 &  1.03-1.07 & 1.3 - 2.0\\
           &   & R & 2700   &  1.03-1.07 & 1.3 - 2.0\\
           & 75 & B & 2700  &  1.14-1.19 & 0.6\\
           &   & R & 2700   &  1.14-1.19 & 0.6\\
           & 75* & B & 3600 &  1.27-1.46 & 0.7\\
           &   & R & 3600   &  1.27-1.46 & 0.7\\
15462-0450 & 10 & B & 2700  &  1.2 & 1.6\\
           &   & R & 2700   &  1.2 & 1.6\\
16156+0146 & 313 & B & 2700 &  1.16-1.21 & 1.2 - 1.7\\
           &   & R & 2700   &  1.16-1.21 & 1.2 - 1.7\\
16474+3430 & 166 & B & 2700 &  1.01 & 1.0\\
           &   & R & 2700   &  1.01 & 1.0\\
16487+5447NE & 110 & B & 2700 &1.12 & 1.2\\
           &   & R & 2700   &  1.12 & 1.2\\
16487+5447NW & 165 & B & 2700 &1.26-1.32 & 1.1\\
           &   & R & 2700   &  1.26-1.32 & 1.1\\
\end{tabular} 
\caption{Log of the spectroscopic observations. In the case of 
  Arp 220 (IRAS 15327+2340), a slit in PA 75 was also used with 
  10 arsec offset to the  north, with respect to the one referred 
  as PA75 in the table: this is labelled as 
  PA75*. The seeing was estimated from DIMM seeing monitor 
  measurements.}
\label{tab:log_observ}
\end{table*}
\addtocounter{table}{-1}
\begin{table*}
\centering
\begin{tabular}{ccccccc}
\hline\hline
Object name & PA & Arm & Exposure & Airmass & Seeing\\
IRAS& $\circ$ & & (s) & & (arcsec) \\ 
(1) & (2) & (3) & (4) & (5) & (6)\\
\hline
17028+5817NE & 130 & B & 2700 &  1.23-1.28 & 1.2\\
           &   & R & 2700         & 1.23-1.28 & 1.2\\
17028+5817NW & 110 & B & 2700 &  1.31-1.39 & 0.9\\
           &   & R & 2700         & 1.31-1.39 & 0.9\\
17044+6720 & 160 & B & 2700     &  1.20-1.29 & 1.4 - 1.9\\
          &   & R & 2700          & 1.20-1.29 & 1.4 - 1.9\\
17179+5444 & 122 & B & 2700     & 1.22-1.28 & 0.9\\
           &   & R & 2700         & 1.22-1.28 & 0.9\\
20414-1651 & 5 & B & 2700         &  1.42 & 1.4 - 1.7\\
           &   & R & 2700         & 1.42 & 1.4 - 1.7\\
21208-0519 & 15 & B & 2700       &  1.20 & 1.2 \\
           &   & R & 2700         & 1.20 & 1.2\\
21219-1757 & 333 & B & 3600   &  1.54-1.67 & 1.5\\
           &   & R & 3600       & 1.54-1.67 & 1.5\\
22491-1808 & 338 & B & 3600   & 1.54-1.58 & 1.0 - 1.5\\ 
           &   & R & 3600       & 1.54-1.58 & 1.0 - 1.5\\
23060+0505 & 338 & B & 3600   &  1.09-1.13 & 0.7\\
           &   & R & 3600       &  1.09-1.13 & 0.7\\
23233+2817 & 168 & B & 2100   &  1.00 & 1.2\\
           &   & R & 2100       & 1.00 & 1.2\\
23234+0946 & 115 & B & 2700   & 1.06 & 1.1\\
           &   & R & 2700       & 1.06 & 1.1\\
23327+2913 & 175 & B & 2700   & 1.00 & 1.7\\
           &   & R & 2700       &  1.00 & 1.7\\
23389+0300 & 24& B & 3600       & 1.12 & 0.7\\
           &   & R & 3600       &  1.12 & 0.7\\
\end{tabular} 
\caption{{\it Continued}}
\end{table*}

The data were reduced (bias subtracted, flat field corrected, cleaned
of cosmic rays, wavelength calibrated and flux calibrated) and
straightened before extraction of the individual spectra using the
standard packages in {\it IRAF} and the {\it STARLINK} packages {\it
FIGARO} and {\it DIPSO}. The wavelength calibration accuracy, measured
as the mean shift between the measured and published
\citep{Osterbrock96} wavelength of night-sky emission lines, is $\sim$
0.35 \AA~for the blue spectra and $\sim$ 0.25 for the red spectra. No
significant variations of these values were found for the observations
of the objects comprising the sample used in this paper. The spectral
resolutions, calculated using the widths of the night-sky emission
lines (FWHM), are in the ranges of 3.4 -- 6.03 \AA~and 4.44 -- 5.84
\AA~for the blue and red arms, respectively.

Accurate relative flux calibration is vital for the purpose of this
project, and therefore several (4 -- 6) photometric standard stars
were observed during each night of the observations. The relative flux
calibration accuracy is estimated to be $\pm$ 5\% over the entire
useful spectral range ($\sim$3300 -- 7800 \AA). This is confirmed
by the excellent match between the fluxes in the overlapping region of
the blue and red arm spectra.

\section{Analysis}

\subsection{Aperture extraction}

\begin{figure*}
\begin{minipage}{\textwidth}
\begin{tabular}{ccc}
\hspace*{0cm}\psfig{file=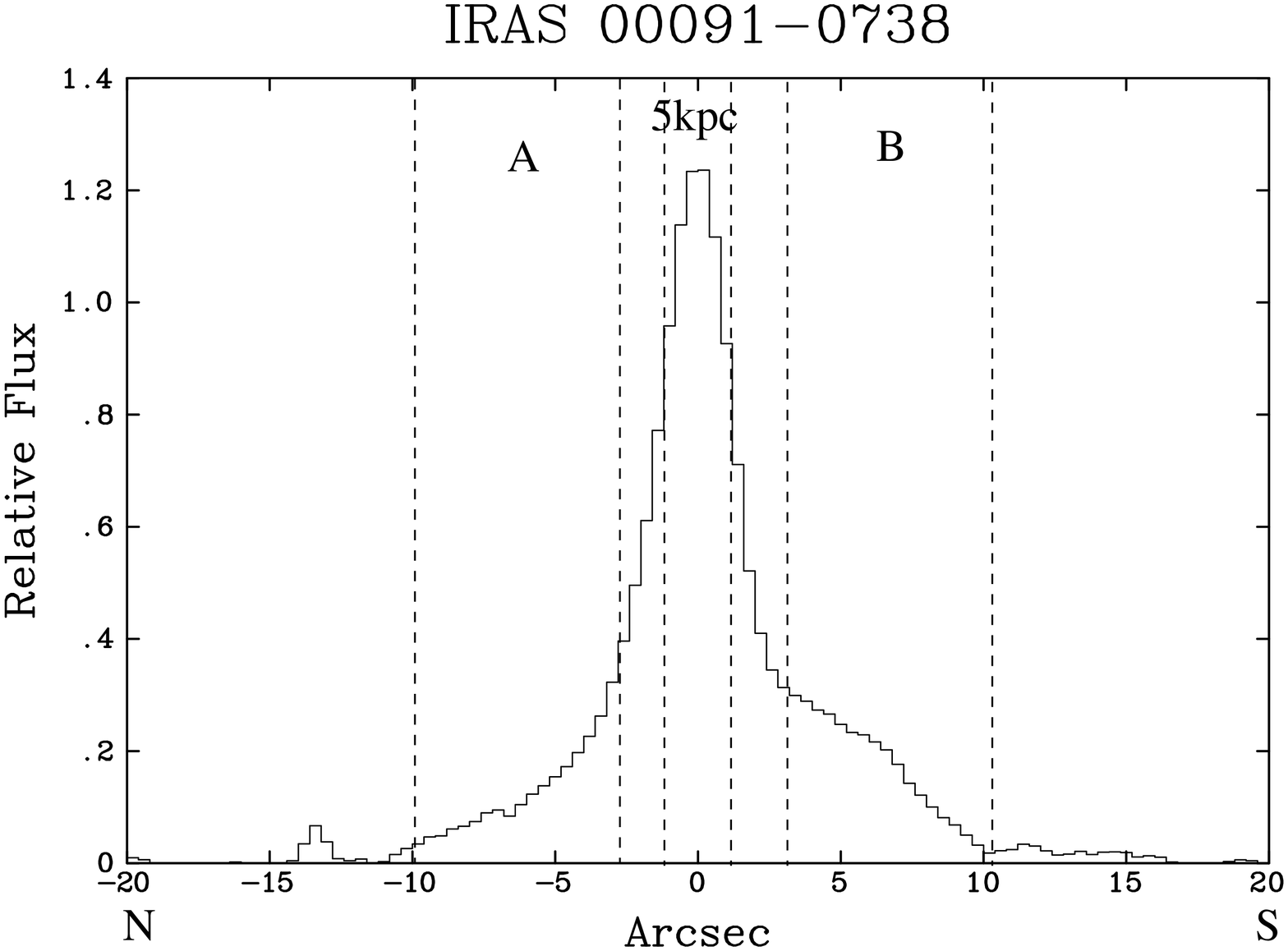,width=5.0cm,angle=0.}&
\psfig{file=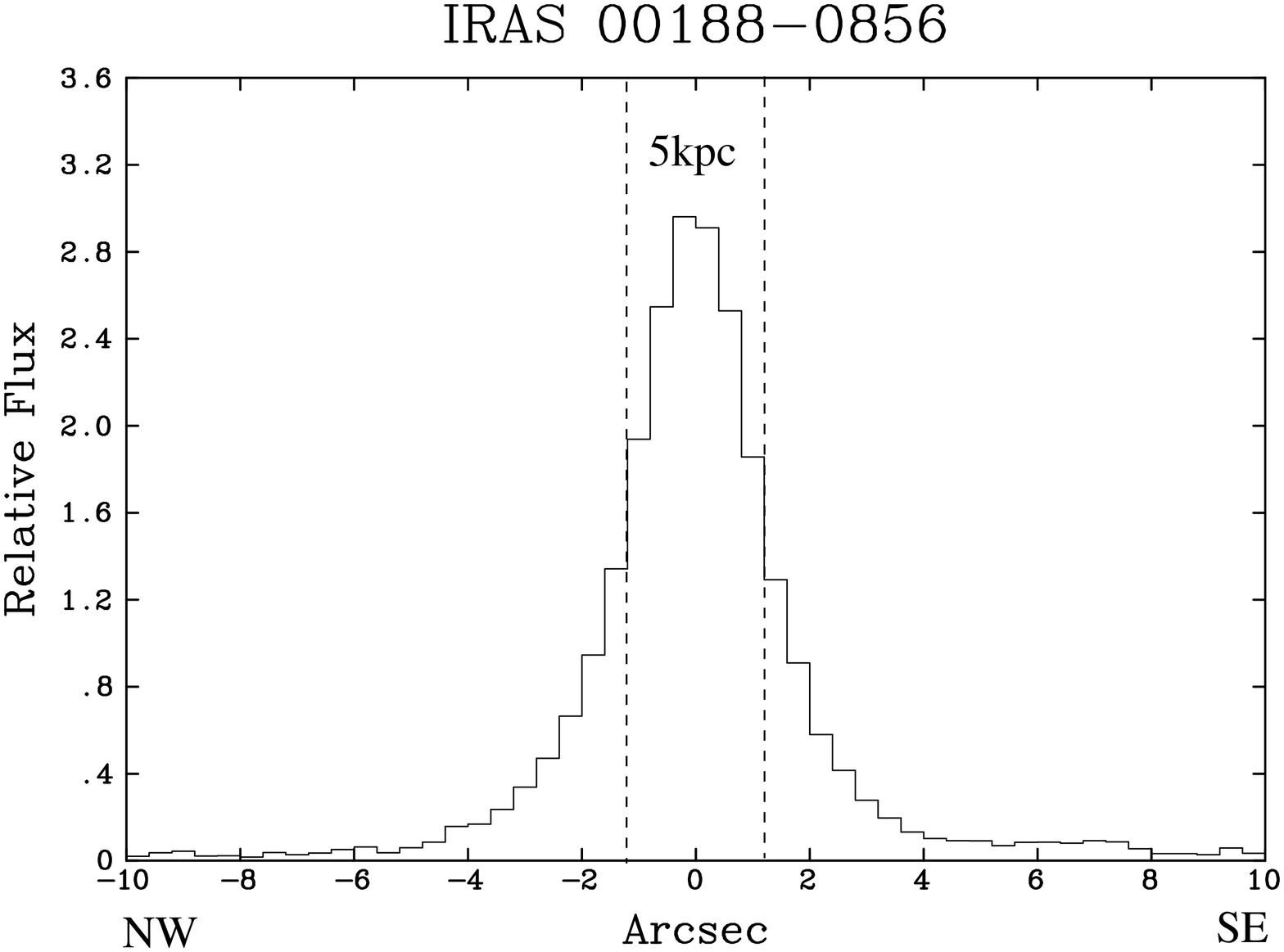,width=5.0cm,angle=0.}&
\psfig{file=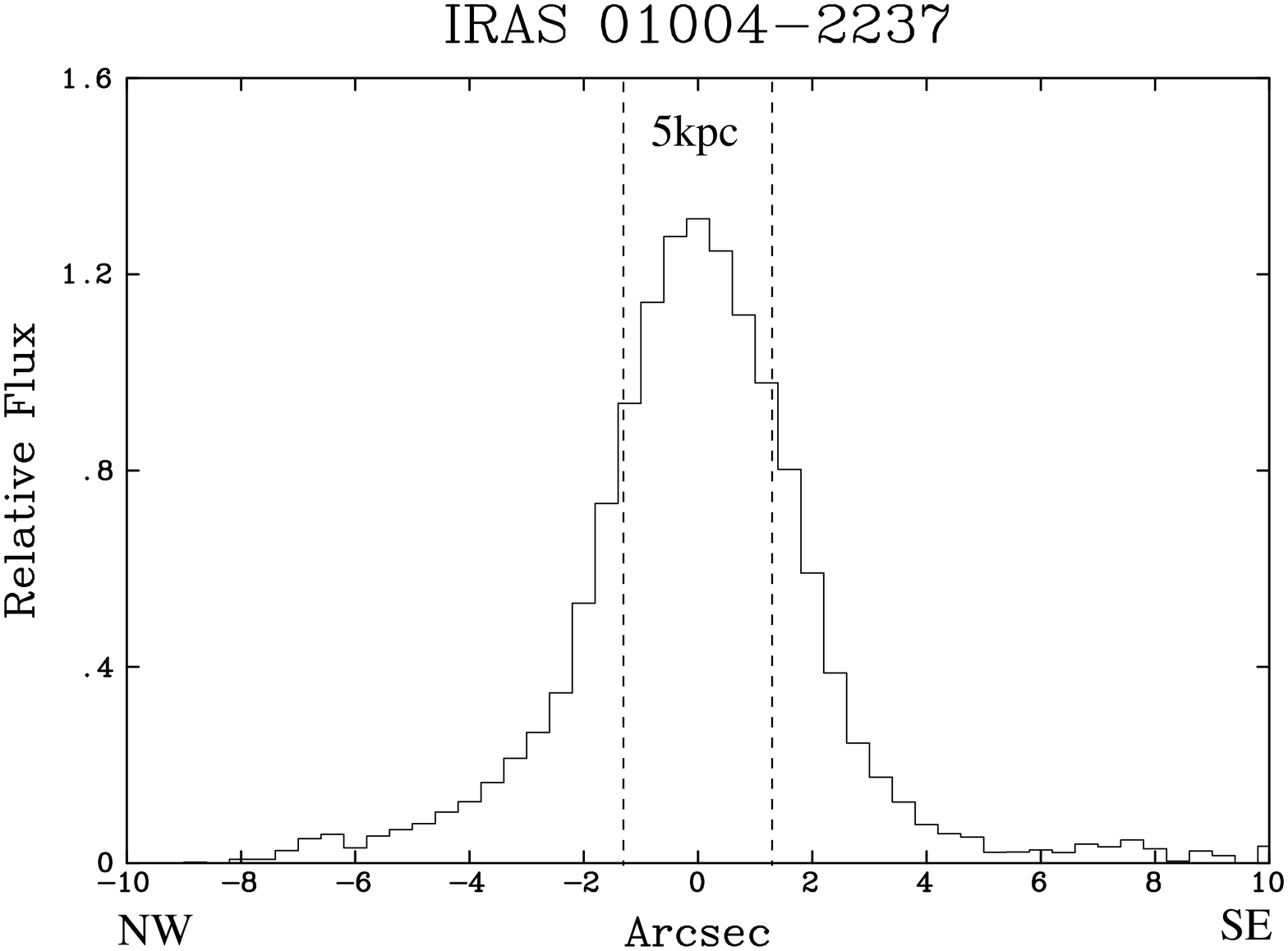,width=5.0cm,angle=0.}\\
\hspace*{0cm}\psfig{file=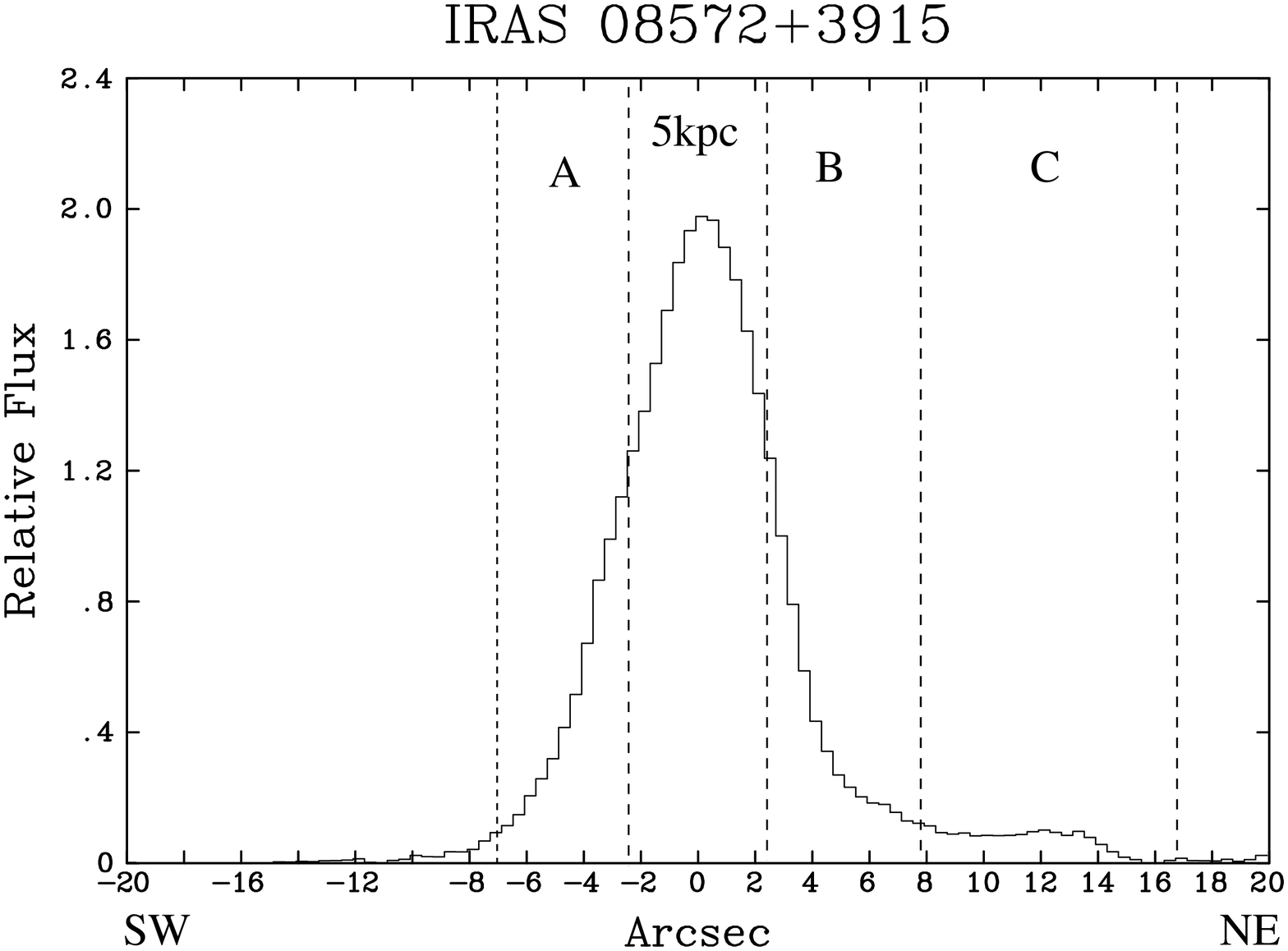,width=5.0cm,angle=0.}&
\psfig{file=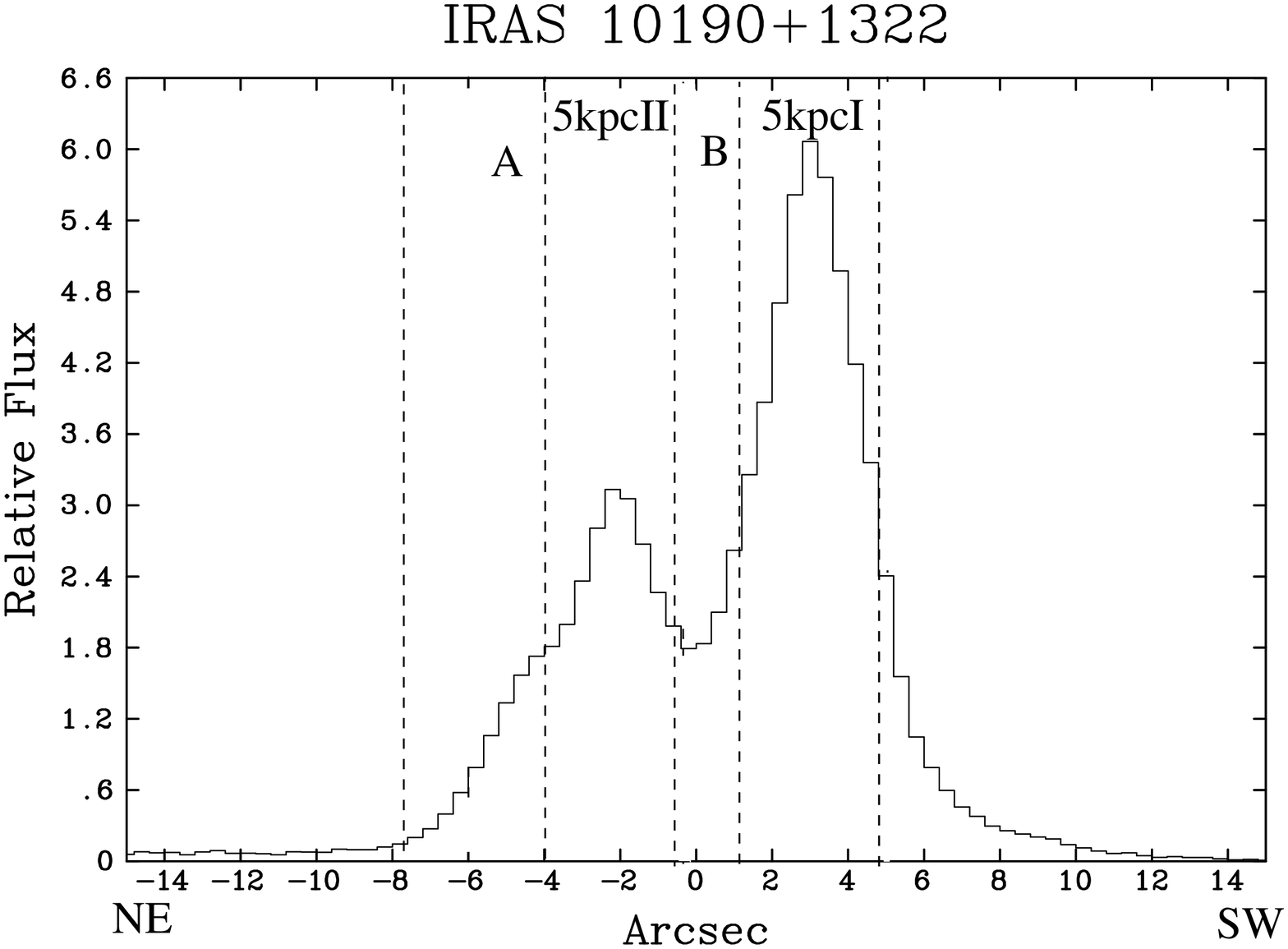,width=5.0cm,angle=0.}&
\psfig{file=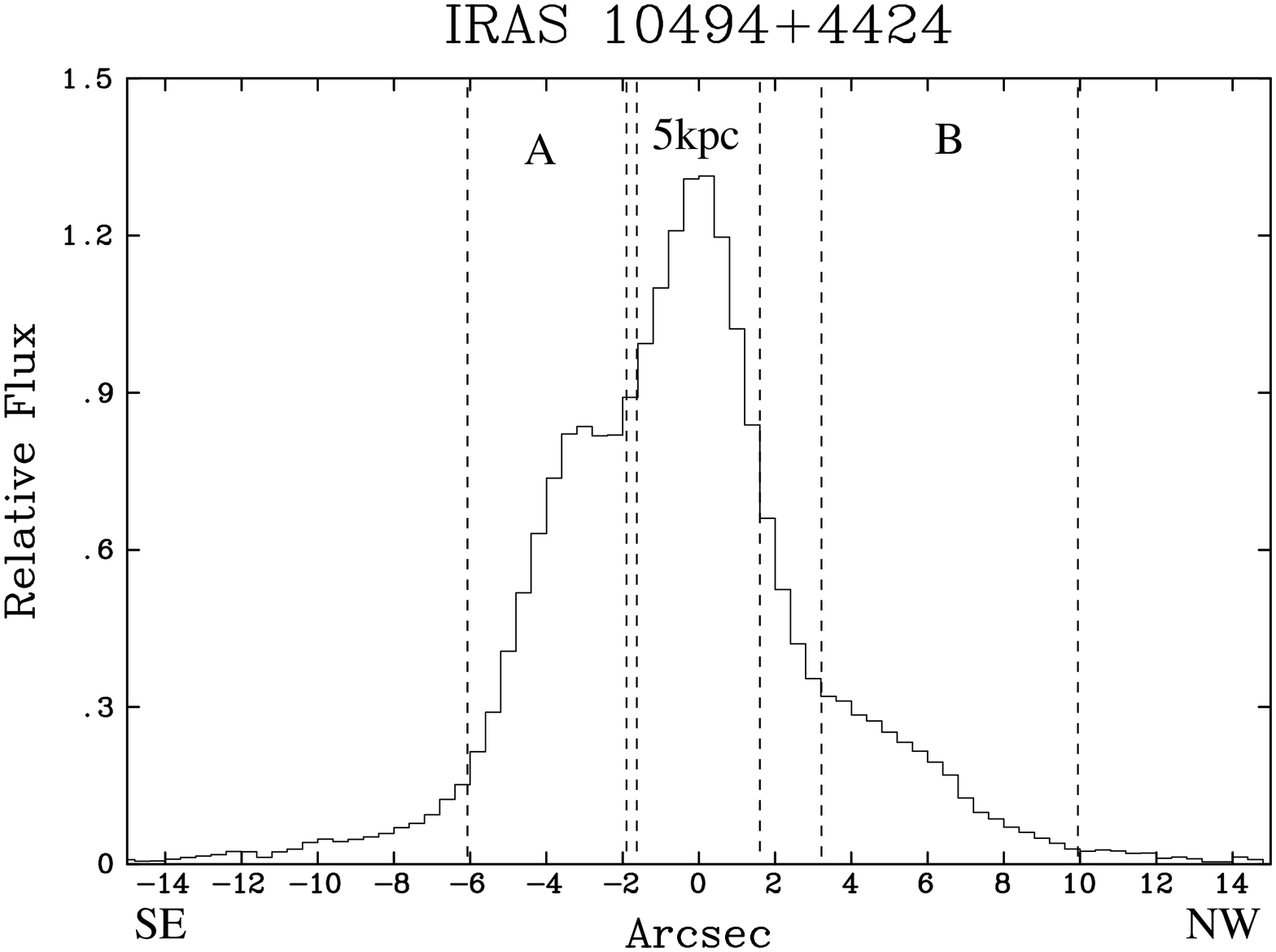,width=5.0cm,angle=0.}\\
\hspace*{0cm}\psfig{file=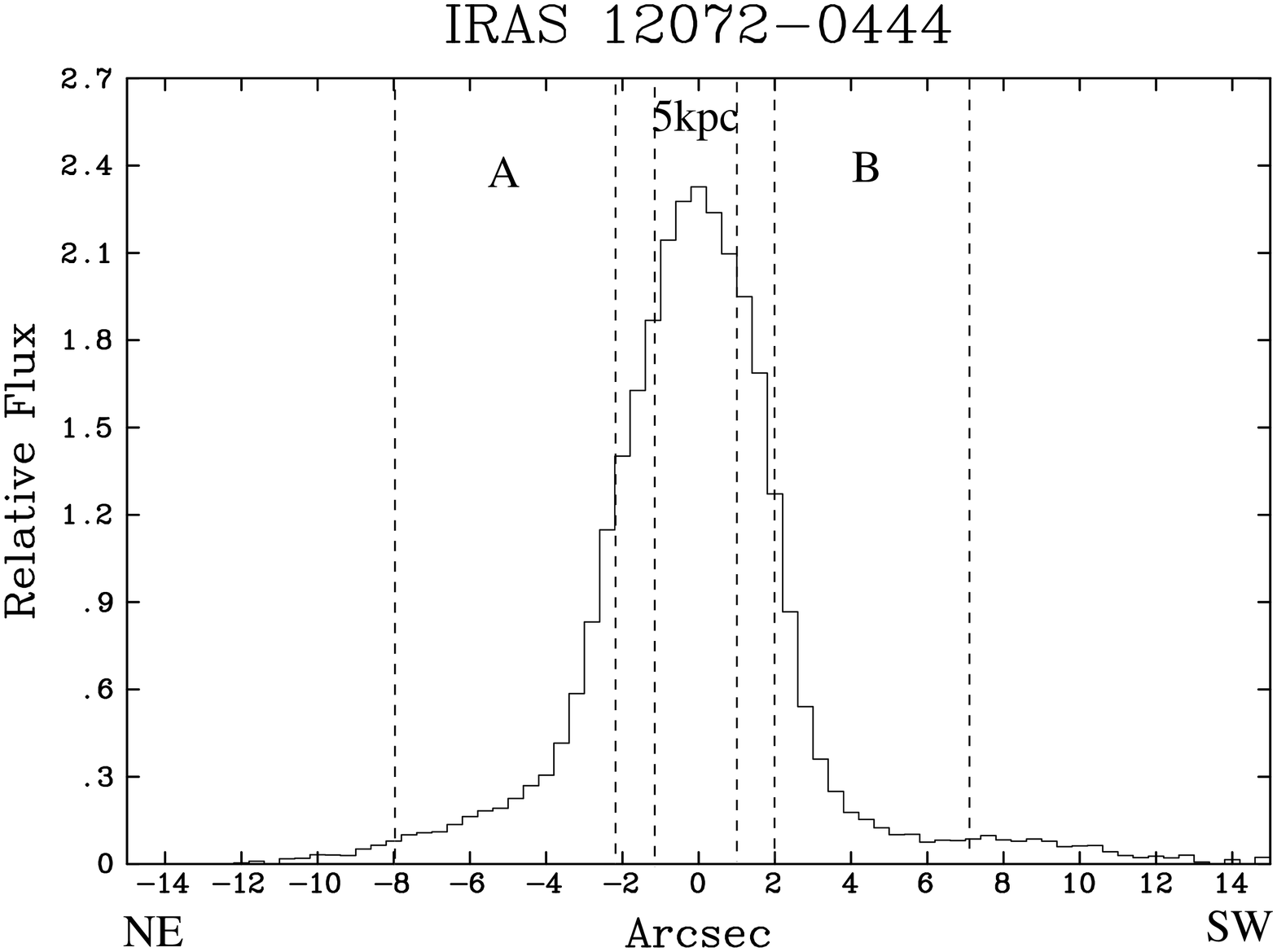,width=5.0cm,angle=0.}&
\psfig{file=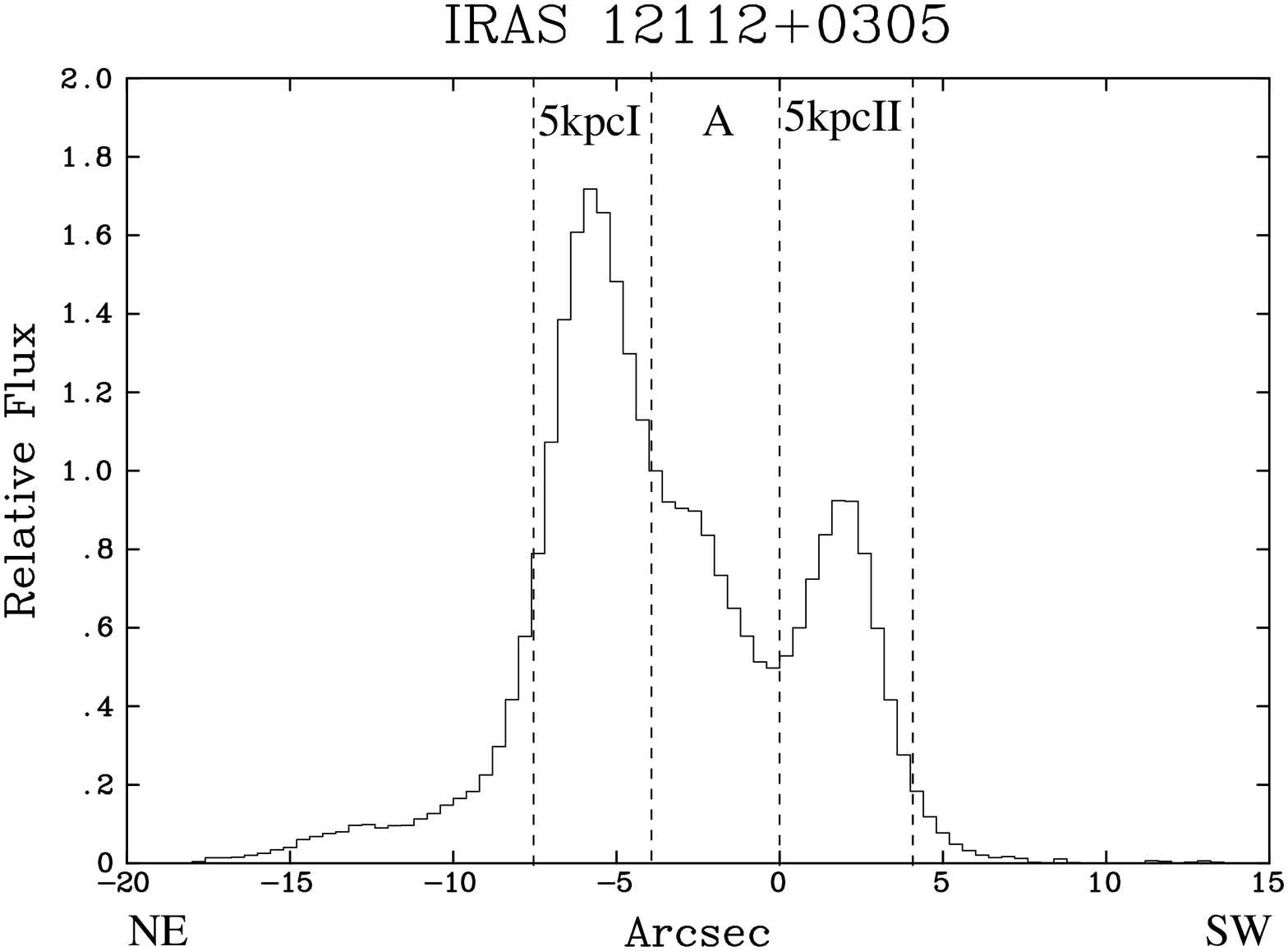,width=5.0cm,angle=0.}&
\psfig{file=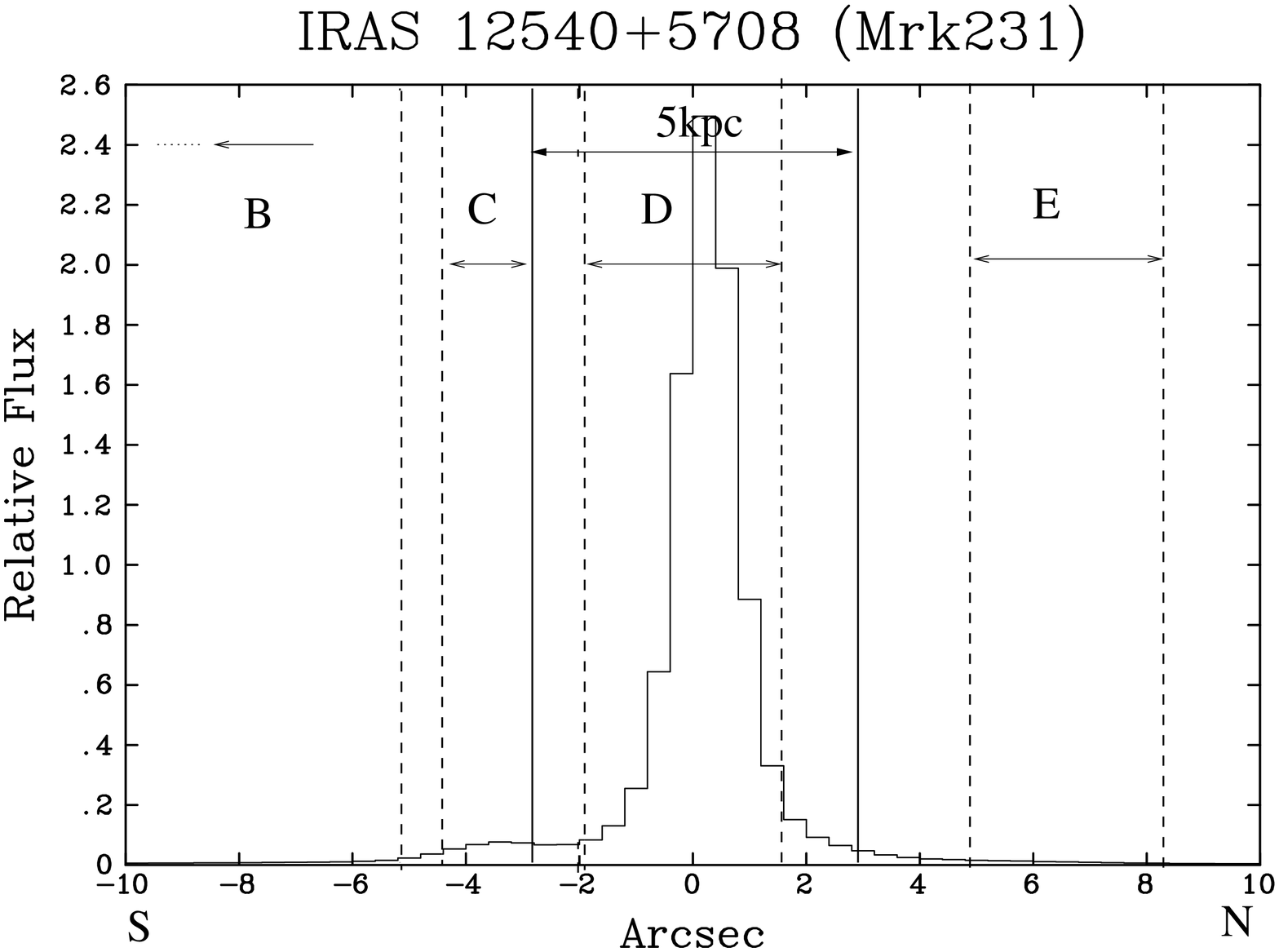,width=5.0cm,angle=0.}\\
\hspace*{0cm}\psfig{file=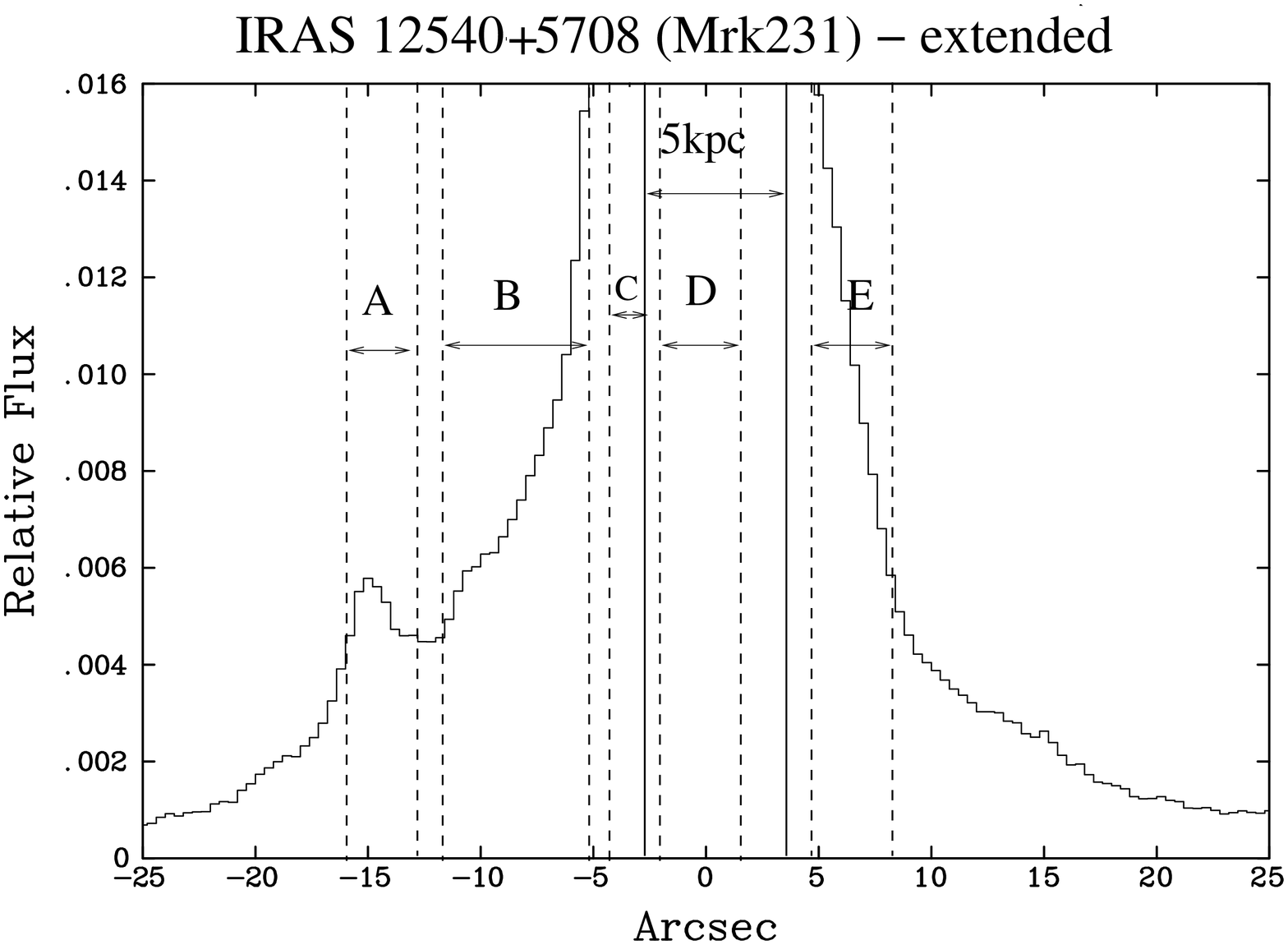,width=5.0cm,angle=0.}&
\psfig{file=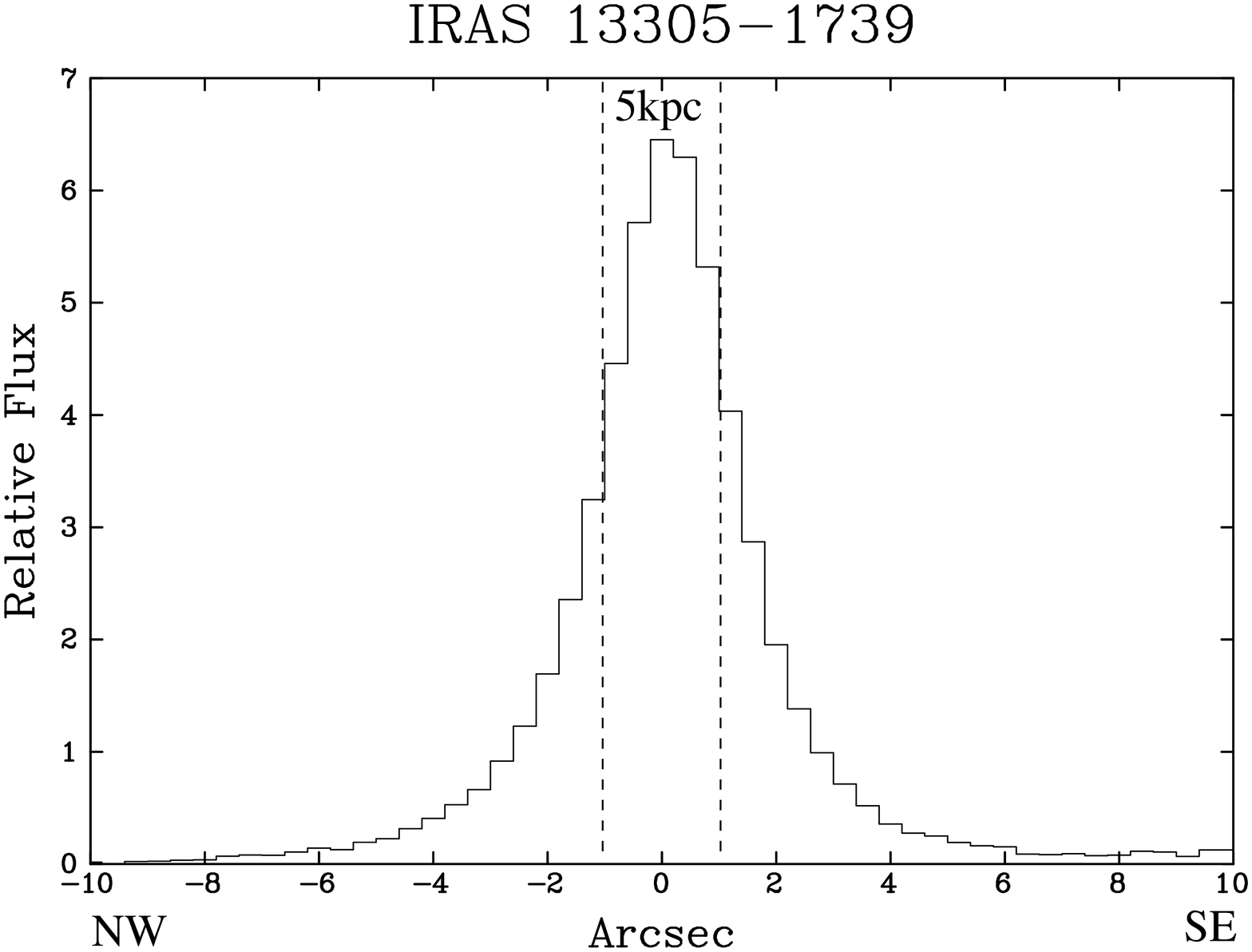,width=5.0cm,angle=0.}&
\psfig{file=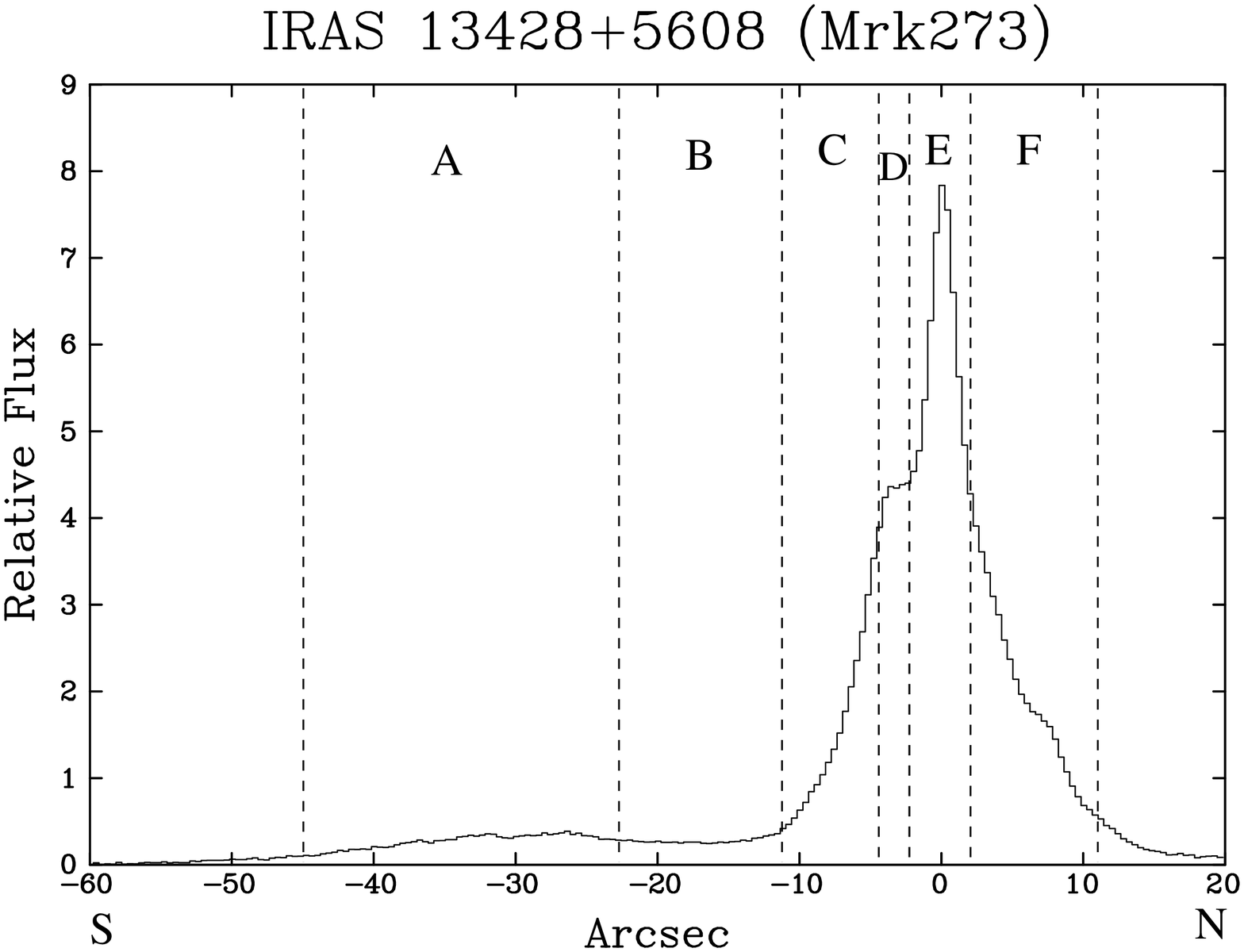,width=5.0cm,angle=0.}\\
\hspace*{0cm}\psfig{file=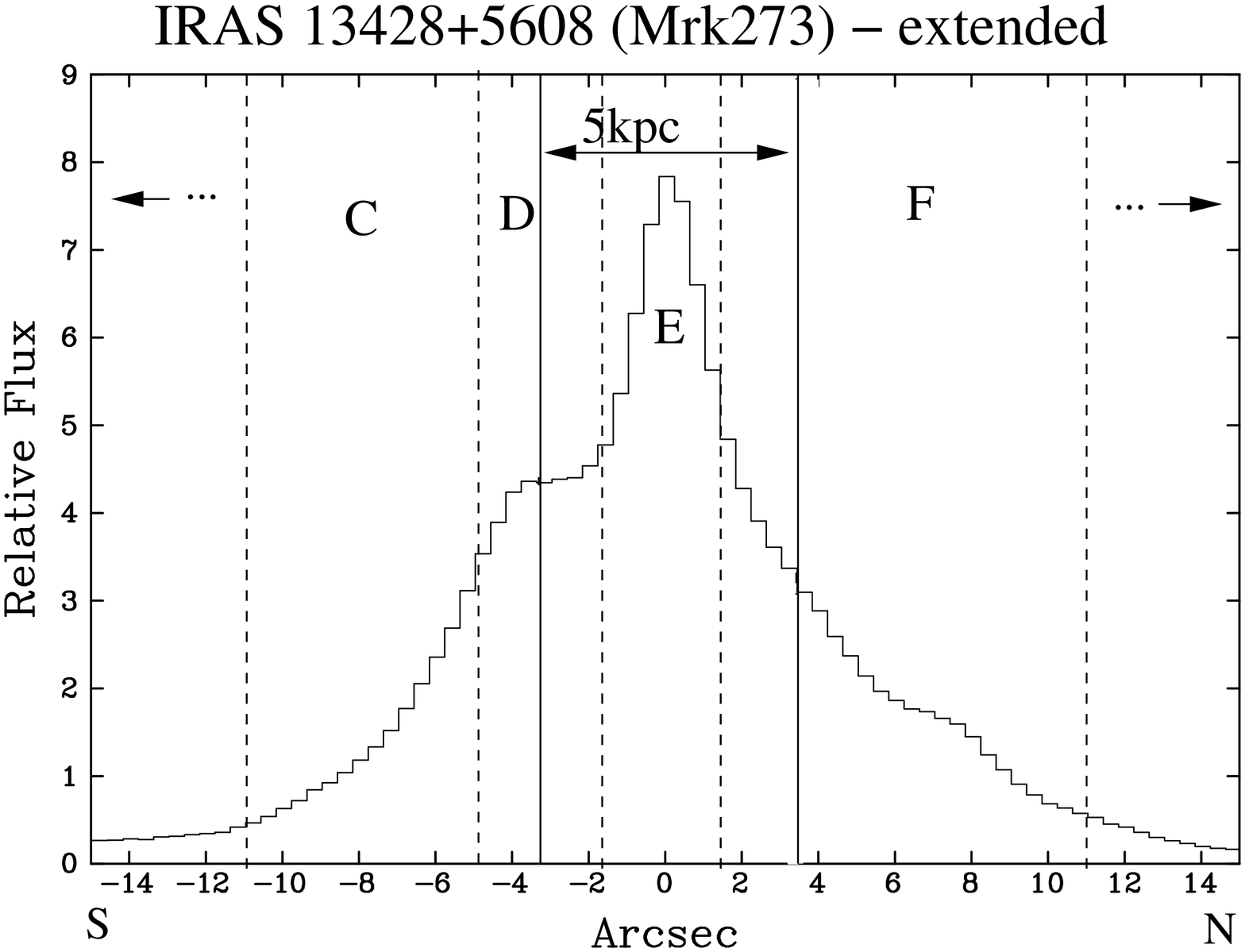,width=5.0cm,angle=0.}&
\psfig{file=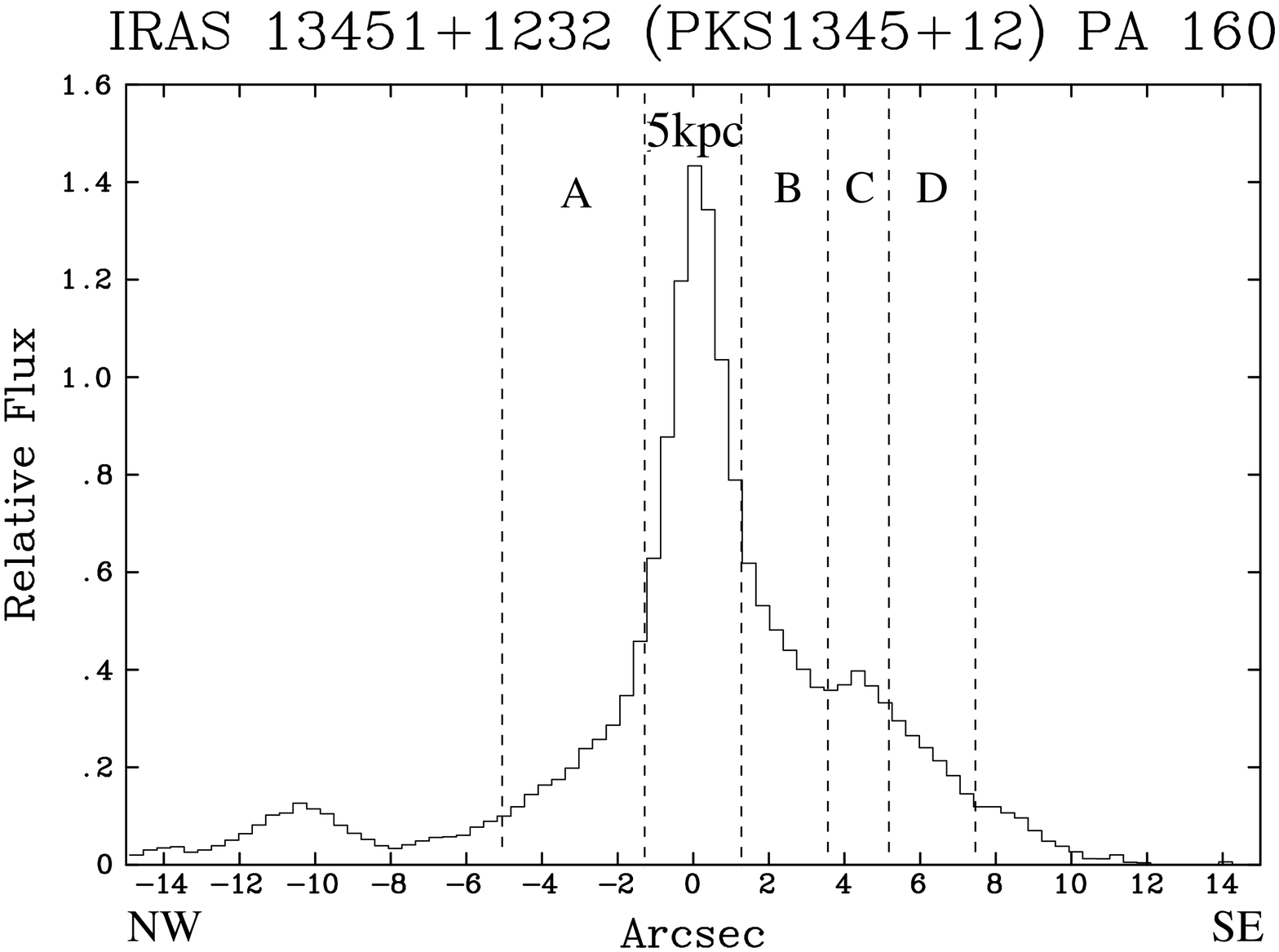,width=5.0cm,angle=0.}&
\psfig{file=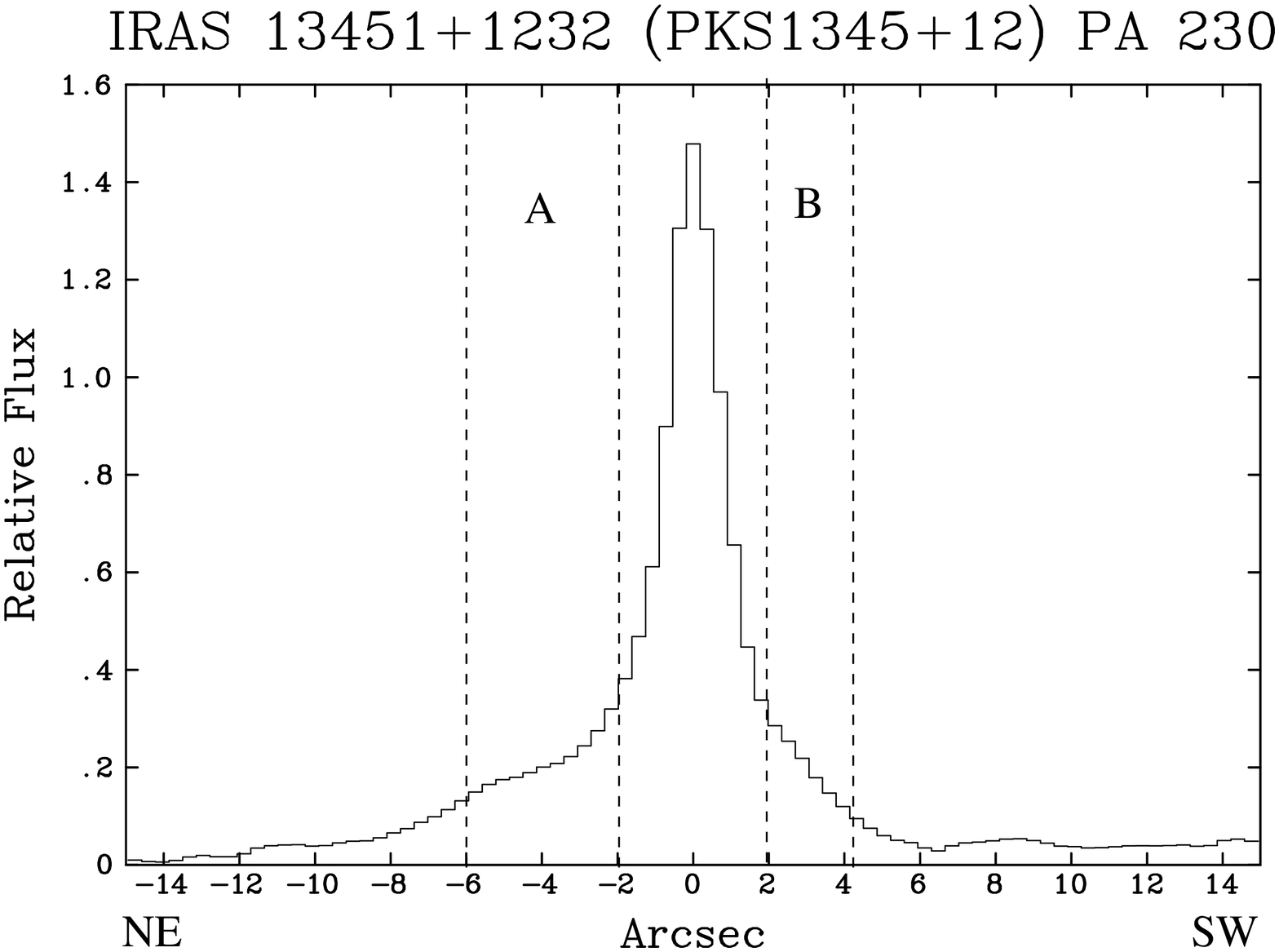,width=5.0cm,angle=0.}\\
\hspace*{0cm}\psfig{file=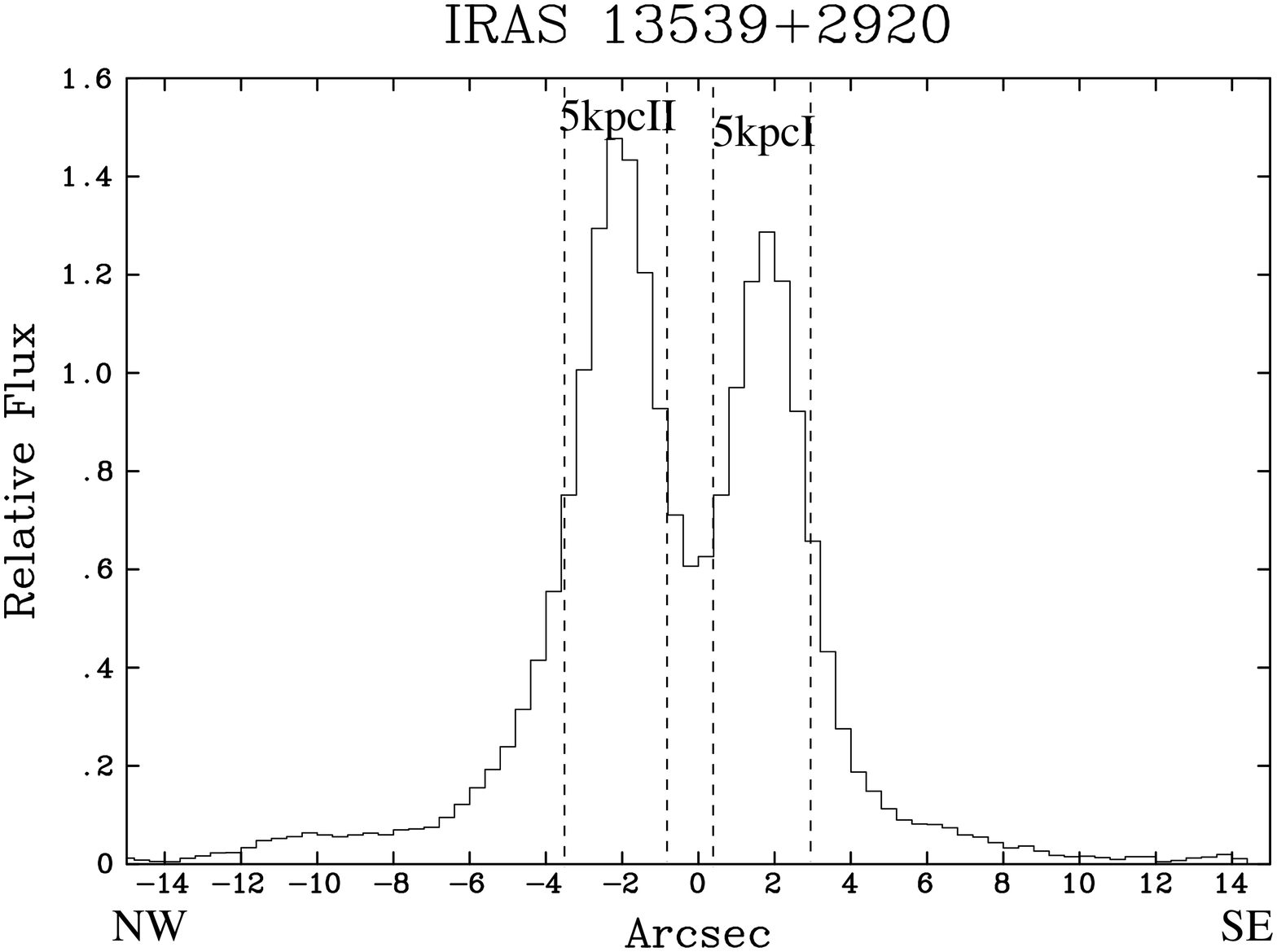,width=5.0cm,angle=0.}&
\psfig{file=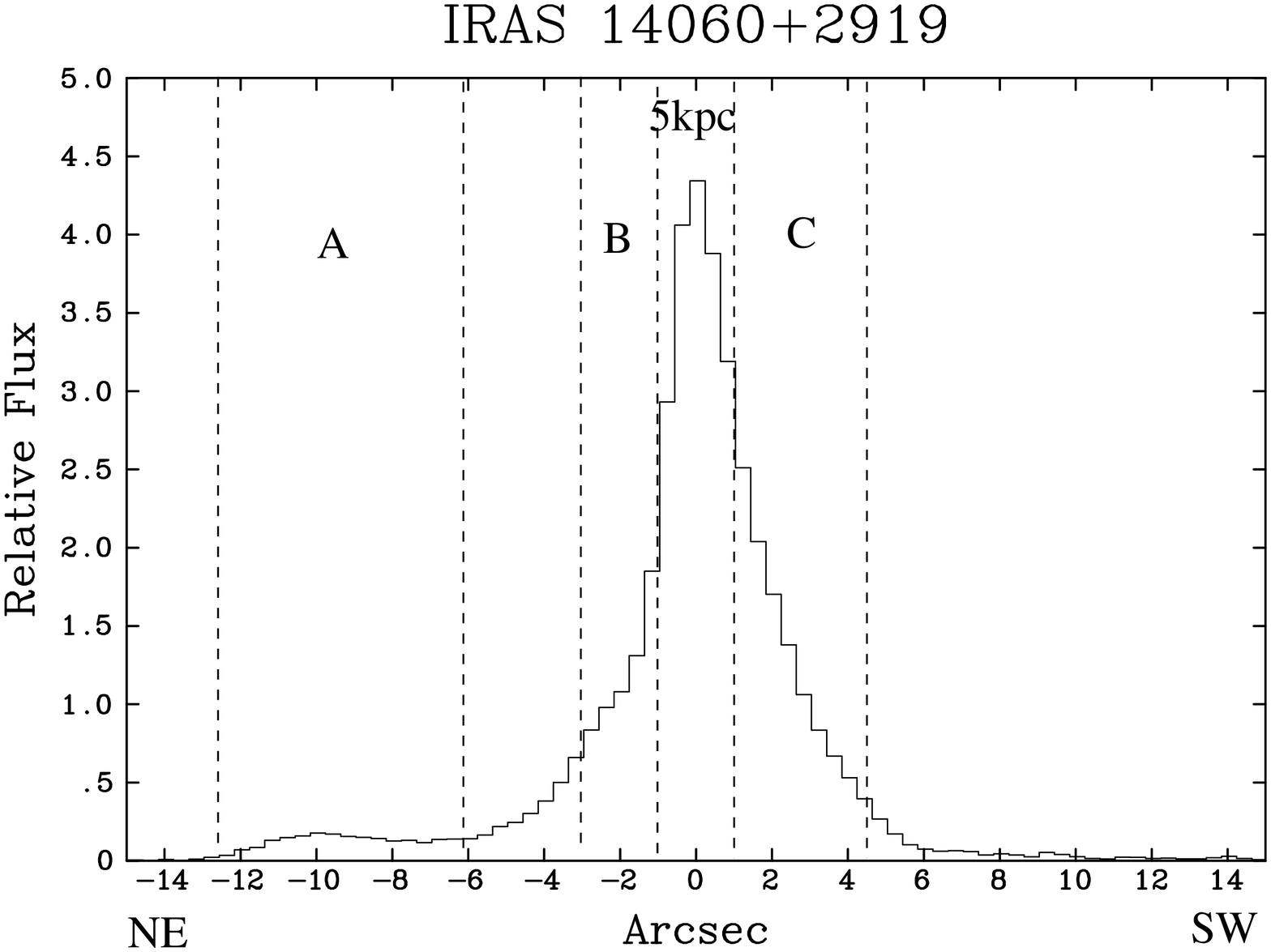,width=5.0cm,angle=0.}&
\psfig{file=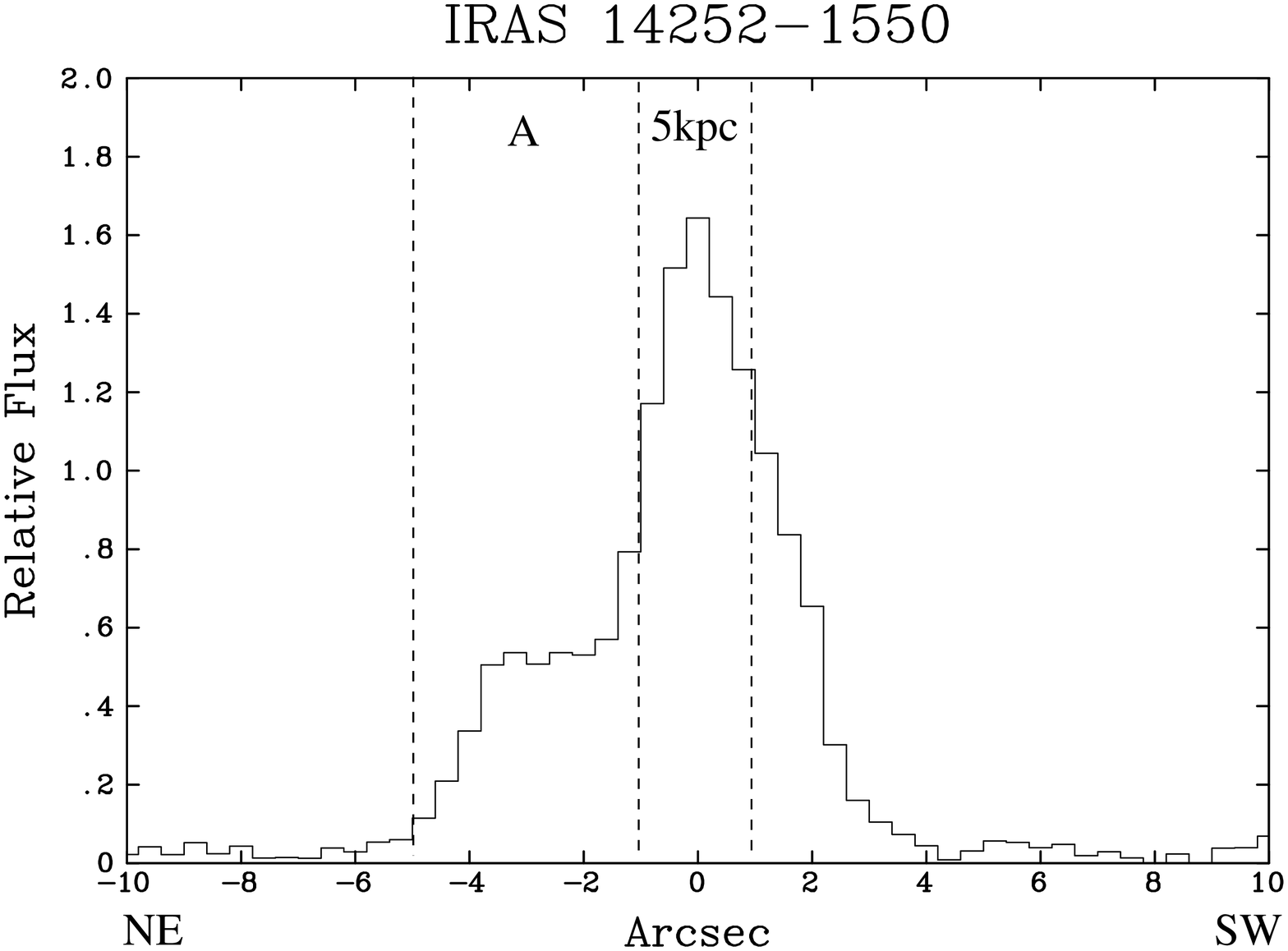,width=5.0cm,angle=0.}\\
\end{tabular}
\caption[Spatial profiles of the 2-D spectra.]{Spatial profiles of the
2-D spectra for the wavelength range 4400 -- 4600~\AA~showing the
positions of the extraction apertures. The symbols N, S, E and W at
the positive and negative ends of the x-axis represent the directions
of the slit on the sky. For example, in the case of IRAS 00091-0738,
positive x corresponds to south, negative x corresponds to north.}
\label{fig:spatial_cuts}
\end{minipage}
\end{figure*}

The relatively low redshifts of the objects in the ES, the wide
spectral coverage, intermediate spectral resolution, excellent flux
calibration and high S/N of the spectra, make it possible to perform
detailed studies of their stellar populations. The extraction
apertures were selected from spatial cuts of the 2-D frames in the
line-free continuum wavelength range 4400 -- 4600~\AA, based on the
visible extended structures and the requirement that the apertures are
large enough to have a sufficiently high S/N ratio for further
analysis. In order to compare the stellar populations between the
objects in our sample, apertures with a metrical scale of 5kpc centred
on the main nuclei were extracted for all the objects in the extended
sample (ES), including separate extractions for multiple nuclei in
individual sources. As mentioned before, a detailed analysis of the
particular cases of IRAS 1341+1232 (PKS 1345+12) and IRAS 15327+2740
(Arp 220) can be found in RZ07 and RZ08 respecively. In the case of
PKS 1345+12 the aperture labelled as 'NUC AP' in the RZ07 paper has a
metric scale of 5.2 kpc and, hereafter, we will refer to this as the
5kpc aperture for this galaxy. In the case of Arp 220, the closest
object in the sample (z = 0.018, resulting in a scale of 0.363 kpc
arcsec$^{-1}$ and a distance of 77.6 Mpc), a 5 kpc aperture covers a
large region of the slit and, therefore, was not selected a priori
when performing the study presented in RZ08. In order to compare Arp
220 with the other ULIRGs included in the sample used for the work
presented here, we used the slit with PA 160 and extracted a 5kpc
aperture, as shown in Figure \ref{fig:spatial_cuts}, sampling the
central region of the galaxy (see Figure 1 in RZ08). A second set of
apertures was then selected to sample the spatial features of those
objects in the ES showing tails, bridges and other diffuse
structures. A total of 133 apertures was extracted for the ULIRGs in
the ES. Figure \ref{fig:spatial_cuts} shows the spatial profiles of
the 2-D frames in the wavelength range 4400 -- 4600~\AA~along with the
extraction apertures.

The extracted spectra, in the rest frame, are shown in Figure
\ref{fig:all1dspectra}. In the particular case of Arp220 we extracted
a total of 24 apertures. As discussed in RZ08, the morphologies of the
extracted spectra (and the properties of the stellar populations) are
similar across the entire extent of the galaxy. Therefore, in Figure
\ref{fig:all1dspectra}, we decided to show a representative sample of
the whole set of extracted apertures. These are the apertures labelled
as AP$_{TOTAL}$ in Figure \ref{fig:spatial_cuts} for each of the slit
PAs used for this galaxy (see RZ08 for more details).

\subsection{Modelling technique}

To study in detail the properties of the stellar populations we have
modelled the full spectral range of the extracted spectra using the
stellar population synthesis model templates of
\cite{Bruzual03}. These spectra were created assuming an instantaneous
burst of star formation, solar metallicity and using the
\cite{Salpeter55} initial mass function (IMF) with lower and upper
mass cut offs: m$_{L}$ = 0.1 M$_{\odot}$ and m$_{U}$ = 100 M$_{\odot}$
\cite[see][for a justification of these assumption]{Tadhunter05}. On
the other hand, ULIRGs have large gas and dust contents. Therefore, it
is important to carefully account for reddening effects. Reddened
synthetic spectra were created using the \cite{Calzetti00} reddening
law, appropriate for starburst galaxies, and also the \cite{Seaton79}
reddening law, representing the Galactic extinction case. During this
process, the intrinsic extinction was modelled using a foreground
screen geometry. Overall, there is good consistency between results
obtained with the two reddening laws, suggesting that the main results
for the ULIRGs in the CS and the ES samples are not sensitive to the
details of the reddening law assumed at optical wavelengths

Of the 133 extraction apertures, four sample the nuclear regions of
Sy1 galaxies (5kpc and Ap D for Mrk 231, 5kpc for IRAS 15462-0405, and
5kpc in the case of IRAS 21219-1757) and, therefore, it was not
possible to model the extracted spectra due to the strong AGN
contamination.  In addition, Ap A for IRAS 23233+2817 samples a region
to the south of the nucleus, including the bright knot detected in the
K-band image shown in \cite{Kim02} (their Figure 1). Our extracted
spectrum for this aperture shows that it is strongly contaminated by a
Galactic M star (see Figure 3). Therefore no attempt was made to model
the spectrum for this aperture.

Prior to the modelling, the spectra were corrected for Galactic
reddening using the far-IR based maps of extinction by
\cite{Schlegel98}. An additional potential contaminant of the stellar
emission in the optical is the nebular continuum emission
\citep[e.g.][]{Dickson95}. In order to account for this effect, we subtracted
a nebular continuum for those apertures for which strong emission
lines were detected, following the technique described in RZ07,
i.e. we used H$\alpha$ to generate the nebular continuum and
considered two extreme cases: (i) maximum nebular continuum assuming
no reddening of the emission-line region; (ii) zero nebular continuum,
corresponding to high reddening.  The final nebular continua were
subtracted from the spectra prior to performing the modelling. The
modelling was carried out for both the nebular-corrected and
uncorrected spectra.

Overall, a total of 128 extraction apertures were modelled. To perform
the fit, we have used the CONFIT code (see \citealt{Robinson00} and
RZ08 for details) which assumes two stellar components plus a
power-law in some cases. To summarize, the CONFIT approach consists of
a direct fit to the overall continuum shape of the extracted spectra
using a minimum $\chi^{2}$ technique. For each spectroscopic aperture,
the flux is measured in several wavelength bins (typically $\sim$50 --
70) chosen to be as evenly distributed in wavelength as possible, and
to avoid strong emission lines and atmospheric absoption features. A
relative flux calibration error of 5\% was assumed during the
modelling. Models with $\chi^2_{red}$ $\lsim$ 1 are deemed to provide
an acceptable fit to the overall continua \cite[see discussion
in][]{Tadhunter05}. From these, the best fitting models were selected
based on a visual inspection of the fits to the detailed absorption
features that have relatively little emission line contamination, such
as high order Balmer lines, He~{\small I} lines, CaII~K~$\lambda$3934,
G-band~$\lambda$4305, MgIb~$\lambda$5173 band. In addition, for those
objects classified as HII-galaxies, we used the H$\alpha$ equivalent
widths to further constrain the ages of the VYSP\footnote{Note that
since the continua of the spectra are affected by dilution due to the
emission of older stellar components, the H$\alpha$ equivalent width
measurements provide upper limits for the age of the VYSP.}.

The model fit results are quantified in terms of the percentage
contribution of the different stellar components in a normalising
bin. Due to the different spectral morphologies of the extracted
spectra, it was not possible to select exactly the same normalising
bin for all the extracted apertures. However, the normalising bin was
always selected to be located within the wavelength range 4400 --
4800~\AA, and usually extends $\sim$~100~\AA.

Since CONFIT allows for a maximum of two stellar components plus a
power-law in some cases and the stellar contents of the galaxy are not
known a priori, we used various combinations of spectra during the
modelling process in order to determine the properties of the dominant
stellar populations. We would like to emphasize that the techniques
and the modelling philosophy used in this paper are based on the
results of RZ08 for the ULIRG Arp 220, and to a lesser extent on those
of RZ07 for the case of PKS1345+12. 

\addtocounter{figure}{-1}
\begin{figure*}
\begin{minipage}{170mm}
\begin{tabular}{ccc}
\hspace*{0cm}\psfig{file=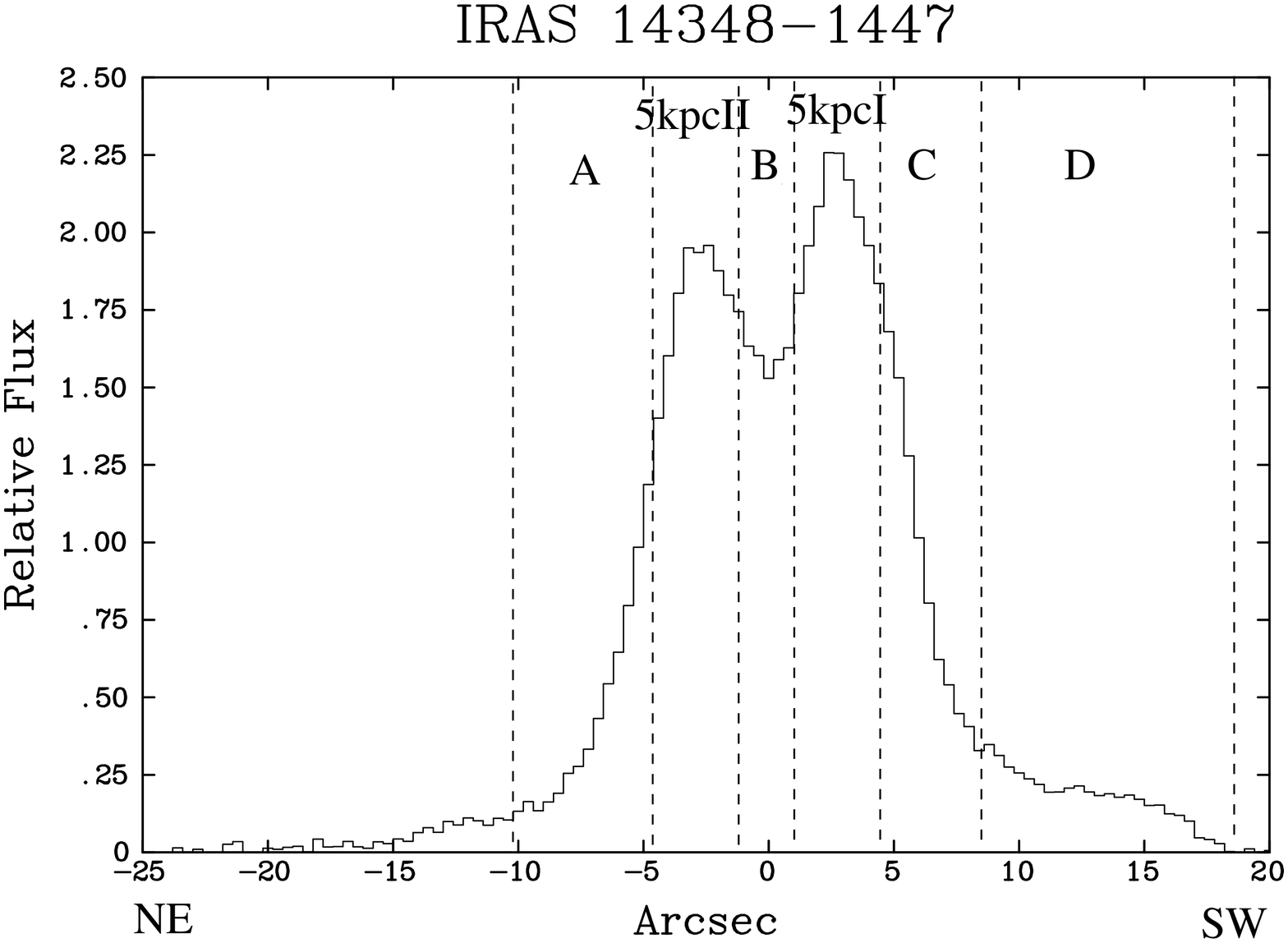,width=5.0cm,angle=0.}&
\psfig{file=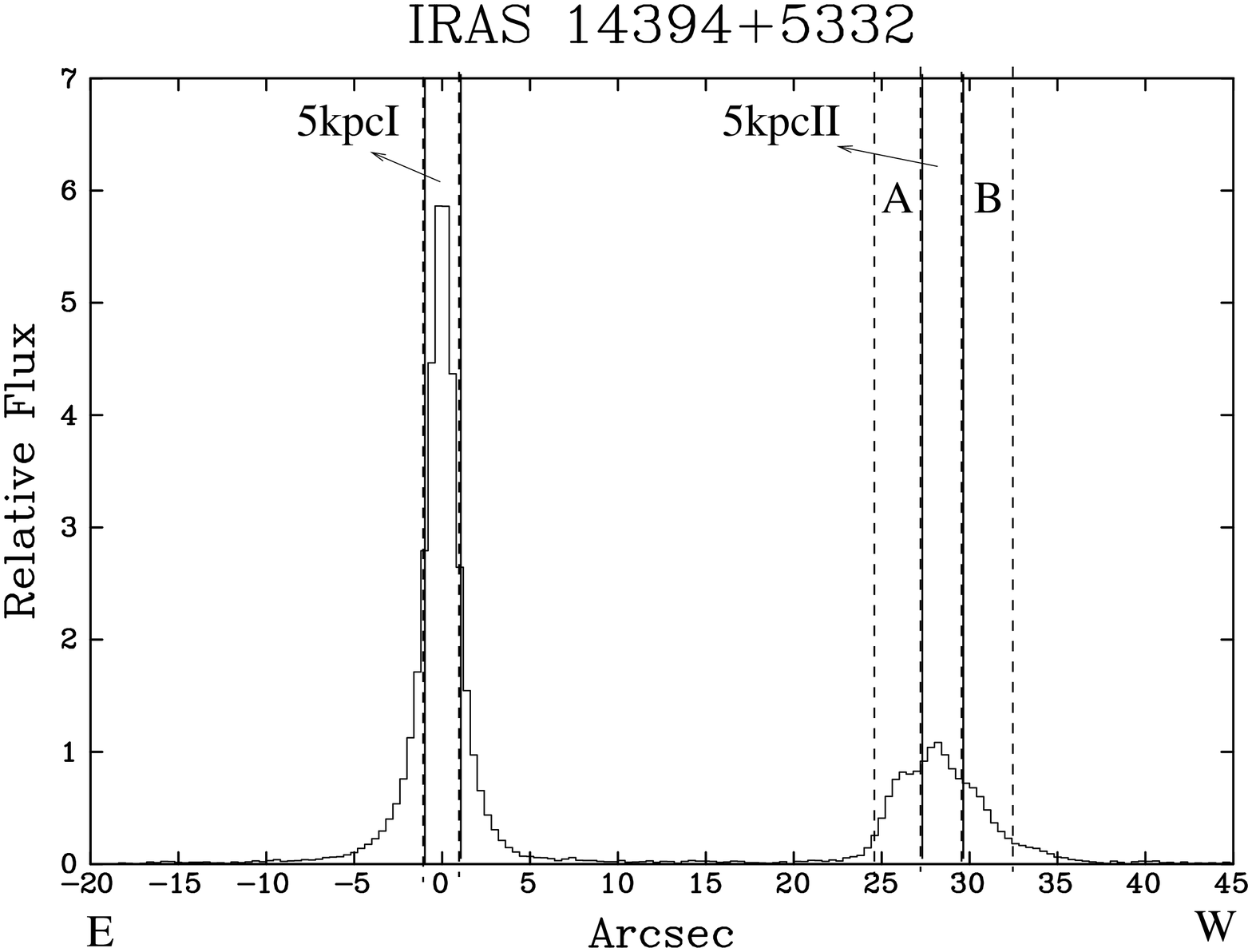,width=5.0cm,angle=0.}&
\psfig{file=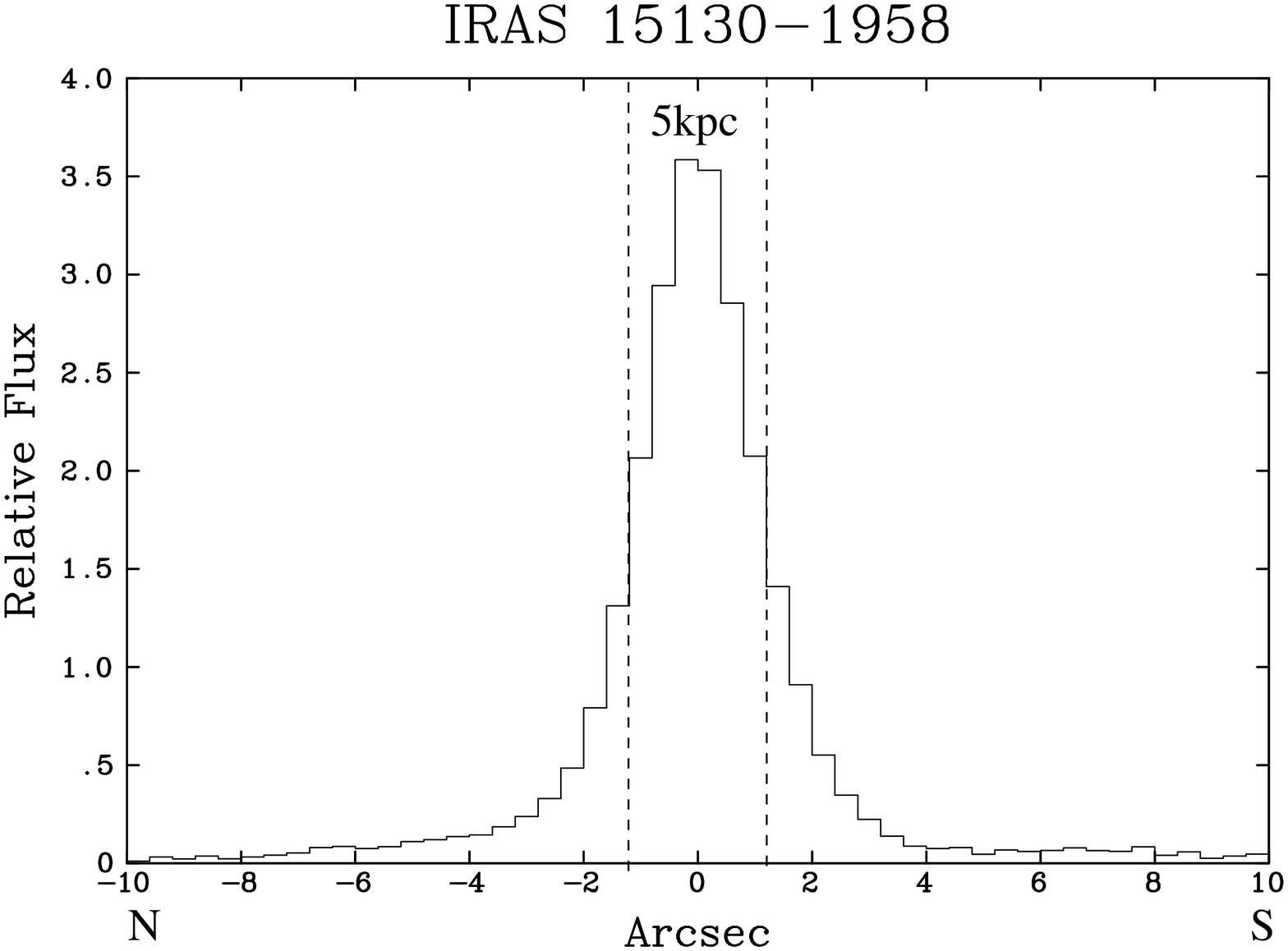,width=5.0cm,angle=0.}\\
\hspace*{0cm}\psfig{file=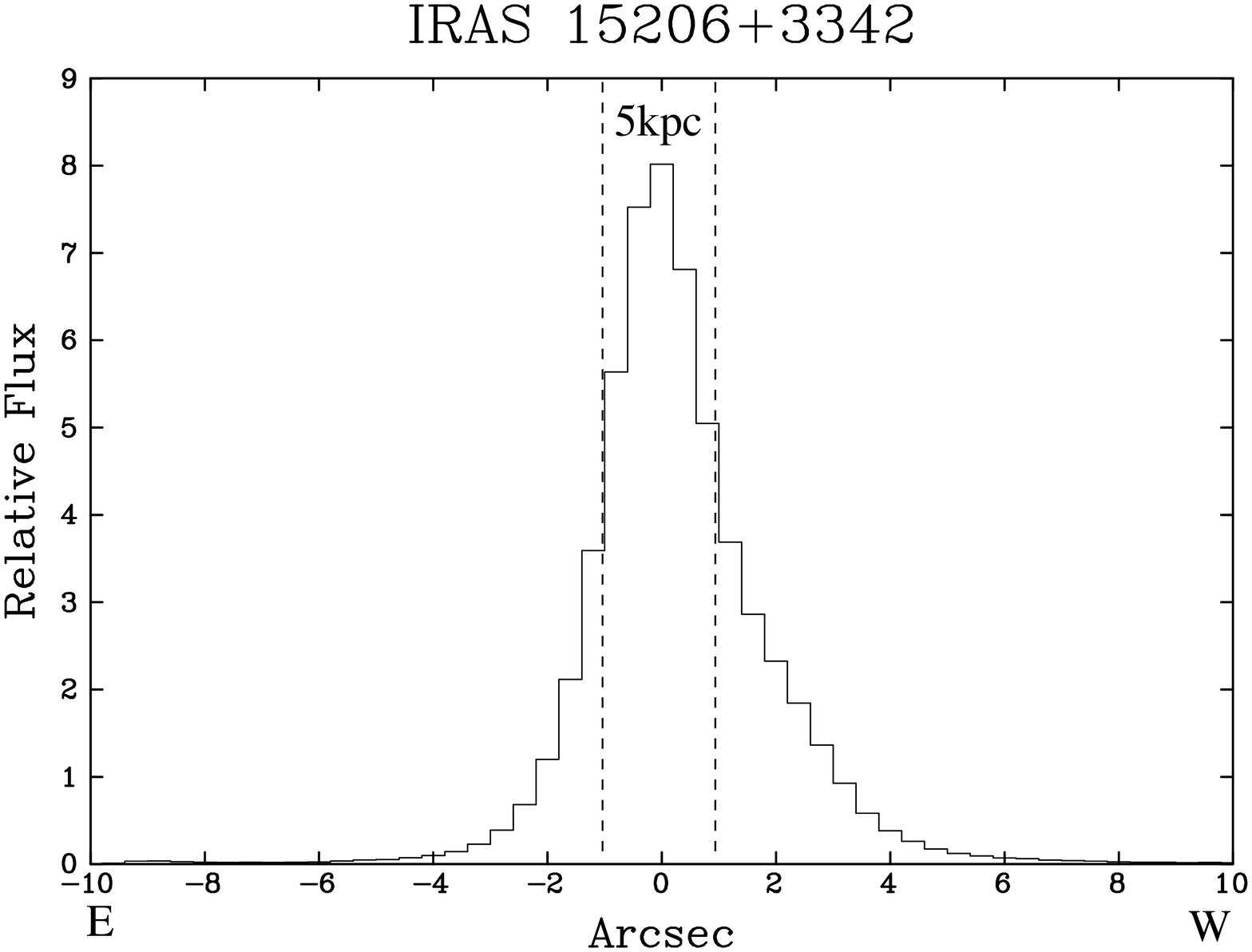,width=5.0cm,angle=0.}&
\psfig{file=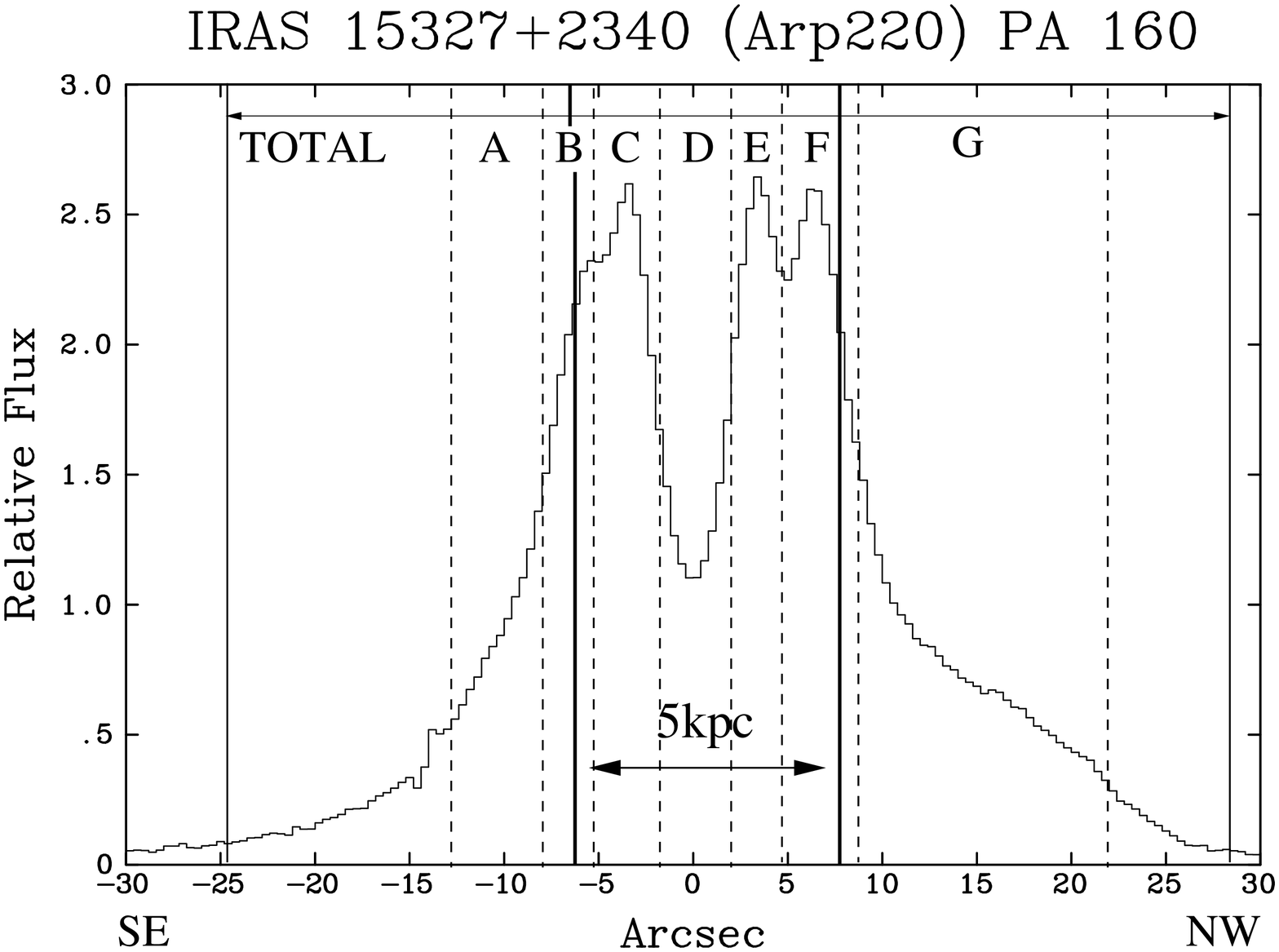,width=5.0cm,angle=0.}&
\psfig{file=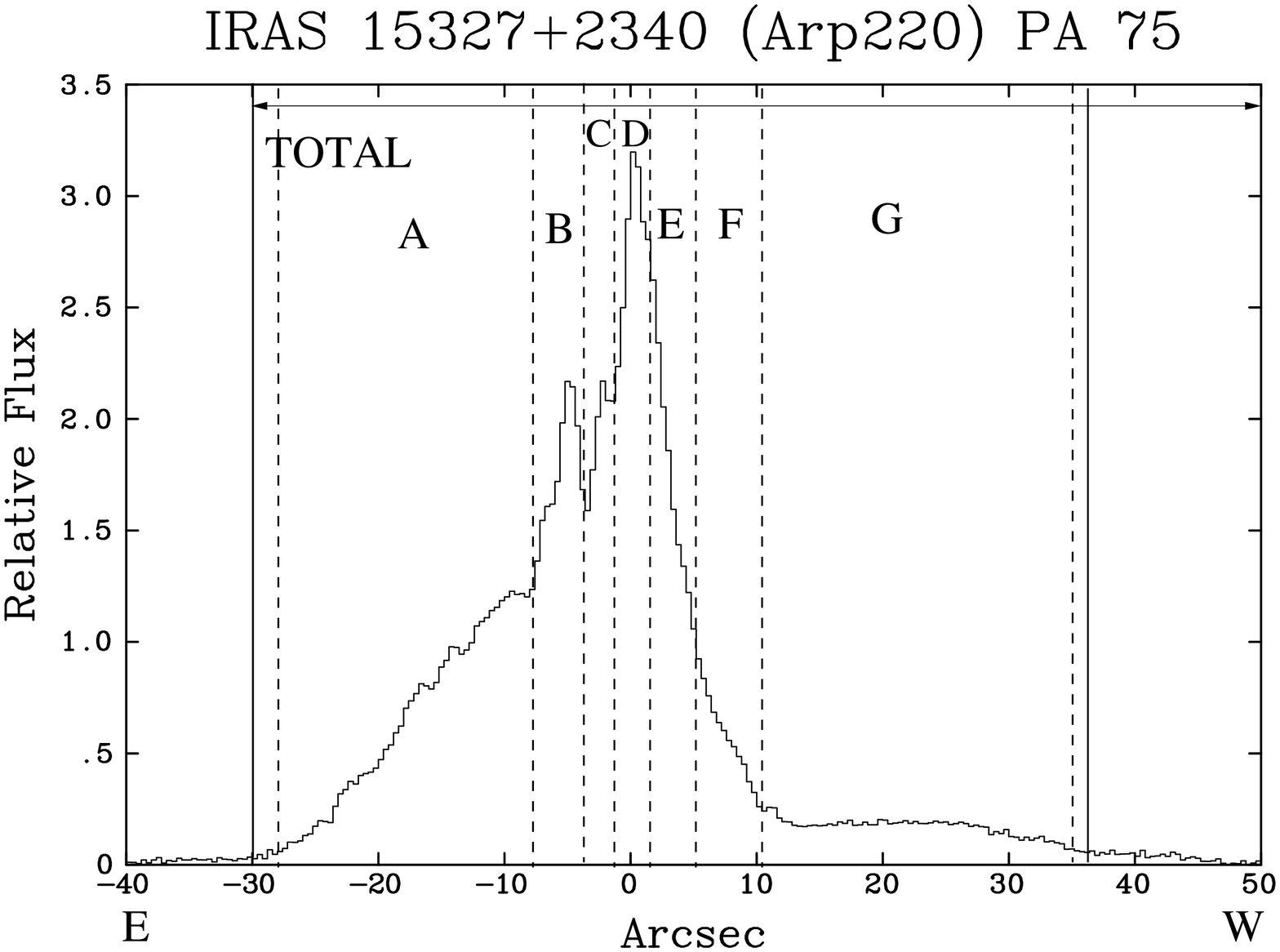,width=5.0cm,angle=0.}\\
\hspace*{0cm}\psfig{file=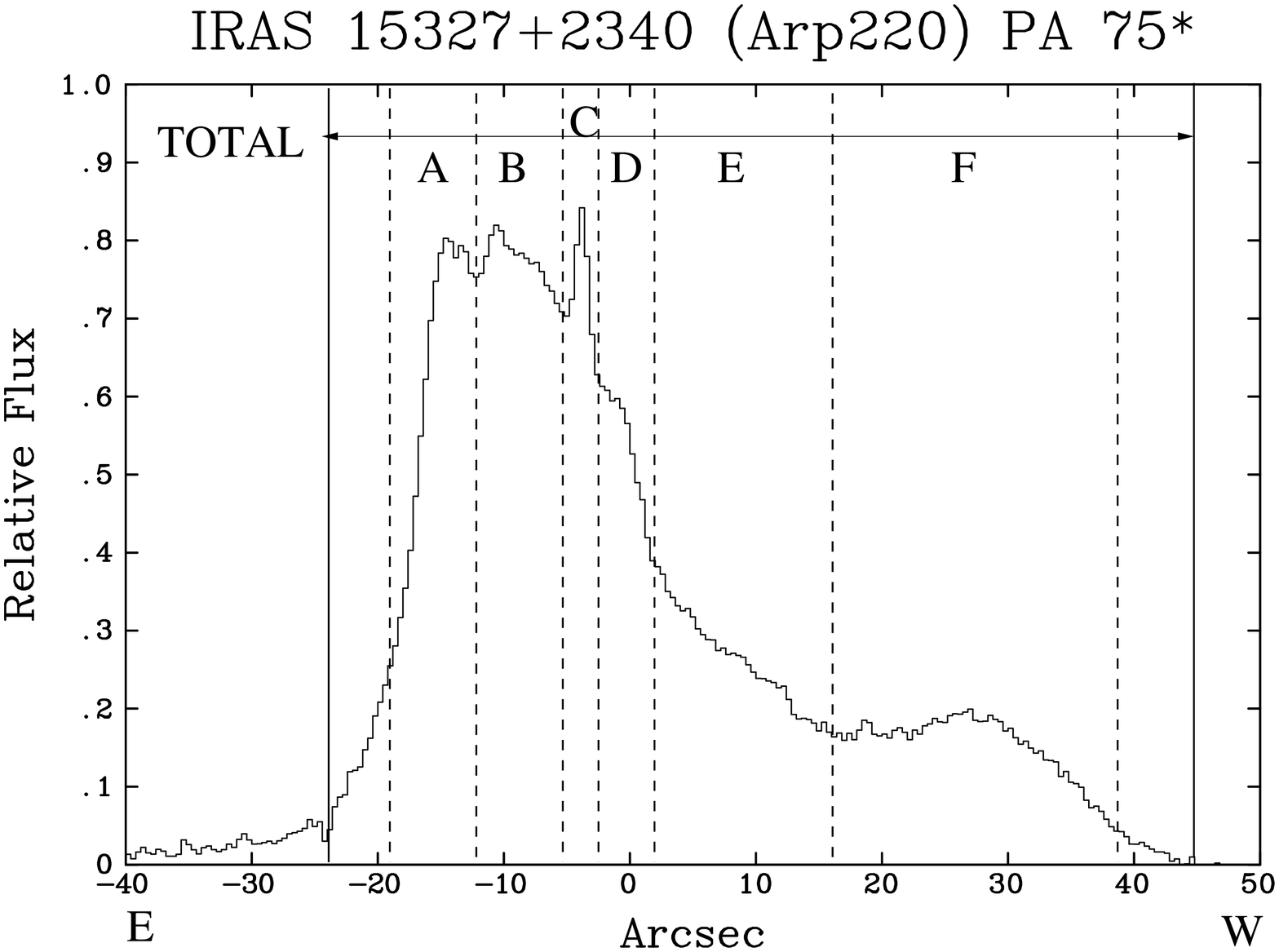,width=5.0cm,angle=0.}&
\psfig{file=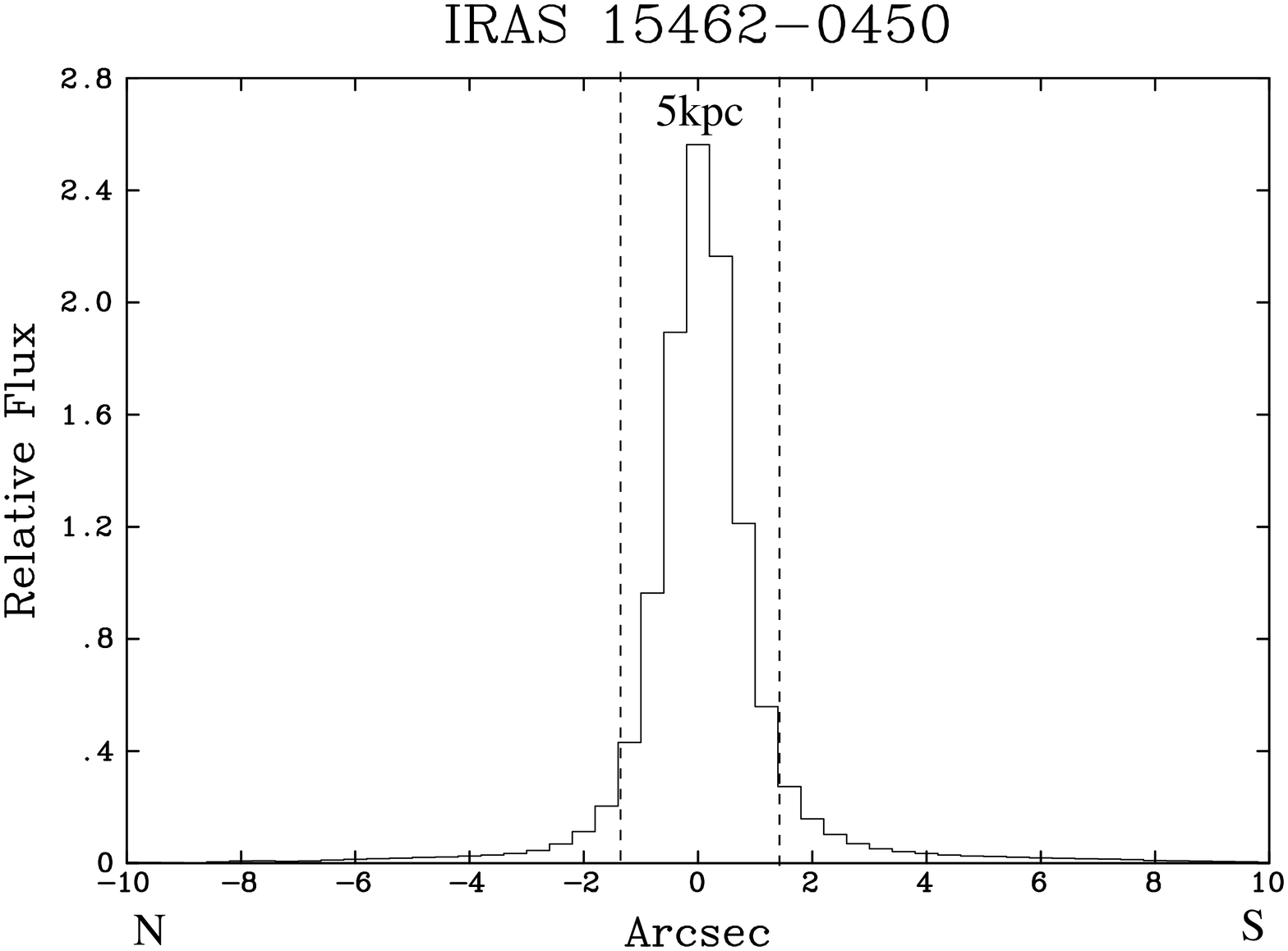,width=5.0cm,angle=0.}&
\psfig{file=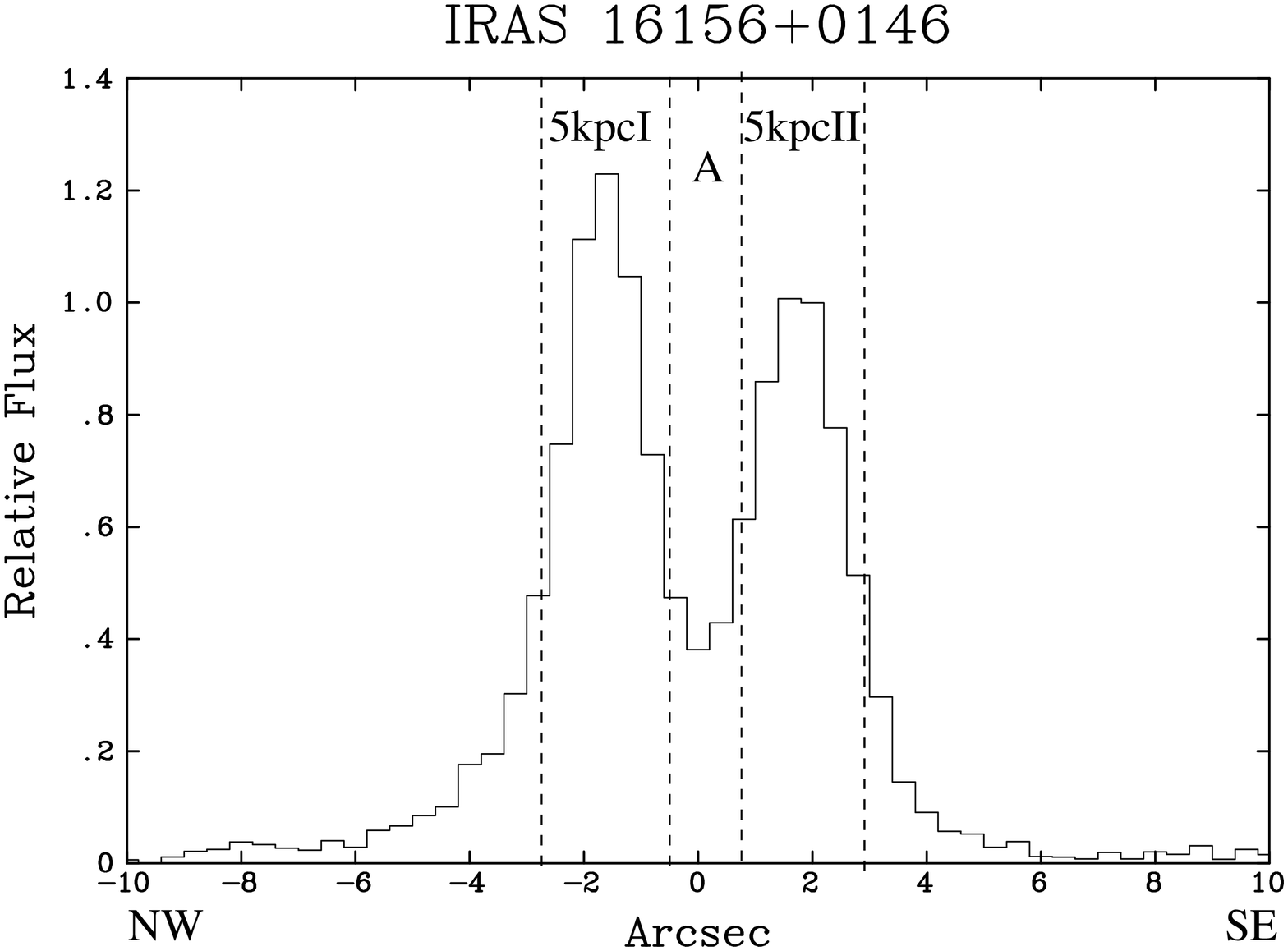,width=5.0cm,angle=0.}\\
\hspace*{0cm}\psfig{file=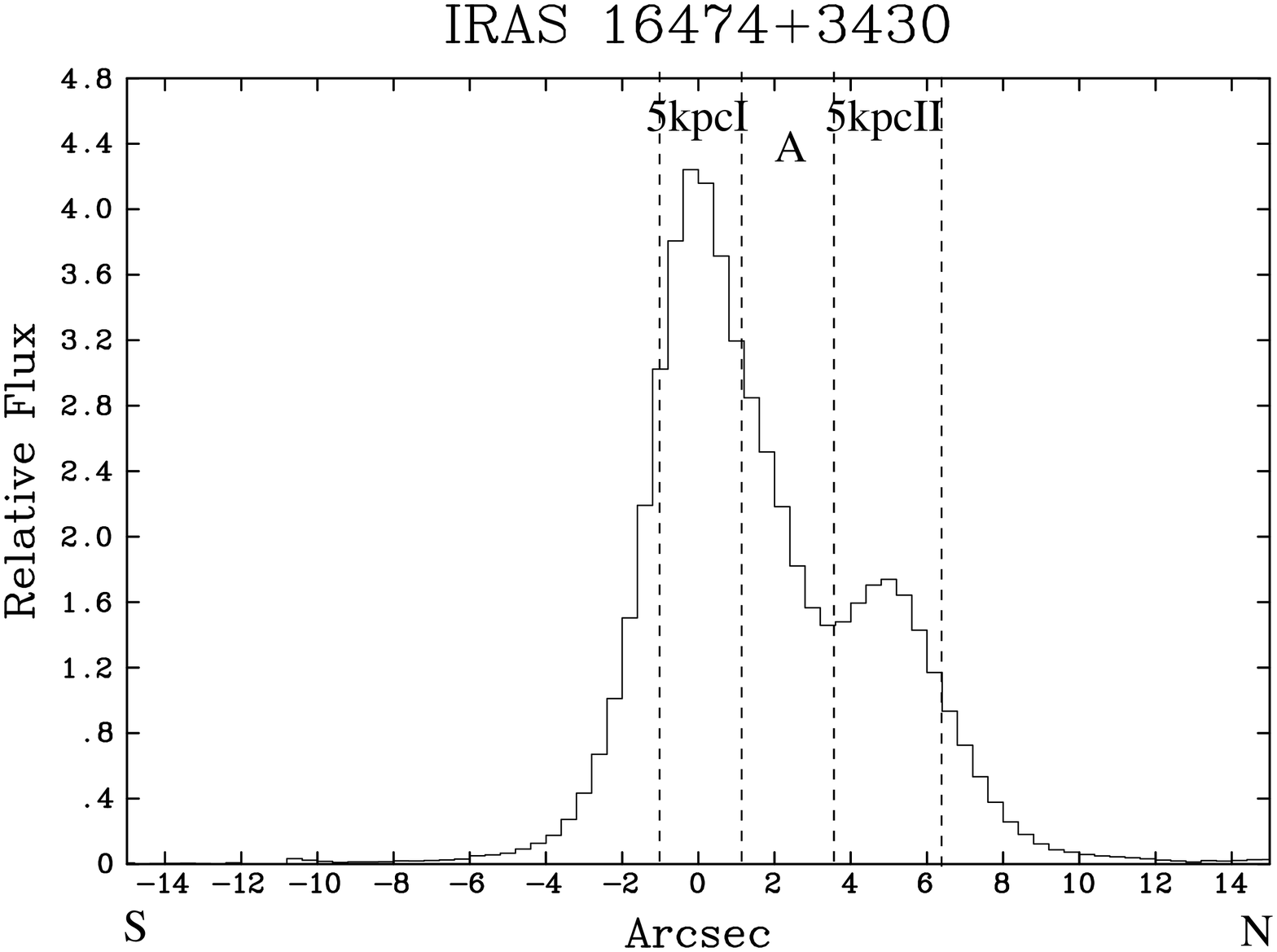,width=5.0cm,angle=0.}&
\psfig{file=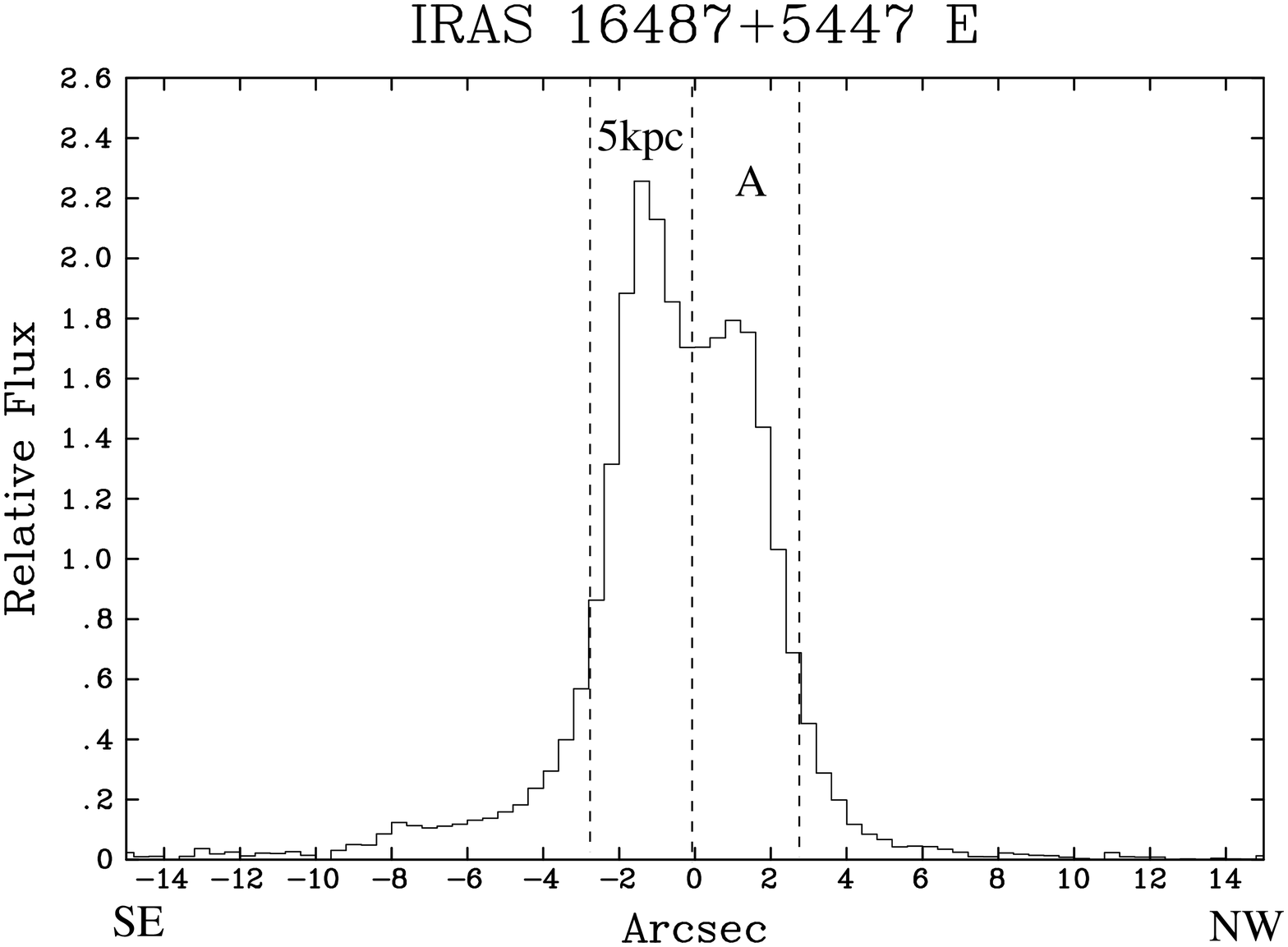,width=5.0cm,angle=0.}&
\psfig{file=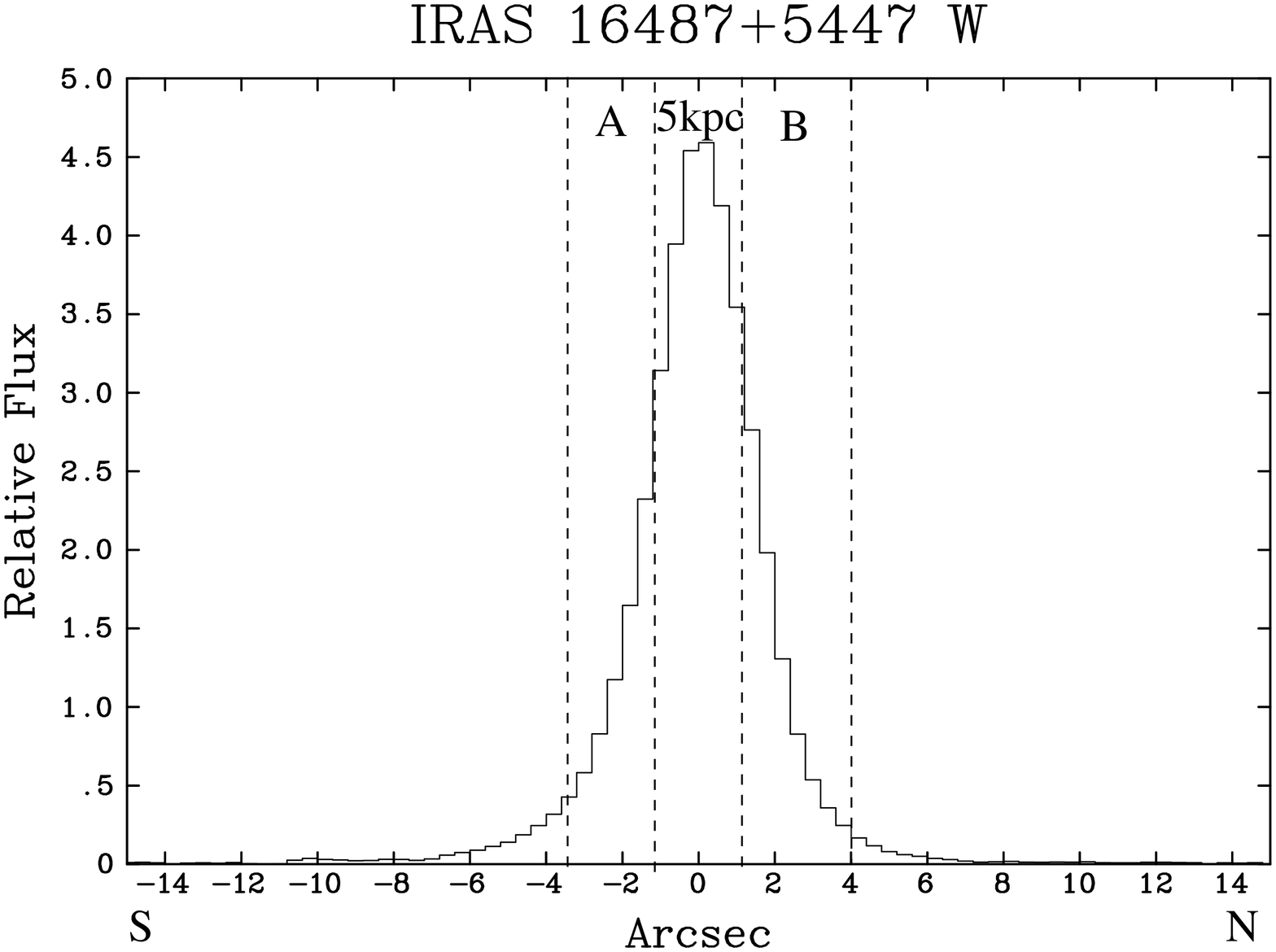,width=5.0cm,angle=0.}\\
\hspace*{0cm}\psfig{file=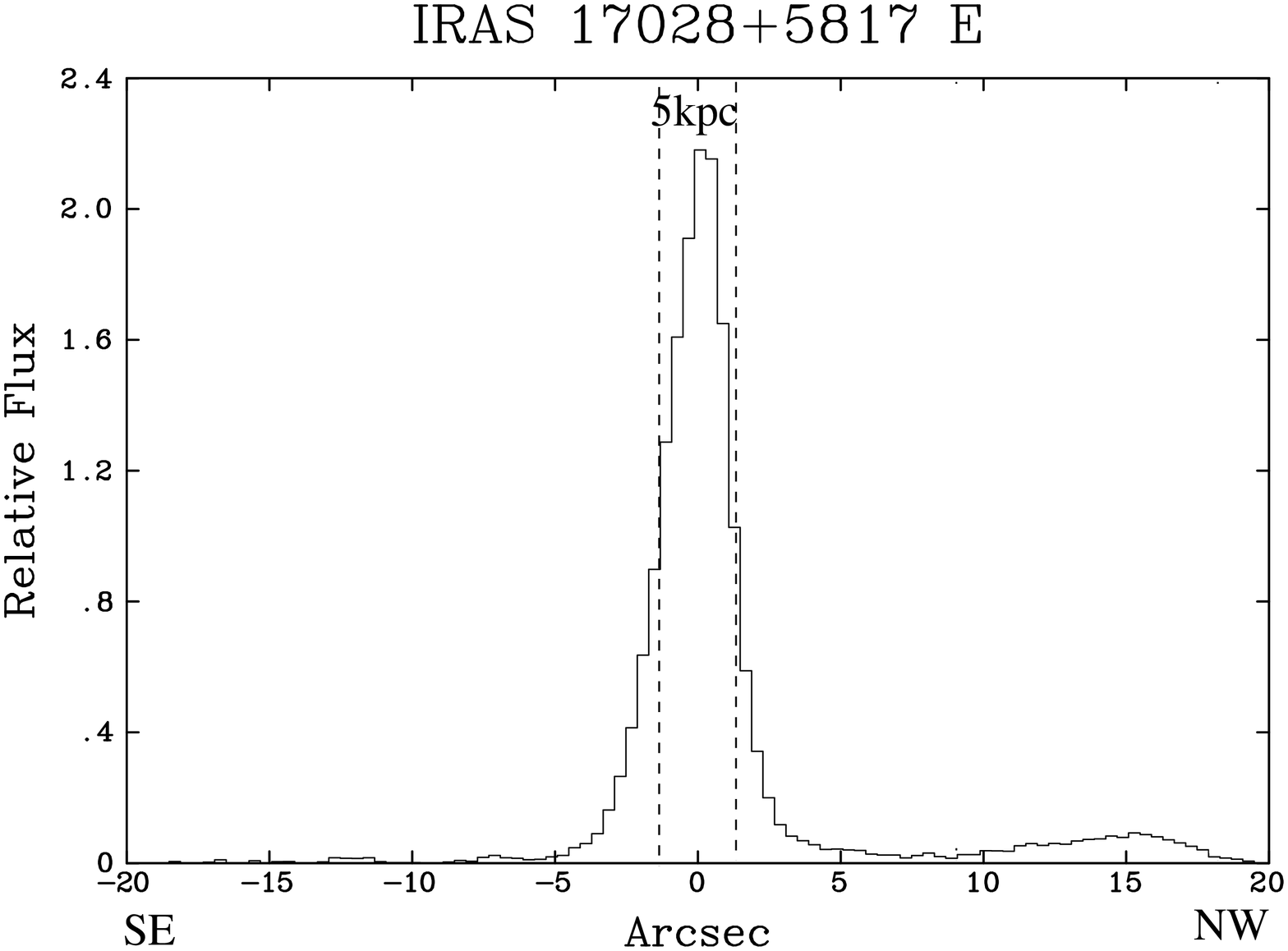,width=5.0cm,angle=0.}&
\psfig{file=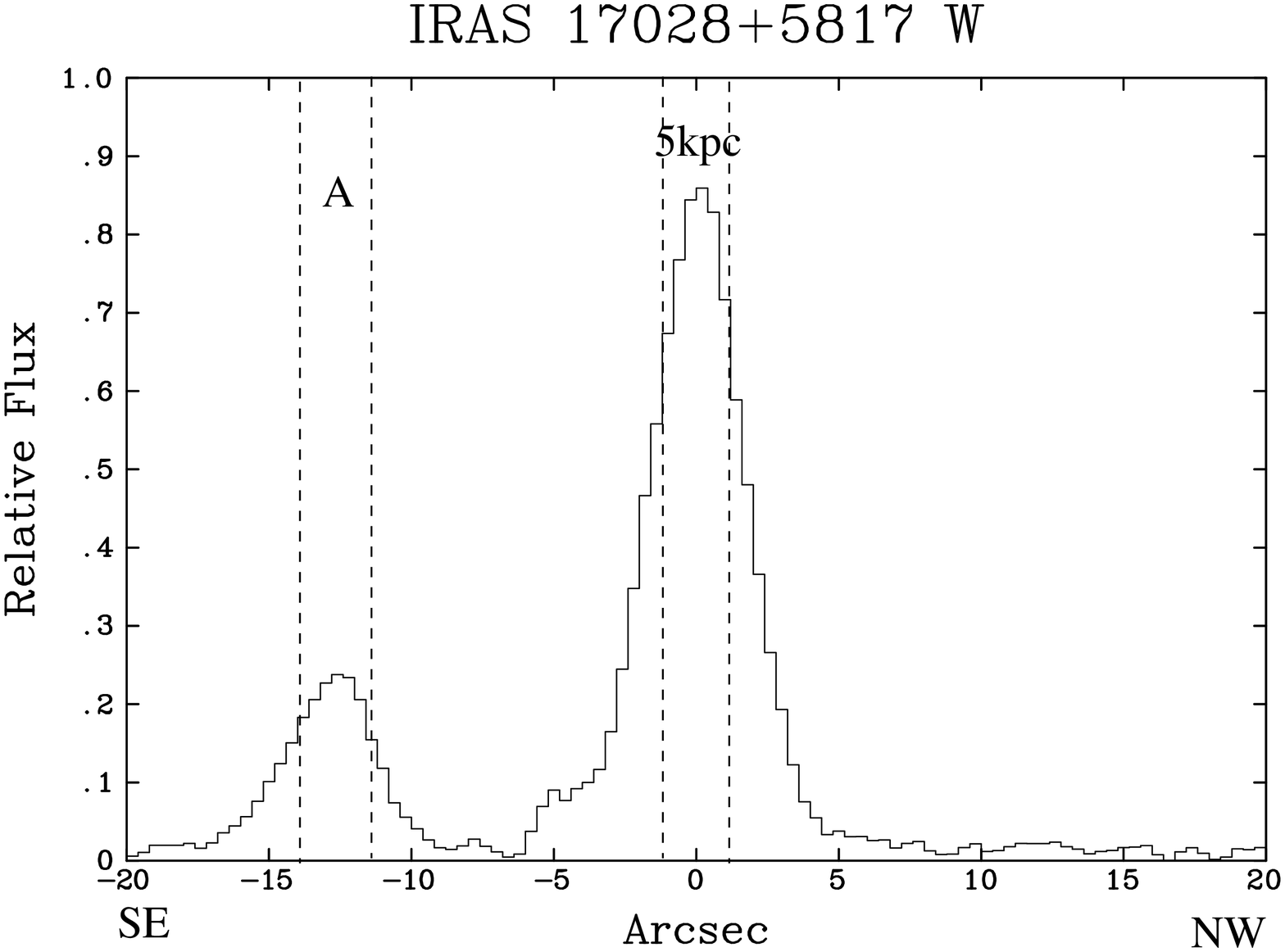,width=5.0cm,angle=0.}&
\psfig{file=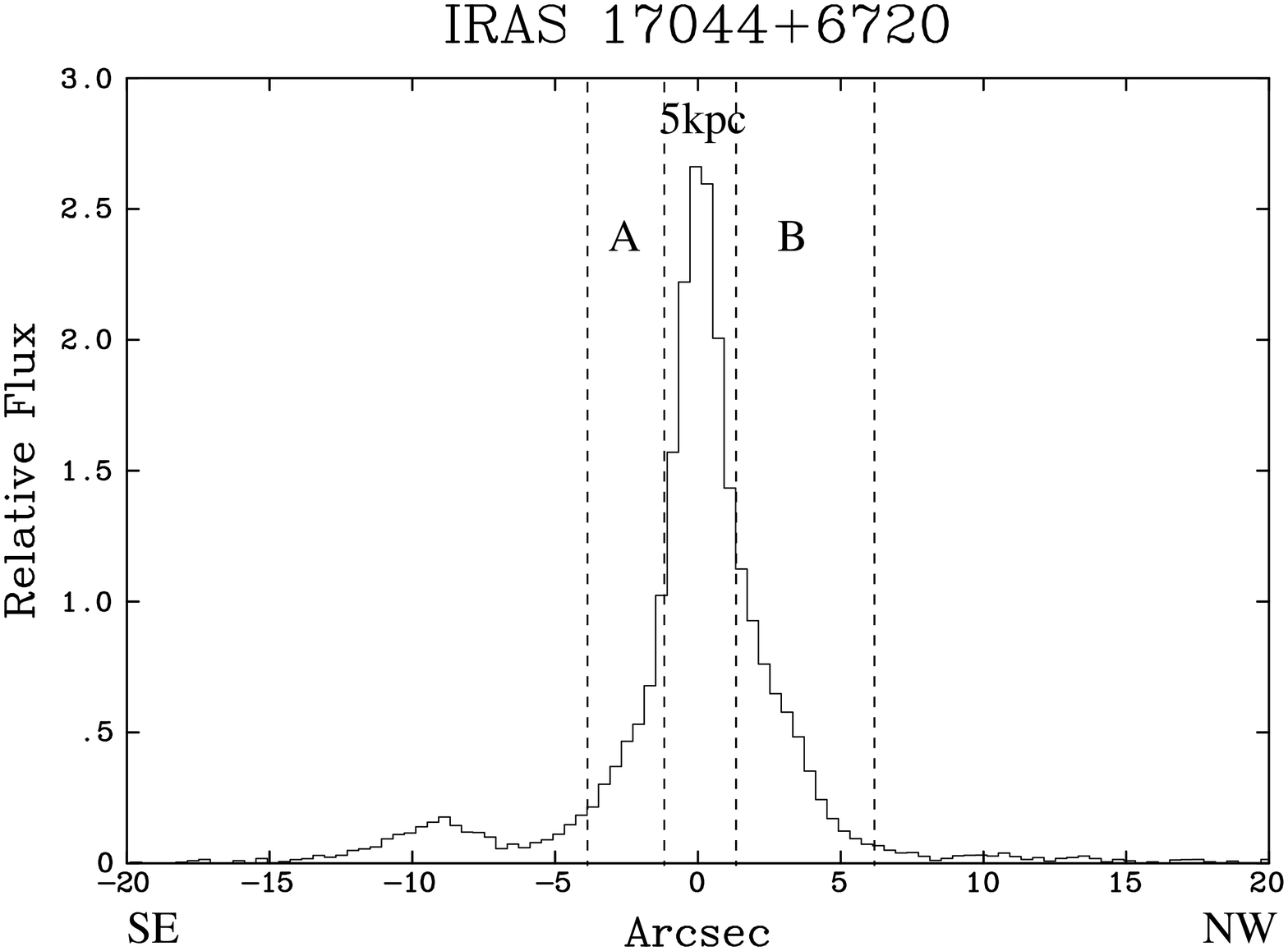,width=5.0cm,angle=0.}\\
\hspace*{0cm}\psfig{file=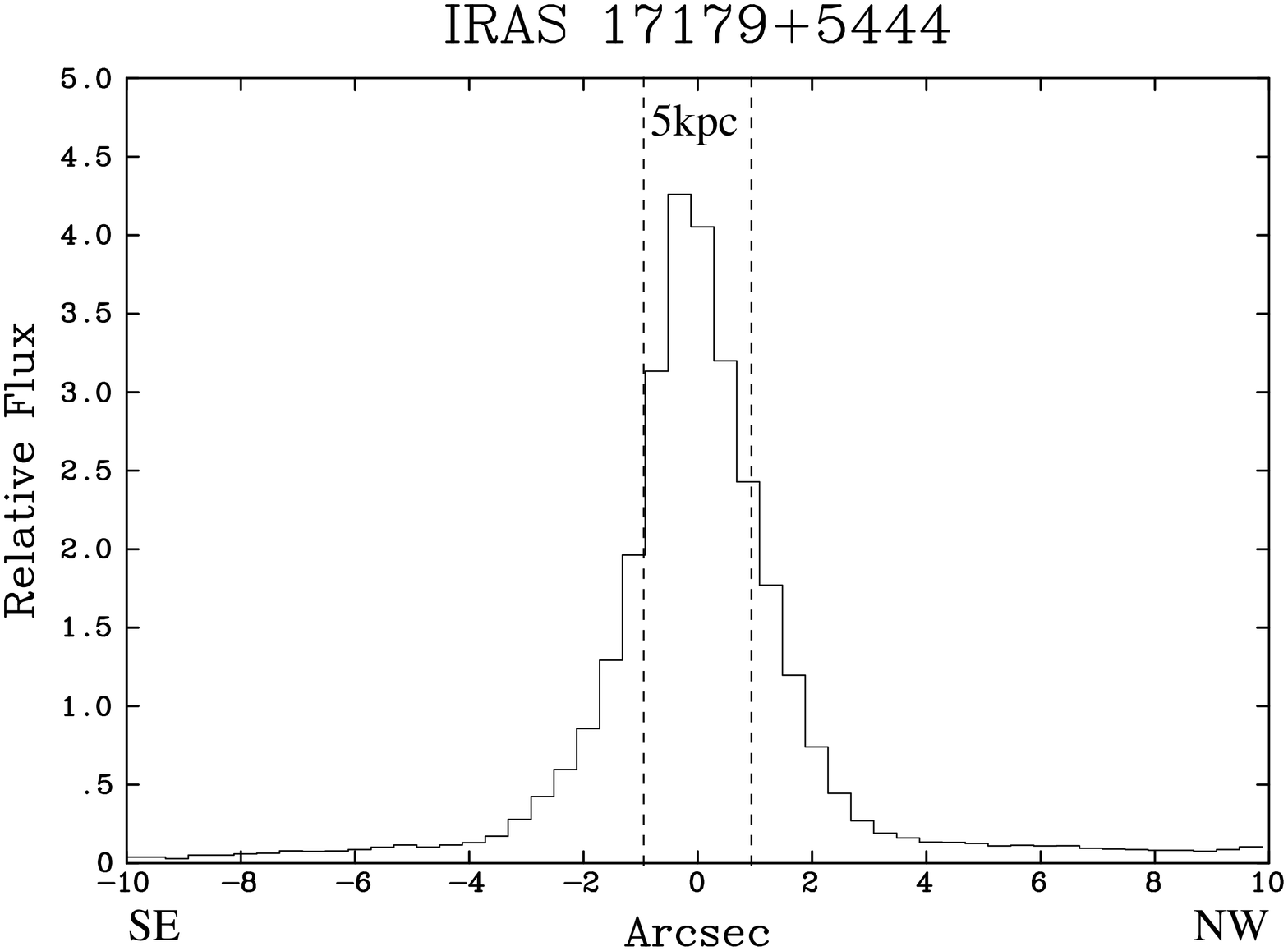,width=5.0cm,angle=0.}&
\psfig{file=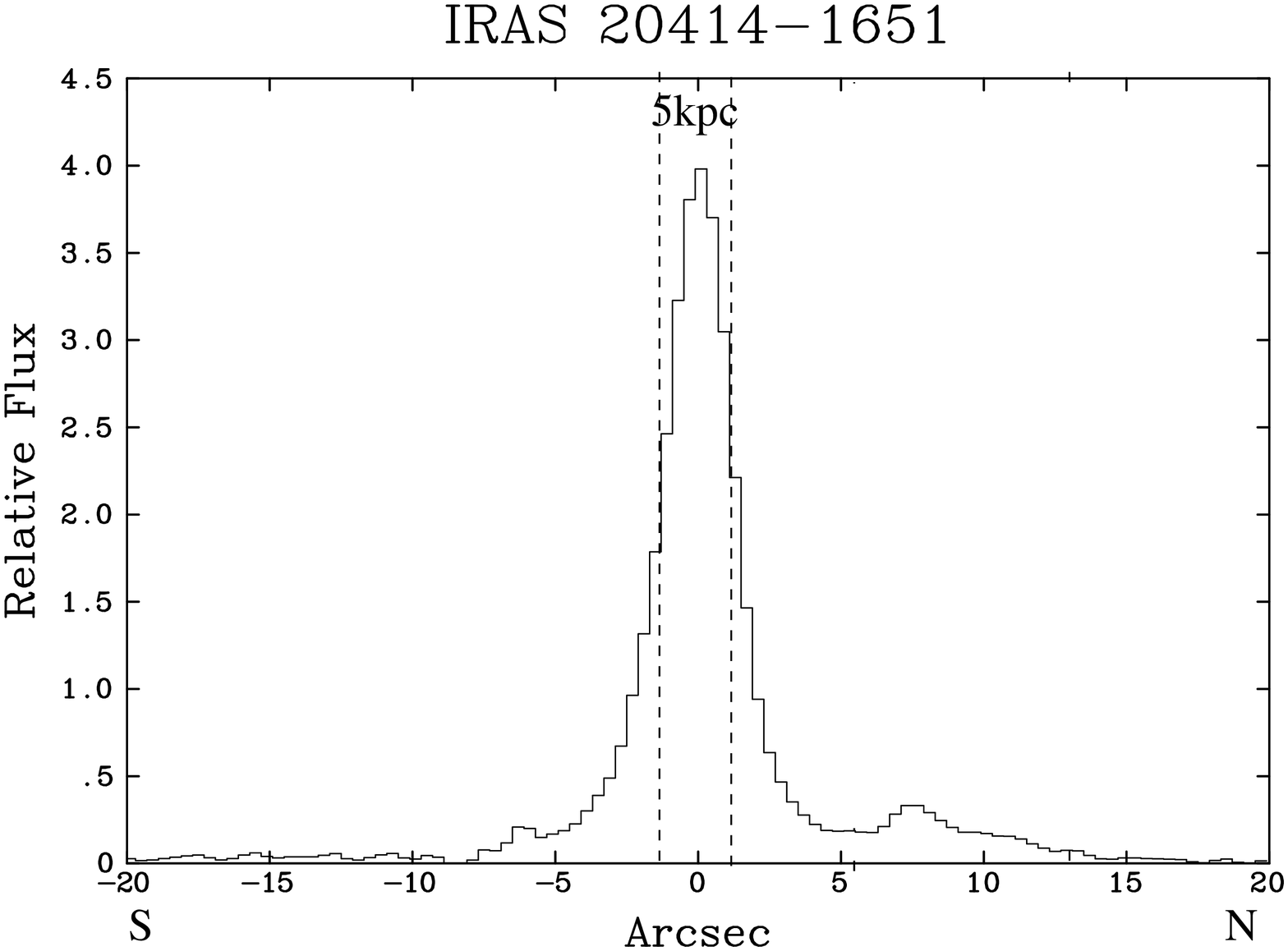,width=5.0cm,angle=0.}&
\psfig{file=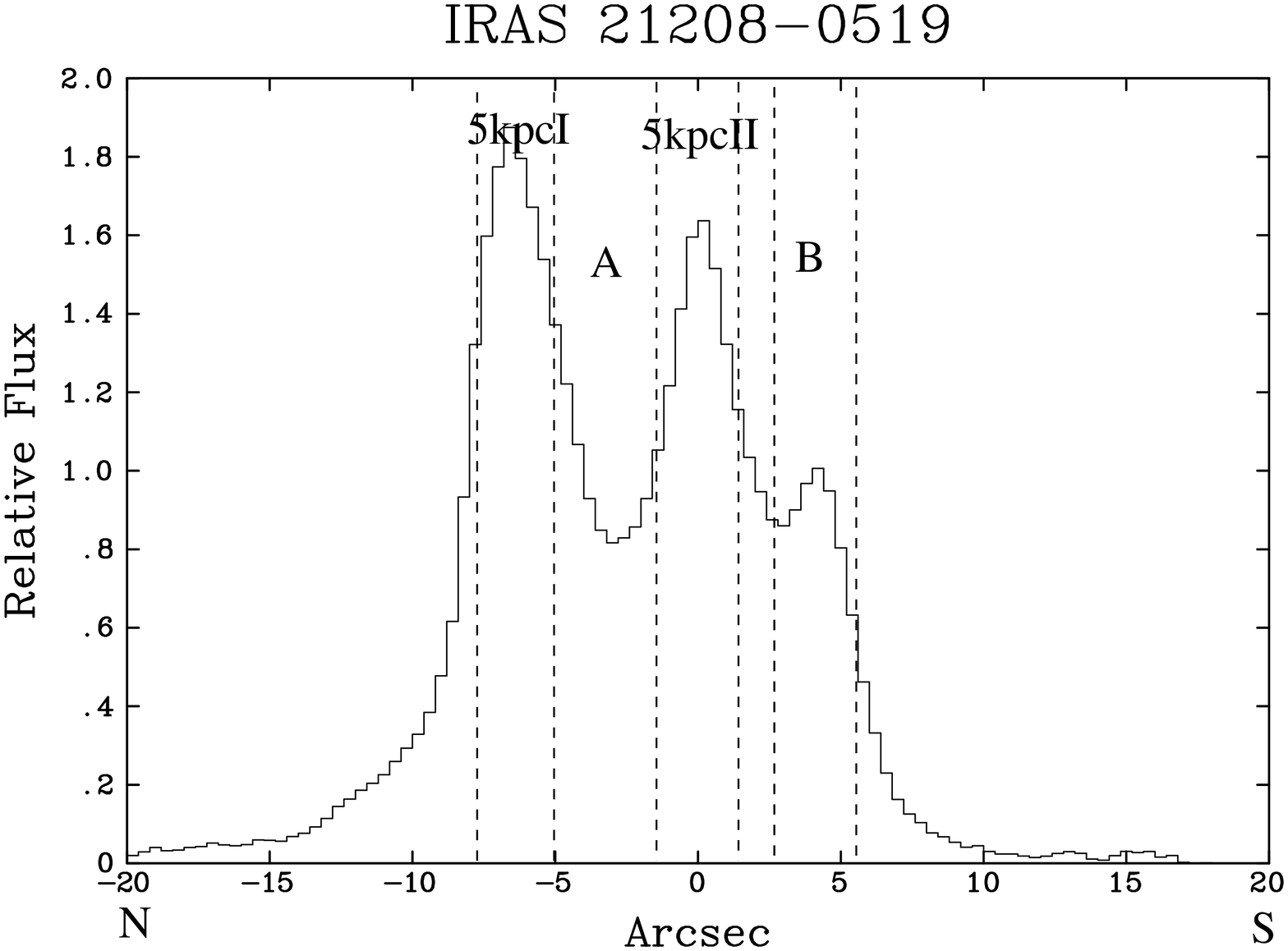,width=5.0cm,angle=0.}\\
\end{tabular}
\caption[{\it Continued}]{}
%\label{fig:SED}
\end{minipage}
\end{figure*}
\addtocounter{figure}{-1}
\begin{figure*}
\begin{minipage}{170mm}
\begin{tabular}{ccc}
\hspace*{0cm}\psfig{file=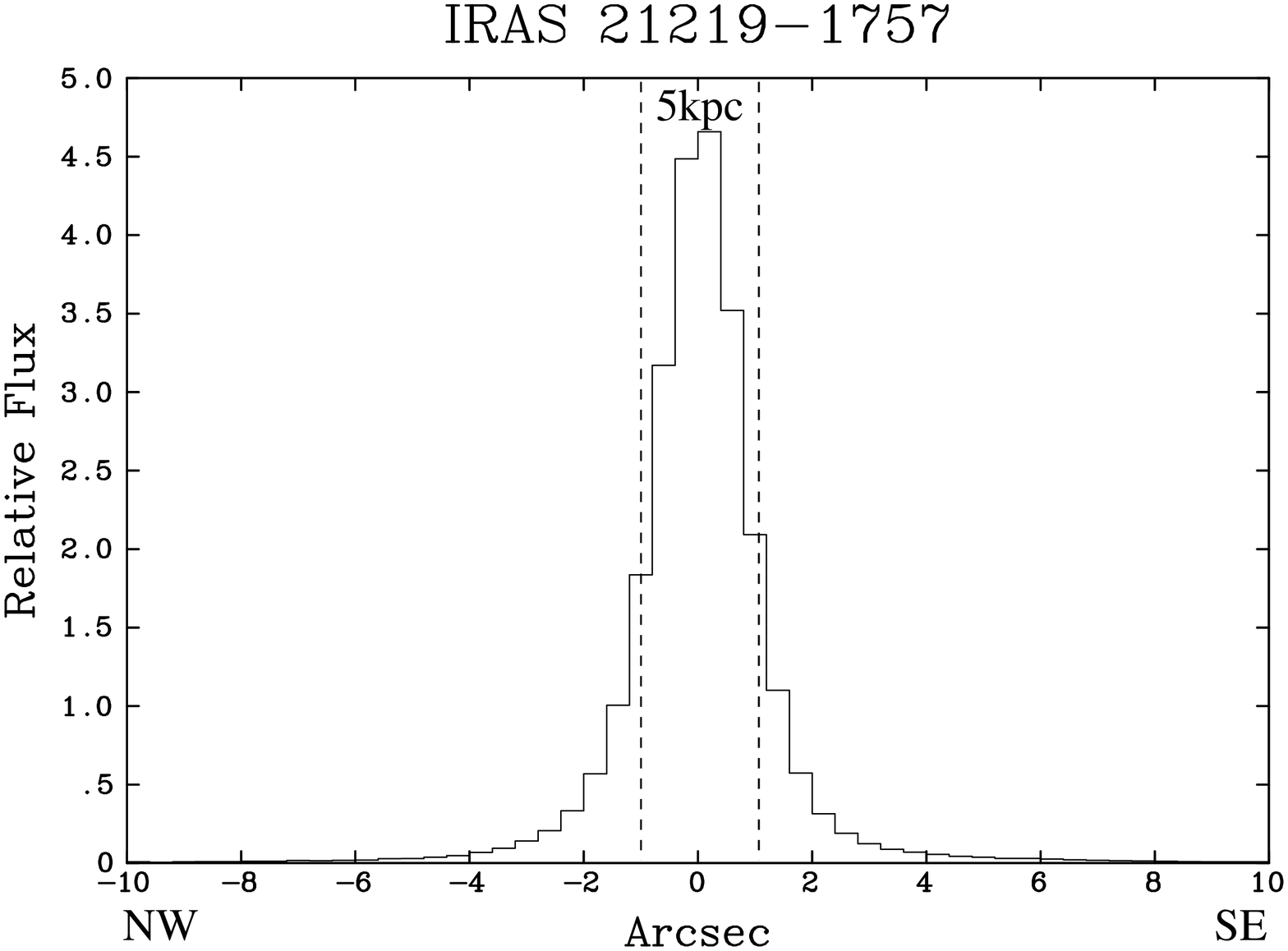,width=5.0cm,angle=0.}&
\psfig{file=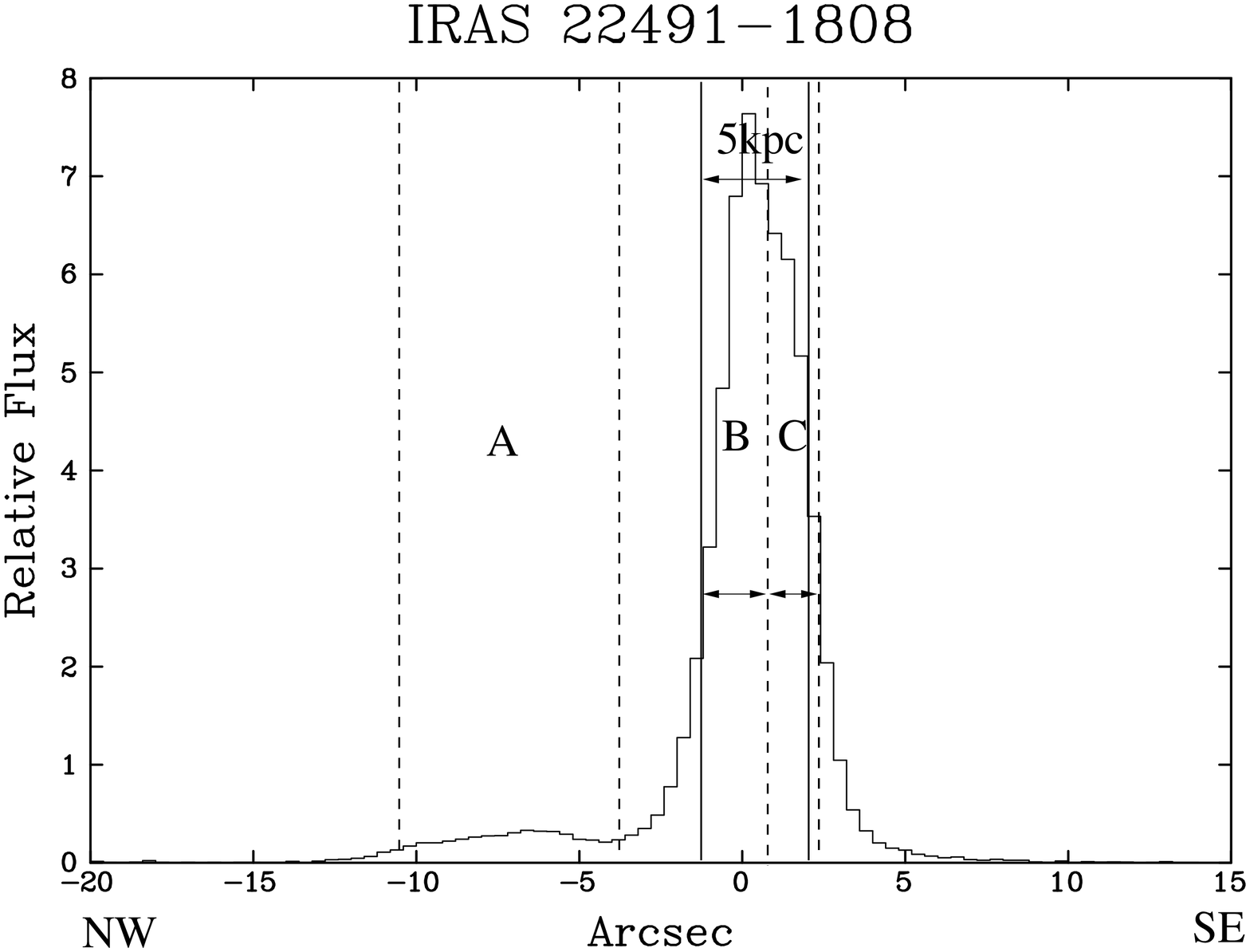,width=5.0cm,angle=0.}&
\psfig{file=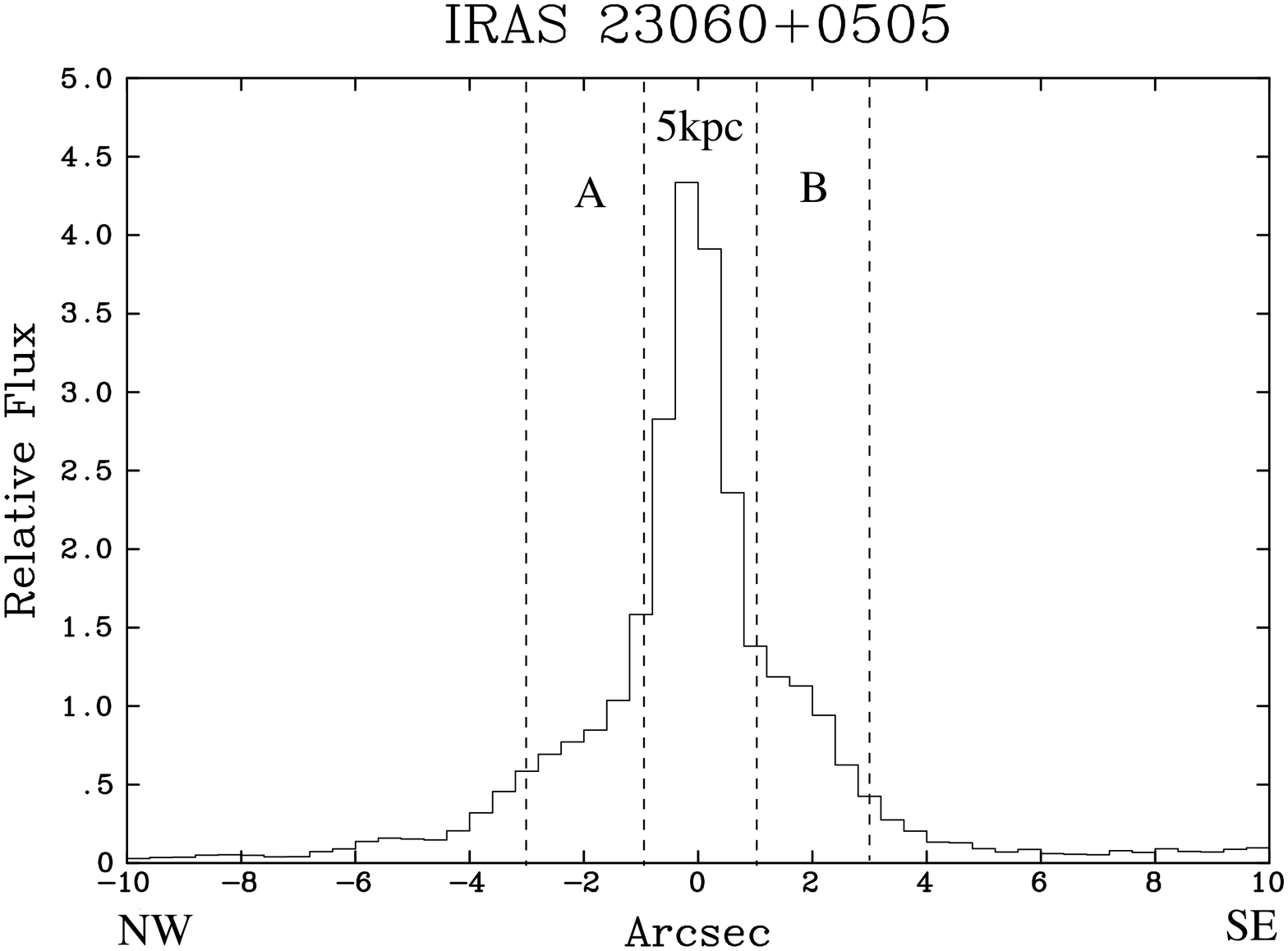,width=5.0cm,angle=0.}\\
\hspace*{0cm}\psfig{file=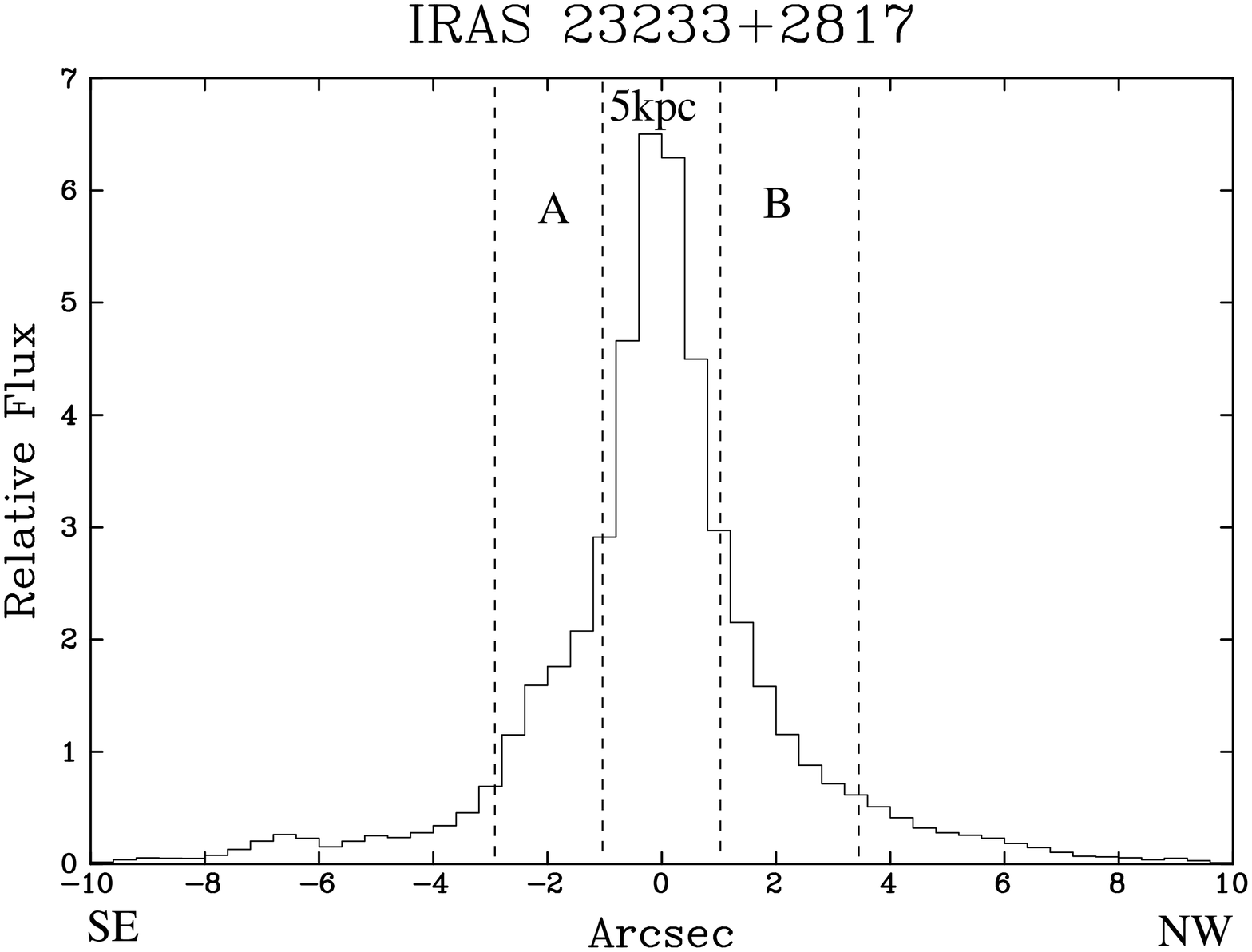,width=5.0cm,angle=0.}&
\psfig{file=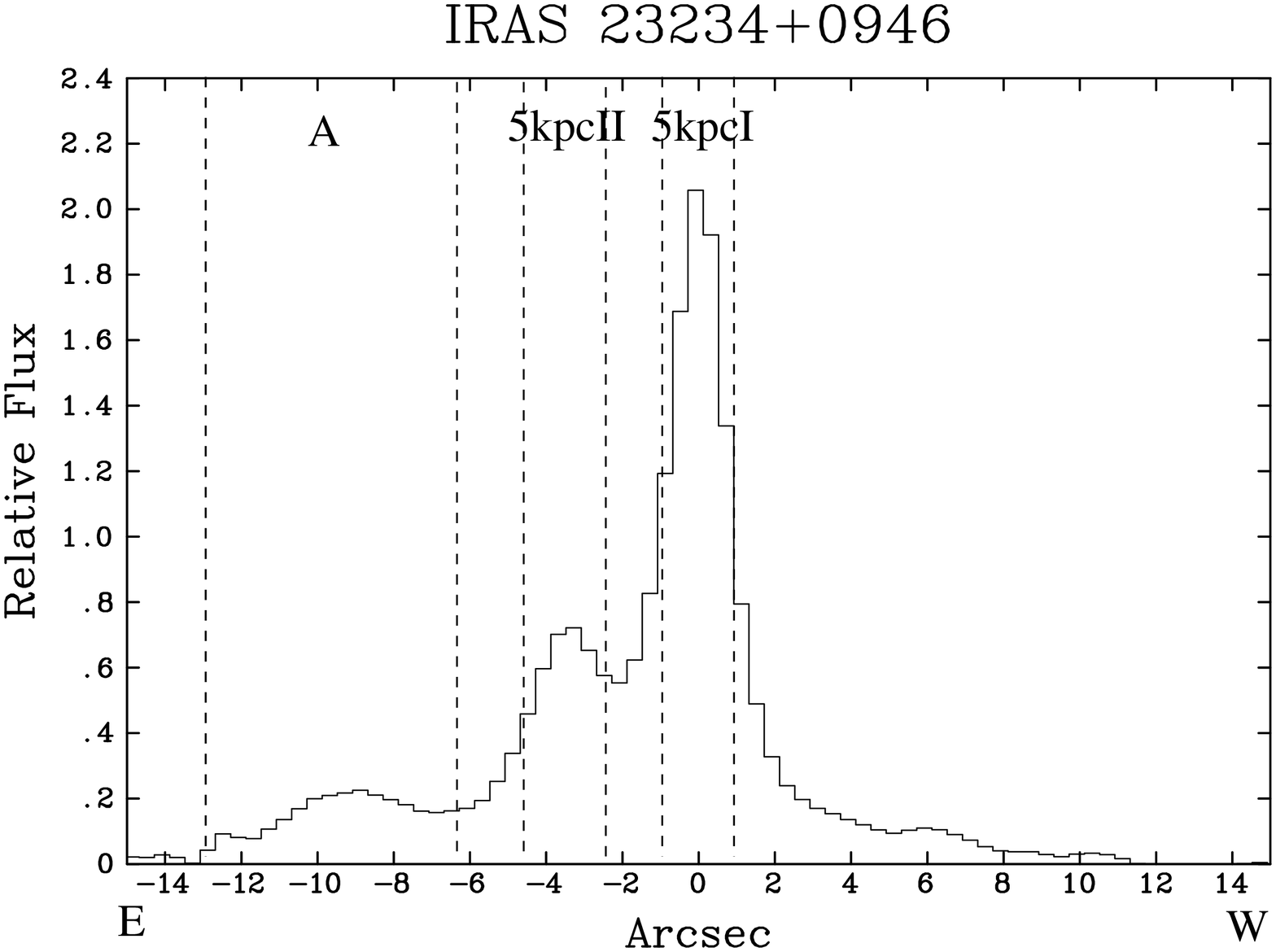,width=5.0cm,angle=0.}&
\psfig{file=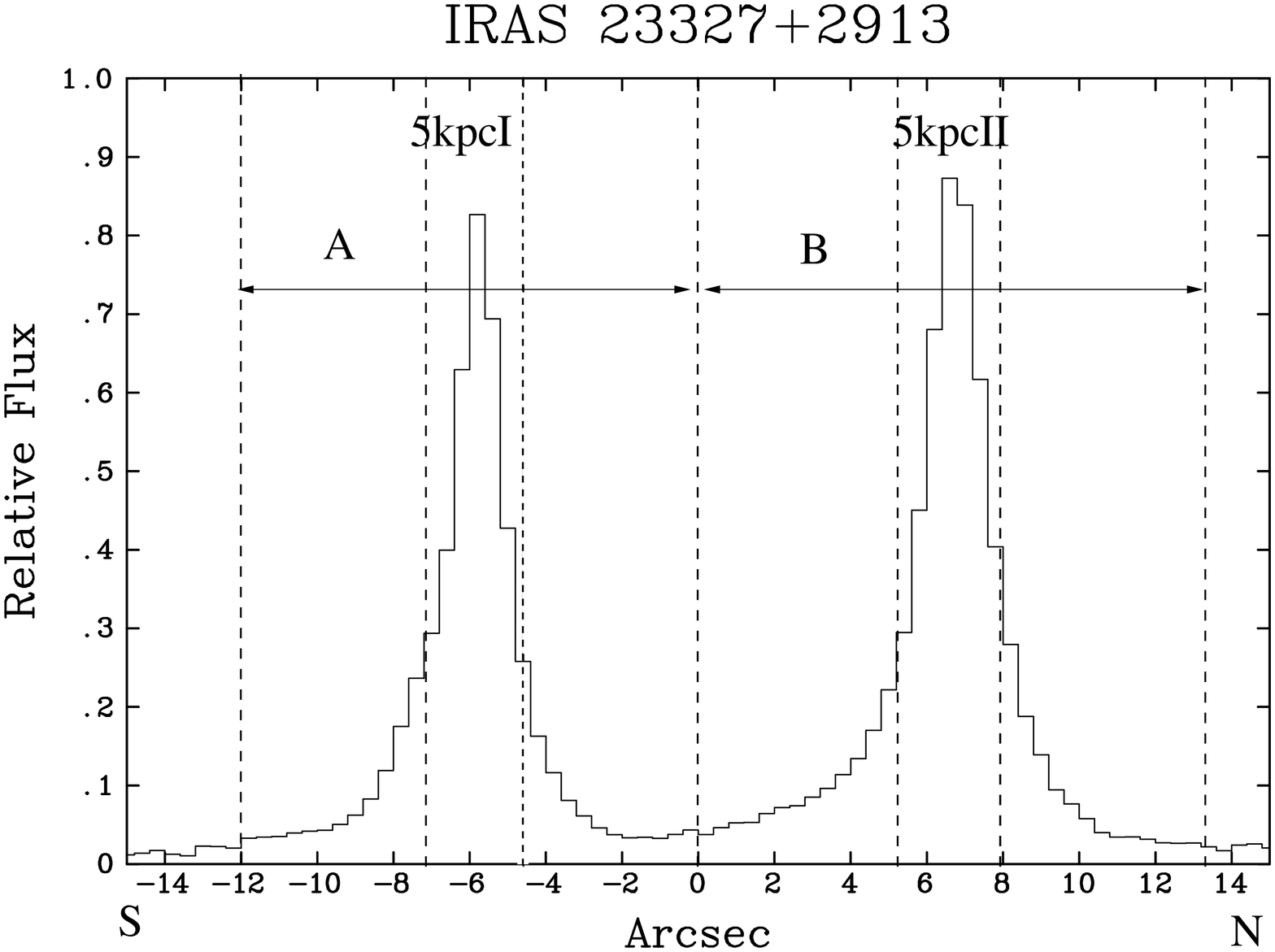,width=5.0cm,angle=0.}\\
\hspace*{0cm}\psfig{file=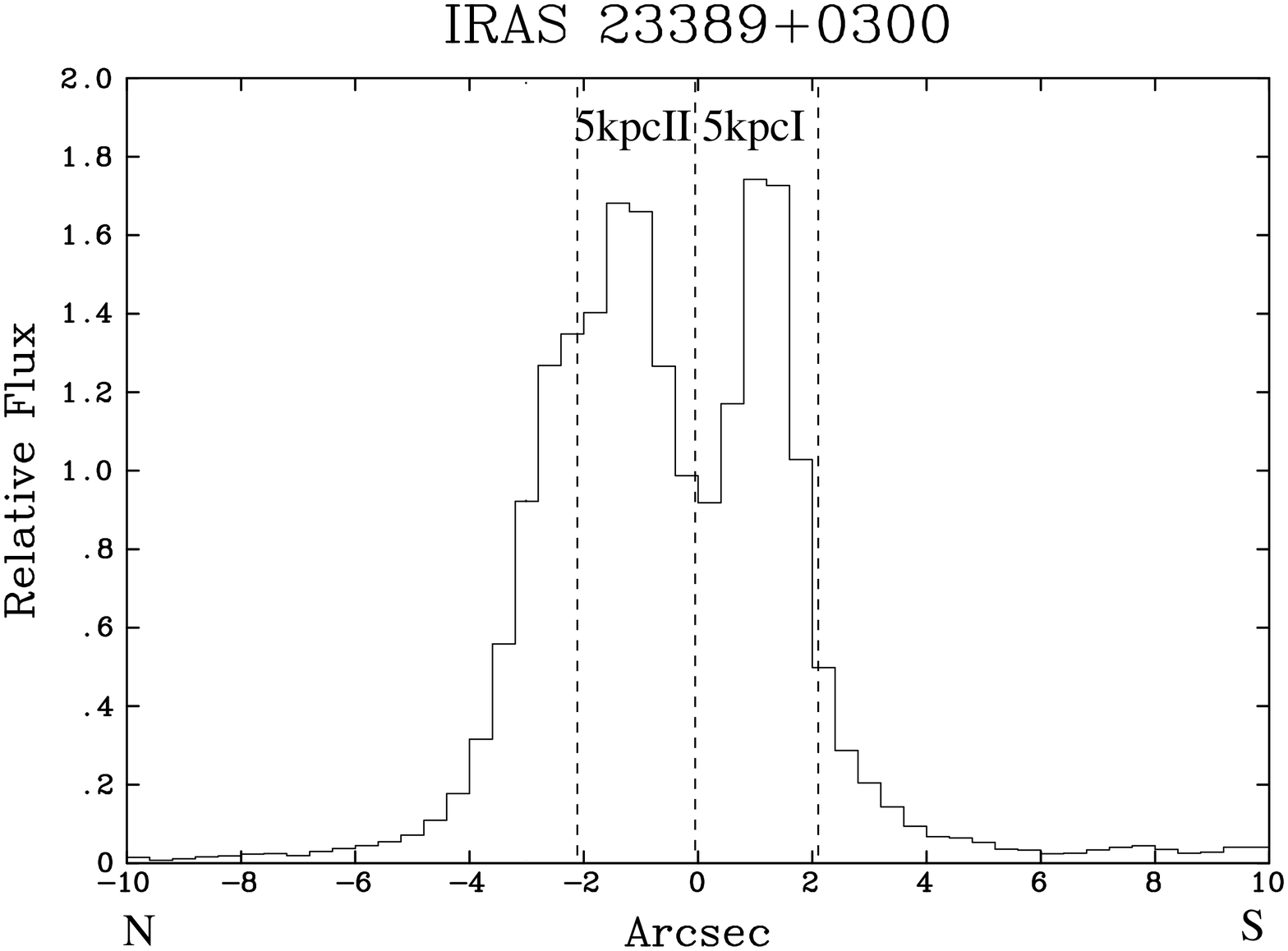,width=5.0cm,angle=0.}&
\psfig{file=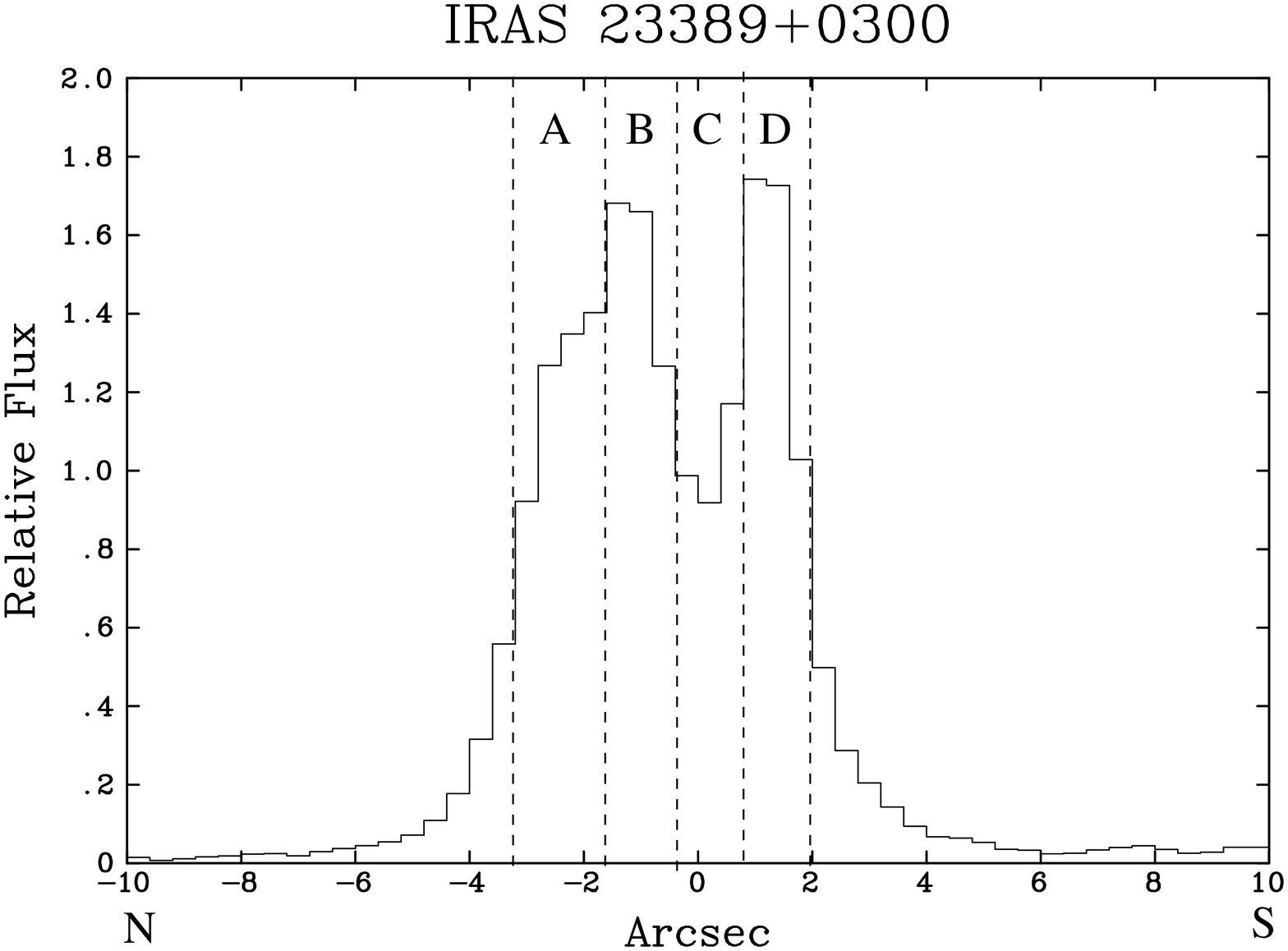,width=5.0cm,angle=0.}\\
\end{tabular}
\caption[{\it Continued}]{}
%\label{fig:SED}
\end{minipage}
\end{figure*}

Throughout this paper, we define {\it young stellar populations}
(YSPs) as stellar components with ages t$_{YSP}$$\leq$ 2 Gyr, and {\it
  old stellar populations} (OSP) as components with ages t$_{OSP}$$>$2
Gyr. With the aim of better describing each combination, it is
convenient to divide the YSPs into two groups:

\begin{itemize}

\item {\it very young stellar populations} (VYSP): stellar components with
ages t$_{\rm VYSP}$~$\leq$~0.1~Gyr;
\item {\it intermediate-age young stellar populations} (IYSP): stellar
components with ages in the range of 0.1 $<$ t$_{\rm IYSP}$ $\leq$ 2 Gyr.

\end{itemize}

Generally, for models that include an old plus a young and/or
an intermediate stellar component it is not possible to distiguish
between OSP ages in the range 2 $<$ t$_{OSP}$ $\leq$ 12 Gyr, and
therefore, we decided not to use such stellar populations during the
modelling analysis described below. Instead, we fixed the age of the
OSP at 12.5 Gyr. The different combinations used during the modelling
were:
\begin{itemize}

\item {\bf Combination I:} as a first approach we used a two component
  model comprising a YSP (t$_{\rm YSP} \leq$ 2 Gyr) with varying
  reddening (0.0 $\leq E(B - V)\leq$ 2.0, increasing in steps of 0.1),
  along with an unreddened\footnote{Although we cannot entirely rule
  the idea that these stellar populations are significantly reddened,
  assuming zero reddening for OSP is reasonable since such populations
  are likely to be situated in the extended bulge/halo in the
  foreground of the system (along the line of sight), or else
  completely obscured by the circum-nuclear dust in the background of
  the system} OSP of age 12.5 Gyr. Accounting for an old component
  covers the case in which one or more of the merging galaxies is
  early-type or bulge-dominated galaxy. It is worth mentioning that
  this combination is almost identical to the one used in RZ07 for the
  ULIRG PKS1345+12. However, the version of the fitting code used here
  represents an updated version of the one used in that
  paper. Therefore, in order to compare the results obtained or this
  galaxy with those of the other ULIRGs in our sample, we decided to
  model again the extracted spectra for PKS1345+12 using this
  combination and the updated version of the code. The results
  obtained here are consistent with those of RZ07. For consistency,
  the spectra of this source were also modelled using the other two
  combinations described below.

\item {\bf Combination II:} this combination consists of three
  components: an OSP of age 12.5 Gyr and zero reddening, along with a
  YSP with variable reddening (0.0 $\leq E(B - V)\leq$ 2.0, increasing
  in steps of 0.1), and a power-law ($F_{\lambda} \propto
  \lambda^{\alpha}$) with a spectral index in the range $-15 < \alpha
  < 15$. The power-law is included to represent either scattered or
  direct AGN continuum component \citep{Tadhunter02}, or a highly
  reddened VYSP.

\item {\bf Combination III:} for ULIRGs as a class, there is no
  reason, a priori, to expect a large contribution to the optical
  emission from a 12.5 Gyr OSP. Therefore, in order to explore the
  possibility of YSPs dominating the optical light from the objects in
  our sample, we used a combination of two YSP components: a IYSP with
  ages in the range 0.3 -- 2.0 Gyr, and $E(B - V)$ values of 0.0, 0.2
  or 0.4, plus a VYSP with age in the range 1 -- 100 Myr and
  reddenings (0.0 $\leq E(B - V) \leq$ 2.0, increasing in steps of
  0.1). Note that for this combination, we assume that the IYSPs have
  relatively low reddening ($E(B - V) \leq$ 0.4).

\end{itemize} 

The modelling results found for all the ULIRGs in the CS and the ES
are shown in Tables \ref{tab:CS_combI}, \ref{tab:ES_combI},
\ref{tab:CS_combII}, \ref{tab:ES_combII}, \ref{tab:CS_combIII} and
\ref{tab:ES_combIII} for Combination I, Combination II and Combination
III (hereafter Comb I, Comb II and Comb III). For the majority of the
objects in our sample, the nebular correction mentioned before did not
lead to a significant change in the modelling results for any of the
combinations, and the uncertainties in the ages, reddenings and
percentage contributions presented in the tables already account for
such minor changes. Therefore we show only results for zero nebular
correction in the tables for all objects with the exception of
PKS1345+12. For this galaxy, the nebular correction changes the
modelling results when using Comb I for Ap A and to a lesser extent Ap
B. In this case both the nebular corrected and uncorrected results are
shown in Table \ref{tab:CS_combI}. We now give a general overview of
the results obtained for Comb I, Comb II and Comb III.

\begin{figure}
\centering
\hspace{0.0cm}
\psfig{figure=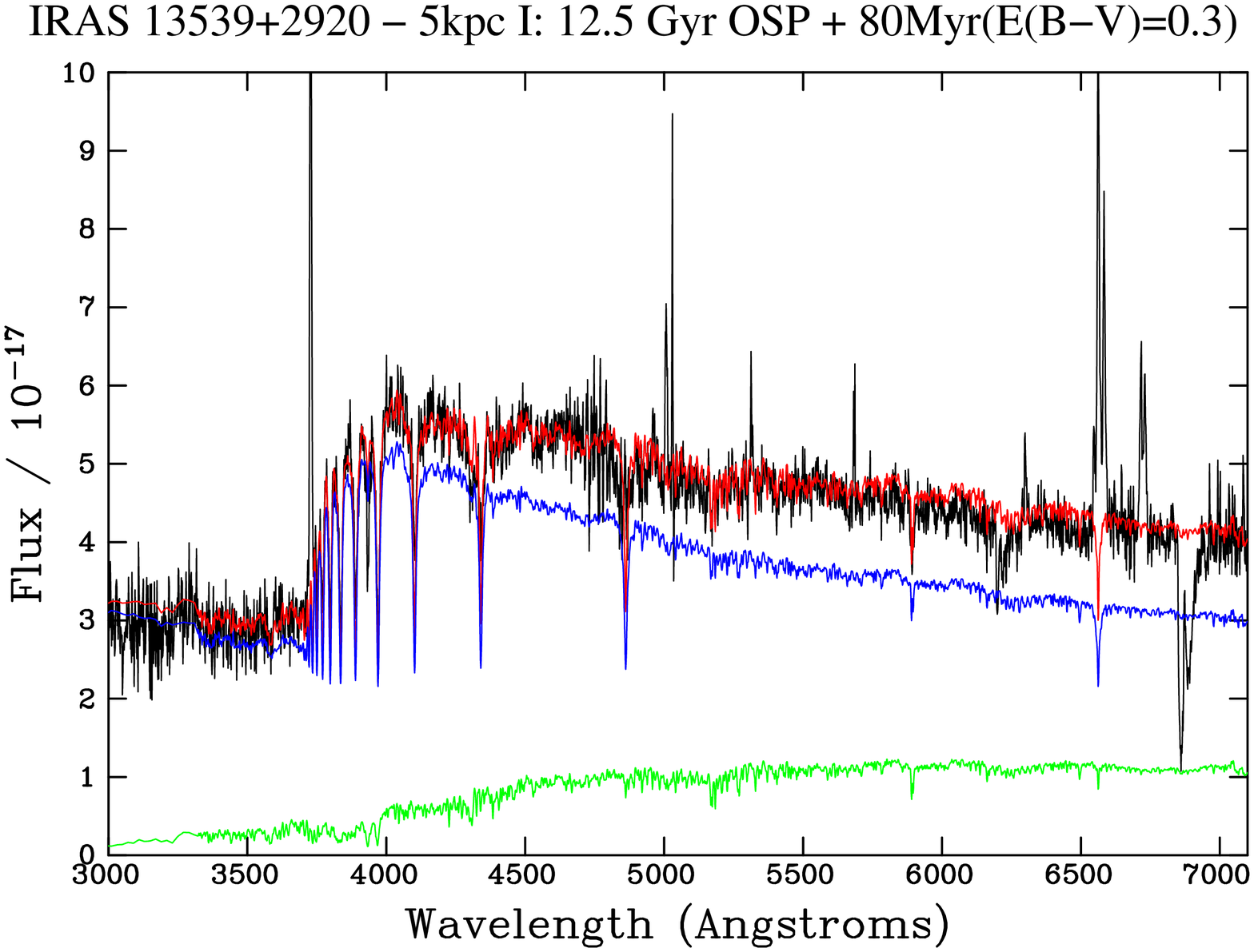,width=8.5cm,angle=0.}\\
\hspace{0.0cm}\psfig{figure=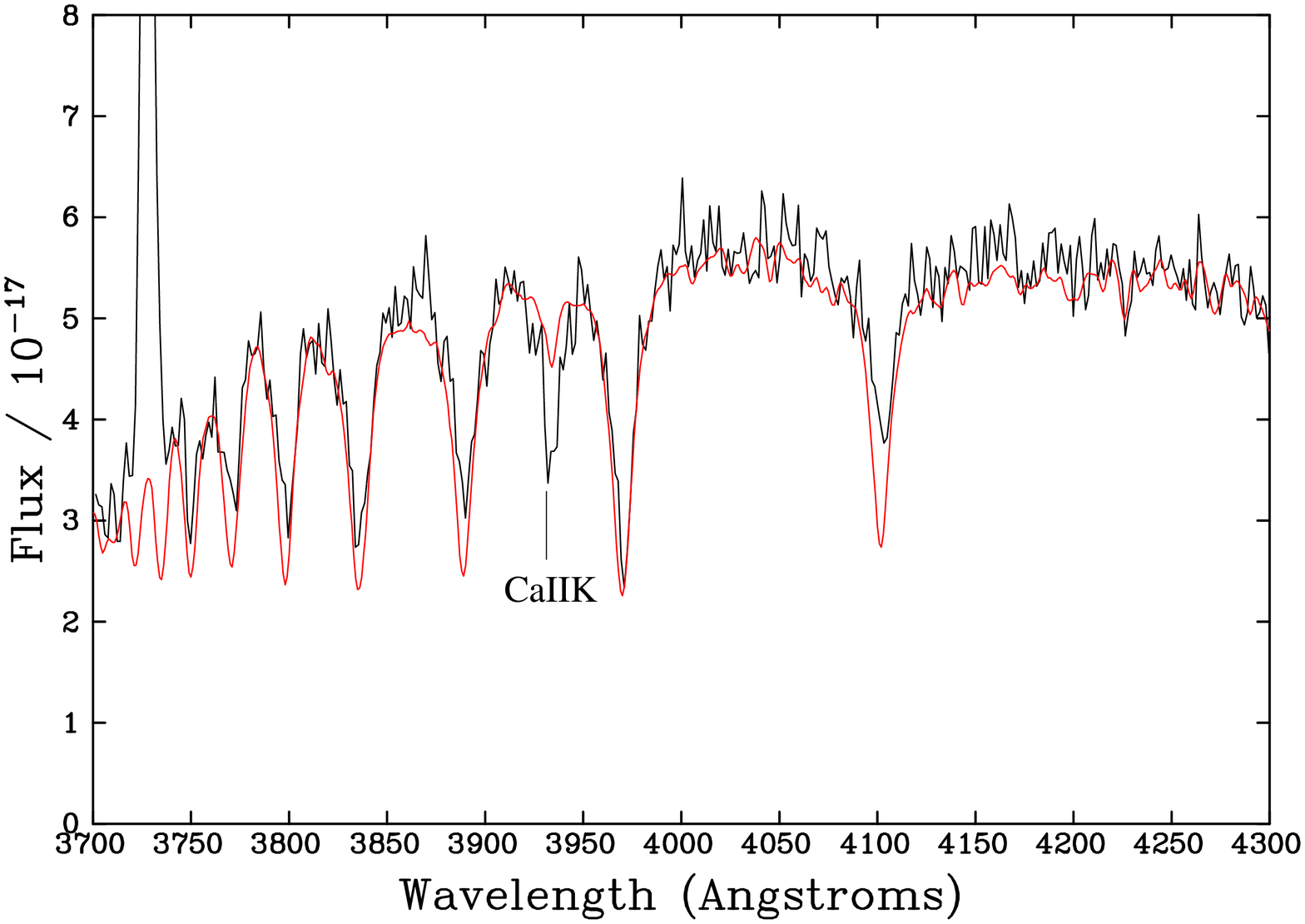,width=6.cm,angle=-90.}\\
\hspace{0.0cm}\psfig{figure=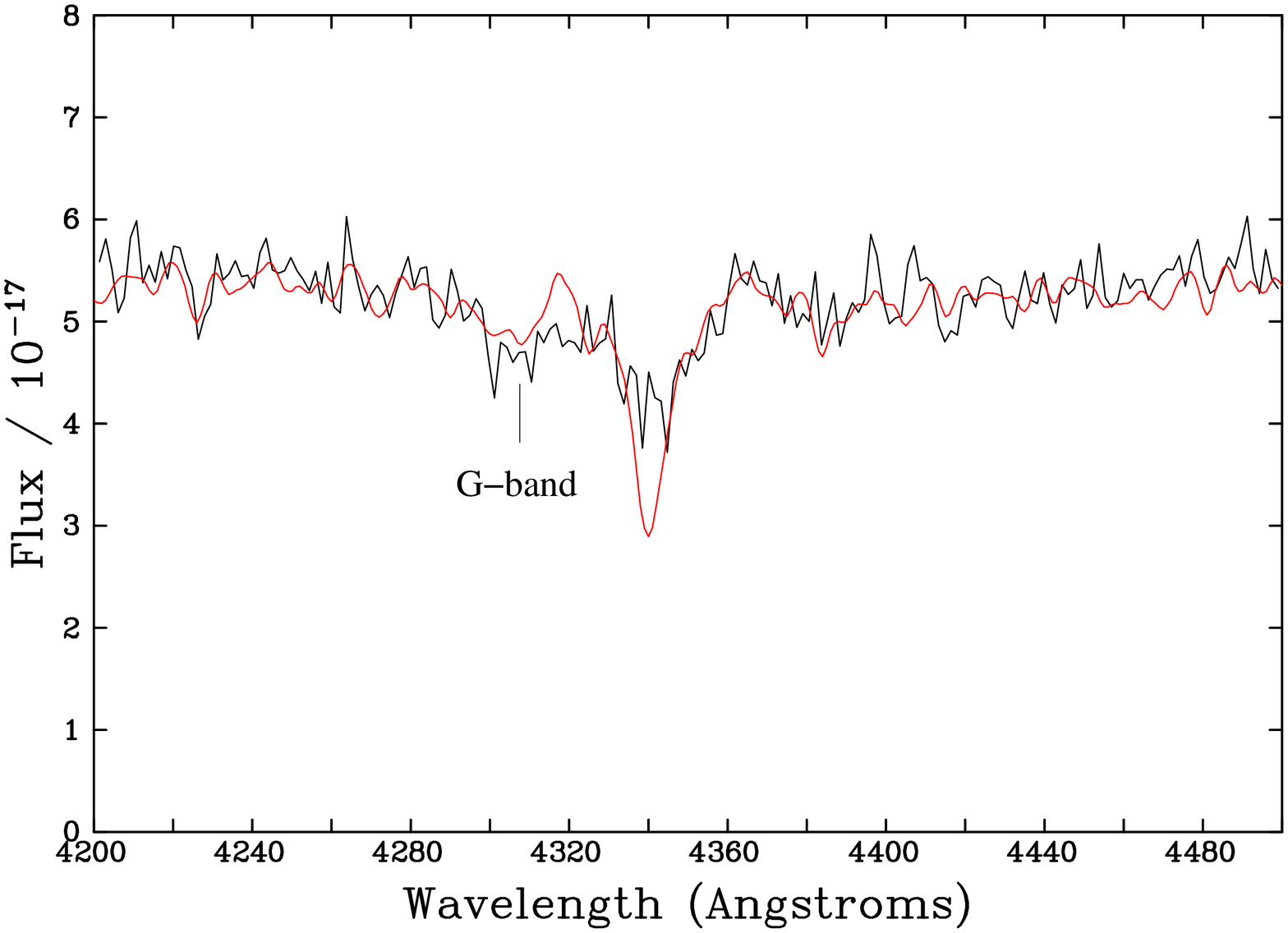,width=6.cm,angle=-90.}\\
\caption[IRAS 13539+2920: an example of the fits obtained using Comb I
for the 5 kpc-I aperture]{An example of the fits obtained using Comb I
for the south-western nucleus (aperture 5 kpc I) in the case of IRAS
13539+2920. The model shown in the figure comprises a 12.5 Gyr OSP
plus an 80 Myr VYSP ($E(B - V)$ = 0.3) which contribute 17\% and 79\%
respectively to the flux in the normalising bin (4600 --
4700~\AA). The green, blue and red spectra correspond to the OSP, the
VYSP, and the sum of the two components respectively. It is clear from
the figure that the shape of the continuum is adequately
fitted. However the fit to the high order Balmer lines and the G-band
is not entirely satisfactory (although some overprediction is expected
for the Balmer lines, since the spectrum shown in the figure has not
been corrected for nebular emission). Moreover, there is a clear
underprediction of the CaII~K absorption line.}
\label{fig:13539_combI}
\end{figure}
\begin{figure}
\centering
\hspace{0.0cm}
\psfig{figure=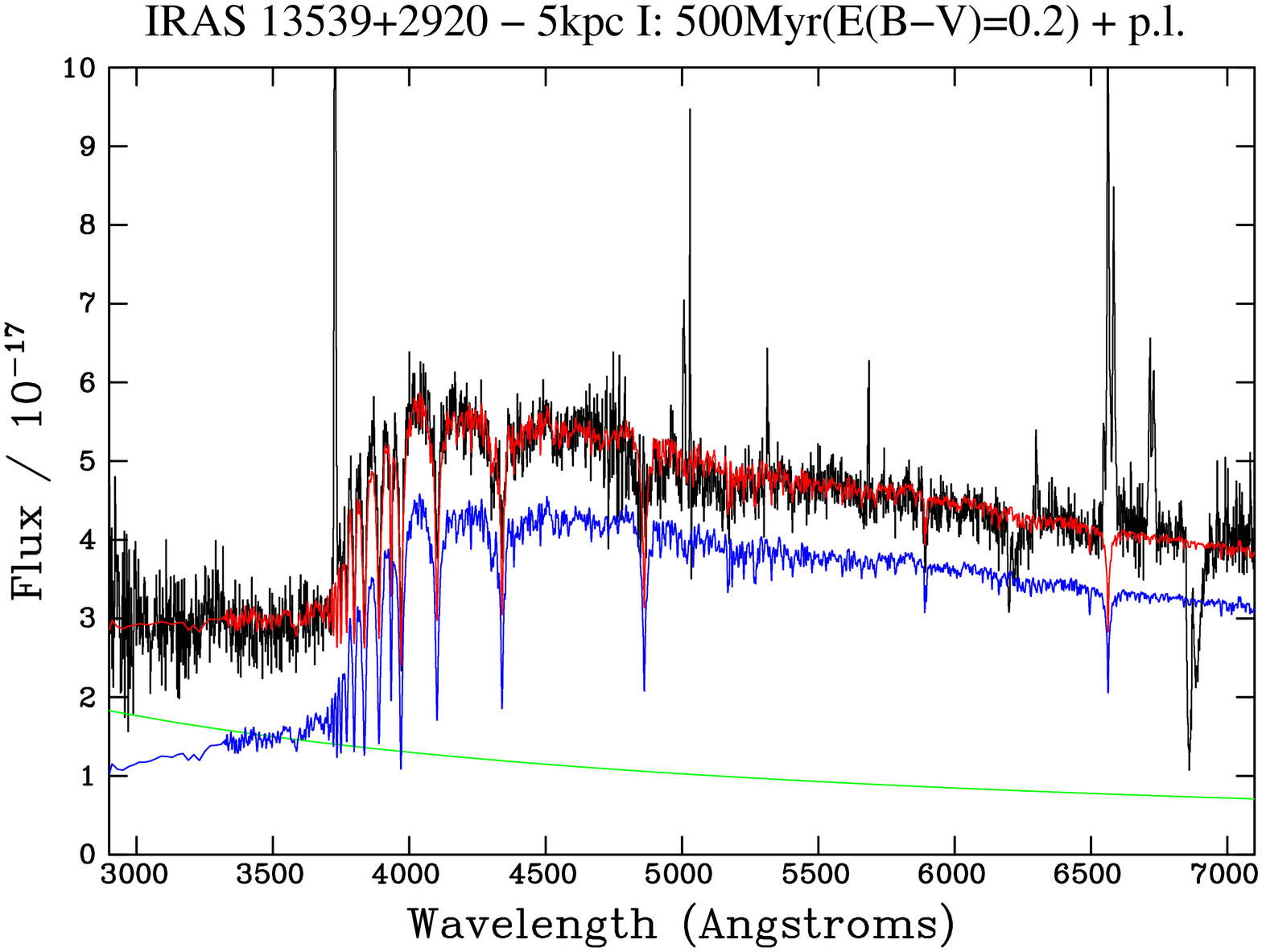,width=8.5cm,angle=0.}\\
\hspace{0.0cm}\psfig{figure=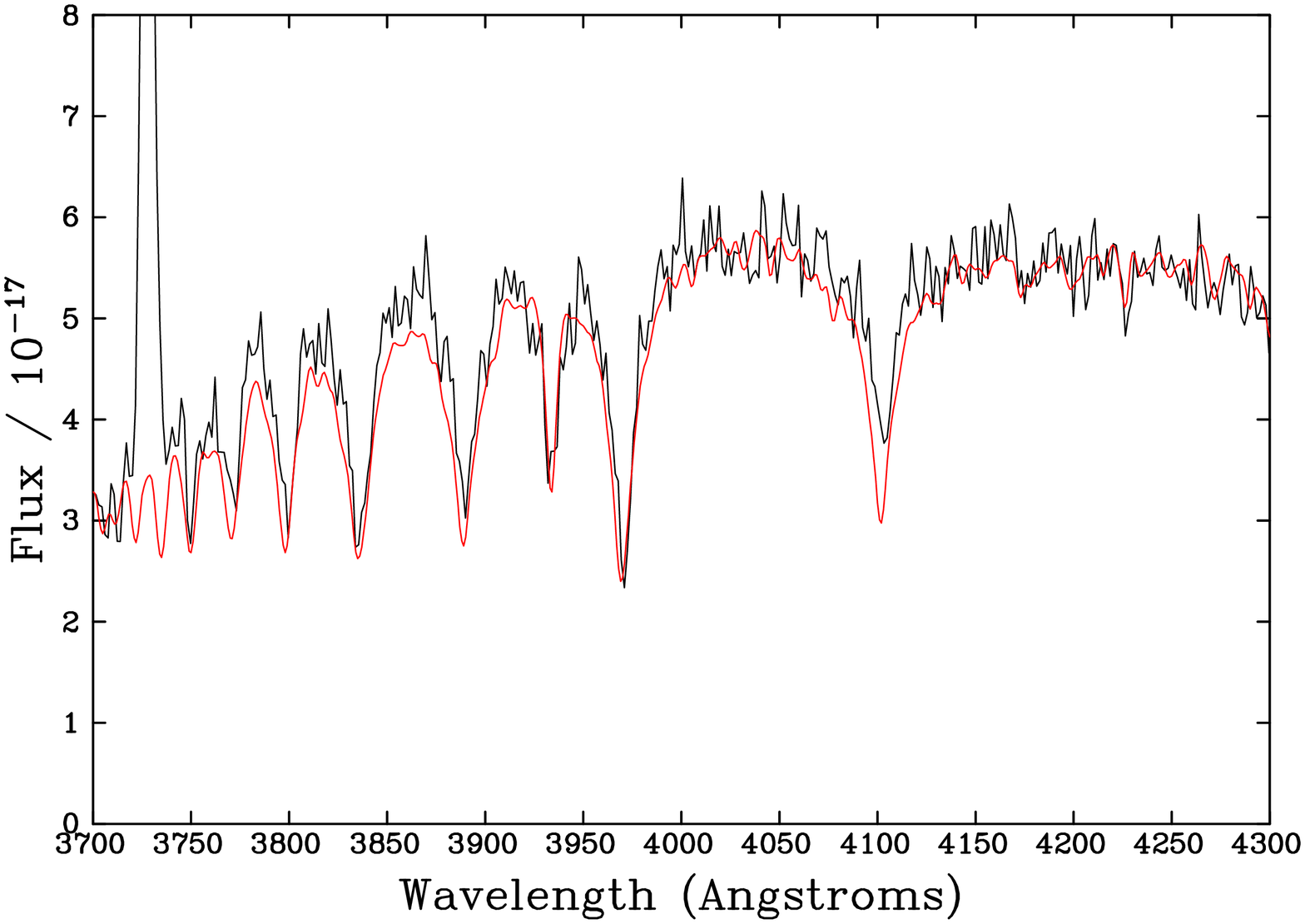,width=6.cm,angle=-90.}\\
\hspace{0.0cm}\psfig{figure=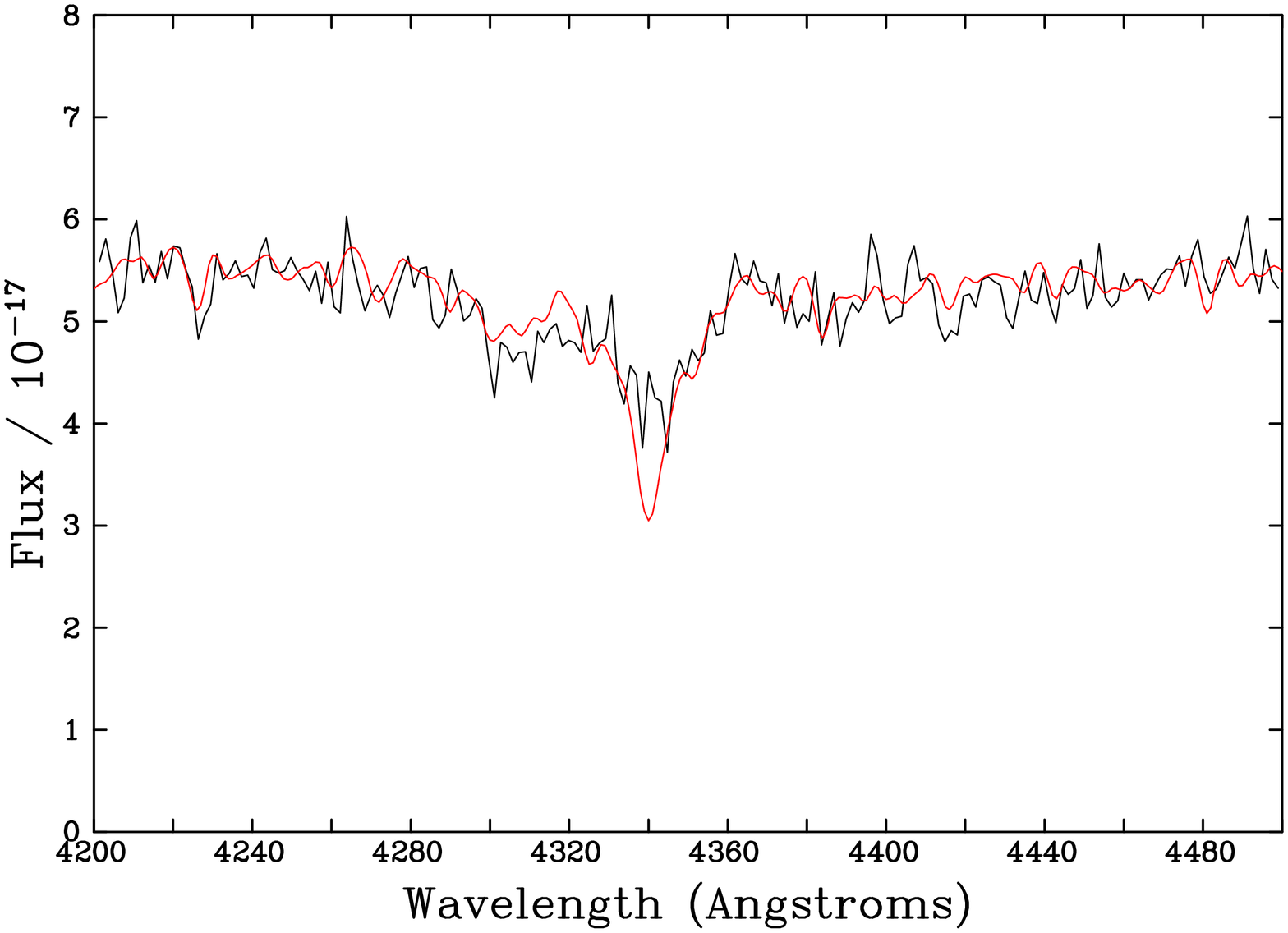,width=6.cm,angle=-90.}\\
\caption[IRAS 13539+2920: an example of the fits obtained using Comb
II for the 5 kpc-I aperture]{Same as Figure \ref{fig:13539_combI}
using Comb II. The figure shows a model comprising a 500 Myr IYSP
($E(B - V)$ = 0.2) and a power-law (power-law index $\alpha$ = -1.06)
which contribute 78\%, 21\% respectively to the flux in the
normalising bin (4600 -- 4700~\AA), with no need of a contribution
from an 12.5 Gyr OSP. The green, blue and red spectra correspond to
the power-law, the IYSP, and the sum of the two components
respectively. It is clear from the figure that the underprediction of
the CaII~K absorption line disppears when a power-law is included. A
small improvement in the fit to the G-band is also present using this
combination. }
\label{fig:13539_combII}
\end{figure}
\begin{figure}
\centering
\hspace{0.0cm}
\psfig{figure=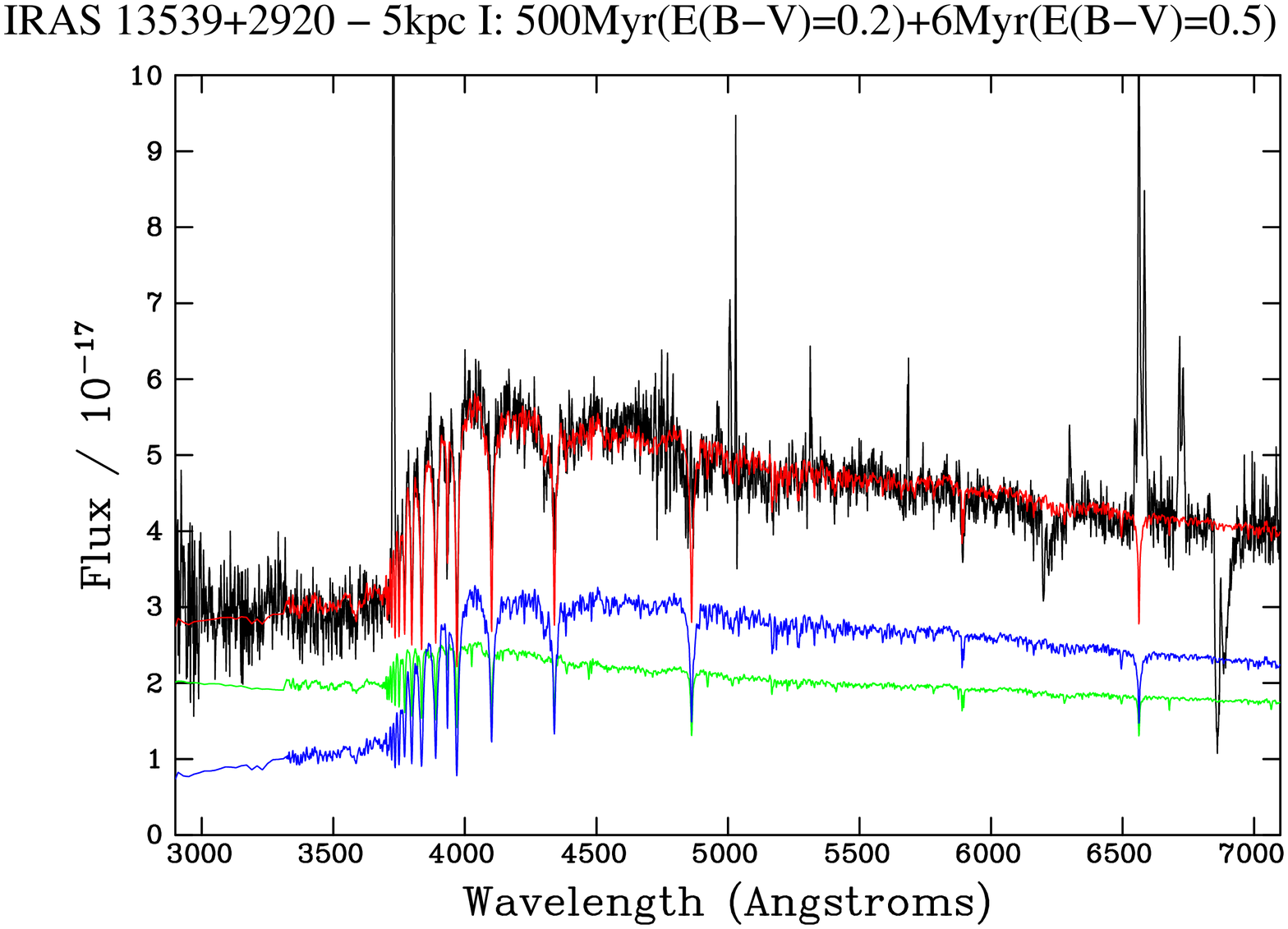,width=8.5cm,angle=0.}\\
\hspace{0.0cm}\psfig{figure=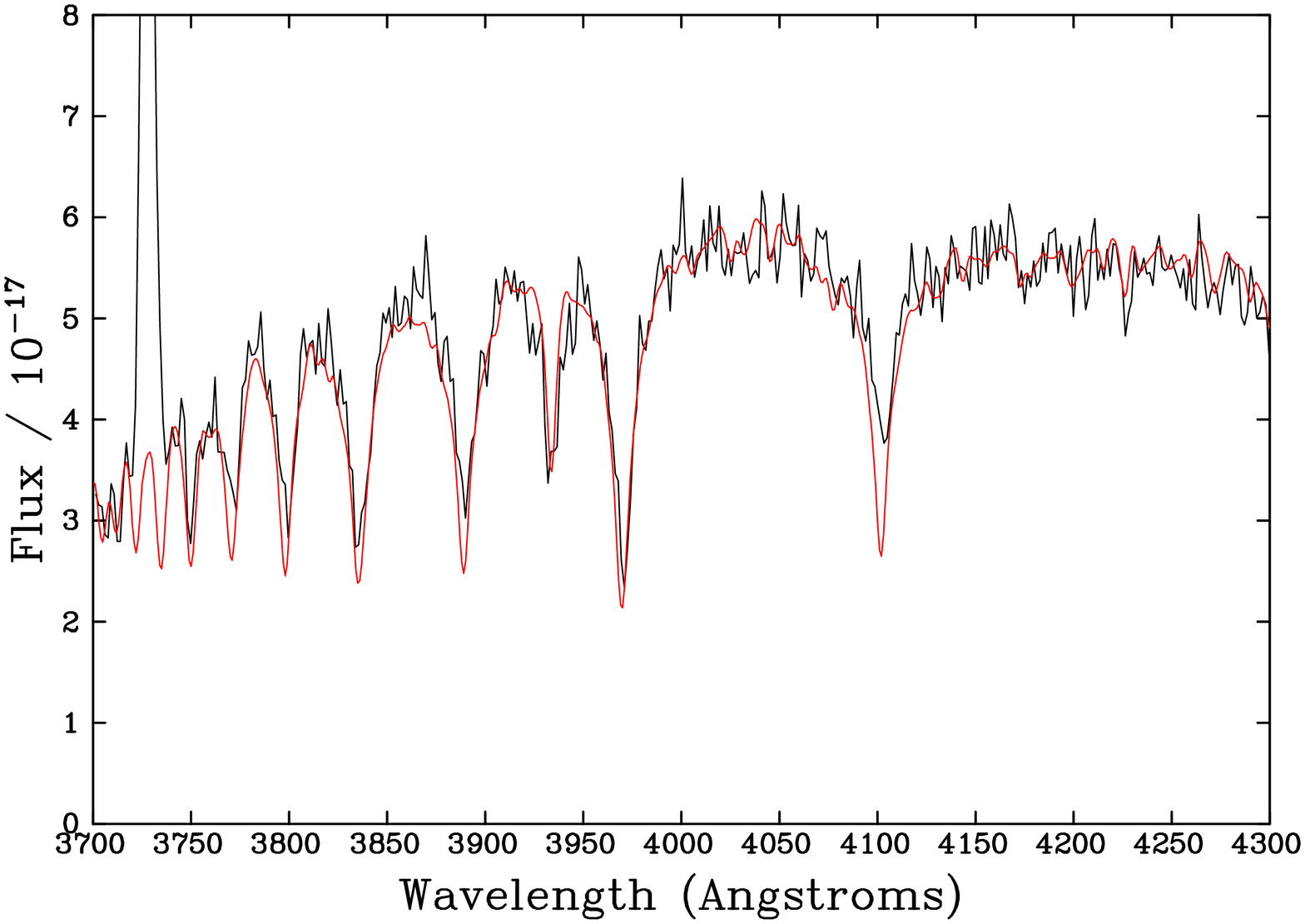,width=6.cm,angle=-90.}\\
\hspace{0.0cm}\psfig{figure=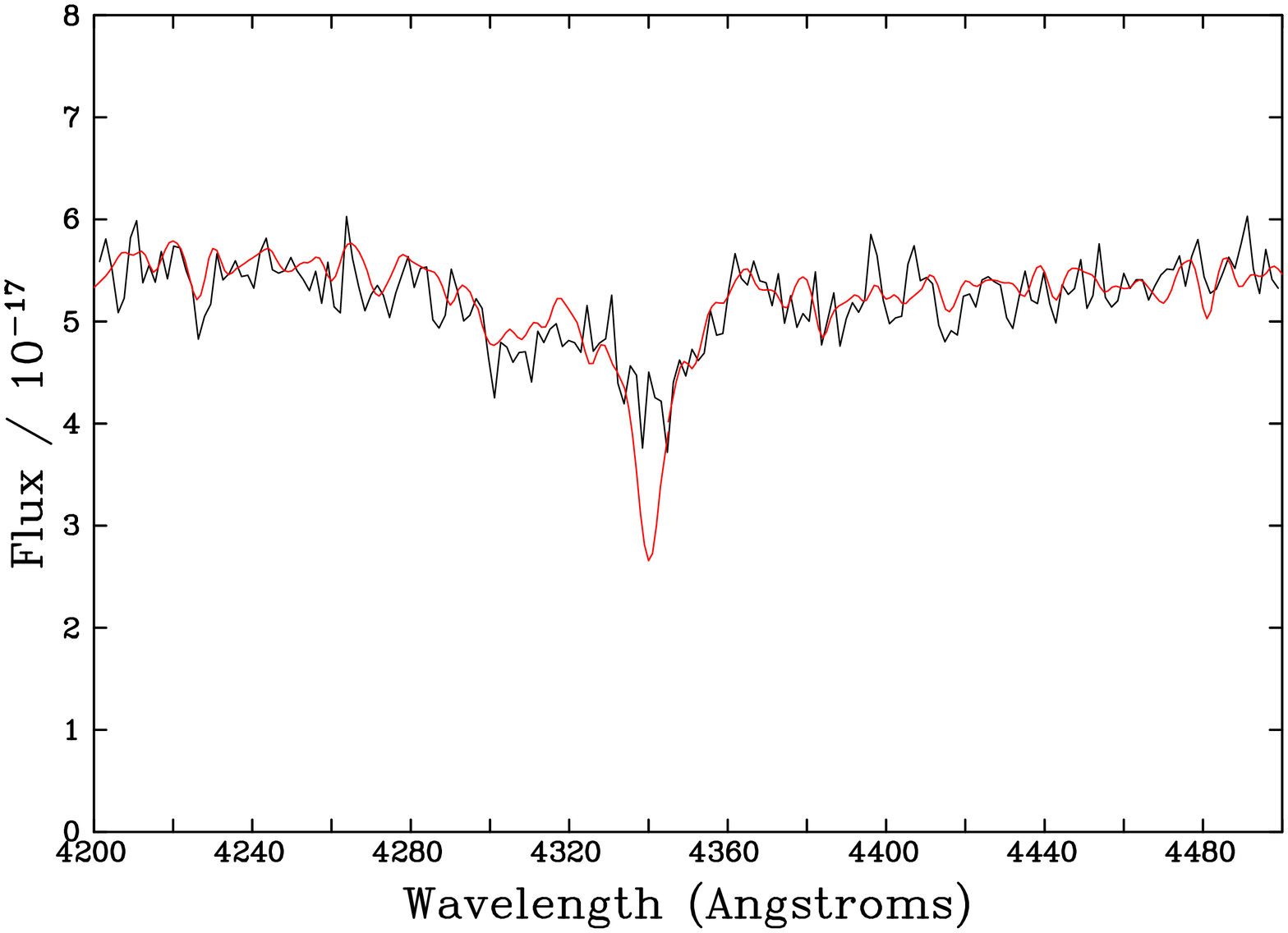,width=6.cm,angle=-90.}\\
\caption[IRAS 13539+2920: an example of the fits obtained using Comb
III for the 5 kpc-I aperture]{Same as Figure \ref{fig:13539_combI} but
using Comb III. The model shown in the figure is comprised of 500 Myr
IYSP ($E(B - V)$ = 0.2) plus a 6 Myr VYSP ($E(B - V)$ = 0.5), which
contribute 56\% and 40\% respectively to the flux in the normalising
bin (4600 -- 4700~\AA).  The green, blue and red spectra correspond to
the VYSP, the IYSP, and the sum of the two components respectively. In
the case of Comb III, both the continuum and the absorption features
(high-order Balmer lines, CaII~K and the G-band) are better fitted
using this combination than in the case of Comb I and Comb II. }
\label{fig:13539_combIII}
\end{figure}

\section{Modelling results.}
\subsection{Combination I}
In the case of Comb I, adequate fits ($\chi^{2} \leq$ 1.0) to the
overall shape of the SED were obtained for all the extraction
apertures modelled, with the exception of the 5 kpc aperture in IRAS
23060+0505, for which a power-law is required. However, although this
combination fits the overall shape of the continuum, it fails to fit
the absoption features (high-order Balmer lines, CaII~K, G-band) for
$\sim$ 42\% (54) of the extraction apertures modelled. Figure
\ref{fig:13539_combI} shows a representative example of the fits
obtained using this combination. The aperture selected for the figure
samples the south-eastern nucleus (5kpcI) of IRAS 13539+2920. The
figure shows the best fitting model to both the continuum and the
detailed absorption feartures, obtained for that aperture. It is clear
from the figure that Comb I fits the optical continuum SED adequately,
but there is an underprediction of the CaII~K absorption feature and,
also, the fit to the Balmer lines and the G-band is not entirely
satisfactory. This result suggests that the simple Comb I models may
not be adequate for modelling the detailed absorption features for a
subset of the ULIRGs in our sample. However, note that, the greatest
disagreement between the Comb I fits and the data often occurs for the
CaII~K line, which potentially has a contribution from absorption by
the ISM in the host galaxy\footnote{$\sim$ 46\% of the apertures for
which Comb I does not perfectly fit the CaII~K absorption feature show
an NaI~D excess. The NaI~D line is also potentially affected by the
ISM of the host galaxy.}. Therefore, for many of the 42\% without
adequate fits to the absorption lines, it is difficult to rule out the
idea that Comb I provides adequate fits to the stellar continuum.

Overall, even if the Comb I fits are not always perfect, they give
single-valued estimates of the luminosity-weighted age, reddening and
percentage contribution of the YSPs within each aperture. Therefore
the results of Comb I -- presented in Table \ref{tab:CS_combI} and
\ref{tab:ES_combI} -- are still useful when performing a
statistical analysis of the results.

\subsection{Combination II}

For those ULIRGs in our sample also classified as Sy2 galaxies, it is
possible that the optical spectrum is a composite of stellar and AGN
continuum emission \cite[e.g. scattered light or weak direct AGN
light: see][]{Tadhunter02}. In this context, we note that significant
starburst activity has already been detected in the nuclear regions in
Sy2 galaxies. \cite{Gonzalez01} found high-order Balmer lines and
He~{\small I} absorption lines, strong indicators of the presence of
YSPs, in their optical spectroscopic study of a sample of 20 Sy2
galaxies. Most of the objects in the \cite{Gonzalez01} sample are
classified as LIRGs, and one, Mrk 273, is classified as a ULIRG and
included in the work presented here. Therefore, it is straightforward
to understand that Comb II, including a power law, is particularly
useful for those ULIRGs in our sample also classified as Sy2 galaxies
at optical wavelegths.

Since the power law also represents a highly reddened VYSP, comb II
gives an idea of the contribution level of the OSPs in each
galaxy. Therefore, we decided to use it for the whole sample of
ULIRGs. Adequate fits ($\chi^{2} \leq$ 1.0) to the overall SED shapes
were found for all of the apertures of all objects using this
combination. On the other hand, this combination fails to fit the
detailed absorption features for 10\% (13 apertures) of the apertures
modelled. Figure \ref{fig:13539_combII} shows an example of a fit
obtained for IRAS 13539+2920 using Comb II for the same extraction
aperture shown in Figure \ref{fig:13539_combI}. It is clear from the
figure that the underprediction of the CaII~K absorption line
disappears when a power-law is included. Finally, as shown in Tables
\ref{tab:CS_combII} and \ref{tab:ES_combII} the minimum precentage
contribution of the OSP component is less that 10\% (due to the
uncertainties inherent in the modelling techniques percentage
contributions $<$ 10\% are considered negligible) for most of the
apertures. This result suggest that an OSP component is not required
to fit the data in the majority of the objects in the ES.

\subsection{Combination III}

A drawback of Comb II is that, for those cases in which the power-law
is likely to represent a VYSP component, it gives no information about
the detailed properties (age and reddening) of such a
component. Therefore, in order to better investigate the detailed
properties of the VYSPs in the galaxies, we used Comb III. Adequate fits to
both the continuum and the detailed absorption features are obtained
for all but 9 ($\sim$ 7\%) of the extraction apertures modelled. In
the minority of the cases for which no adequate fits could be obtained
using Comb III, either an OSP or a power-law component was required
in addition to a YSP. Figure \ref{fig:13539_combIII} shows an example
of a fit obtained using Comb III for the same extraction aperture
shown in Figures \ref{fig:13539_combI} and \ref{fig:13539_combII} in
the case of IRAS 13539+2920. The figure clearly demonstrates that Comb
III provides the best fit to all of the detailed absorption features.

At this stage, it is worth re-iterating that there is no reason, a
priori, to expect a major contribution to the optical emission from a
12.5 Gyr OSP in ULIRGs. The lack of an OSP is also suggested by the
results of Combination II and, to a lesser extent, those of
Combination I. Therefore, in the following, we will concentrate on the
modelling results for Comb III and use those for Comb I and II as
additional information, unless otherwise specified.

\section{Estimating masses and bolometric luminosities}

\begin{table*}
%\hspace*{2.0cm}
\centering
{\small
%\centering
\begin{tabular}{@{}lcccccc@{}}
\hline
Object Name& Mass  & Mass & Mass & L$_{\rm bol}$ & L{\tiny IR} & \% of L{\tiny IR}\\    
IRAS       & VYSP  & IYSP & OSP  & & &\\ 
           & 10$^{9}$ M$_{\odot}$ & 10$^{10}$ M$_{\odot}$ & 10$^{11}$ M$_{\odot}$  & 10$^{12}$ L$_{\odot}$ & 10$^{12}$ L$_{\odot}$ & \\
(1)        & (2)                   &  (3)                  &  (4)                  &  (5) & (6) & (7)\\    
\hline 
00091--0738 & 1.8 & 4.4 & - & 0.56 & 1.69 &33\\
00188--0856 & 7.8 & 9.3 & - & 0.24 & 2.34 &10\\
01004--2237 & 0.4 & 1.5 & - & 0.07 & 1.90 &4\\
08572+3915  & 3.9 & 2.4 & - & 0.11 & 1.41 &8\\
10190+1322  & 3.0 & 3.6 & - & 0.83 & 1.10 &75\\
10494+4424  & 31.0 & 2.1 & - & 0.48 & 1.48&32\\ 
12072--0444 & 17.4& 1.9 & - & 0.18 & 2.45 &7\\
12112+0305  & 0.4 & 7.2 & - & 0.21 & 2.10 &10\\
%12540+5708$^a$&  -  &  -  & - &  -  & 3.4 &\\
13305--1739 & 4.9 & 3.4 & - & 1.80 & 1.78 &101\\
13428+5608  & 4.3 & 7.0 & - & 0.58 & 1.08 &54\\
13451+1232  & 4.5 & 50.0& - & 2.10 & 2.09 &100\\
13539+2920  & 7.4 & 2.7 & - & 0.88 & 1.10 &80\\
14060+2919  & 2.1 & 5.4 & - & 0.53 & 1.17 &45\\
14252--1550 & 46.0& 0.7 & - & 0.96 & 1.55 &62\\
14348--1447 & 13.0& 9.5 & - & 0.36 & 2.09 &17\\
14394+5332E & 10.0& 1.2 & - & 0.80 & 1.20 &67\\
14394+5332W & 18.0& 2.2 & - & 0.53 & -    & -\\
15130--1958 & 0.5 & 4.0 & - & 0.20 & 1.23 &16\\
15206+3342  & 2.2 & 8.2 & - & 1.69 & 1.66 &102\\
15327+2340  & 0.5 & 3.5 & - & 0.16 & 1.62 &10\\
%15462-0450$^a$  &  -  &   - & - & -  & 1.62\\
16156+0146  & 1.0 & 3.1 & - & 0.27 & 1.20 &23\\
16474+3430  & 1.1 & 8.4 & - & 0.33 & 1.41 &23\\
16487+5447  & 3.3 & 9.0 & - & 0.17 & 1.44 &12\\
17028+5817E & 1.3 & 0.5 & - & 0.22 & -    &\\
17028+5817W & 41.4& 0.9 & - & 0.65 & 1.38 &47\\
17044+6720  & 3.4 & 2.4 & - & 0.77 & 1.48 &52\\
17179+5444  & 7.5 & 16.6& - & 2.0 & 1.74  &114\\
20414-1651  & 0.8 & 3.4 & - & 0.32 & 1.51 &21\\
21208--0519 & 3.8 & 9.2 & - & 0.77 & 1.12  &69\\
%21219--1757$^a$ &  -  & -	& - & - & 1.25\\
22491--1808 & 2.7 & 2.5 & - & 0.8 & 1.35   &59\\
23060+0505  & 19.0& -	& 0.5 & 2.2 & 3.01 &73\\
23233+2817  & 10.0& -	& 2.2  & 0.16 & 1.1&15\\
23234+0946  & 5.2 & 4.3 & - & 0.65 & 1.23  &53\\
23327+2913N & -   & -	& 3.5 & - & -      &\\
23327+2913S & 1.8 & -	& 1.2 & 0.23 & 1.25&18\\
23389+0303  & 2.1 & 1.9 &  -  & 0.49 & 1.35&36\\
\end{tabular}}
\caption[Estimated masses and bolometric luminosities associated with
the stellar components detected at optical wavelengths]{Estimated
total masses and bolometric luminosities associated with the stellar
components detected at optical wavelengths. Col (1): object name. Col
(2): estimated mass associated with the VYSP component. Col (3):
estimated masses associated with the IYSP component. Col (4):
estimated masses associated with the OSP component (if present). Col
(5): estimated bolometric luminosities associated with the stellar
components. Col(6): mid- to far IR luminosity of the sources from
\cite{Kim98a}. Col (7): estimated bolometric luminosities of the YSPs
detected in the optical expressed as percentages of the mid- to far-IR
luminosities.\newline Note that since it was not possible to model the
stellar populations in the nuclear regions of the three ULIRGs
classified as Sy1 galaxies (IRAS 12540+5708,IRA 15462--0450 and IRAS
21219--1757), no results for these objects are presented in the table
(see Apendix A for details).}
\label{tab:Mass_Lbol}
\end{table*}

It is also possible to estimate the masses and bolometric luminosities
associated with the stellar populations detected at optical
wavelengths using the modelling results. For this purpose, we
extracted and modelled a large aperture including all the structure
visible in the 2-D frames (and in the spatial cut in Figure
\ref{fig:spatial_cuts}) for each object. Assuming that the stellar
population mix remains constant across the entire visible extents of
the galaxies, we chose a model from all the adequate fits obtained,
selected to be as representative as possible of all the stellar
populations in the galaxy. Once the model was selected, we estimated
the mass contribution of each of the stellar components in the large
aperture. Since the slit does not cover the full extent of each
galaxy, we then used the \cite{Kim02} total magnitudes of the galaxies
in the R-band to scale the results and estimate the masses in the
various stellar components. However, due to the diversity of the
stellar populations found, it is sometimes hard to select a particular
representative model for some ULIRGs. In such cases, the selection is
biased towards models consistent with the results obtained for the
nuclear region, since it is likely that most of the mid- to far-IR
luminostity arises from these regions. The estimated total masses of
the YSPs (VYSPs plus IYSPs, results presented in Table 10) fall in the
range 0.18$\times 10^{10} \leq M_{\rm YSP}\leq$ 50 $\times10^{10}$
M$_{\odot}$. In the cases where both VYSPs and IYSPs are present, the
IYSP is, in most cases, more massive than the VYSP (the exceptions are
IRAS 10494+4424, IRAS 14252-15550 and IRAS 17028+5817W)

To estimate the bolometric luminosities associated with the stellar
populations detected in the optical, we assumed the same combination
of stellar populations as used for the mass measurements, scaling the
unreddened stellar templates using the mass values previously
obtained. We then integrated the scaled synthetic spectra over their
entire wavelength range. The results are presented in Table
\ref{tab:Mass_Lbol}. We find stellar bolometric luminosities in the
range of 0.07 $\times$ 10$^{12}$ $<$ L$_{\rm bol} <$ 2.2 $\times$
10$^{12}$ L$_{\odot}$. Assuming that most of the optical light is
absorbed and then re-processed by dust, we can compare the estimated
bolometric luminosities with the mid- to far-IR luminosities of the
sources, presented in Table \ref{comp_sample}. Column 7 in the table
shows the estimated bolometric luminosities of the YSPs detected in the
optical expressed as percentages of the measured mid- to far-IR luminosities.

Finally, we would like to emphasize that the results in Table
\ref{tab:Mass_Lbol} are indicative rather than exact, and they must be
interpreted carefully due to the various uncertainties attached to the
process. Above all, the major source of uncertainity arises from the
selection of the model; different combinations of stellar populations
may change the results significantly.

\section{Summary of general patterns and results}

In order to give an overview of the properties of the stellar
populations of the ULIRGs in our sample as a whole, the general results
and patterns obtained from the modelling can be summarized as follows.

\begin{itemize}

\item{\bf Modelling technique}: three different combinations of
  stellar populations have been used for the analysis presented here:
  {\bf combination I} (OSP + YSP), {\bf combination II} (OSP + YSP +
  p.l.) and {\bf combination III} (VYSP + IYSP). In general, adequate
  fits to the overall shapes of the continua in individual apertures
  are obtained using all of the combinations. However, Comb I fails to
  fit the detailed absorption features for 42\% (54) of the extraction
  apertures. This percentage is reduced down to 10\% (13 apertures)
  and 7\% (9 apertures) in the cases of Comb II and Comb III,
  respectively. Note that adequate fits for both the continuum and the
  detailed absorption features are found for 60\% of the extraction
  apertures using {\it all} of the combinations. This clearly
  demonstrates the difficulty in obtaining a unique solution when
  using combinations of different stellar populations to model the
  optical spectra of starburst systems.

\item{\bf Young stellar populations}: young stellar populations (YSPs,
  t$_{\rm YSP} \leq$ 2.0 Gyr) are present in {\it all} apertures in
  {\it all} sources, with the exception of IRAS 23327+2913
  5kpc-II. Furthermore, very young stellar populations (VYSPs, t$_{\rm
  VYSP}\leq$ 100 Myr) are essential to obtain adequate fits in {\it
  all but} 20 ($\sim$ 15\%) of the extraction apertures, 14 of which
  sample the extended regions of Arp220 and the radio galaxy IRAS
  13451+1232 (PKS1345+12).

  At this stage we would like to add a caveat about the properties of
  the so-called intermediate-age young stellar populations (IYSPs, 0.1
  $<$ t$_{\rm IYSP} \leq$ 2.0 Gyr). For this component, we have assumed
  a limited range of ages and, in particular, reddening values, in
  comparison with those of the VYSPs. Therefore, the ranges of age and
  reddening presented in Tables \ref{tab:CS_combIII} and
  \ref{tab:ES_combIII} for the IYSPs are partially a consequence of
  these assumptions.
 
\item{\bf Old stellar populations}: old stellar populations (OSPs,
  12.5 Gyr) do not make a major contribution to the optical light in
  most apertures for most sources. The modelling results show that
  OSPs are essential for fitting the optical continuum in only 7
  ($\sim$ 5\%) of all the apertures considered; in the majority of the
  cases, adequate fits can be obtained using combinations of IYSPs and
  VYSPs. Only in the cases of IRAS 23327+2913 (5kpc-II) and IRAS
  21208-0519 (5kpc-II) we find that the OSPs dominate. In addition, in
  the cases of IRAS 13451+1232, IRAS 14394+5332 5kpc-II, IRAS
  16487+5447E and IRAS 23234+0946 5kpc-II either OSPs, or dominant
  ``old'' IYSPs (1 - 2 Gyr), are required to model the data.

\item{\bf Reddening}: the study presented here further emphasizes the
importance of carefully accounting for dust reddening effects when
modelling the stellar populations of star-forming galaxies.

\item{\bf Double nucleus systems}: 20 of the 36 ULIRGs in our sample 
are classified as double nucleus systems and one, IRAS 14394+5322, is
classified as a multiple ($>$ 2 nucleus) system. For 14 of these
systems, it was possible to study the stellar populations within the
different nuclei individually. In 71\% of the cases (10 of the 14
objects) the results are consistent with the stellar populations being
similar in the two nuclei in terms of ages, reddenings and percentage
contributions. However, in four cases (IRAS 14394+5332, IRAS
16487+5447, IRAS 21208-0519 and IRAS 23327+2913) we find marked
differences between the mix of ages of the stellar populations in the
two nuclei.

\item{\bf Extended and nuclear apertures}: in the majority of the
objects we find that we require the same stellar components (IYSP and
VYSP) at all locations of the systems to model the spectra
adequatelly.

\item{\bf Masses of the YSP}: the estimated total masses of the YSPs
(VYSPs plus IYSPs) fall in the range 0.18$\times 10^{10} \leq M_{\rm
YSP}\leq$ 50 $\times10^{10}$ M$_{\odot}$. In the cases where both
VYSPs and IYSPs are present, the IYSP is, in most cases, more massive
than the VYSP (the exceptions are IRAS 10494+4424, IRAS 14252-15550
and IRAS 17028+5817W).
 
\item{\bf Bolometric luminosities}: we have also estimated the
bolometric luminosities associated with these stellar populations. For
48\% of the objects in the ES sample (16 of the 33 objects, excluding the
Sy1 galaxies), the bolometric luminosities of the YSPs detected in the
optical can account for a large fraction ($\gsim$ 50\%) of the mid- to
far-IR luminosity of the source.

More detailed information about individual sources is presented in
Appendix A.
%Given that the YSP in most such cases are
%highly reddened, the detected stellar populations may provide a
%significant fraction of the luminosity required to heat the mid-to-far
%IR emitting dust. Therefore, the main source for the mid- to far-IR is
%not always hidden in ULIRGs.
\end{itemize}

\section{Conclusions}

This is the first of two papers presenting a detailed study of the
stellar populations in a large sample of ULIRGs, based on long-slit
optical spectroscopic observations. In this paper we have introduced
the sample and described the analysis techniques and the
general results obtained from the modelling. In addition, we present
estimated values for the total masses of the stellar populations
within the objects and the bolometric luminosities associated with
them. 

Overall, the results for the sample in general are consistent with
those obtained in the detailed studies of RZ07 and RZ08 for PKS1345+12
and Arp 220 respectively, in terms of:

\begin{itemize}
\item
the requirement for a mixture of VYSPs and IYSPs, with the
proportional contribution of the VYSP component increasing towards the
nuclei in most cases; and 
\item
the importance of taking into account the reddening of the YSP
component, with clear evidence of increasing reddening towards the
nuclei in a number of the objects.
\end{itemize}

The work presented here clearly shows the utility of spectral
synthesis modelling techniques for studying the properties of the
stellar populations in star forming galaxies. In addition, it also
shows the difficulty in obtaining a unique solution when using
combinations of different stellar populations to model spectra,
especially on the basis of fits to the SED alone (i.e. not examining
the detailed absorption features). We have shown that some results,
and therefore the conclusions based on them, can change significantly
if different models are used.

The results described in this paper will be analyzed in more detail
in a forthcoming paper (Rodr\'iguez Zaur\'in et al. 2009b), where
they will be discussed in the context of evolutionary models for
merging systems, and the results obtained for similar systems in the
high redshift universe.

\section*{Acknowledgments} 
We thank the referee for useful comments that have helped to improve
the manuscript. JRZ acknowledges financial support from the STFC in
the form of a PhD studentship. JRZ also acknowledges financial support
from the spanish grant ESP2007-65475-C02-01. RGD is supported by the
Spanish Ministerio de Educaci\'on y Ciencia under grant AYA
2007-64712. We also thank support for a joint CSIC-Royal Astronomy
Society bilateral collaboration grant. The William Herschel Telescope
is operated on the island of La Palma by the Isaac Newton Group in the
Spanish Observatorio del Roque de los Muchachos of the Instituto de
Astrofisica de Canarias.

\bibliographystyle{mn2e} 
\bibliography{Thesis_paper_I_ref_reply}

\begin{thebibliography}{}

\bibitem[\protect\citeauthoryear{{Armus}, {Charmandaris}, {Bernard-Salas},
  {Spoon} H.W.W.and~{Marshall}, {Higdon}, {Desai}, {Teplitz}, {Hao}, {Devost},
  {Brandl}, {Wu}, {Sloan}, {Soifer}, {Houck} \& {Herter}}{{Armus}
  et~al.}{2007}]{Armus07}
{Armus} L.,  {Charmandaris} V.,  {Bernard-Salas} J.,  {Spoon}
  H.W.W.and~{Marshall} J.,  {Higdon} S.,  {Desai} V.,  {Teplitz} H.,  {Hao} L.,
   {Devost} D.,  {Brandl} B.,  {Wu} Y.,  {Sloan} G.,  {Soifer} B.,  {Houck} J.,
     {Herter} T.,  2007, ApJ, 656, 148

\bibitem[\protect\citeauthoryear{{Armus}, {Heckman} \& {Miley}}{{Armus}
  et~al.}{1990}]{Armus90}
{Armus} L.,  {Heckman} T.,    {Miley} G.,  1990, ApJ, 364, 471

\bibitem[\protect\citeauthoryear{{Arribas} \& {Colina}}{{Arribas} \&
  {Colina}}{2002}]{Arribas02}
{Arribas} S.,  {Colina} L.,  2002, ApJ, 573, 576

\bibitem[\protect\citeauthoryear{{Arribas}, {Colina} \& {Borne}}{{Arribas}
  et~al.}{2000}]{Arribas00}
{Arribas} S.,  {Colina} L.,    {Borne} K.,  2000, ApJ, 542, 228

\bibitem[\protect\citeauthoryear{{Bruzual} \& {Charlot}}{{Bruzual} \&
  {Charlot}}{2003}]{Bruzual03}
{Bruzual} G.,  {Charlot} S.,  2003, MNRAS, 344, 1000

\bibitem[\protect\citeauthoryear{{Calzetti}, {Armus}, {Bohlin}, {Kinney},
  {Koornneef} \& {Storchi-Bergman}}{{Calzetti} et~al.}{2000}]{Calzetti00}
{Calzetti} D.,  {Armus} L.,  {Bohlin} R.,  {Kinney} A.,  {Koornneef} J.,
  {Storchi-Bergman} T.,  2000, ApJ, 533, 682

\bibitem[\protect\citeauthoryear{{Canalizo} \& {Stockton}}{{Canalizo} \&
  {Stockton}}{2000a}]{Canalizo00a}
{Canalizo} G.,  {Stockton} A.,  2000a, AJ, 528, 201

\bibitem[\protect\citeauthoryear{{Canalizo} \& {Stockton}}{{Canalizo} \&
  {Stockton}}{2000b}]{Canalizo00b}
{Canalizo} G.,  {Stockton} A.,  2000b, AJ, 120, 1750

\bibitem[\protect\citeauthoryear{{Canalizo} \& {Stockton}}{{Canalizo} \&
  {Stockton}}{2001}]{Canalizo01}
{Canalizo} G.,  {Stockton} A.,  2001, ApJ, 555, 719

\bibitem[\protect\citeauthoryear{{Cole}, {Pedlar}, {Holloway} \&
  {Mundell}}{{Cole} et~al.}{1999}]{Cole99}
{Cole} G.,  {Pedlar} A.,  {Holloway} A.,    {Mundell} C.,  1999, MNRAS, 310,
  1033

\bibitem[\protect\citeauthoryear{{Condon}, {Huang}, {Yin} \& {Thuan}}{{Condon}
  et~al.}{1991}]{Condon91}
{Condon} J.,  {Huang} Z.-P.,  {Yin} Q.,    {Thuan} T.,  1991, ApJ, 378, 65

\bibitem[\protect\citeauthoryear{{Dasyra}, {Tacconi}, {Davies}, {Naab},
  {Genzel}, {Lutz}, {Sturm}, {Baker}, {Veilleux}, {Sanders} \&
  {Burkert}}{{Dasyra} et~al.}{2006a}]{Dasyra06a}
{Dasyra} K.,  {Tacconi} L.,  {Davies} R.,  {Naab} T.,  {Genzel} R.,  {Lutz} D.,
   {Sturm} E.,  {Baker} A.,  {Veilleux} S.,  {Sanders} D.,    {Burkert} A.,
  2006a, ApJ, 638, 745

\bibitem[\protect\citeauthoryear{{Dasyra}, {Tacconi}, {Davies}, {Naab},
  {Genzel}, {Lutz}, {Sturm}, {Baker}, {Veilleux}, {Sanders} \&
  {Burkert}}{{Dasyra} et~al.}{2006b}]{Dasyra06b}
{Dasyra} K.,  {Tacconi} L.,  {Davies} R.,  {Naab} T.,  {Genzel} R.,  {Lutz} D.,
   {Sturm} E.,  {Baker} A.,  {Veilleux} S.,  {Sanders} D.,    {Burkert} A.,
  2006b, ApJ, 651, 835

\bibitem[\protect\citeauthoryear{{Davies}, {Tacconi} \& {Genzel}}{{Davies}
  et~al.}{2004}]{Davies04}
{Davies} R.,  {Tacconi} L.,    {Genzel} R.,  2004, ApJ, 613, 781

\bibitem[\protect\citeauthoryear{{Dickson}, {Tadhunter}, {Shaw}, {Clark} \&
  {Morganti}}{{Dickson} et~al.}{1995}]{Dickson95}
{Dickson} R.,  {Tadhunter} C.~N.,  {Shaw} M.,  {Clark} N.,    {Morganti} R.,
  1995, MNRAS, 273, L29

\bibitem[\protect\citeauthoryear{{Downes} \& {Solomon}}{{Downes} \&
  {Solomon}}{1998}]{Downes98}
{Downes} D.,  {Solomon} P.,  1998, ApJ, 507, 615

\bibitem[\protect\citeauthoryear{{Evans}, {Kim}, {Mazzarella}, {Scoville} \&
  {Sanders}}{{Evans} et~al.}{1999}]{Evans99}
{Evans} A.,  {Kim} D.,  {Mazzarella} J.,  {Scoville} N.,    {Sanders} D.,
  1999, ApJ, 521, L107

\bibitem[\protect\citeauthoryear{{Evans}, {Mazzarella}, {Surace} \&
  {Sanders}}{{Evans} et~al.}{2002}]{Evans02}
{Evans} A.,  {Mazzarella} J.,  {Surace} J.,    {Sanders} D.,  2002, AJ, 580,
  749

\bibitem[\protect\citeauthoryear{{Farrah}, {Afonso}, {Efstathiou},
  {Rowan-Robinson}, {Fox, M}. \& {Clements}}{{Farrah} et~al.}{2003}]{Farrah03}
{Farrah} D.,  {Afonso} J.,  {Efstathiou} A.,  {Rowan-Robinson} M.,  {Fox, M}.
  {Clements} D.,  2003, MNRAS, 343, 585

\bibitem[\protect\citeauthoryear{{Farrah}, {Bernard-Salas}, {Spoon}, {Soifer},
  {Armus}, {Brandl}, {Charmandaris}, {Desai}, {Higdon}, {Devost} \&
  {Houck}}{{Farrah} et~al.}{2007}]{Farrah07}
{Farrah} D.,  {Bernard-Salas} J.,  {Spoon} H.~W.~W.,  {Soifer} B.~T.,  {Armus}
  L.,  {Brandl} B.,  {Charmandaris} V.,  {Desai} V.,  {Higdon} S.,  {Devost}
  D.,    {Houck} J.,  2007, ApJ, 667, 149

\bibitem[\protect\citeauthoryear{{Farrah}, {Surace}, {Veilleux}, {Sanders} \&
  {Vacca}}{{Farrah} et~al.}{2005}]{Farrah05}
{Farrah} D.,  {Surace} J.~A.,  {Veilleux} S.,  {Sanders} D.~B.,    {Vacca}
  W.~D.,  2005, ApJ, 626, 70

\bibitem[\protect\citeauthoryear{{Franceschini}, {Aussel}, {Cesarsky}, {Elbaz}
  \& {Fadda}}{{Franceschini} et~al.}{2001}]{Franceschini01}
{Franceschini} A.,  {Aussel} H.,  {Cesarsky} C.~J.,  {Elbaz} D.,    {Fadda} D.,
   2001, A\&A, 378, 1

\bibitem[\protect\citeauthoryear{{Franceschini}, {Braito}, {Persic}, {Della
  Ceca}, {Bassani}, {Cappi}, {Malaguti}, {Palumbo}, {Risaliti}, {Salvati} \&
  {Severgnini}}{{Franceschini} et~al.}{2003}]{Franceschini03}
{Franceschini} A.,  {Braito} V.,  {Persic} M.,  {Della Ceca} R.,  {Bassani} L.,
   {Cappi} M.,  {Malaguti} P.,  {Palumbo} G.,  {Risaliti} G.,  {Salvati} M.,
  {Severgnini} P.,  2003, MNRAS, 343, 1181

\bibitem[\protect\citeauthoryear{{Genzel} R.~{Lutz}, {Sturm}, {Egami}, {Kunze},
  {Moorwood}, {Rigopoulou}, {Spoon}, {Sternberg}, {Tacconi-Garman}, {Tacconi}
  \& {Thatte}}{{Genzel} et~al.}{1998}]{Genzel98}
{Genzel} R.~{Lutz} D.,  {Sturm} E.,  {Egami} E.,  {Kunze} D.,  {Moorwood} A.,
  {Rigopoulou} D.,  {Spoon} H.,  {Sternberg} A.,  {Tacconi-Garman} L.,
  {Tacconi} L.,    {Thatte} N.,  1998, ApJ, 498, 579

\bibitem[\protect\citeauthoryear{{Gonz\'alez Delgado}, {Heckman} \&
  {Leitherer}}{{Gonz\'alez Delgado} et~al.}{2001}]{Gonzalez01}
{Gonz\'alez Delgado} R.,  {Heckman} T.,    {Leitherer} 2001, ApJ, 546, 845

\bibitem[\protect\citeauthoryear{{Gonz{\'a}lez Delgado}, {Cervi{\~n}o},
  {Martins}, {Leitherer} \& {Hauschildt}}{{Gonz{\'a}lez Delgado}
  et~al.}{2005}]{Gonzalez05}
{Gonz{\'a}lez Delgado} R.~M.,  {Cervi{\~n}o} M.,  {Martins} L.~P.,  {Leitherer}
  C.,    {Hauschildt} P.~H.,  2005, MNRAS, 357, 945

\bibitem[\protect\citeauthoryear{{Hamilton} \& {Keel}}{{Hamilton} \&
  {Keel}}{1987}]{Hamilton87}
{Hamilton} D.,  {Keel} W.,  1987, ApJ, 321, 211

\bibitem[\protect\citeauthoryear{{Houck}, {Sneider}, {Danielson}, {Beichman},
  {Lonsdale}, {Neugebauer} \& B.T.}{{Houck} et~al.}{1985}]{Houck85}
{Houck} J.,  {Sneider} D.,  {Danielson} G.,  {Beichman} C.,  {Lonsdale} C.,
  {Neugebauer} C.,    B.T. S.,  1985, ApJ., 290, L5

\bibitem[\protect\citeauthoryear{{Houck}, {Soifer}, {Neugebauer}, {Beichman},
  {Aumann}, {Clegg}, {Gillett}, {Habing}, {Hauser}, {Low}, {Miley},
  {Rowan-Robinson} \& {Walker}}{{Houck} et~al.}{1984}]{Houck84}
{Houck} J.,  {Soifer} B.~T.,  {Neugebauer} G.,  {Beichman} C.~A.,  {Aumann}
  H.~H.,  {Clegg} P.~E.,  {Gillett} F.~C.,  {Habing} H.~J.,  {Hauser} M.~G.,
  {Low} F.~J.,  {Miley} G.,  {Rowan-Robinson} M.,    {Walker} R.~G.,  1984,
  ApJ., 278, L63

\bibitem[\protect\citeauthoryear{{Imanishi} \& {Dudley}}{{Imanishi} \&
  {Dudley}}{2002}]{Imanishi00}
{Imanishi} M.,  {Dudley} C.,  2002, ApJ, 545, 701

\bibitem[\protect\citeauthoryear{{Imanishi}, {Dudley}, {Maiolino}, {Maloney},
  {Nakagawa} \& {Risaliti}}{{Imanishi} et~al.}{2007}]{Imanishi07}
{Imanishi} M.,  {Dudley} C.,  {Maiolino} R.,  {Maloney} P.,  {Nakagawa} T.,
  {Risaliti} G.,  2007, ApJ, 171, 72

\bibitem[\protect\citeauthoryear{{Imanishi}, {Dudley} \& {Maloney}}{{Imanishi}
  et~al.}{2006}]{Imanishi06}
{Imanishi} M.,  {Dudley} C.,    {Maloney} P.,  2006, AJ, 637, 114

\bibitem[\protect\citeauthoryear{{Kim} \& {Sanders}}{{Kim} \&
  {Sanders}}{1998}]{Kim98a}
{Kim} D.~C.,  {Sanders} D.~B.,  1998, ApJ, 119, 41

\bibitem[\protect\citeauthoryear{{Kim}, {Veilluex} \& {Sanders}}{{Kim}
  et~al.}{2002}]{Kim02}
{Kim} D.~C.,  {Veilluex} S.,    {Sanders} D.~B.,  2002, ApJ, 143, 277

\bibitem[\protect\citeauthoryear{{Knapen}, {Laine}, {Yates} J.A.~{Robinson},
  {Richards}, {Doyon} \& {Nadeau}}{{Knapen} et~al.}{1997}]{Knapen97}
{Knapen} J.,  {Laine} S.,  {Yates} J.A.~{Robinson} A.,  {Richards} A.,  {Doyon}
  R.,    {Nadeau} D.,  1997, ApJ, 490, L29

\bibitem[\protect\citeauthoryear{{Leitherer}, {Schaerer}, {Goldader},
  {Gonz\'alez Delgado}, {Robert}, {Foo Kune}, {de Mello}, {Devost} \&
  {Heckman}}{{Leitherer} et~al.}{1999}]{Leitherer99}
{Leitherer} C.,  {Schaerer} D.,  {Goldader} J.,  {Gonz\'alez Delgado} R.,
  {Robert} C.,  {Foo Kune} D.,  {de Mello} D.,  {Devost} D.,    {Heckman} T.,
  1999, ApJ, 123, 3

\bibitem[\protect\citeauthoryear{{Lutz}, {Veilluex} \& {Genzel}}{{Lutz}
  et~al.}{1999}]{Lutz99}
{Lutz} D.,  {Veilluex} S.,    {Genzel} R.,  1999, ApJ, 517, L13

\bibitem[\protect\citeauthoryear{{Majewski}, {Hereld}, {Koo}, {Illingworth} \&
  {Heckman}}{{Majewski} et~al.}{1993}]{Majewski93}
{Majewski} S.,  {Hereld} M.,  {Koo} D.,  {Illingworth} G.,    {Heckman} T.,
  1993, ApJ, 403, 125

\bibitem[\protect\citeauthoryear{{Murphy}, {Soifer}, {Matthews} \&
  {Armus}}{{Murphy} et~al.}{2001}]{Murphy01}
{Murphy} T.,  {Soifer} B.,  {Matthews} K.,    {Armus} L.,  2001, ApJ, 559, 201

\bibitem[\protect\citeauthoryear{{Nagar}, {Wilson}, {Falcke}, {Veilleux} \&
  {Maiolino}}{{Nagar} et~al.}{2003}]{Nagar03}
{Nagar} N.,  {Wilson} A.,  {Falcke} H.,  {Veilleux} S.,    {Maiolino} R.,
  2003, A\&A, 409, 115

\bibitem[\protect\citeauthoryear{{Osterbrock}, {Fulbright}, {Keane} \&
  {Trager}}{{Osterbrock} et~al.}{1996}]{Osterbrock96}
{Osterbrock} D.,  {Fulbright} J.,  {Keane} M.,    {Trager} S.,  1996, PASP,
  108, 277

\bibitem[\protect\citeauthoryear{{Rigopoulou}, {Spoon}, {Genzel}, {Lutz},
  {Moorwood} \& {Tran}}{{Rigopoulou} et~al.}{1999}]{Rigopoulou99}
{Rigopoulou} D.,  {Spoon} H.,  {Genzel} R.,  {Lutz} D.,  {Moorwood} A.,
  {Tran} Q.,  1999, ApJ, 118, 2625

\bibitem[\protect\citeauthoryear{{Risaliti}, {Maiolino}, {Marconi}, {Sani},
  {Berta}, {Braito}, {Ceca}, {Franceschini} \& {Salvati}}{{Risaliti}
  et~al.}{2006}]{Risaliti06}
{Risaliti} G.,  {Maiolino} R.,  {Marconi} A.,  {Sani} E.,  {Berta} S.,
  {Braito} V.,  {Ceca} R.,  {Franceschini} A.,    {Salvati} M.,  2006, MNRAS,
  365, 303

\bibitem[\protect\citeauthoryear{{Robinson}, {Tadhunter}, {Axon} \&
  {Robinson}}{{Robinson} et~al.}{2000}]{Robinson00}
{Robinson} T.,  {Tadhunter} C.,  {Axon} D.,    {Robinson} A.,  2000, MNRAS,
  317, 922

\bibitem[\protect\citeauthoryear{{Rodr\'iguez Zaur\'in}, {Holt}, {Tadhunter} \&
  {Gonz\'alez Delgado}}{{Rodr\'iguez Zaur\'in}
  et~al.}{2007}]{Rodriguez-Zaurin07}
{Rodr\'iguez Zaur\'in} J.,  {Holt} J.,  {Tadhunter} C.,    {Gonz\'alez Delgado}
  R.,  2007, MNRAS, 375, 1133

\bibitem[\protect\citeauthoryear{{Rodr{\'{\i}}guez Zaur{\'{\i}}n}, {Tadhunter}
  \& {Gonz{\'a}lez Delgado}}{{Rodr{\'{\i}}guez Zaur{\'{\i}}n}
  et~al.}{2008}]{Rodriguez-Zaurin08}
{Rodr{\'{\i}}guez Zaur{\'{\i}}n} J.,  {Tadhunter} C.~N.,    {Gonz{\'a}lez
  Delgado} R.~M.,  2008, MNRAS, 384, 875

\bibitem[\protect\citeauthoryear{{Rupke}, {Veilleux} \& {Sanders}}{{Rupke}
  et~al.}{2002}]{Rupke02}
{Rupke} D.,  {Veilleux} S.,    {Sanders} D.,  2002, ApJ, 570, 588

\bibitem[\protect\citeauthoryear{{Salpeter}}{{Salpeter}}{1955}]{Salpeter55}
{Salpeter} E.,  1955, ApJ, 121, 161

\bibitem[\protect\citeauthoryear{{Schlegel}, {Finkbeiner} \&
  {Davis}}{{Schlegel} et~al.}{1998}]{Schlegel98}
{Schlegel} J.,  {Finkbeiner} D.,    {Davis} M.,  1998, ApJ, 500, 525

\bibitem[\protect\citeauthoryear{{Scoville}, {Evans}, {Thompson}, {Rieke},
  {Hines}, {Low}, {Dinshaw}, {Surace} \& {Armus}}{{Scoville}
  et~al.}{2000}]{Scoville00}
{Scoville} N.~Z.,  {Evans} A.~S.,  {Thompson} R.,  {Rieke} M.,  {Hines} D.~C.,
  {Low} F.~J.,  {Dinshaw} N.,  {Surace} J.~A.,    {Armus} L.,  2000, ApJ, 119,
  991

\bibitem[\protect\citeauthoryear{{Seaton}}{{Seaton}}{1979}]{Seaton79}
{Seaton} M.,  1979, MNRAS, 187, 73

\bibitem[\protect\citeauthoryear{{Soifer}, {Neugebauer}, {Matthews}, {Egami},
  {Becklin}, {Weinberger}, {Ressler}, {Werner}, {Evans}, {Scoville}, {Surace}
  \& {Condon}}{{Soifer} et~al.}{2000}]{Soifer00}
{Soifer} B.,  {Neugebauer} G.,  {Matthews} K.,  {Egami} E.,  {Becklin} E.,
  {Weinberger} A.,  {Ressler} M.,  {Werner} M.,  {Evans} A.,  {Scoville} N.,
  {Surace} J.,    {Condon} J.,  2000, AJ, 119, 509

\bibitem[\protect\citeauthoryear{{Soifer}, {Sanders}, {Madore}, {Neugebauer},
  {Danielson}, {Elias}, {Lonsdale} \& {Rice}}{{Soifer} et~al.}{1987}]{Soifer87}
{Soifer} B.,  {Sanders} D.,  {Madore} B.,  {Neugebauer} C.,  {Danielson} G.,
  {Elias} E.,  {Lonsdale} C.,    {Rice} W.,  1987, ApJ., 320, 74

\bibitem[\protect\citeauthoryear{{Soifer}, {Sanders}, {Neugebauer},
  {Danielson}, {Lonsdale}, {Madore} \& {Persson}}{{Soifer}
  et~al.}{1986}]{Soifer86}
{Soifer} B.,  {Sanders} D.,  {Neugebauer} C.,  {Danielson} G.,  {Lonsdale} C.,
  {Madore} B.,    {Persson} S.,  1986, ApJ., 303, L41

\bibitem[\protect\citeauthoryear{{Soifer}}{{Soifer}}{1984a}]{Soifer84a}
{Soifer} B. e.~a.,  1984a, ApJ., 278, L71

\bibitem[\protect\citeauthoryear{{Soifer}}{{Soifer}}{1984b}]{Soifer84b}
{Soifer} B. e.~a.,  1984b, ApJ., 283, L1

\bibitem[\protect\citeauthoryear{{Surace}, {Sanders} \& {Evans}}{{Surace}
  et~al.}{2000}]{Surace00a}
{Surace} J.,  {Sanders} D.,    {Evans} A.,  2000, ApJ, 529, 170

\bibitem[\protect\citeauthoryear{{Surace} \& {Sanders}}{{Surace} \&
  {Sanders}}{2000}]{Surace00b}
{Surace} J.~A.,  {Sanders} D.,  2000, ApJ, 120, 604

\bibitem[\protect\citeauthoryear{{Surace}, {Sanders}, {Vacca} W. D.~{Veilleux}
  \& {Mazzarella}}{{Surace} et~al.}{1998}]{Surace98b}
{Surace} J.~A.,  {Sanders} D.~B.,  {Vacca} W. D.~{Veilleux} S.,    {Mazzarella}
  J.~M.,  1998, ApJ, 492, 116

\bibitem[\protect\citeauthoryear{{Surace} \& {Sanders}}{{Surace} \&
  {Sanders}}{1999}]{Surace99}
{Surace} J.~A.,  {Sanders} D.~C.,  1999, ApJ, 512, 162

\bibitem[\protect\citeauthoryear{{Tadhunter}, {Dickson}, {Morganti},
  {Robinson}, {Wills}, {Villar-Martin} \& {Hughes}}{{Tadhunter}
  et~al.}{2002}]{Tadhunter02}
{Tadhunter} C.,  {Dickson} R.,  {Morganti} R.,  {Robinson} T.,  {Wills} K.,
  {Villar-Martin} M.,    {Hughes} M.,  2002, MNRAS, 330, 977

\bibitem[\protect\citeauthoryear{{Tadhunter}, {Robinson}, {Gonz\'alez Delgado},
  {Wills} \& {Morganti}}{{Tadhunter} et~al.}{2005}]{Tadhunter05}
{Tadhunter} C.~N.,  {Robinson} T.~G.,  {Gonz\'alez Delgado} R.~M.,  {Wills} K.,
     {Morganti} R.,  2005, MNRAS, 356, 480

\bibitem[\protect\citeauthoryear{{Ulvestad}
  J.S.~{Wilson}}{{Ulvestad}}{1984}]{Ulvestad84}
{Ulvestad} J.S.~{Wilson} A.,  1984, ApJ, 278, 544

\bibitem[\protect\citeauthoryear{{Vald\'es}, {Berta}, {Bressan},
  {Franceschini}, {Rigopoulou} \& {Rodighiero}}{{Vald\'es}
  et~al.}{2005}]{Valdes05}
{Vald\'es} J.,  {Berta} S.,  {Bressan} A.,  {Franceschini} A.,  {Rigopoulou}
  D.,    {Rodighiero} G.,  2005, A\&A, 434, 149

\bibitem[\protect\citeauthoryear{{Veilleux}, {Kim}, {Mazzarella} \&
  {Soifer}}{{Veilleux} et~al.}{1995}]{Veilleux95}
{Veilleux} S.,  {Kim} D.~C.,  {Mazzarella} J.~M.,    {Soifer} B.~T.,  1995,
  ApJ, 98, 171

\bibitem[\protect\citeauthoryear{{Veilleux}, {Kim} \& {Sanders}}{{Veilleux}
  et~al.}{1999}]{Veilleux99a}
{Veilleux} S.,  {Kim} D.-C.,    {Sanders} D.~B.,  1999, ApJ, 522, 113

\bibitem[\protect\citeauthoryear{{Veilleux}, {Kim} \& {Sanders}}{{Veilleux}
  et~al.}{2002}]{Veilleux02}
{Veilleux} S.,  {Kim} D.~C.,    {Sanders} D.~B.,  2002, ApJ, 143, 315

\bibitem[\protect\citeauthoryear{{Veilleux}, {Sanders} \& {Kim}}{{Veilleux}
  et~al.}{1997}]{Veilleux97}
{Veilleux} S.,  {Sanders} D.,    {Kim} D.-C.,  1997, ApJ, 484, 92

\bibitem[\protect\citeauthoryear{{Veilleux}, {Sanders} \& {Kim}}{{Veilleux}
  et~al.}{999b}]{Veilleux99b}
{Veilleux} S.,  {Sanders} D.~B.,    {Kim} D.-C.,  1999b, ApJ, 522, 139

\bibitem[\protect\citeauthoryear{{Veilluex}, {Kim}, {Peng}, {Ho}, {Tacconi},
  {Dasyra}, {Genzel}, {Lutz} \& {Sanders}}{{Veilluex}
  et~al.}{2006}]{Veilleux06}
{Veilluex} S.,  {Kim} D.-C.,  {Peng} C.,  {Ho} L.,  {Tacconi} L.,  {Dasyra} K.,
   {Genzel} R.,  {Lutz} D.,    {Sanders} D.,  2006, ApJ, 643, 707

\bibitem[\protect\citeauthoryear{{Wilson}, {Harris}, {Longden} \&
  {Scoville}}{{Wilson} et~al.}{2006}]{Wilson06}
{Wilson} C.,  {Harris} W.,  {Longden} R.,    {Scoville} N.,  2006, ApJ, 641,
  763

\end{thebibliography}

%\newpage
%\begin{minipage}{160mm}
%\appendix
%\begin{centering}
%\section{Notes on individual objects}
%\end{centering}
%\hspace{6.0cm}{\bf IRAS 00091-0738} 
%\end{minipage}

\clearpage

\begin{figure*}
\begin{minipage}{170mm}
\begin{tabular}{cc}
\hspace*{0cm}\psfig{file=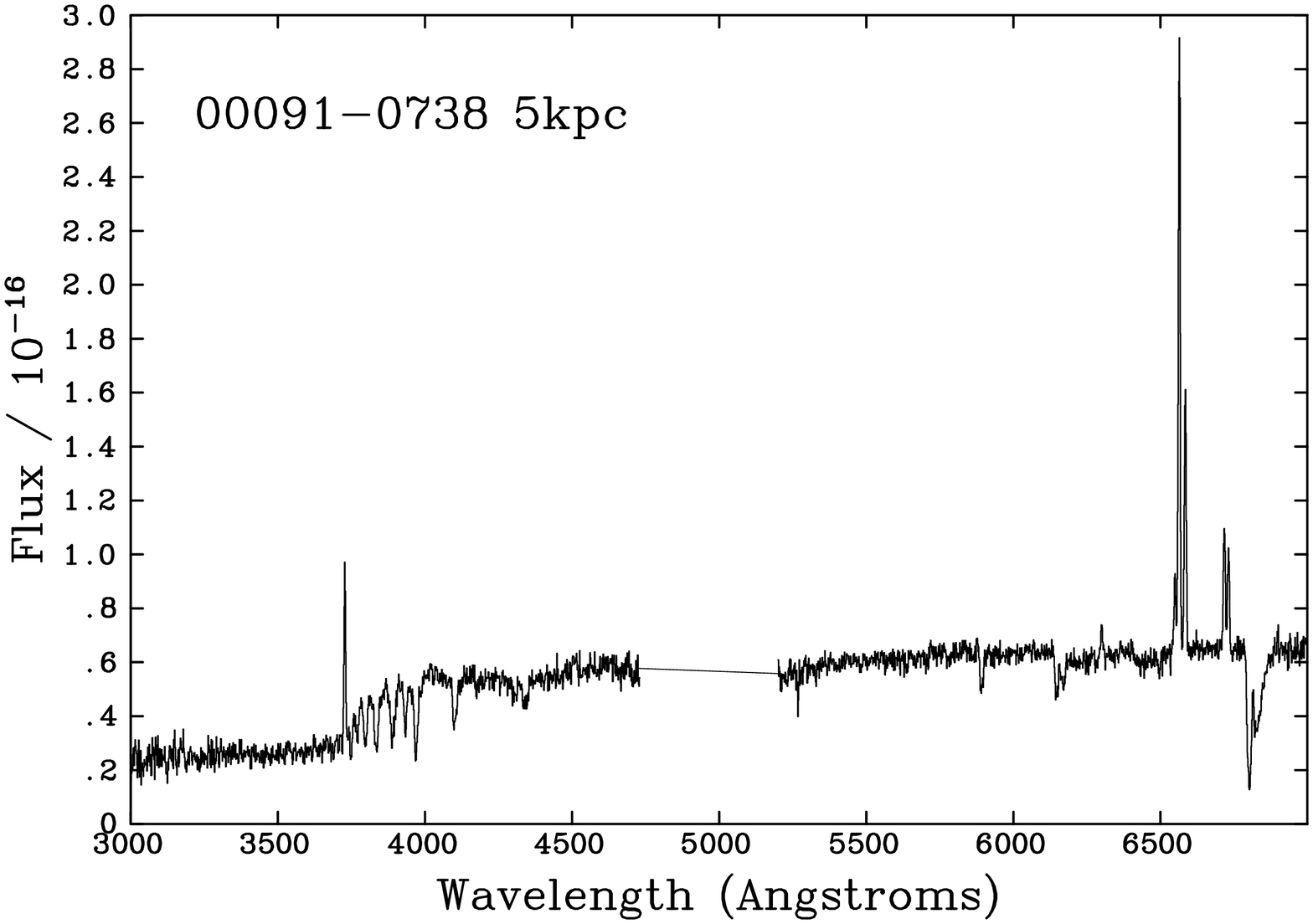,width=7.8cm,angle=0.}&
\psfig{file=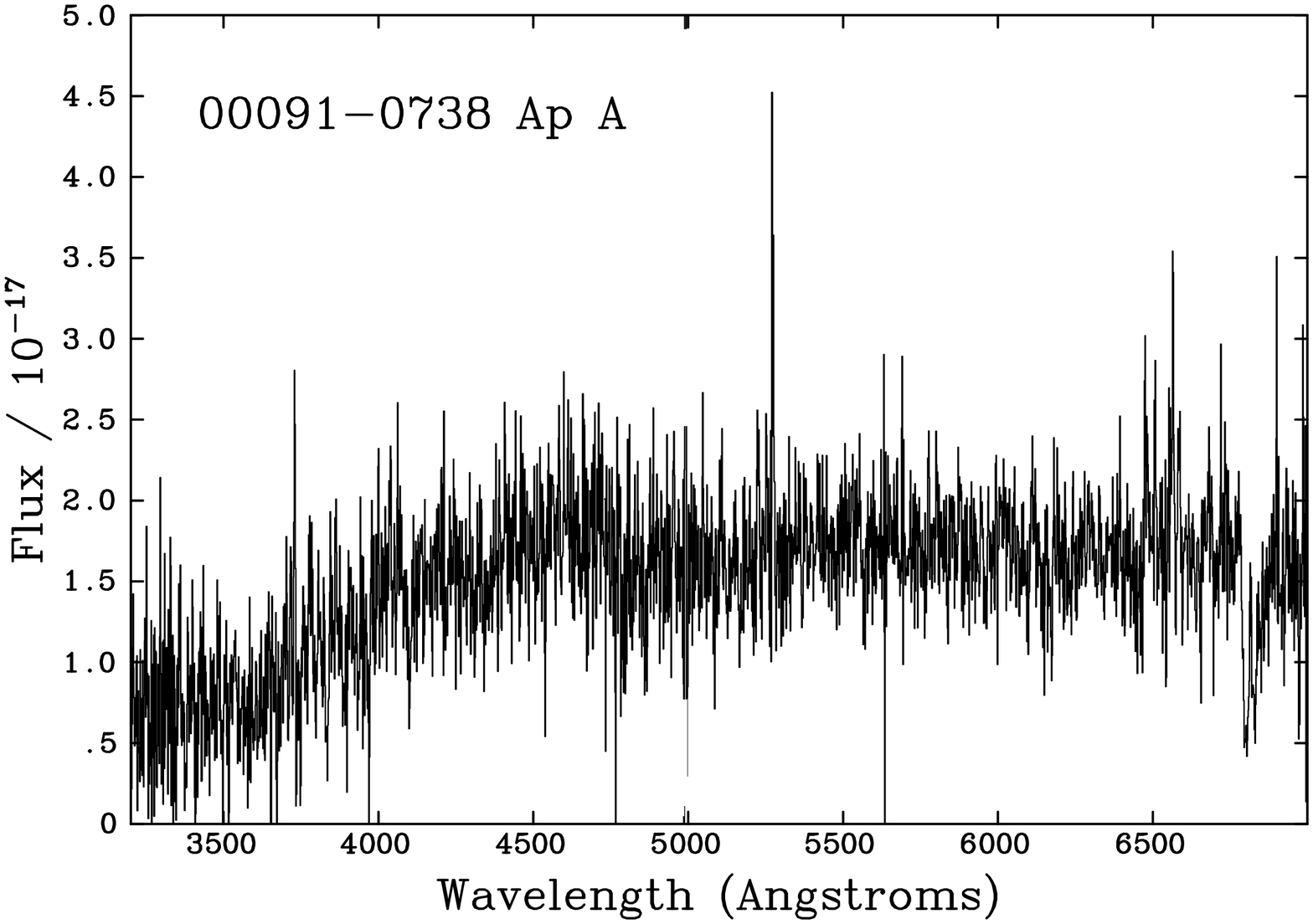,width=7.8cm,angle=0.}\\
\hspace*{0cm}\psfig{file=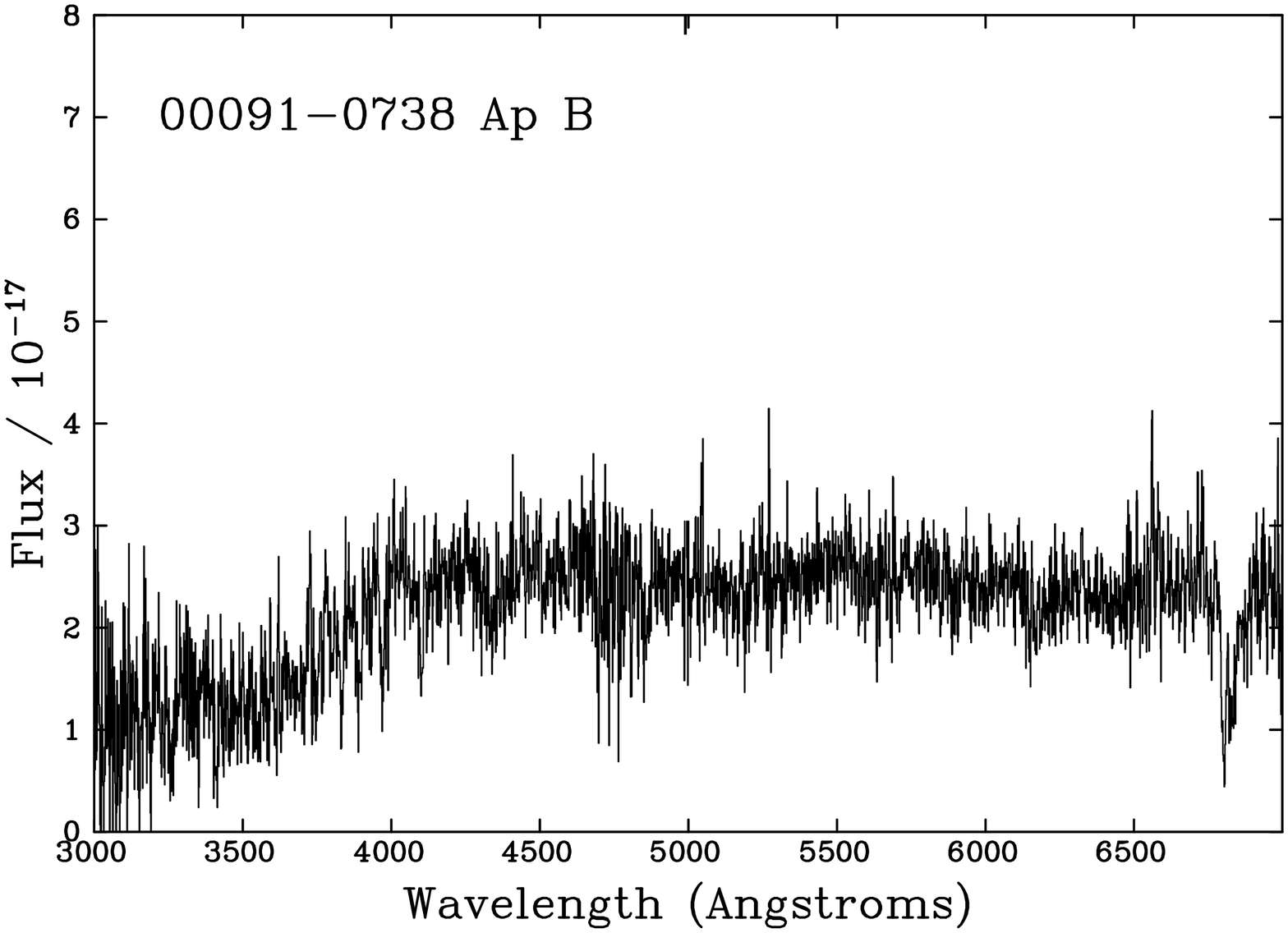,width=7.8cm,angle=0.}&
\psfig{file=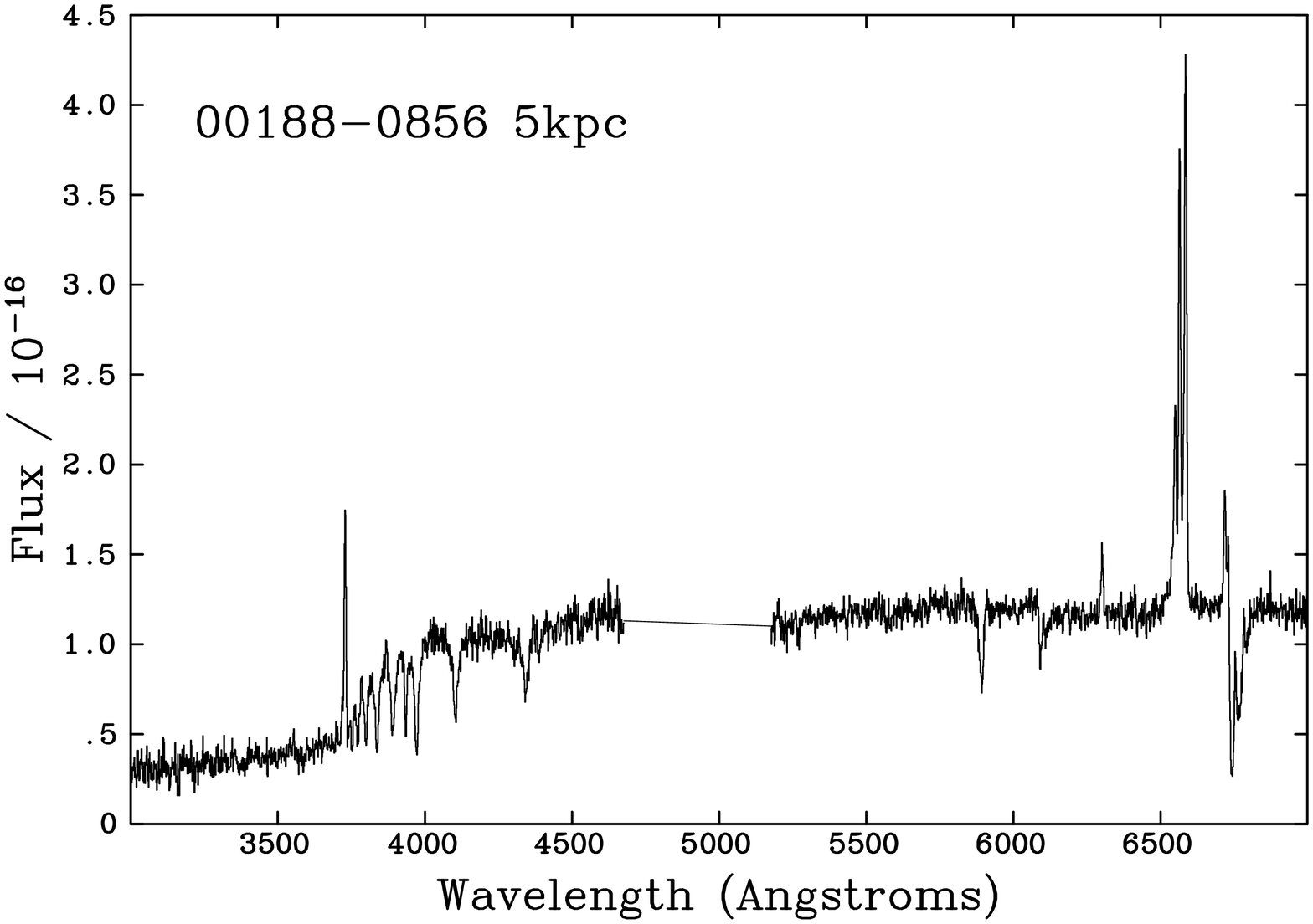,width=7.8cm,angle=0.}\\
\hspace*{0cm}\psfig{file=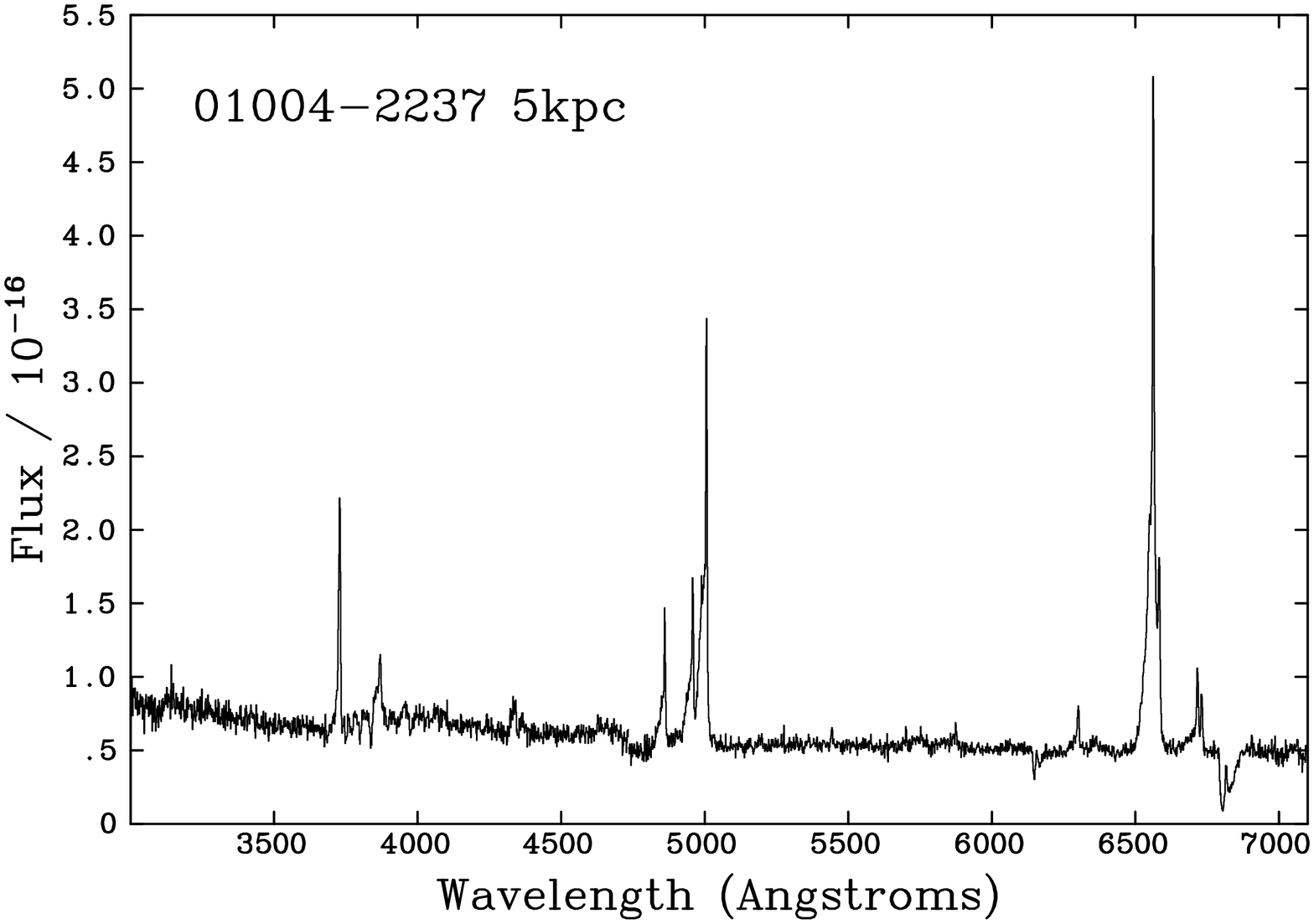,width=5.5cm,angle=-90.}&
\psfig{file=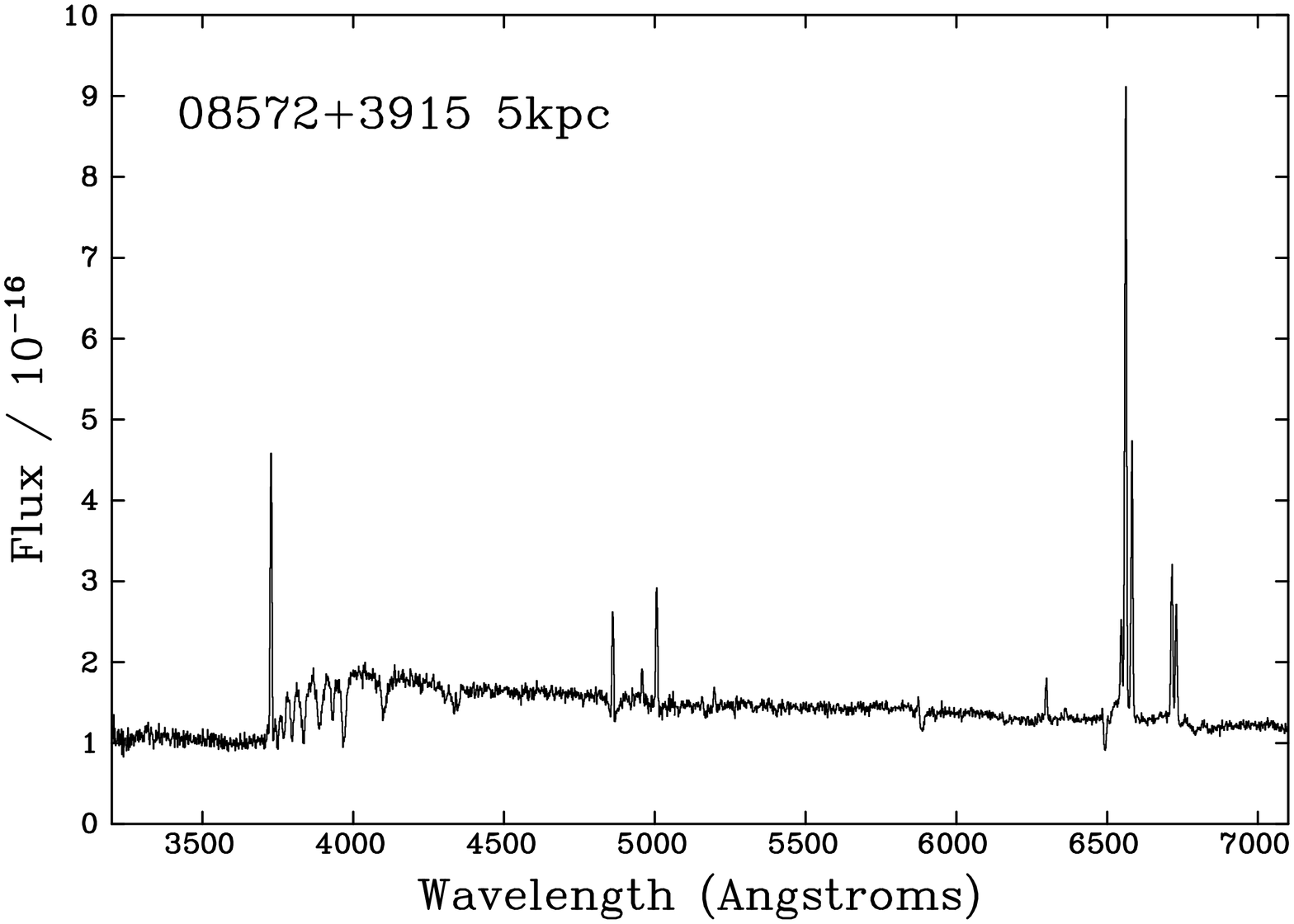,width=5.5cm,angle=-90.}\\
\hspace*{0cm}\psfig{file=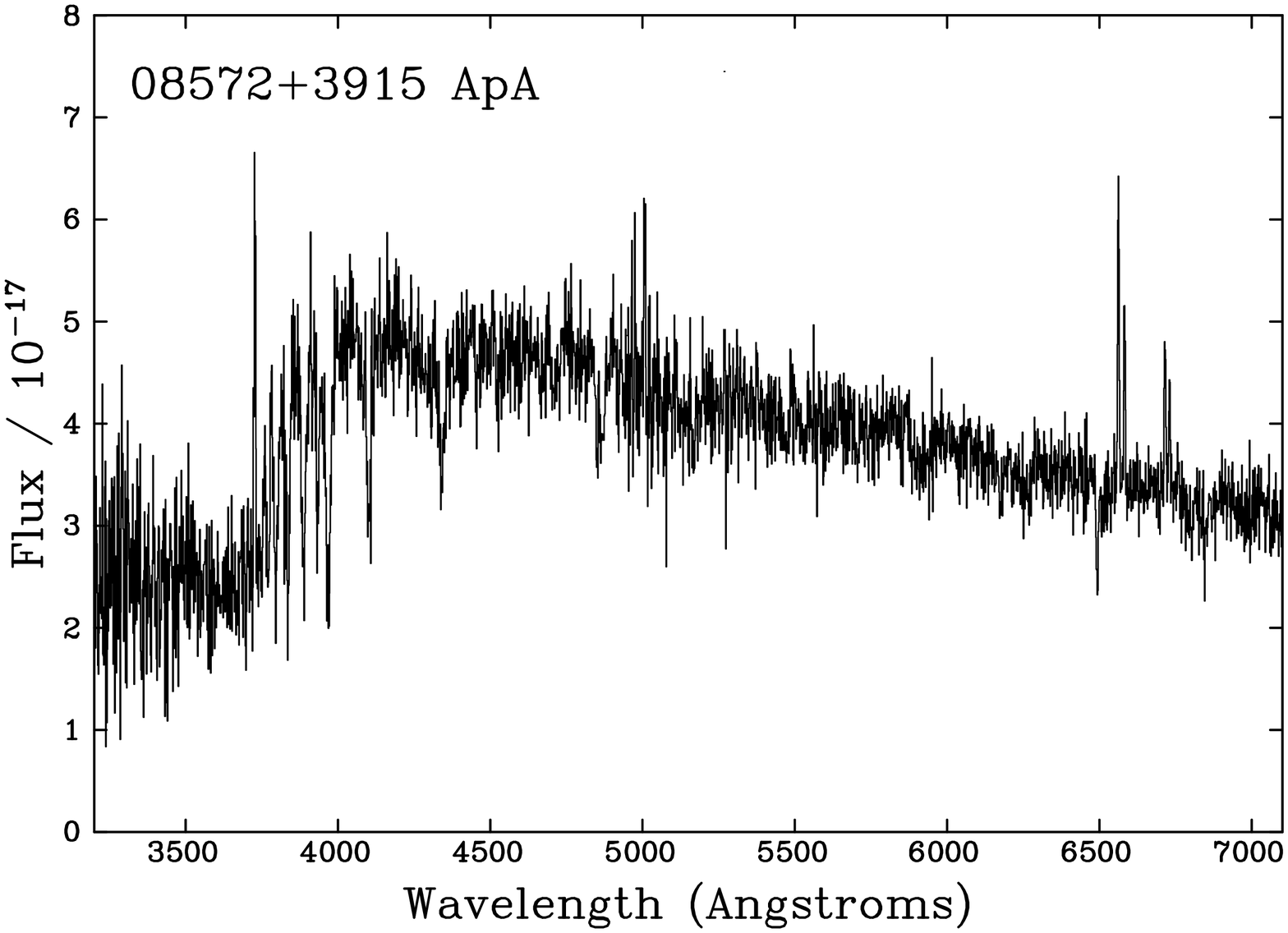,width=7.8cm,angle=0.}&
\psfig{file=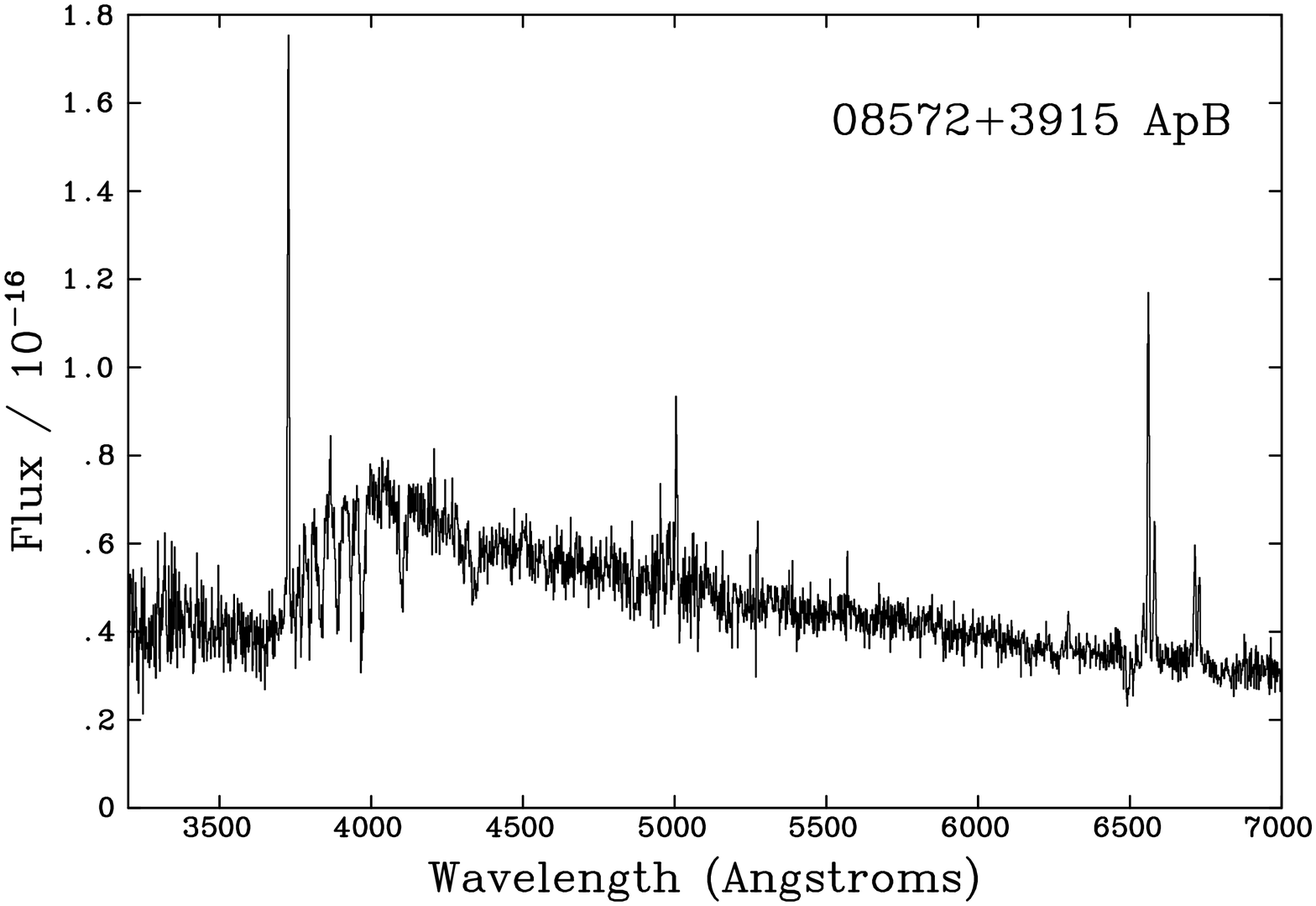,width=7.8cm,angle=0.}\\
\end{tabular}
\caption[Extracted 1D-spectra for all the ULIRGs in the ES] {The
extracted 1D-spectra for all the ULIRGs and apertures in the ES. When
the dichroic effects are important, it is not possible to show an
adequate plot of the entire wavelength range, and the dichroic region
is removed from the plots. The fluxes are presented in wavelength
units.}
\label{fig:all1dspectra}
\end{minipage}
\end{figure*}
\addtocounter{figure}{-1}
\begin{figure*}
\begin{minipage}{170mm}
\begin{tabular}{cc}
\hspace*{0cm}\psfig{file=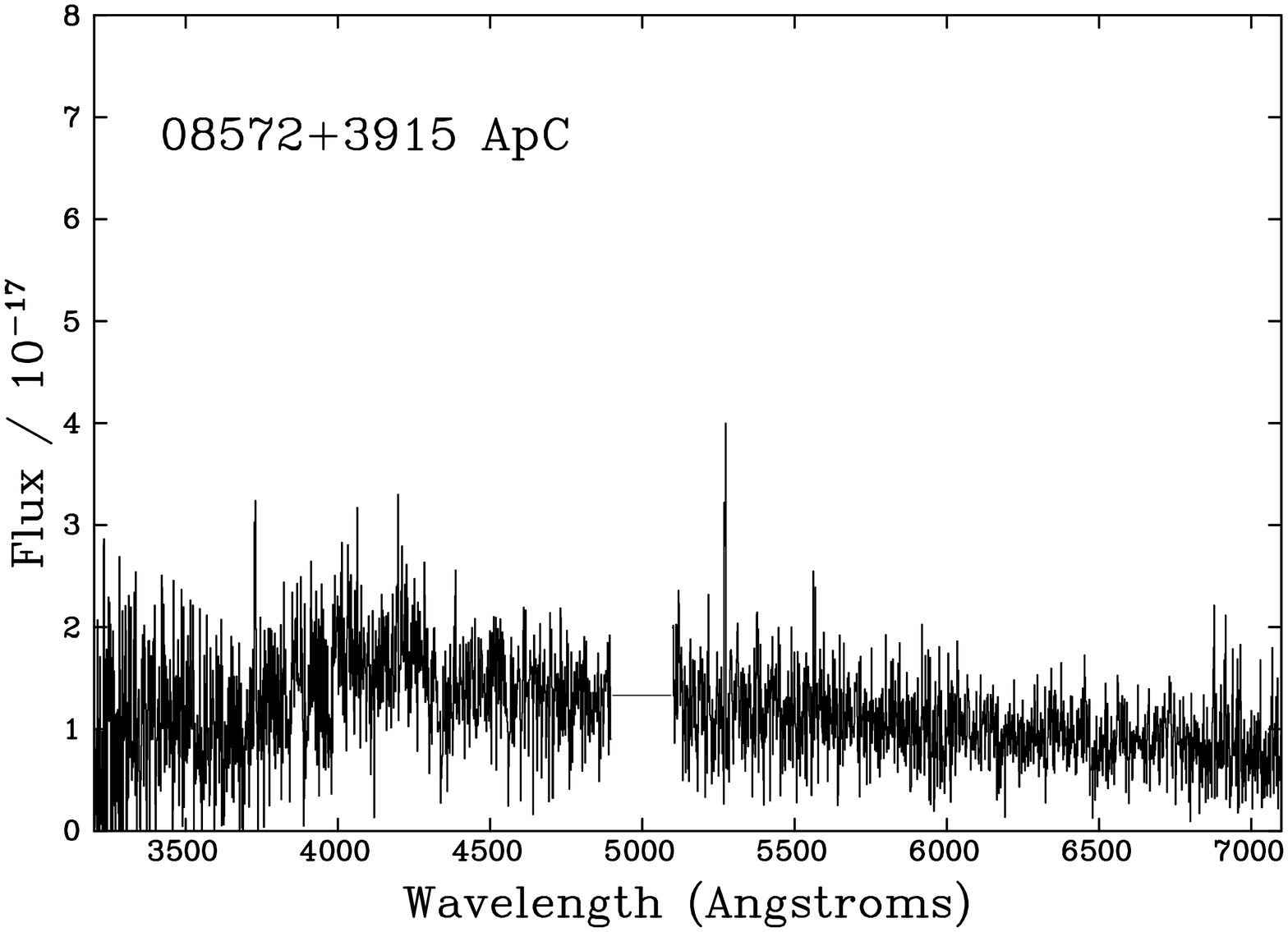,width=7.8cm,angle=0.}&
\psfig{file=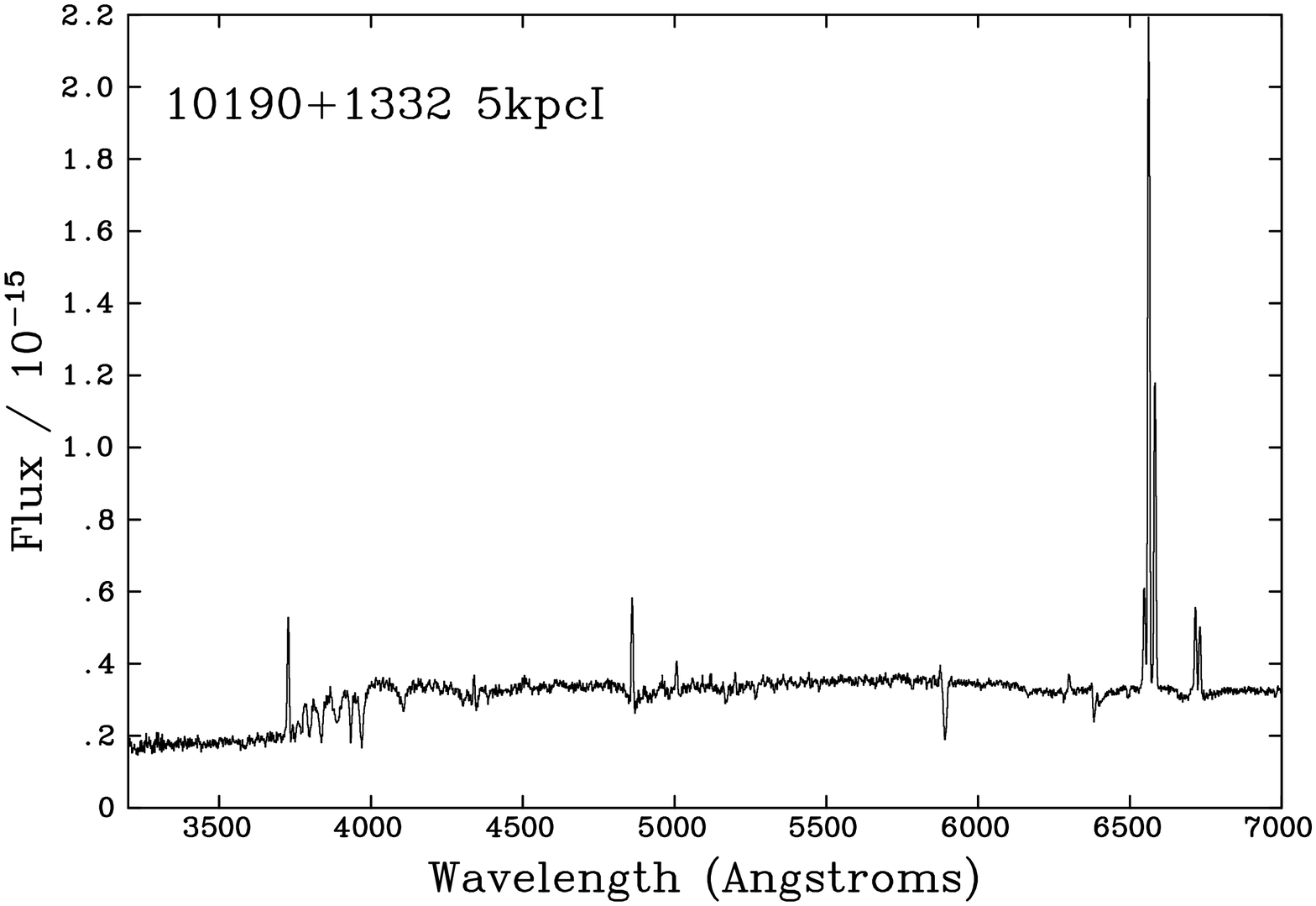,width=7.8cm,angle=0.}\\
\hspace*{0cm}\psfig{file=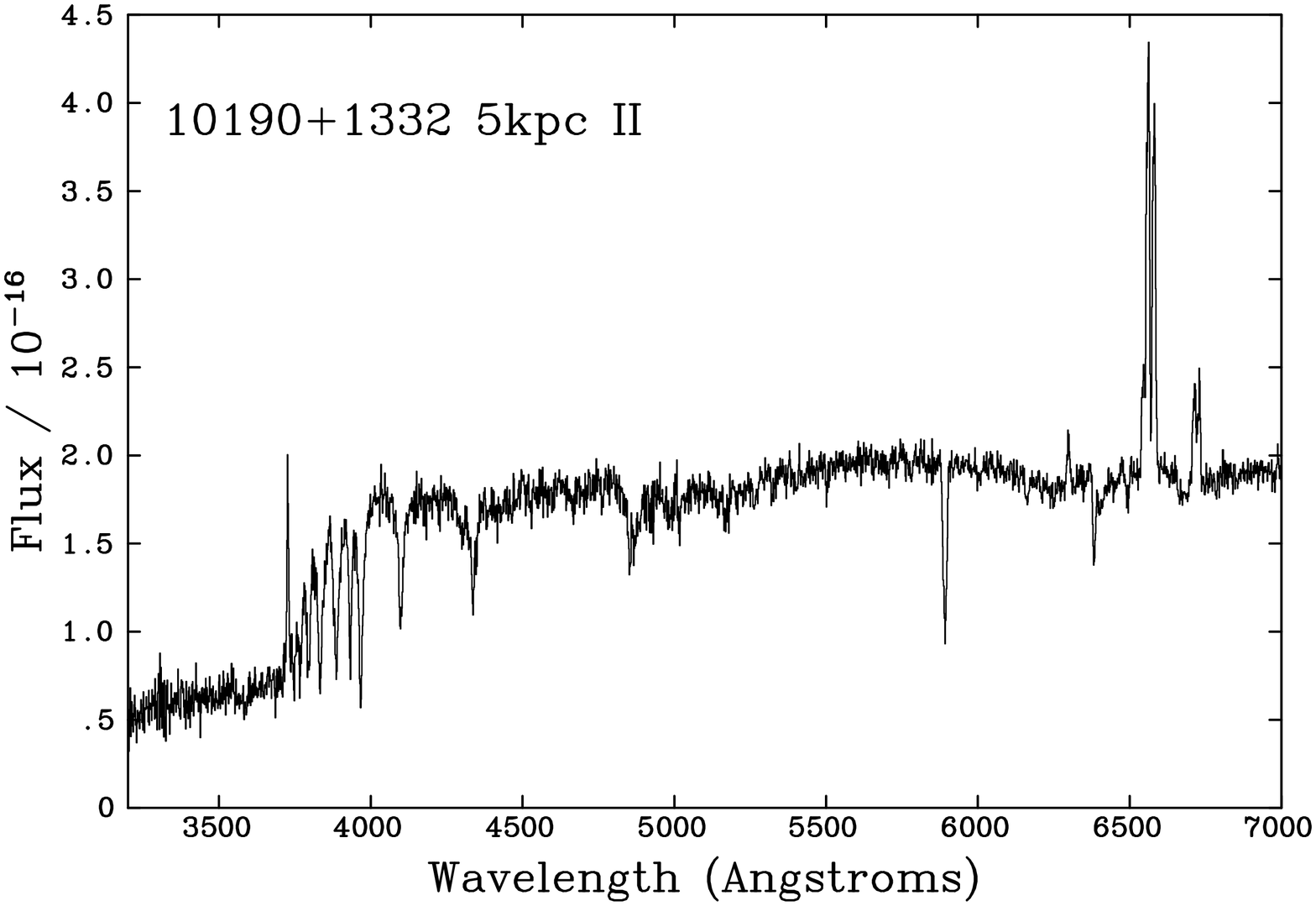,width=5.5cm,angle=-90.}&
\psfig{file=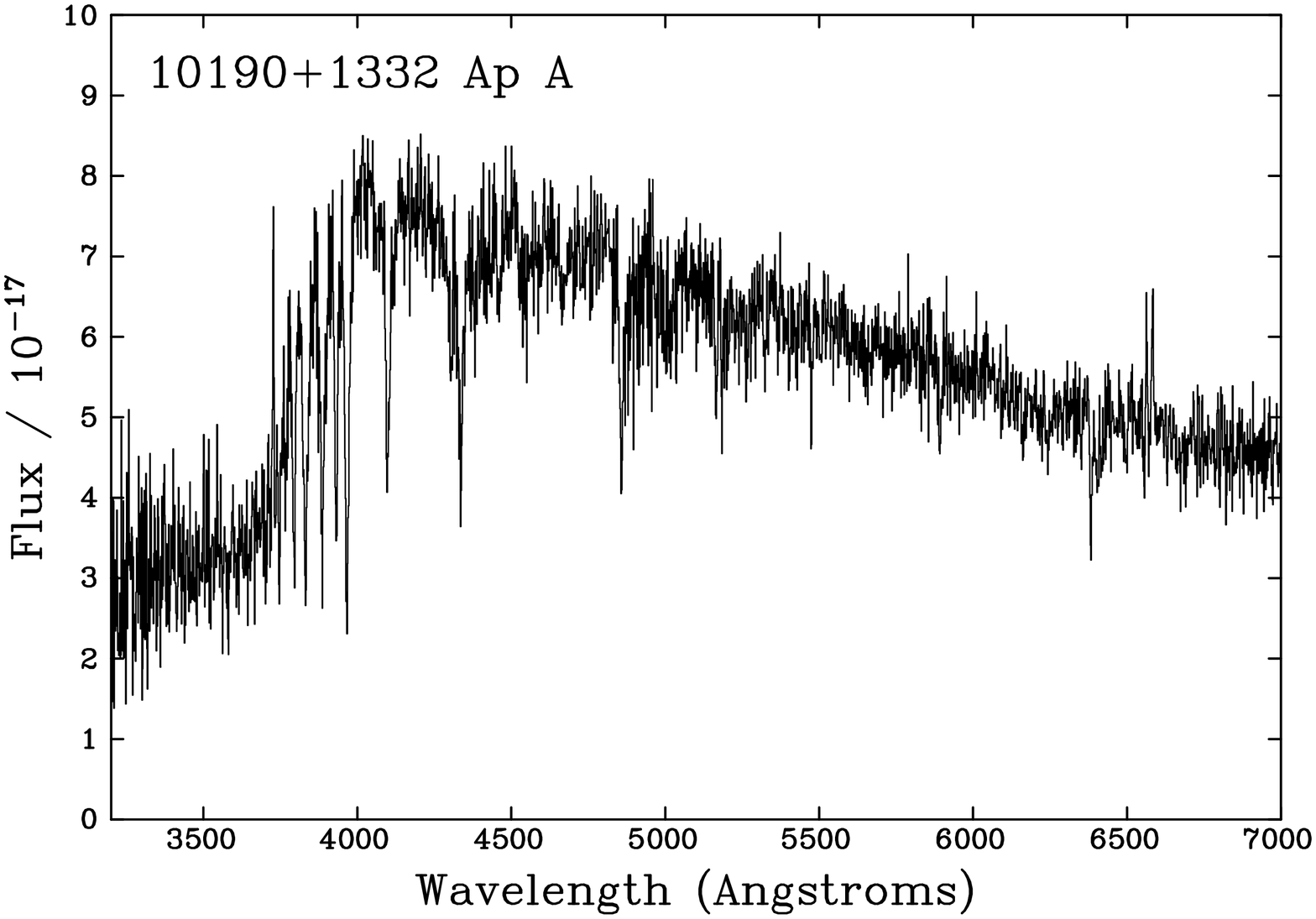,width=5.5cm,angle=-90.}\\
\hspace*{0cm}\psfig{file=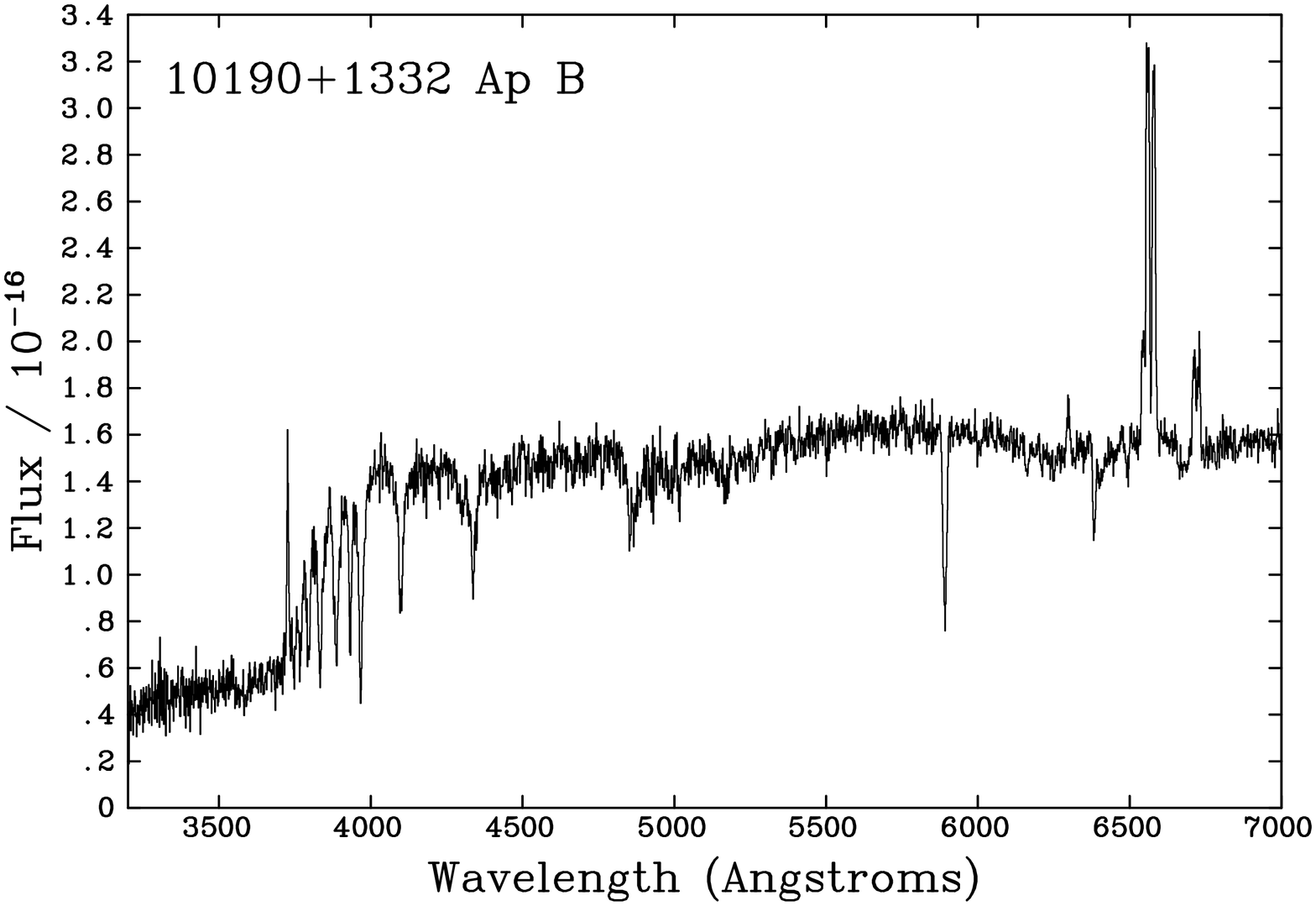,width=5.5cm,angle=-90.}&
\psfig{file=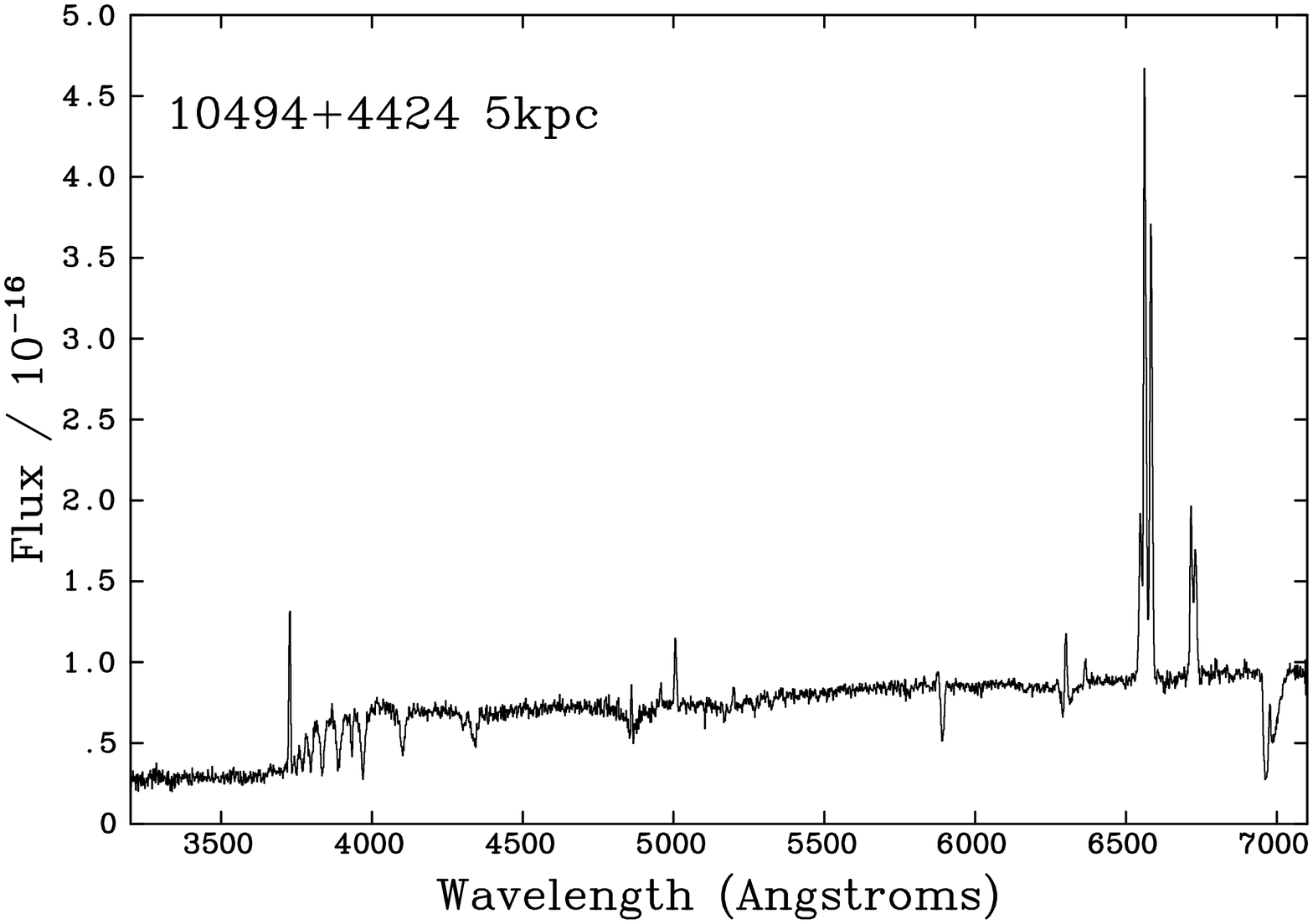,width=5.5cm,angle=-90.}\\
\hspace*{0cm}\psfig{file=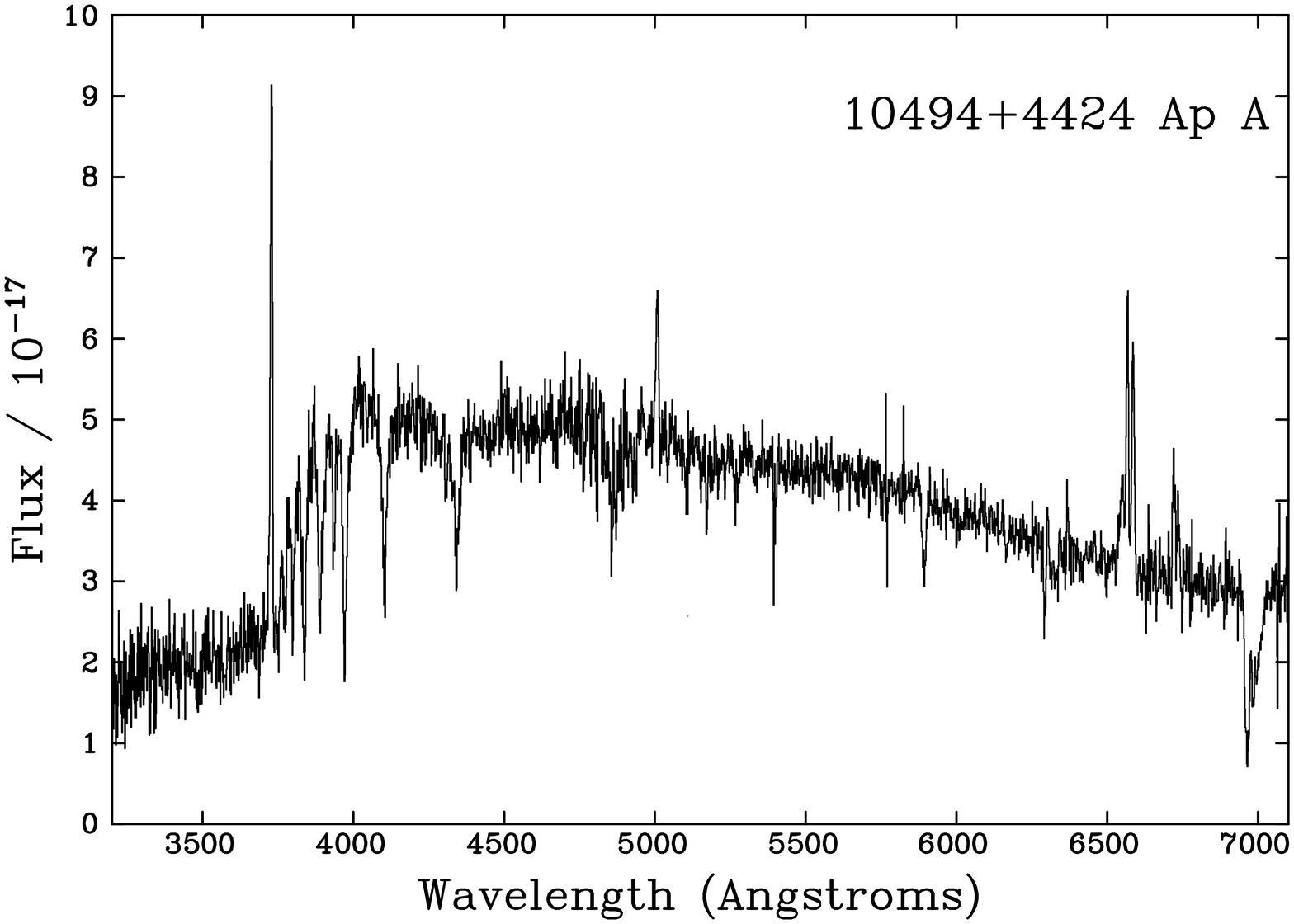,width=7.8cm,angle=0.}&
\psfig{file=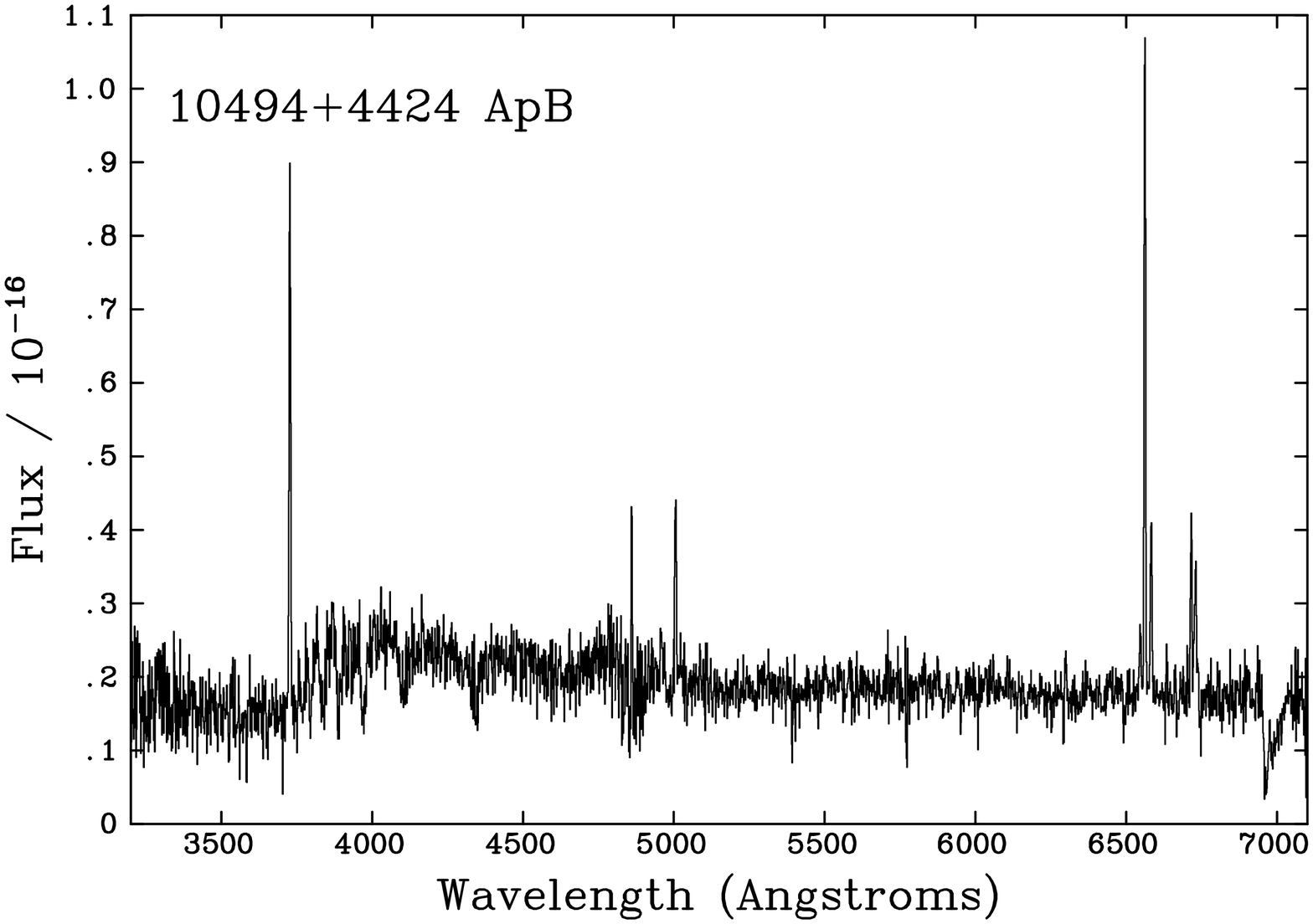,width=7.8cm,angle=0.}
\end{tabular}
\caption[{\it Continued}]{Continued}
%\label{fig:SED}
\end{minipage}
\end{figure*}
\addtocounter{figure}{-1}
\begin{figure*}
\begin{minipage}{170mm}
\begin{tabular}{cc}
\hspace*{0cm}\psfig{file=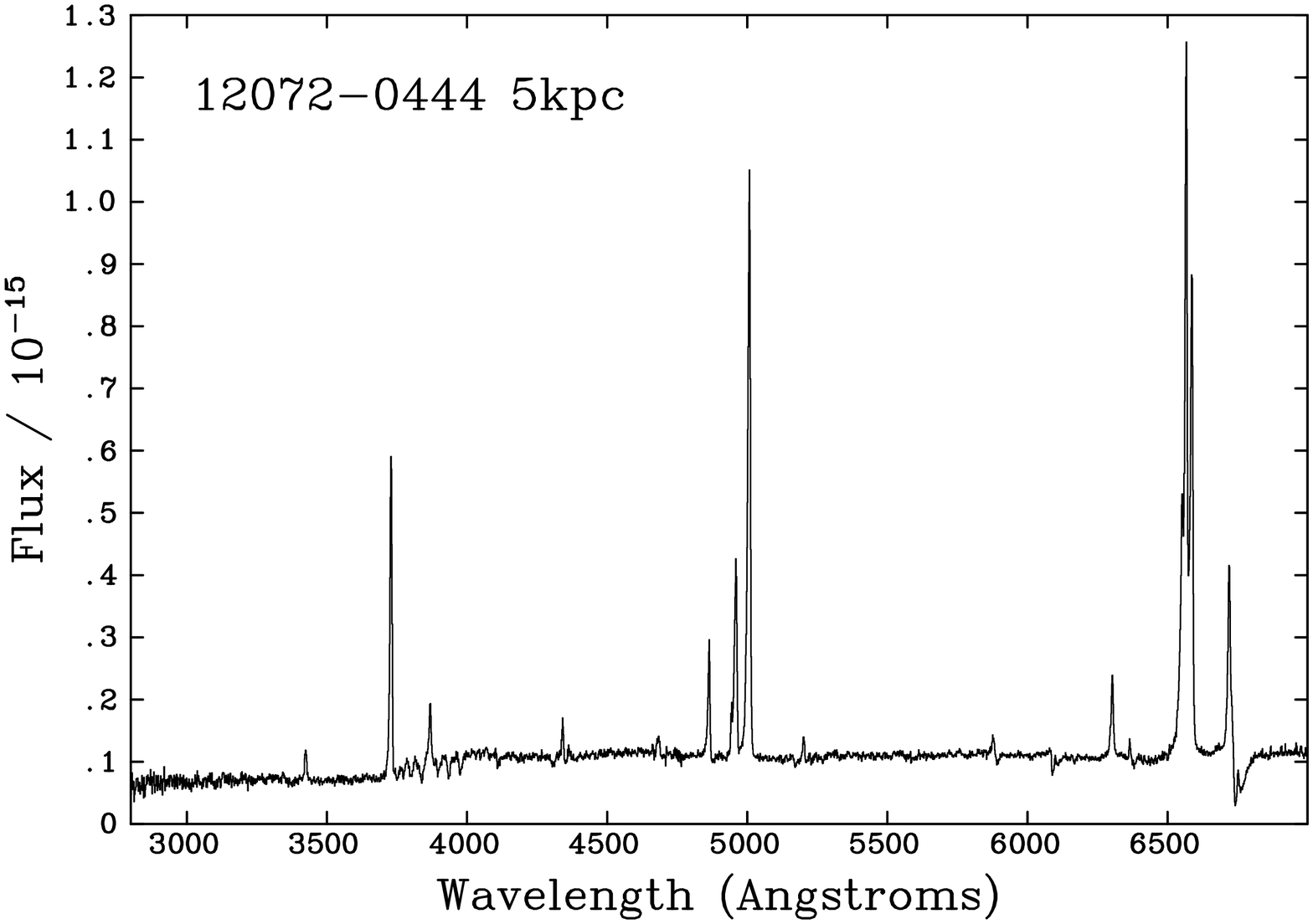,width=7.8cm,angle=0.}&
\psfig{file=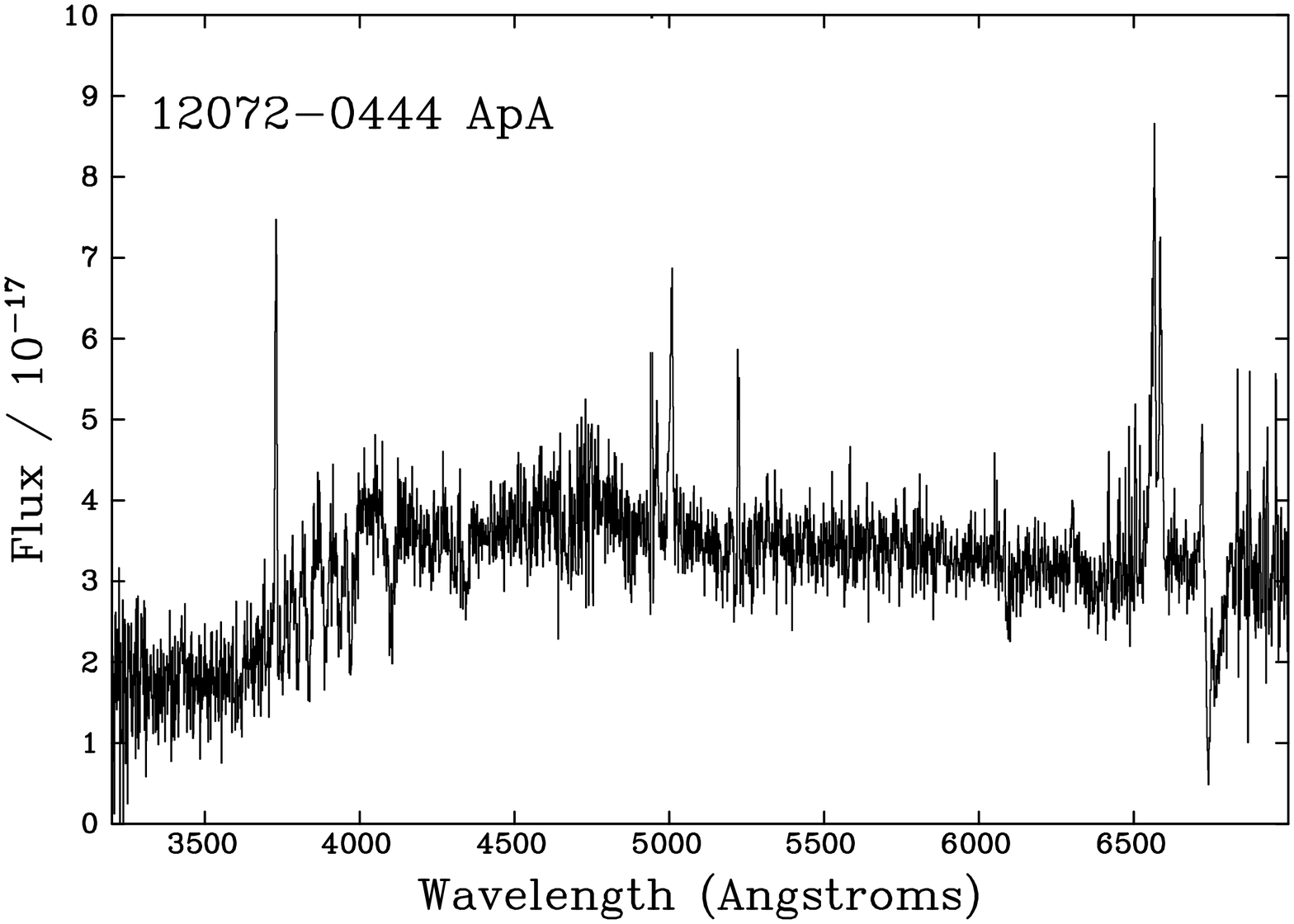,width=7.8cm,angle=0.}\\
\hspace*{0cm}\psfig{file=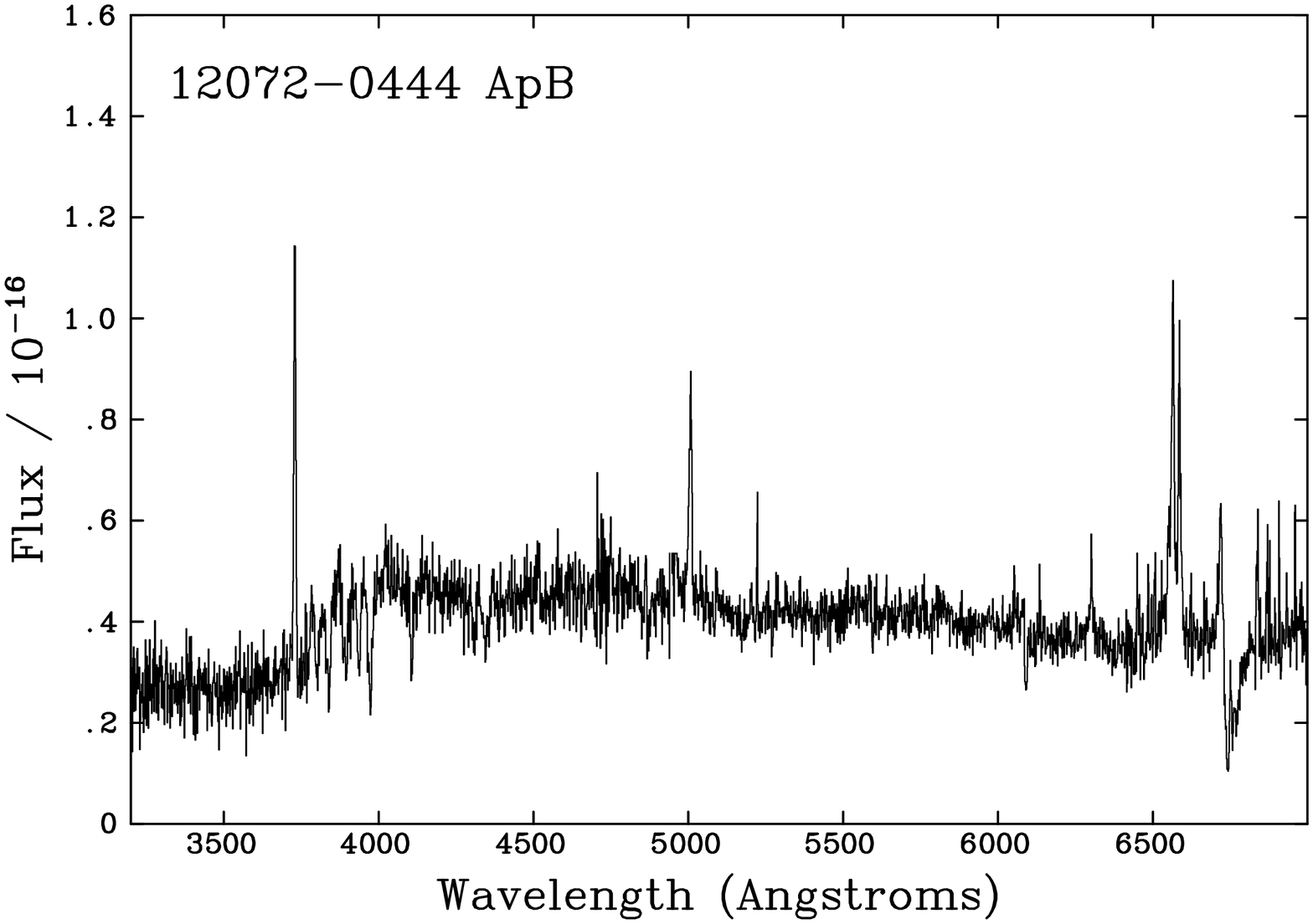,width=7.8cm,angle=0.}&
\psfig{file=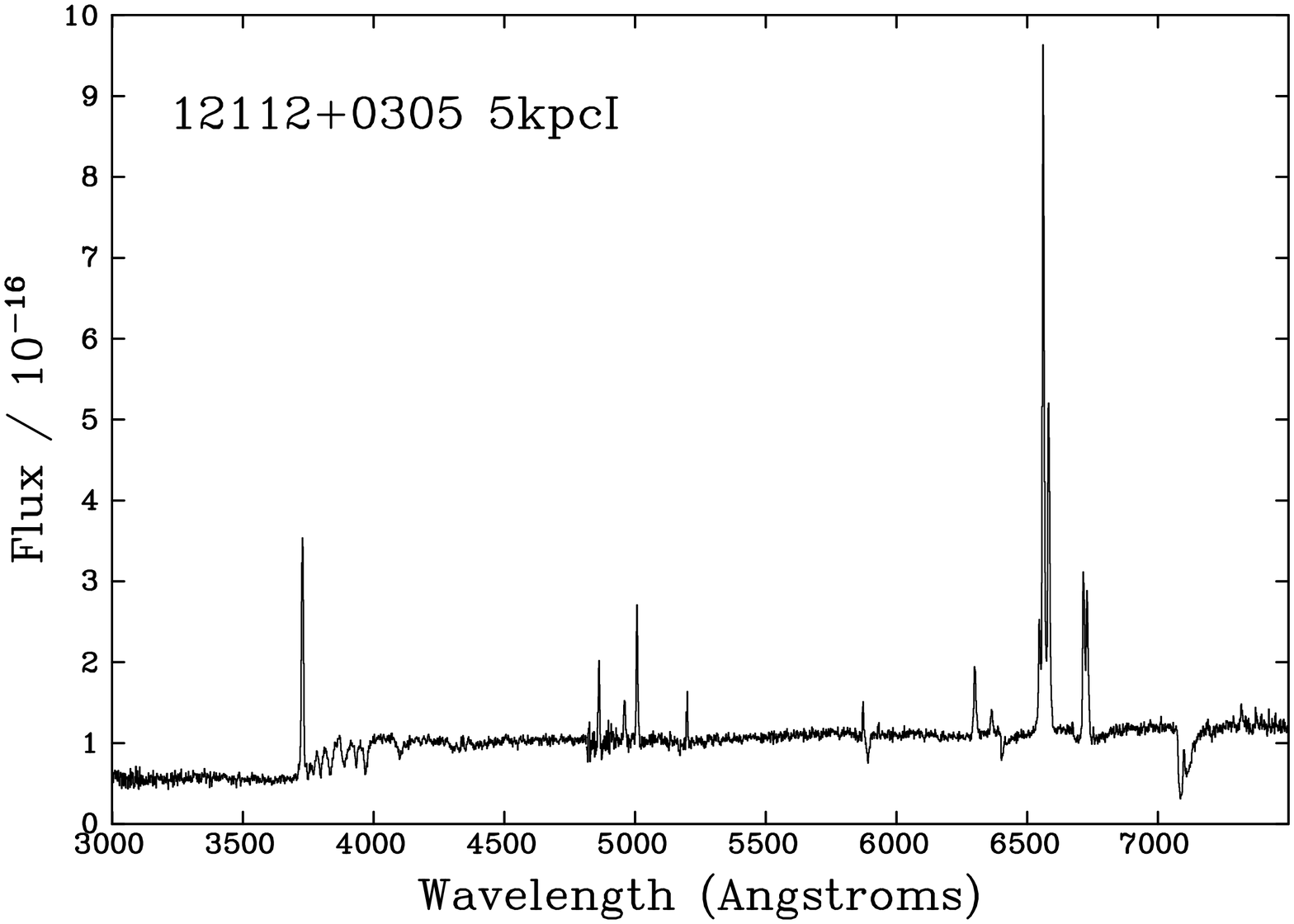,width=7.8cm,angle=0.}\\
\hspace*{0cm}\psfig{file=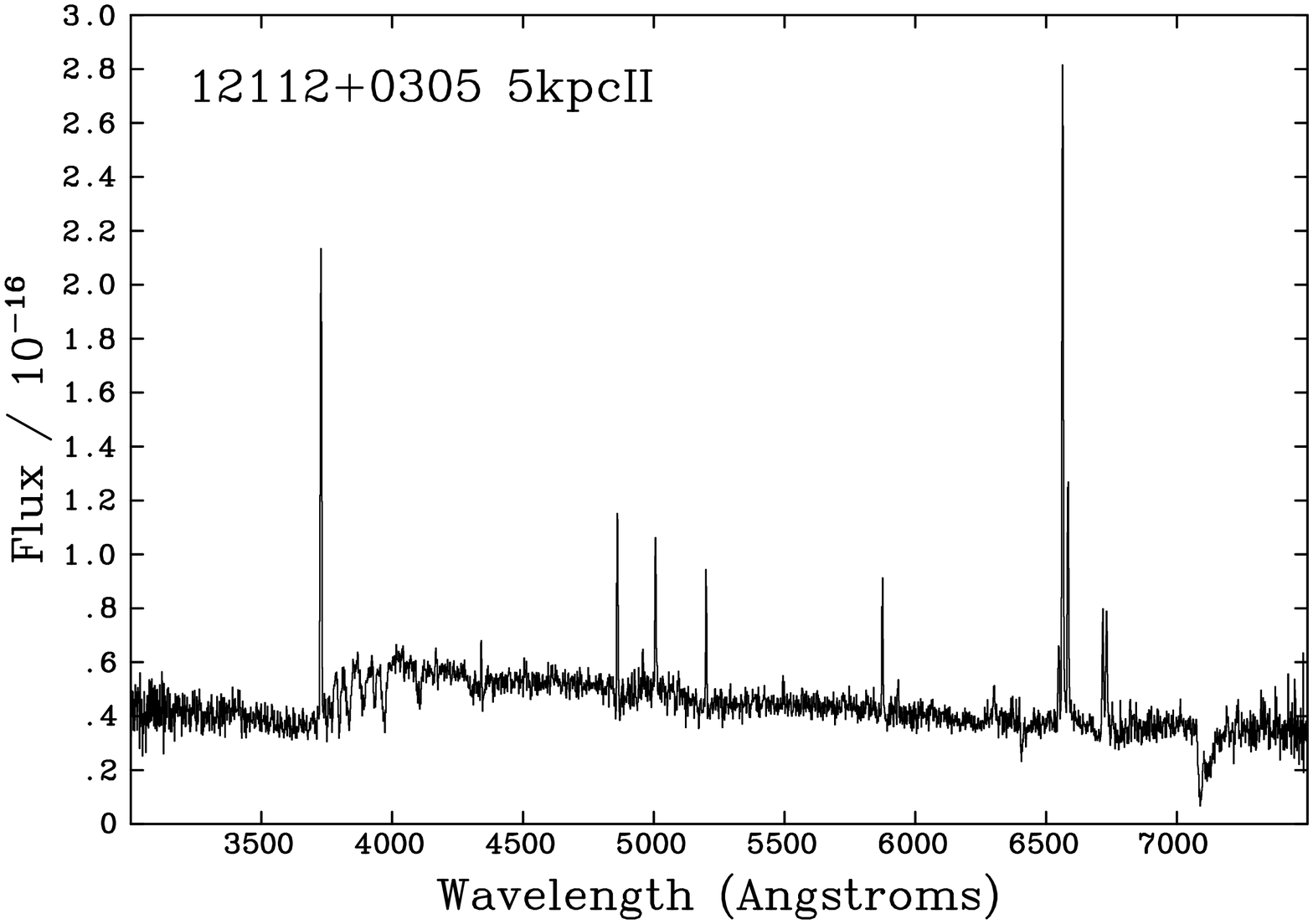,width=7.8cm,angle=0.}&
\psfig{file=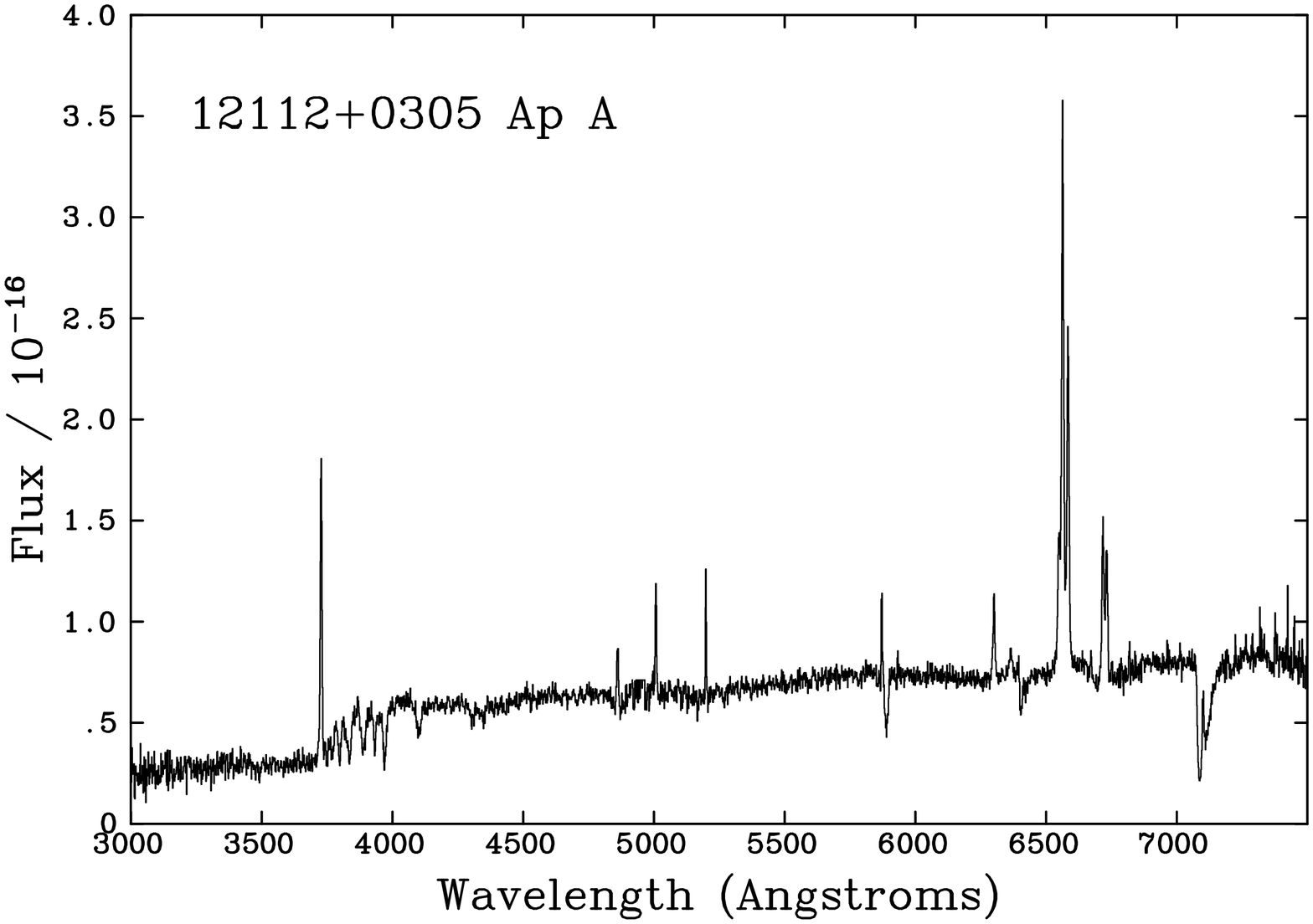,width=7.8cm,angle=0.}\\
\hspace*{0cm}\psfig{file=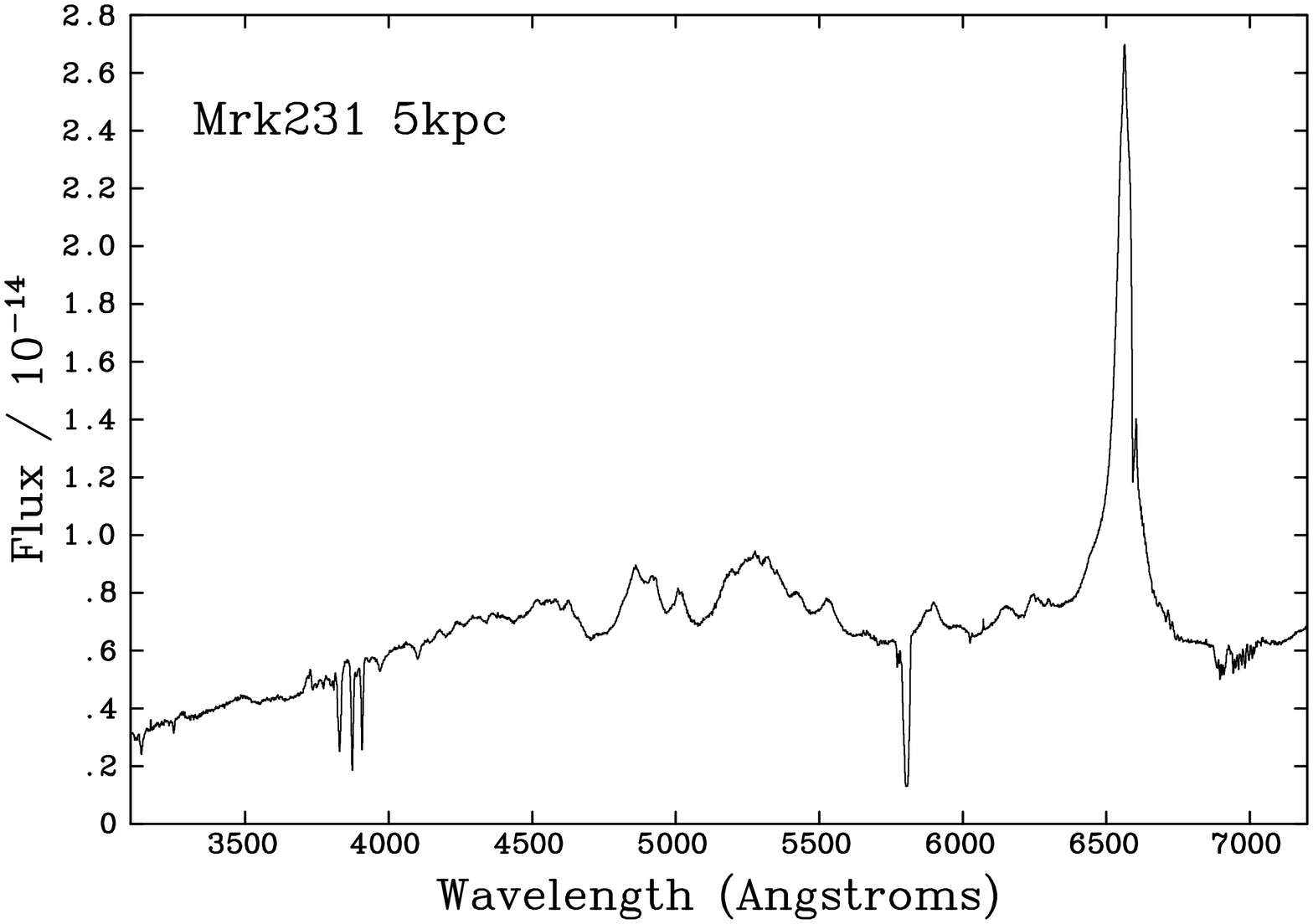,width=5.5cm,angle=-90.}&
\psfig{file=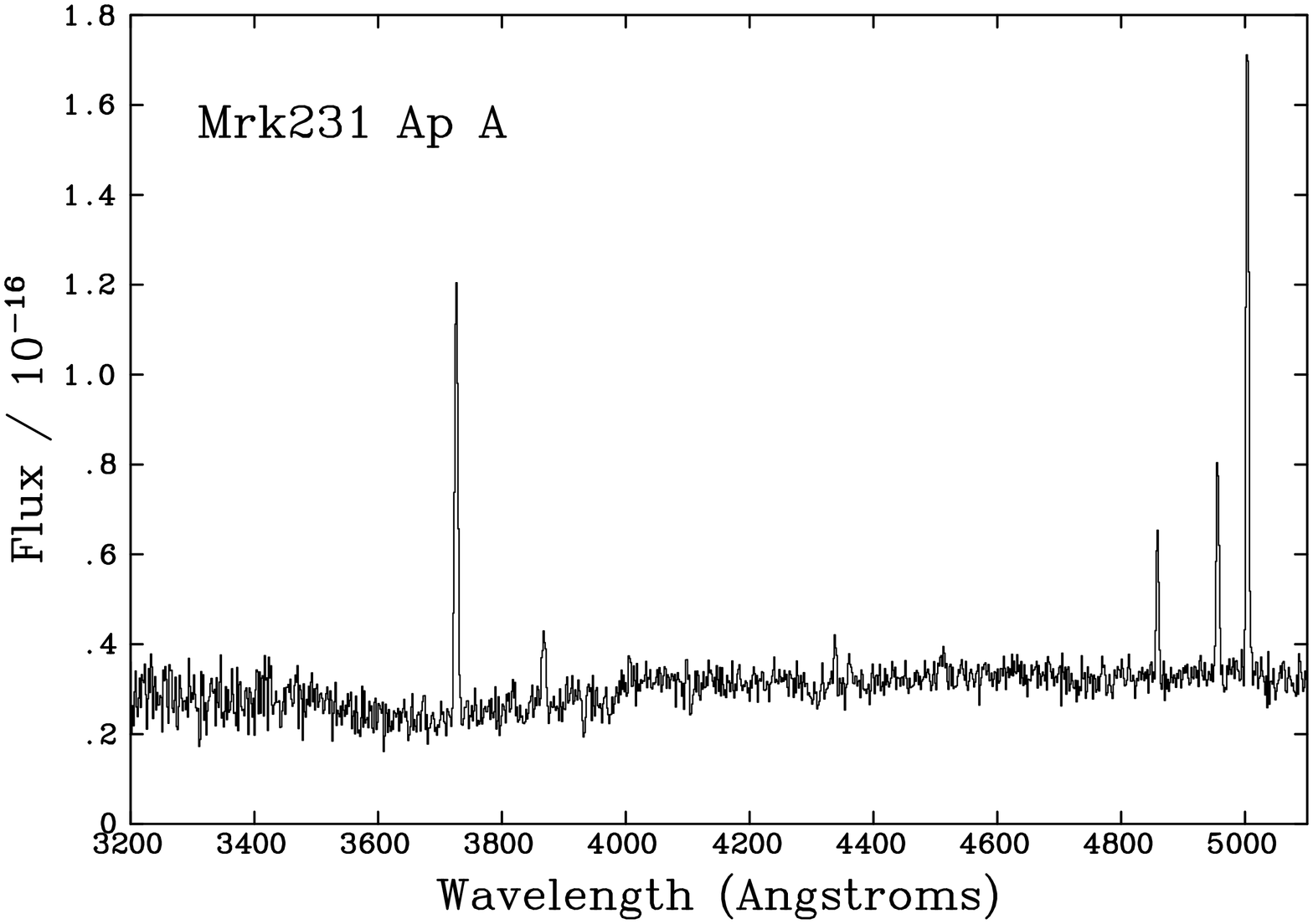,width=5.5cm,angle=-90.}
\end{tabular}
\caption[{\it Continued}]{Continued}
%\label{fig:SED}
\end{minipage}
\end{figure*}
\addtocounter{figure}{-1}
\begin{figure*}
\begin{minipage}{170mm}
\begin{tabular}{cc}
\hspace*{0cm}\psfig{file=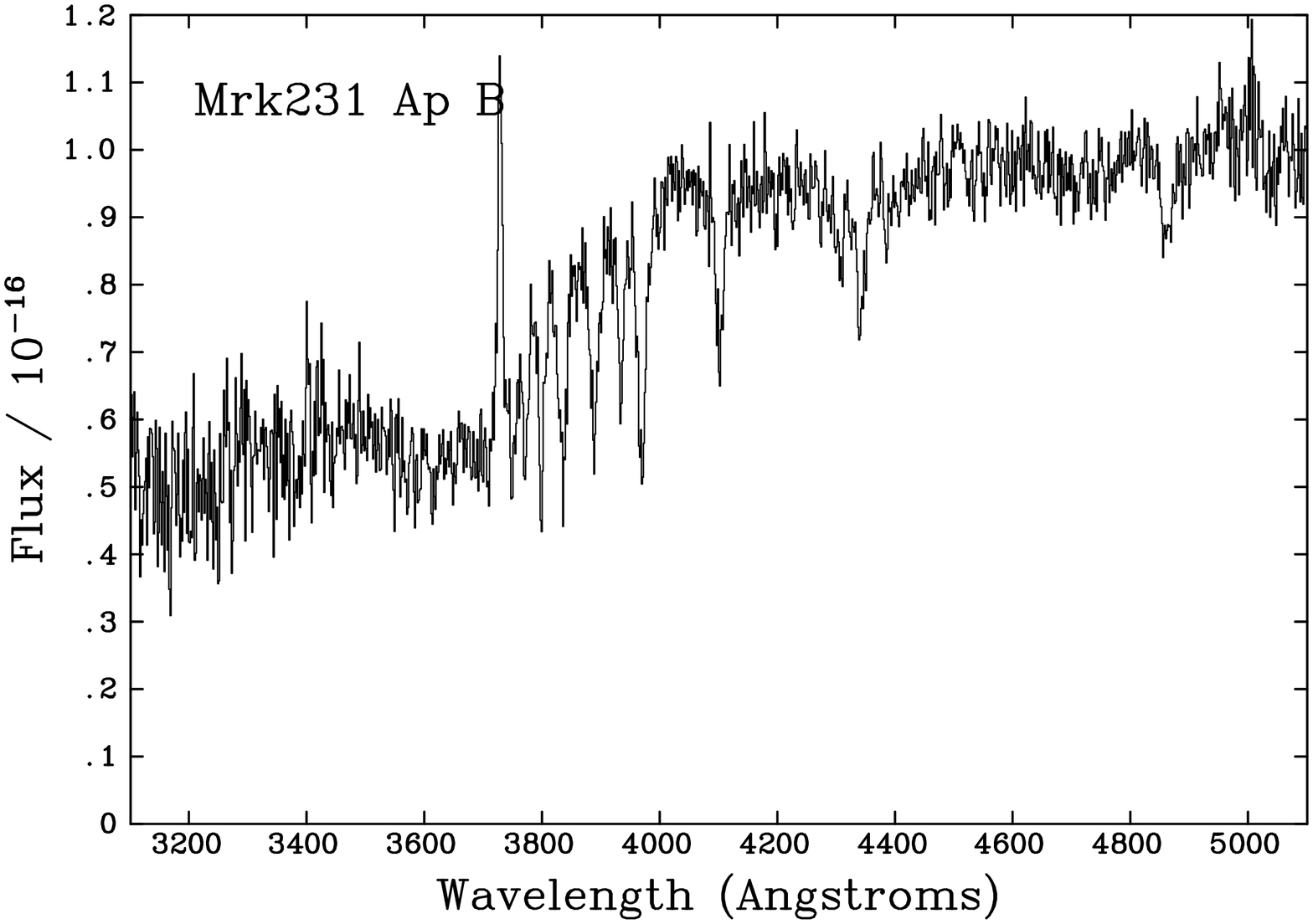,width=5.5cm,angle=-90.}&
\psfig{file=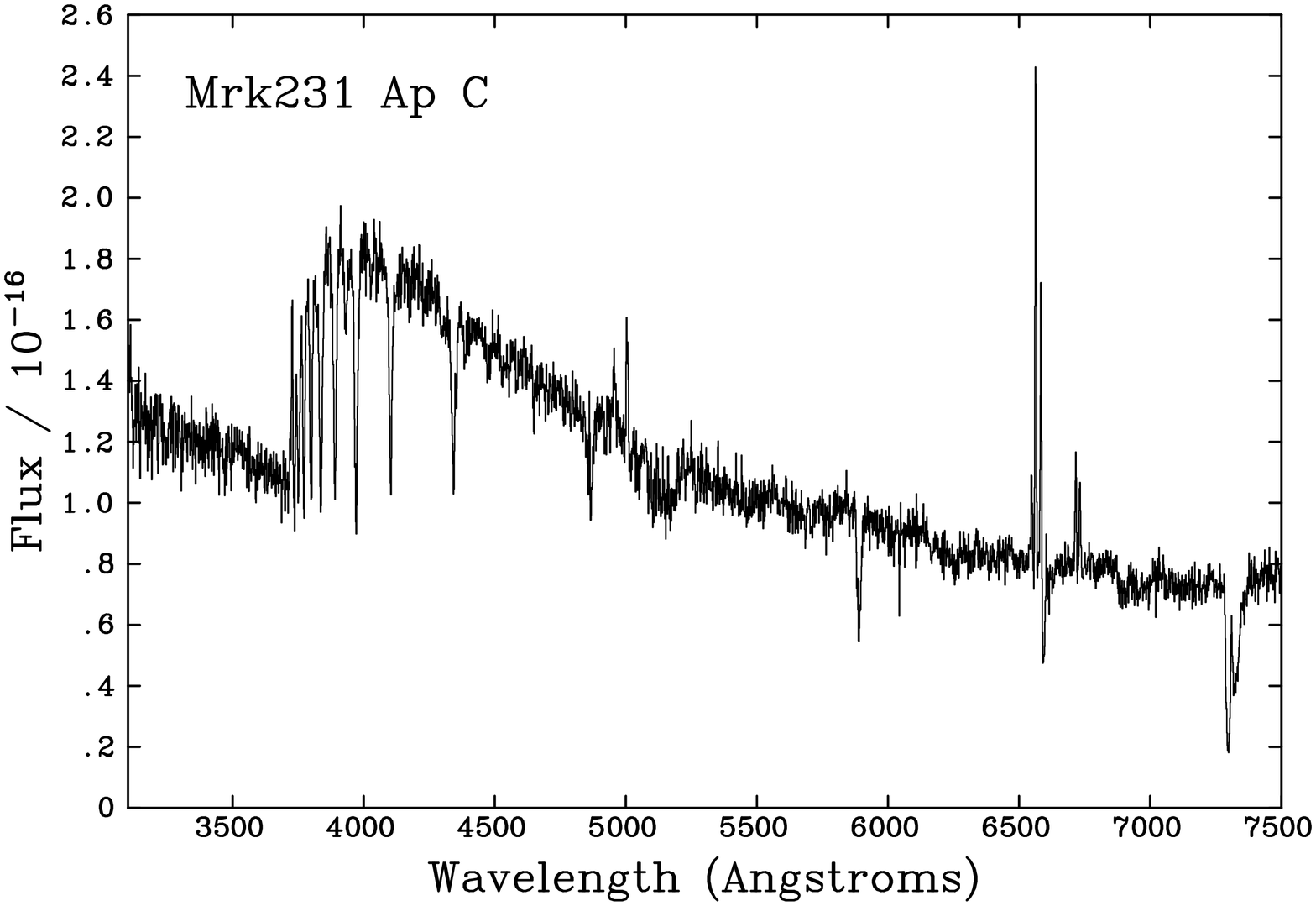,width=5.5cm,angle=-90.}\\
\hspace*{0cm}\psfig{file=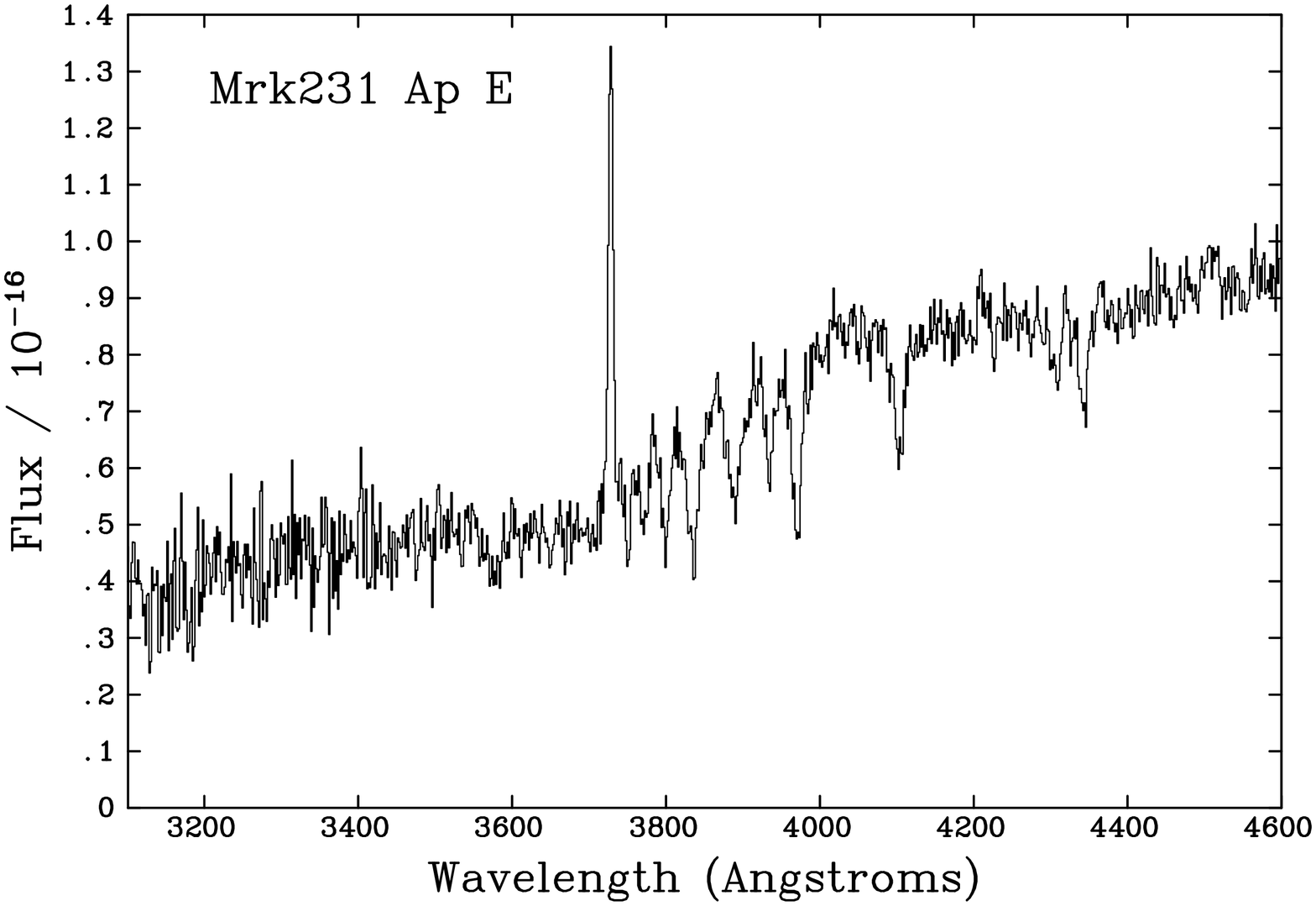,width=5.5cm,angle=-90.}&
\psfig{file=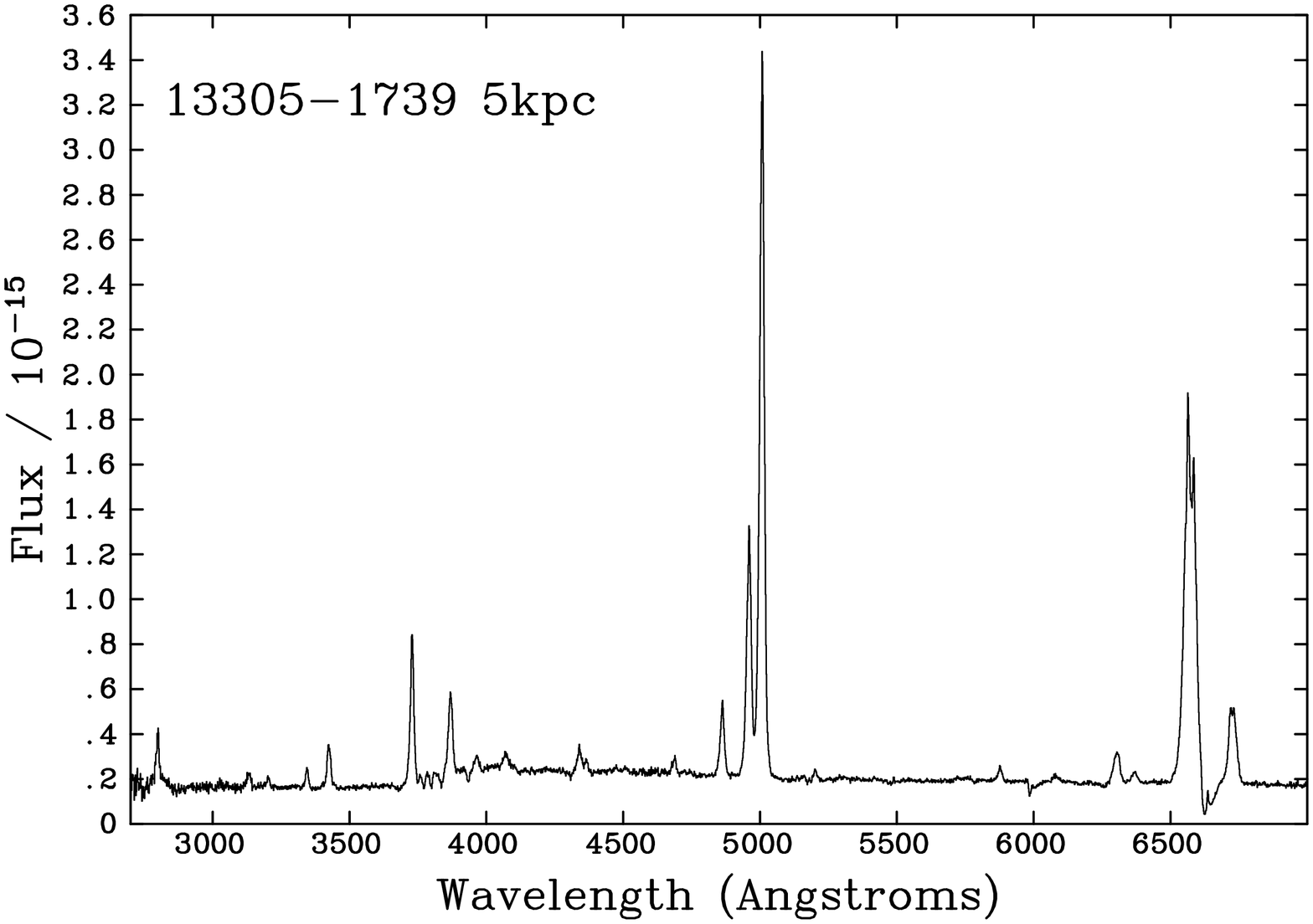,width=5.5cm,angle=-90.}\\
\hspace*{0cm}\psfig{file=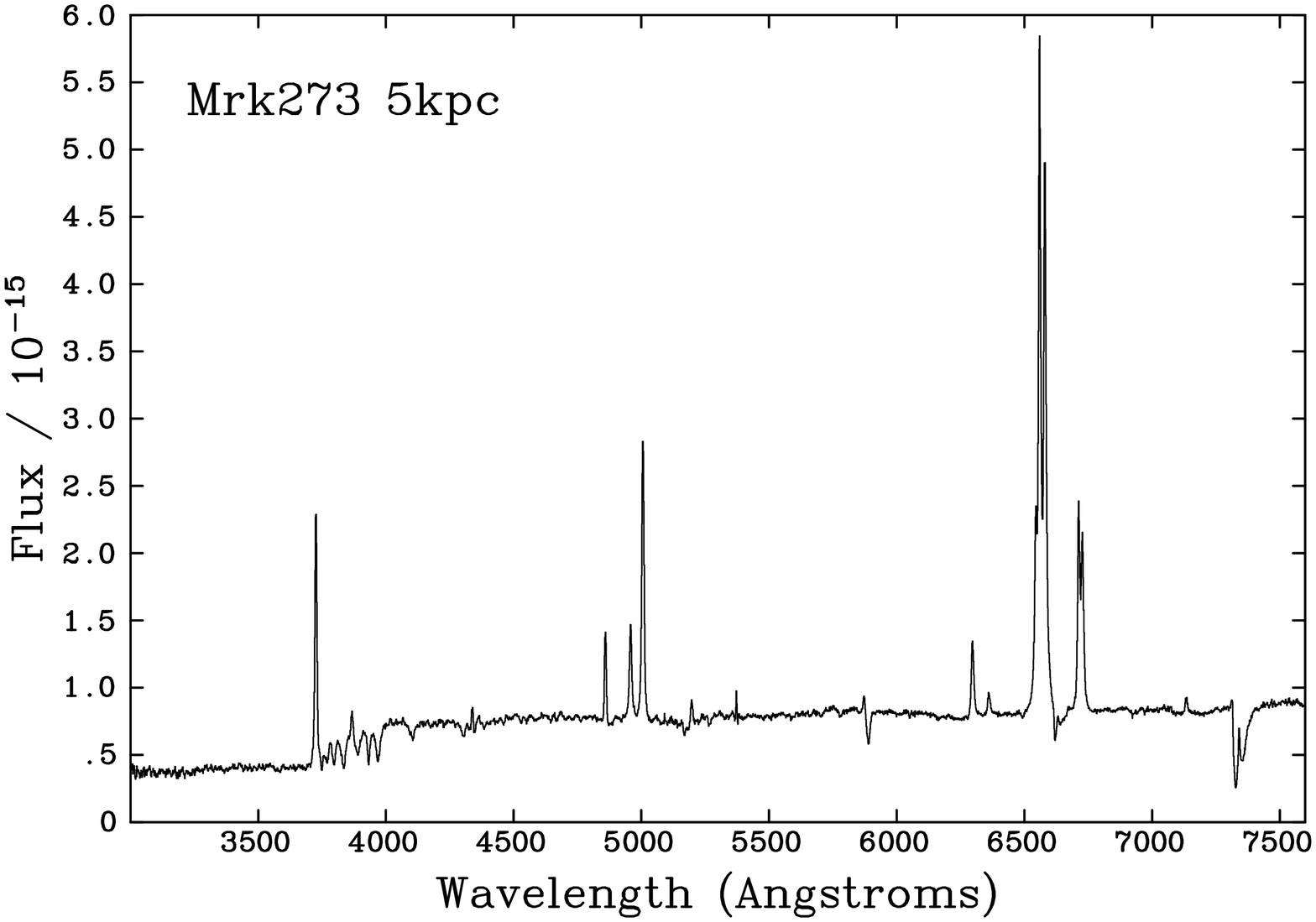,width=7.8cm,angle=0.}&
\psfig{file=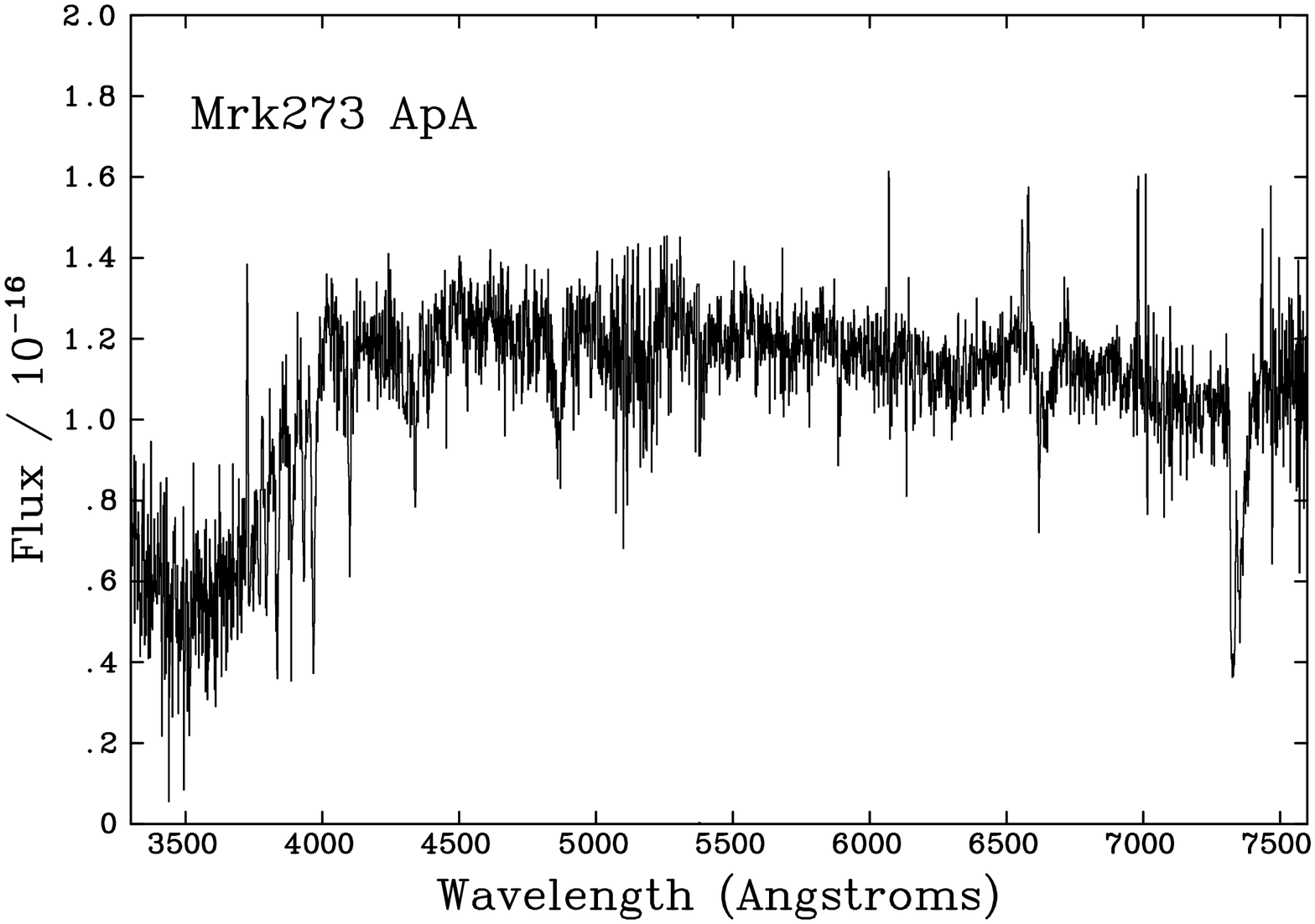,width=7.8cm,angle=0.}\\
\hspace*{0cm}\psfig{file=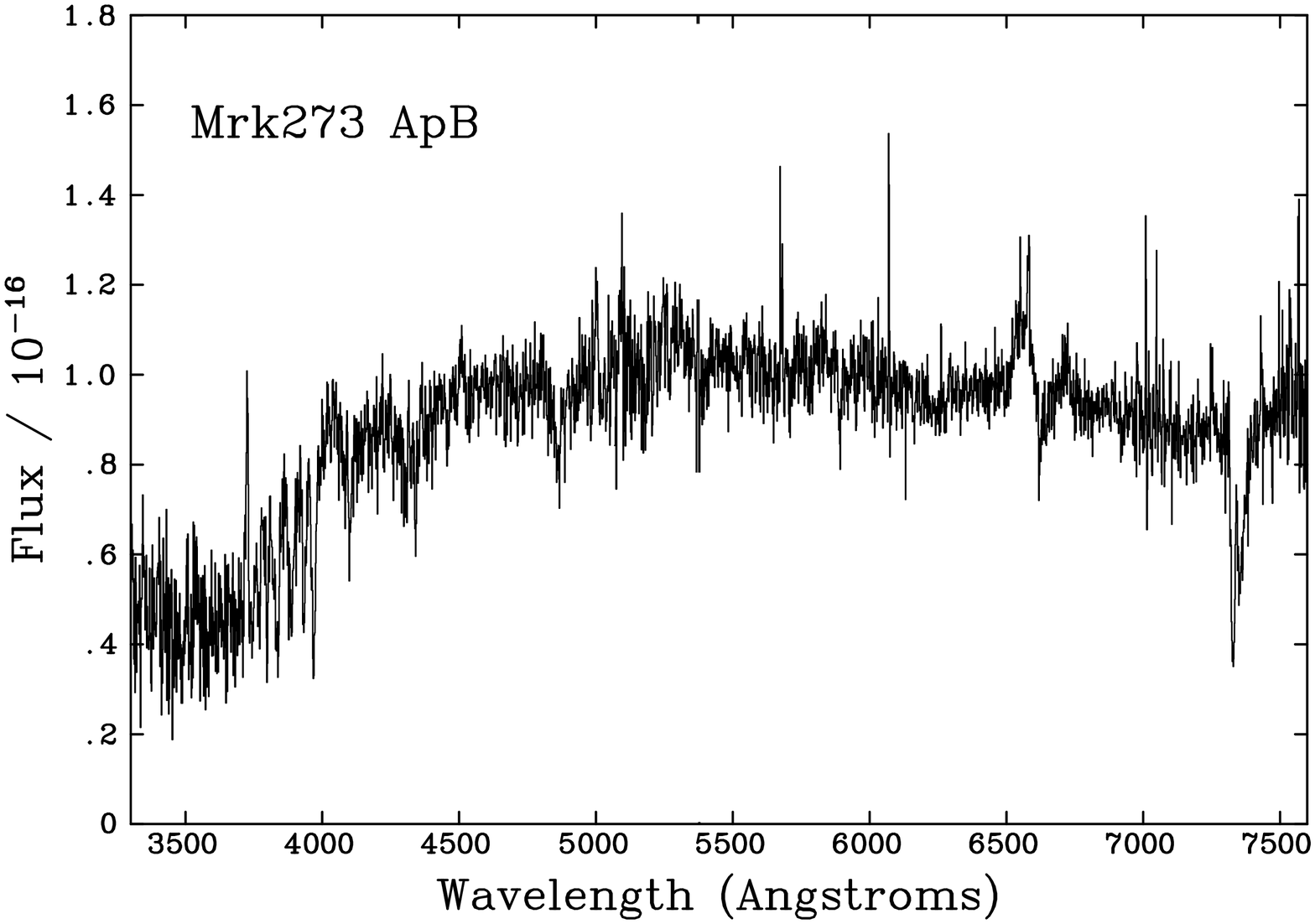,width=7.8cm,angle=0.}&
\psfig{file=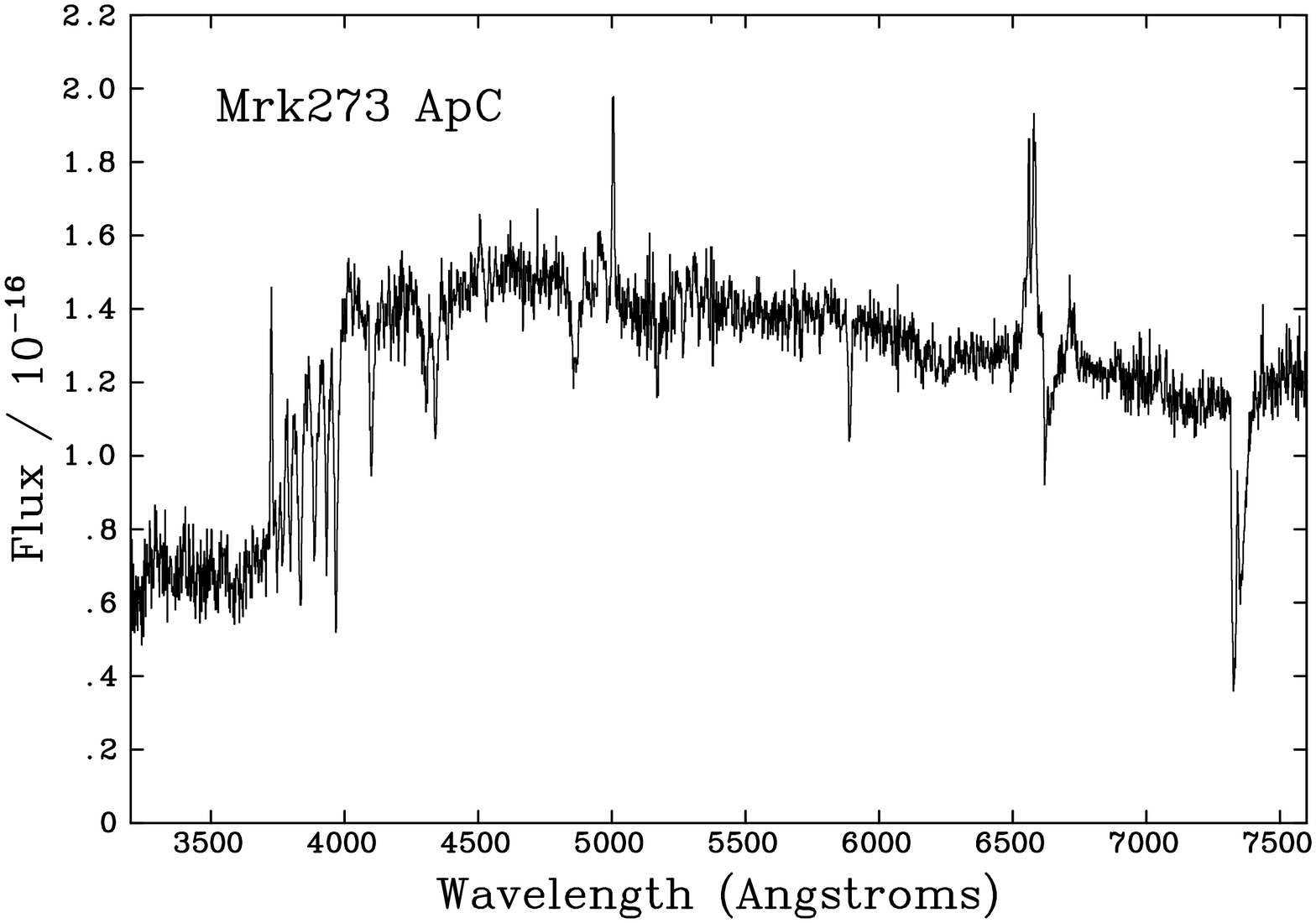,width=7.8cm,angle=-0.}
\end{tabular}
\caption[{\it Continued}]{Continued}
%\label{fig:SED}
\end{minipage}
\end{figure*}
\addtocounter{figure}{-1}
\begin{figure*}
\begin{minipage}{170mm}
\begin{tabular}{cc}
\hspace*{0cm}\psfig{file=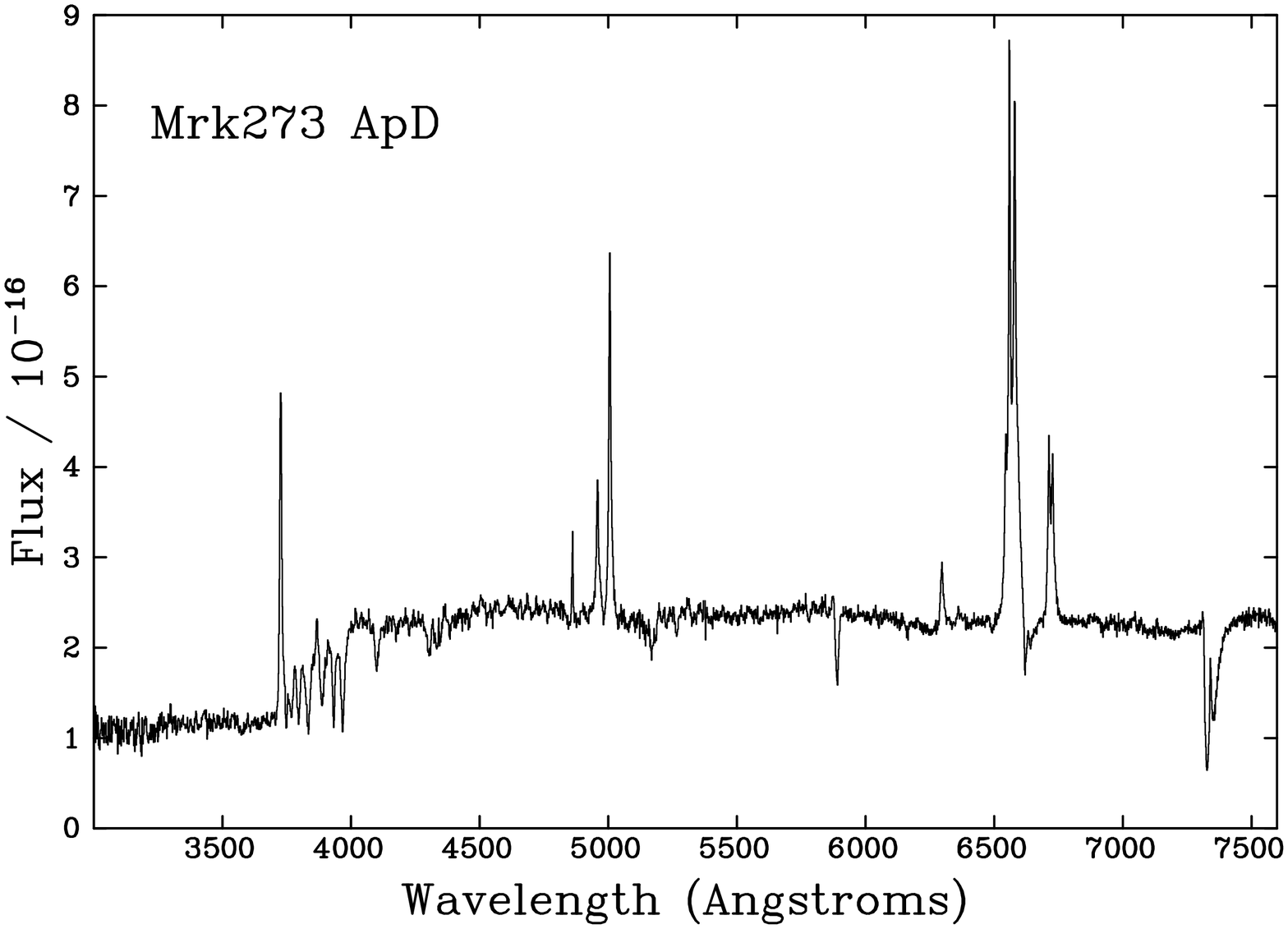,width=7.8cm,angle=0.}&
\psfig{file=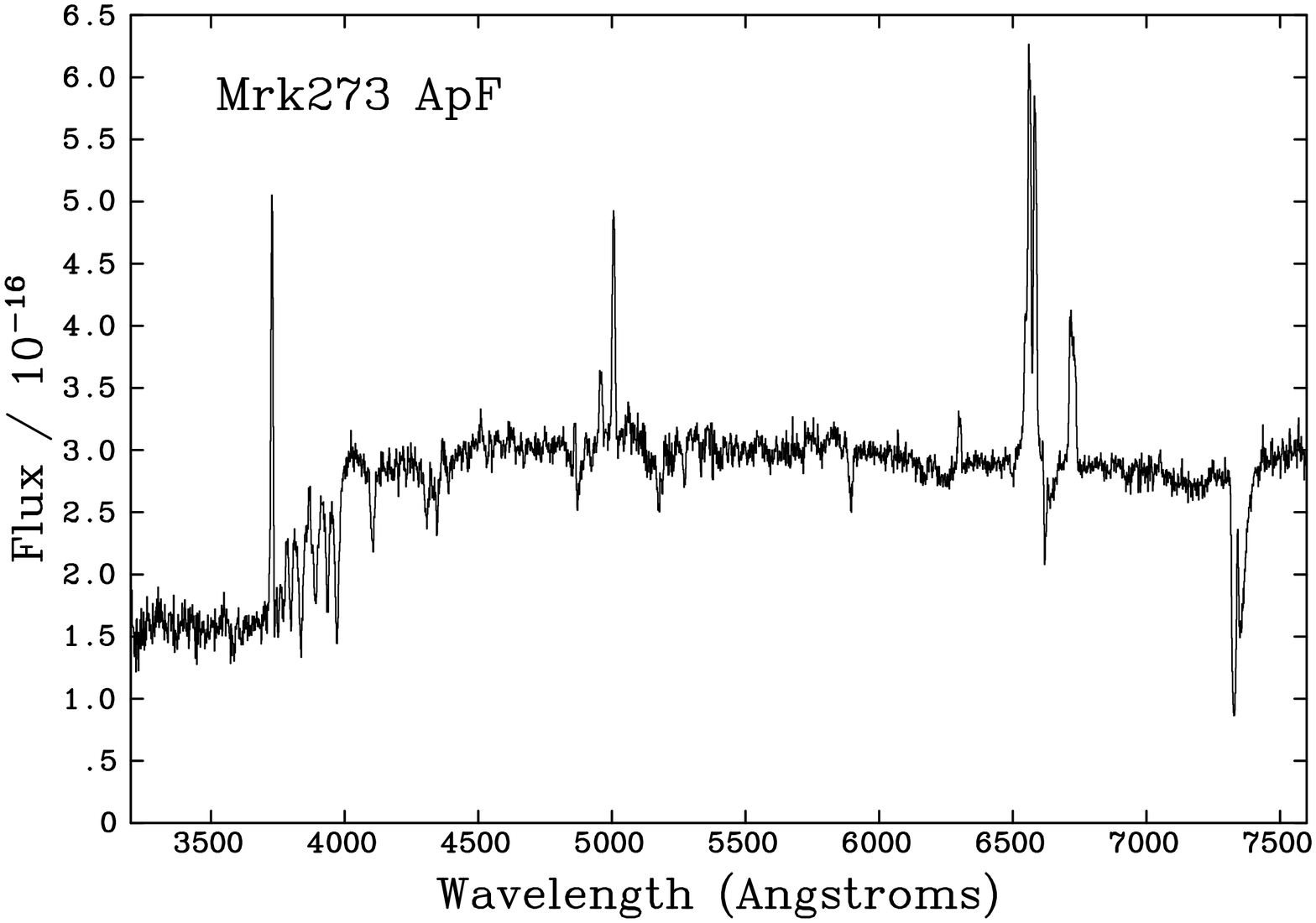,width=7.8cm,angle=0.}\\
\hspace*{0cm}\psfig{file=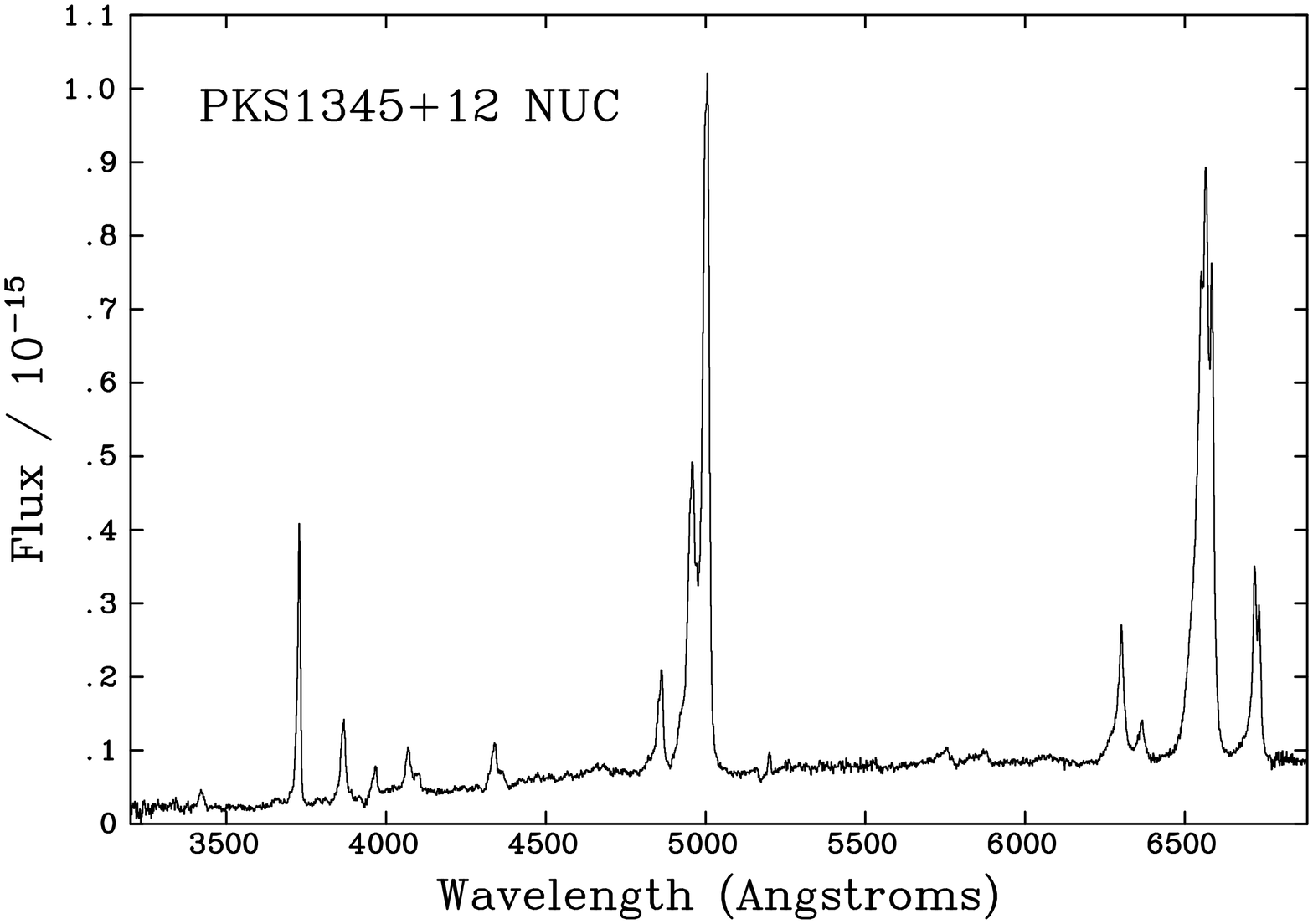,width=5.5cm,angle=-90.}&
\psfig{file=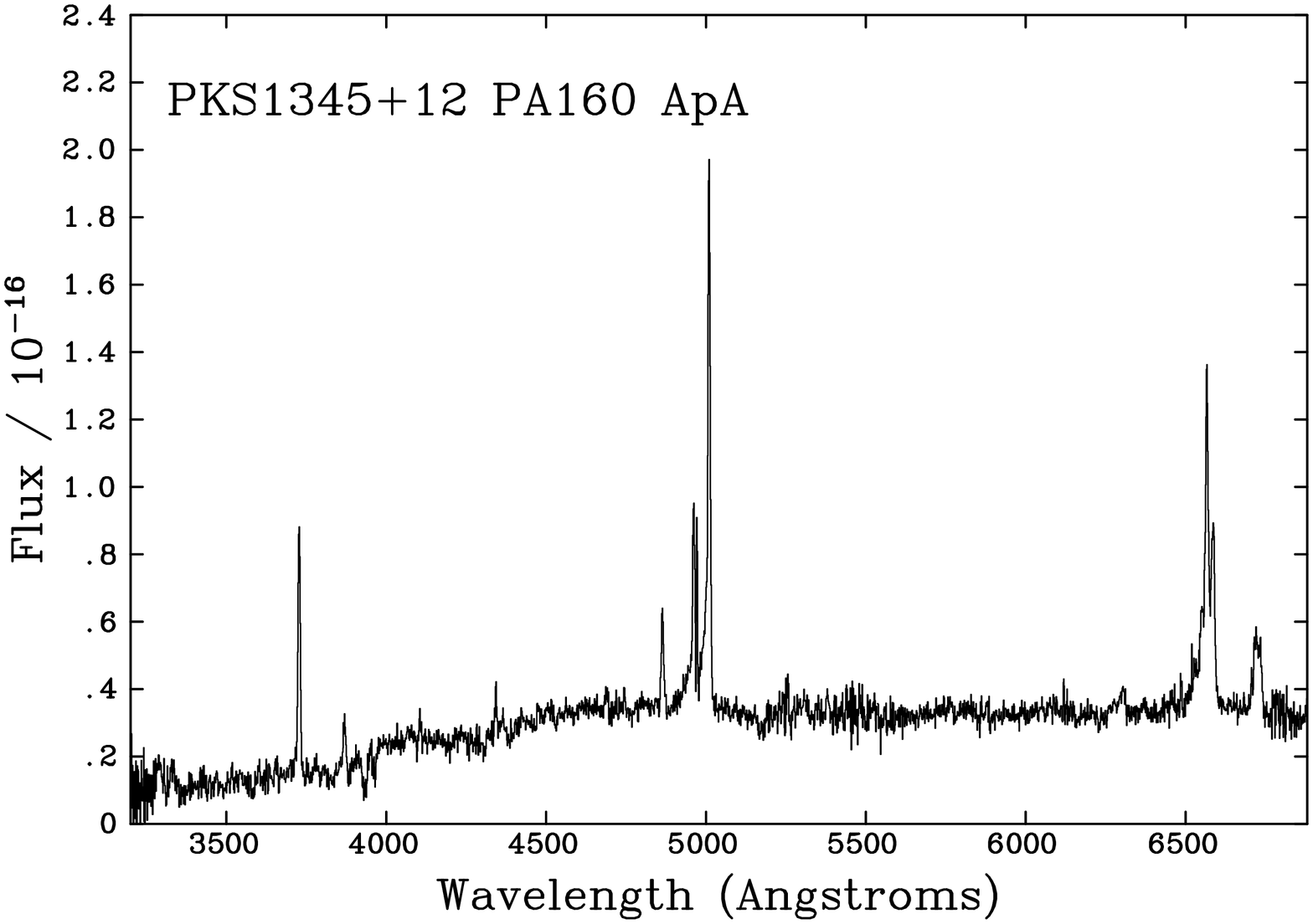,width=5.5cm,angle=-90.}\\
\hspace*{0cm}\psfig{file=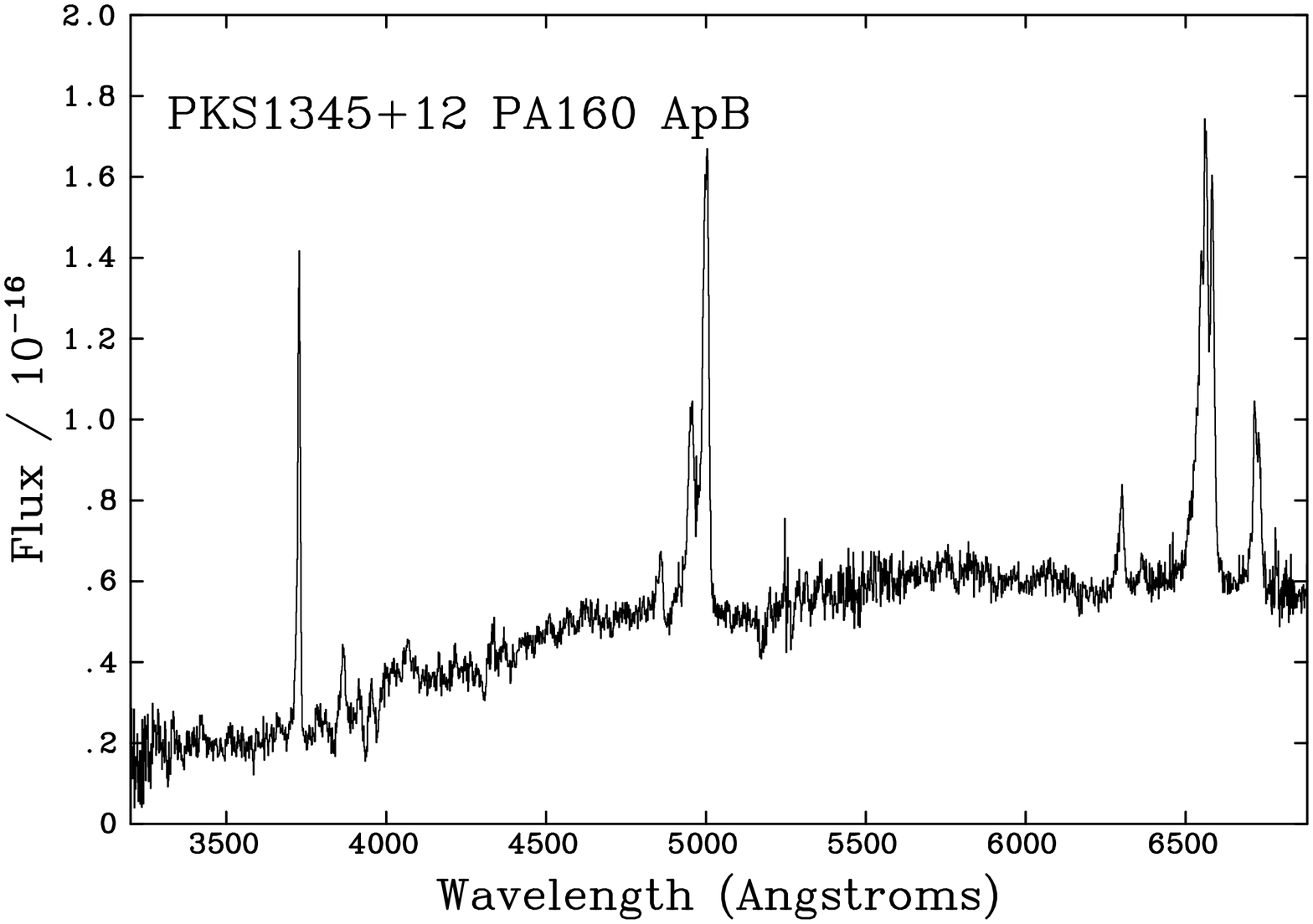,width=5.5cm,angle=-90.}&
\psfig{file=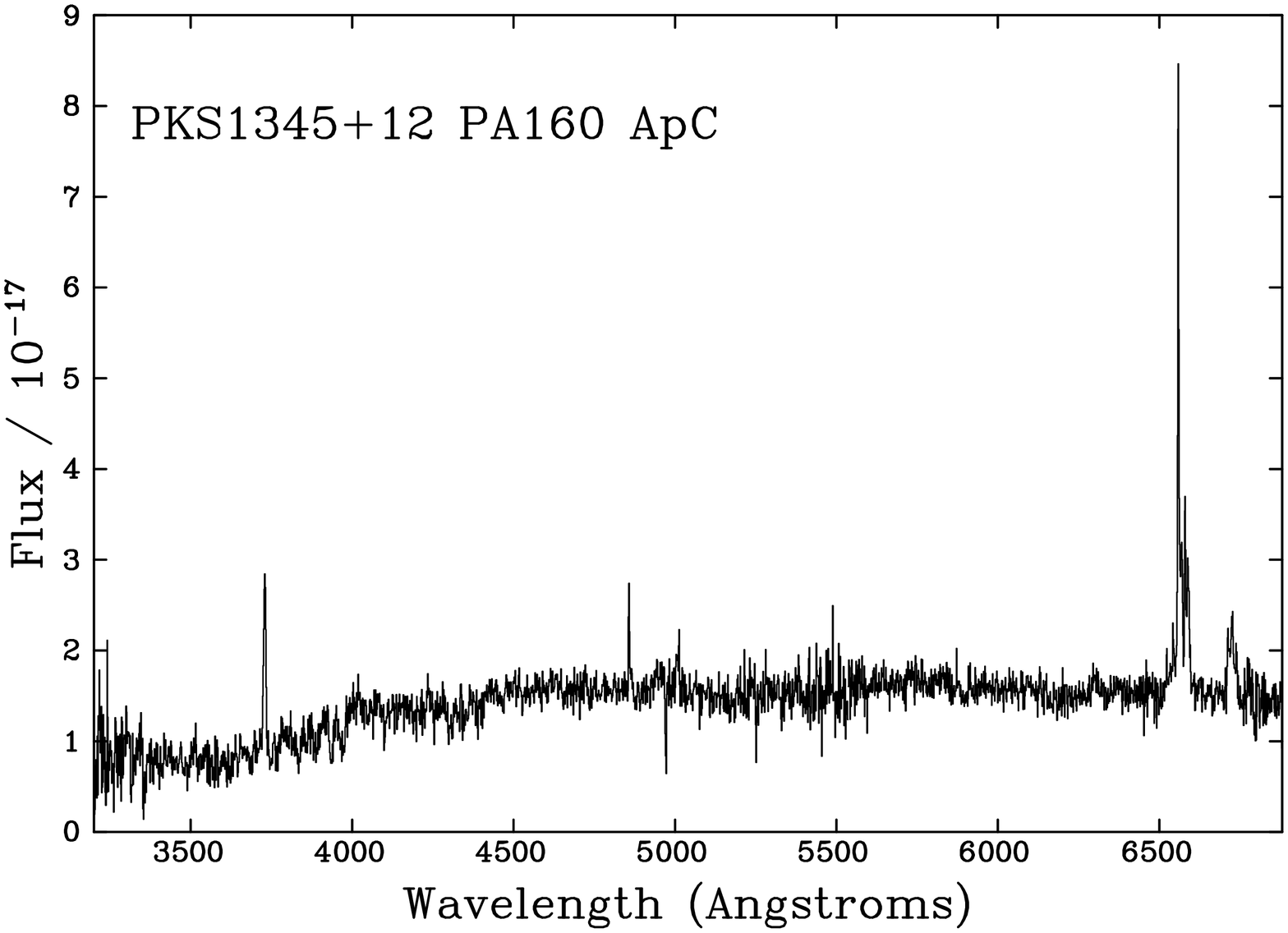,width=5.5cm,angle=-90.}\\
\hspace*{0cm}\psfig{file=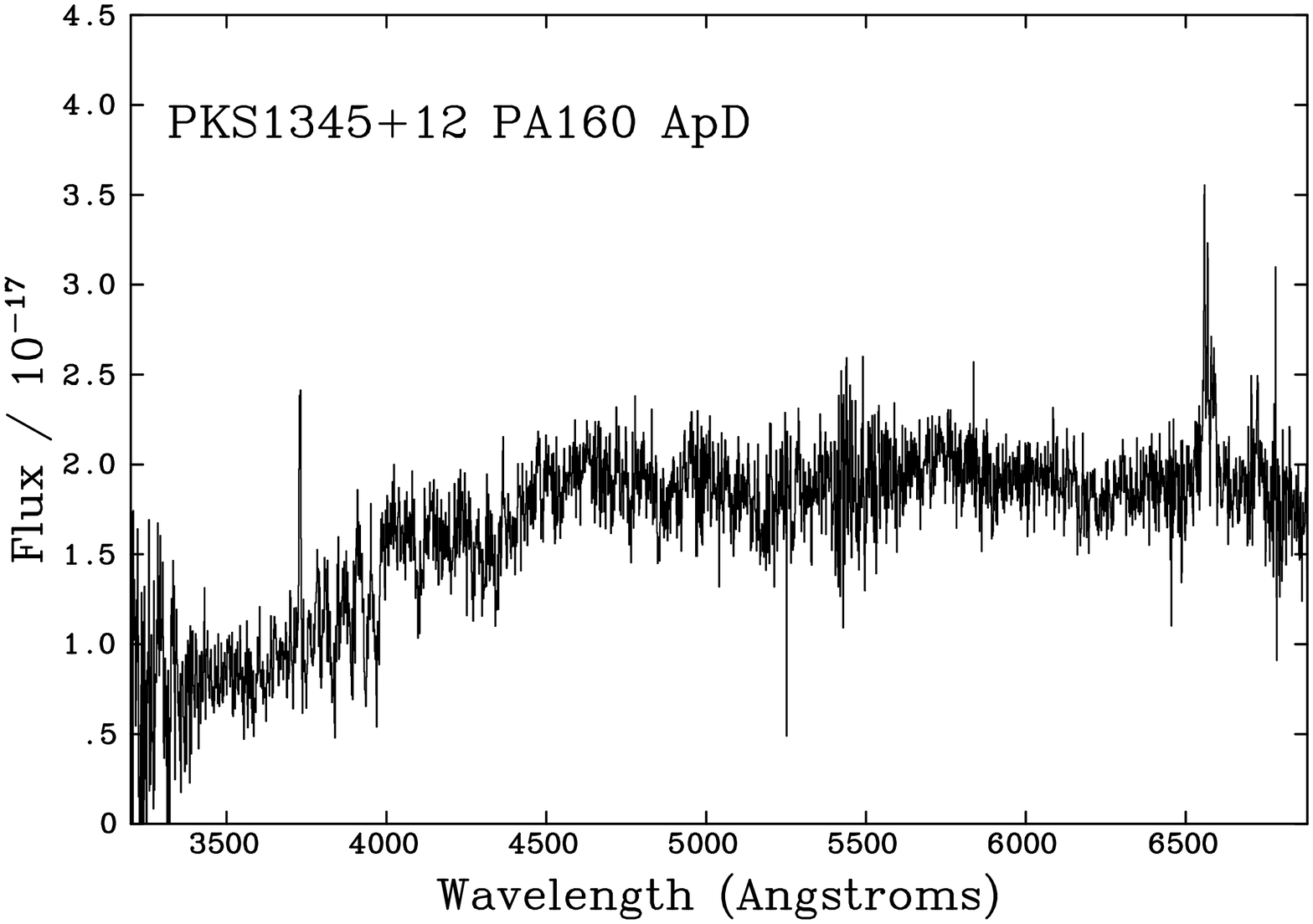,width=7.8cm,angle=0.}&
\psfig{file=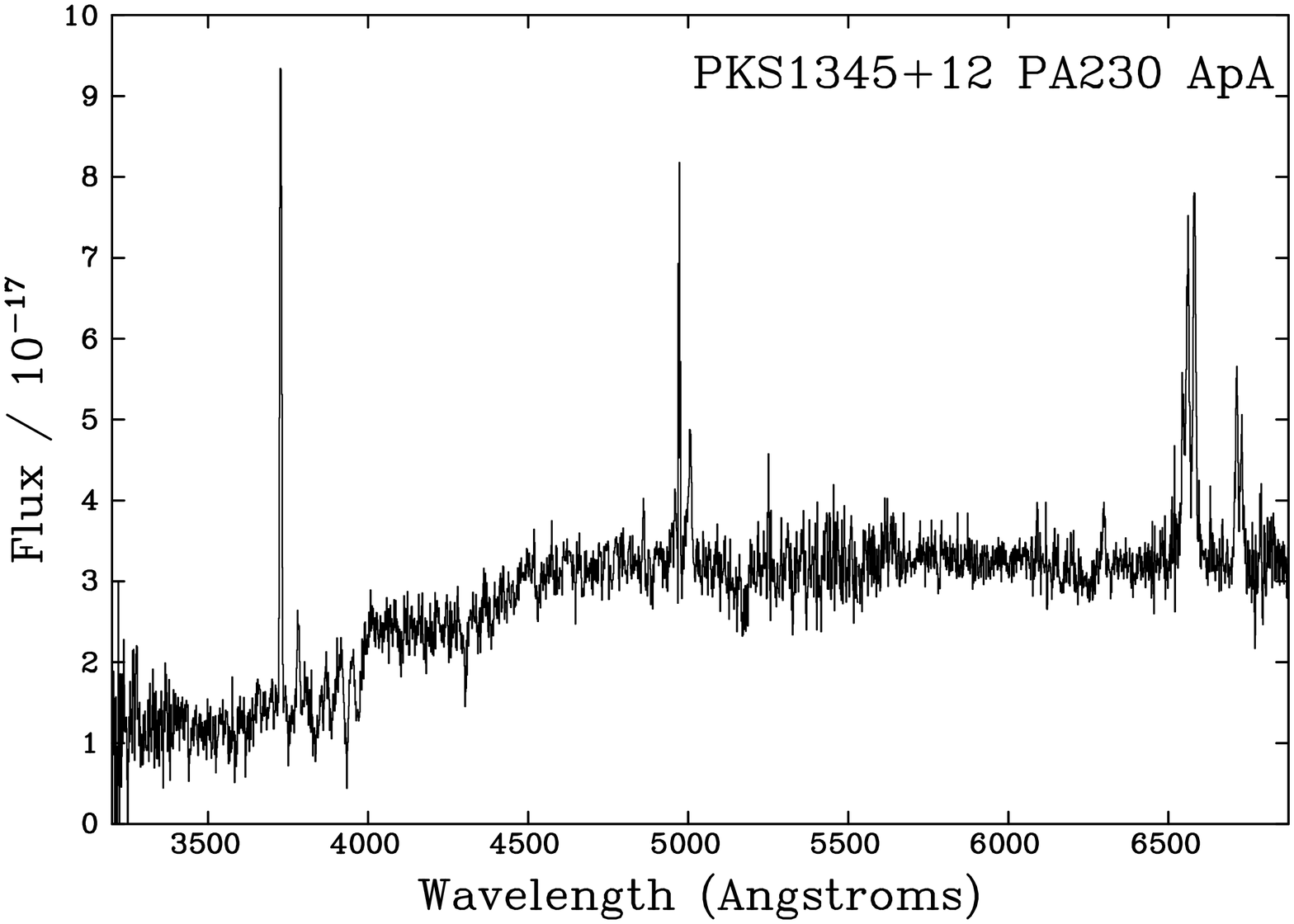,width=7.8cm,angle=0.}
\end{tabular}
\caption[{\it Continued}]{Continued}
%\label{fig:SED}
\end{minipage}
\end{figure*}
\addtocounter{figure}{-1}
\begin{figure*}
\begin{minipage}{170mm}
\begin{tabular}{cc}
\hspace*{0cm}\psfig{file=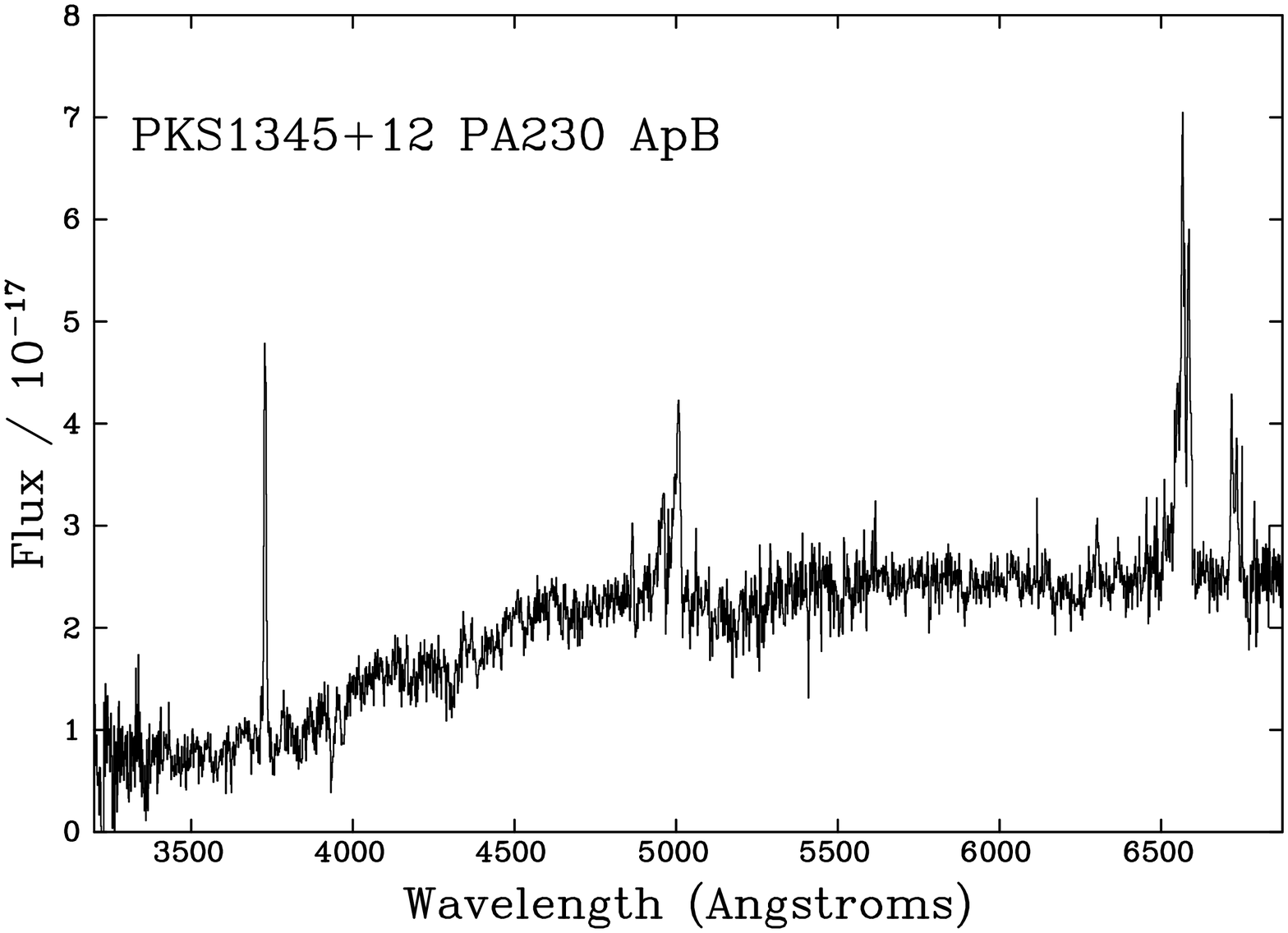,width=7.8cm,angle=0.}&
\psfig{file=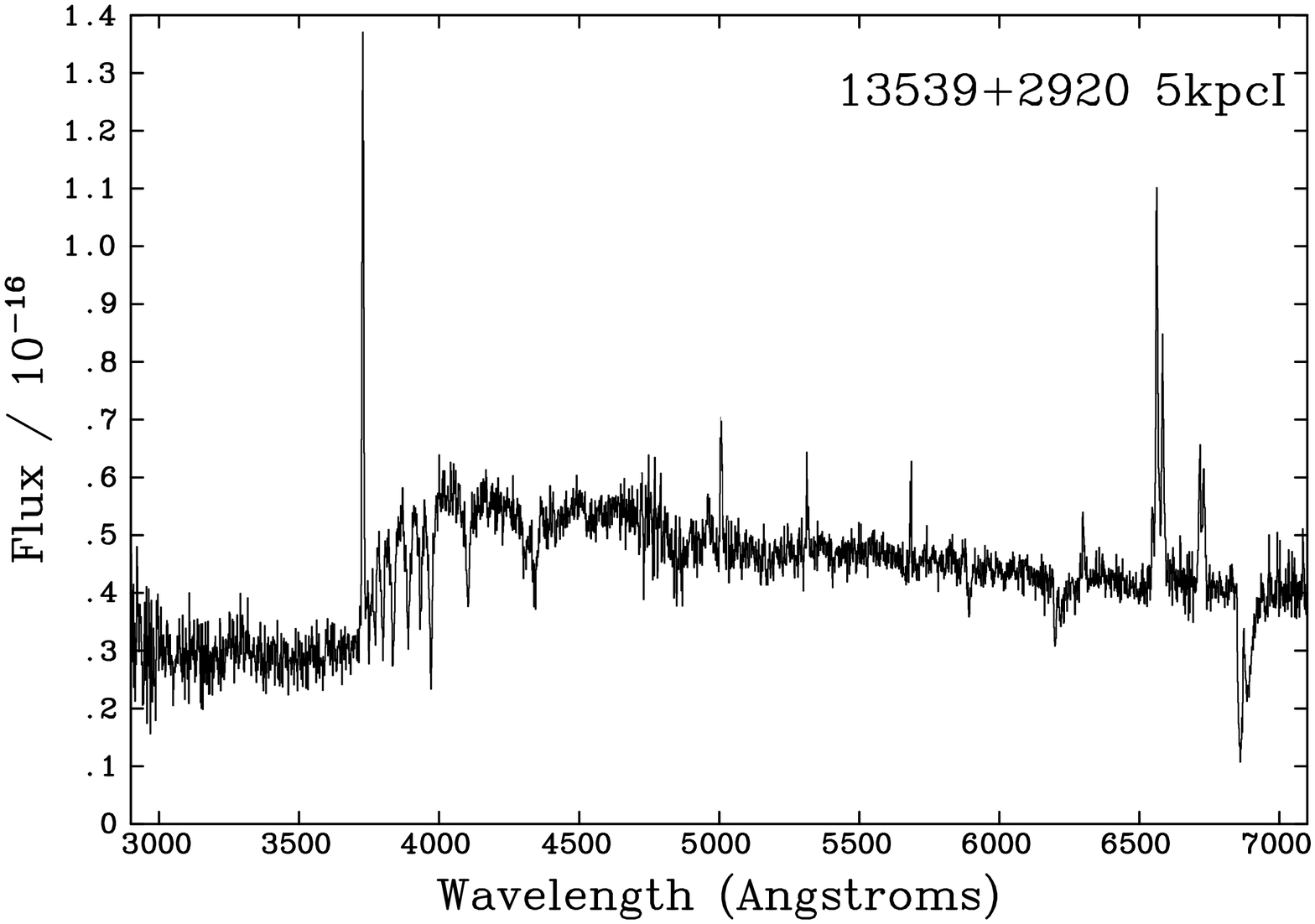,width=7.8cm,angle=0.}\\
\hspace*{0cm}\psfig{file=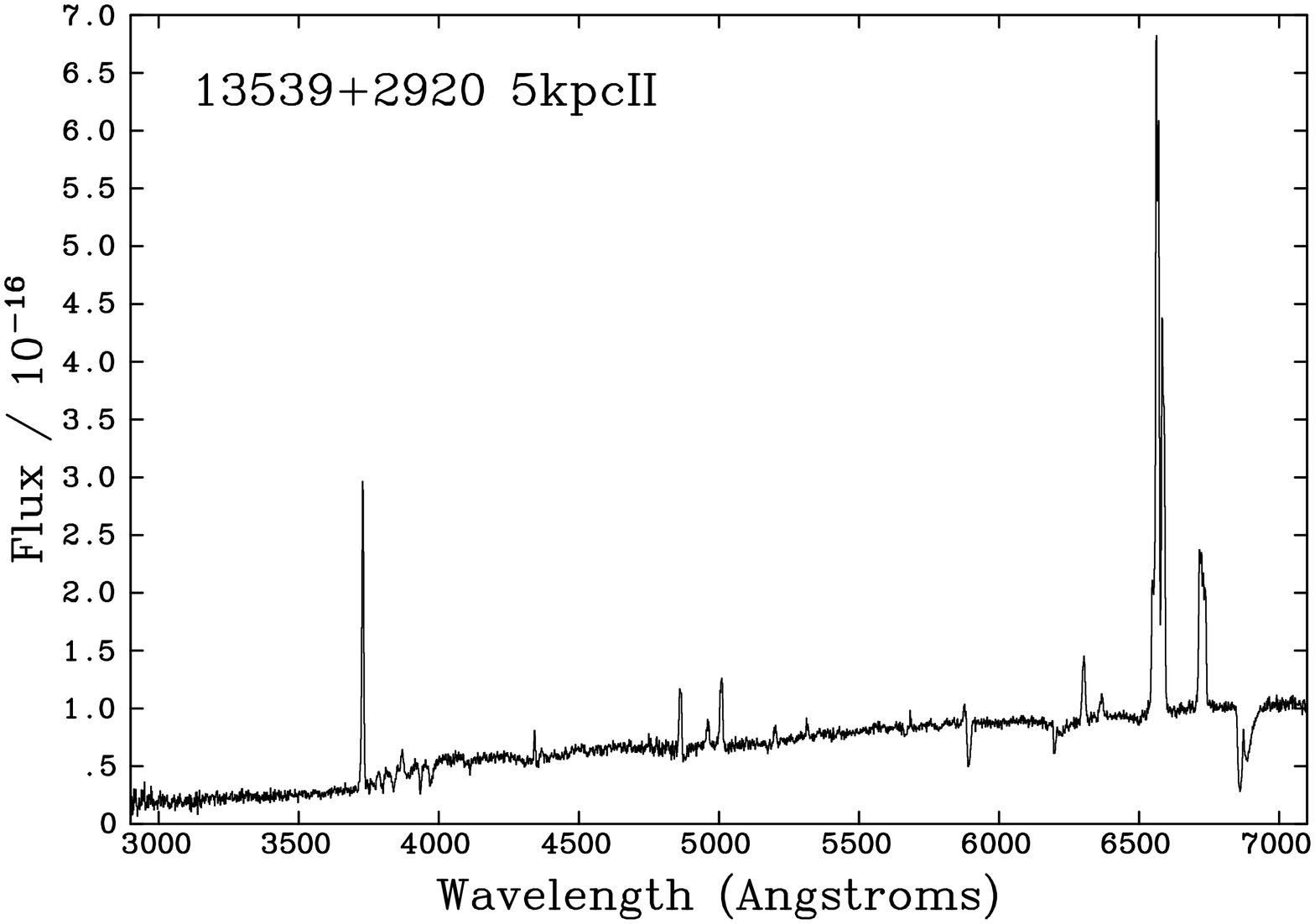,width=5.5cm,angle=-90.}&
\psfig{file=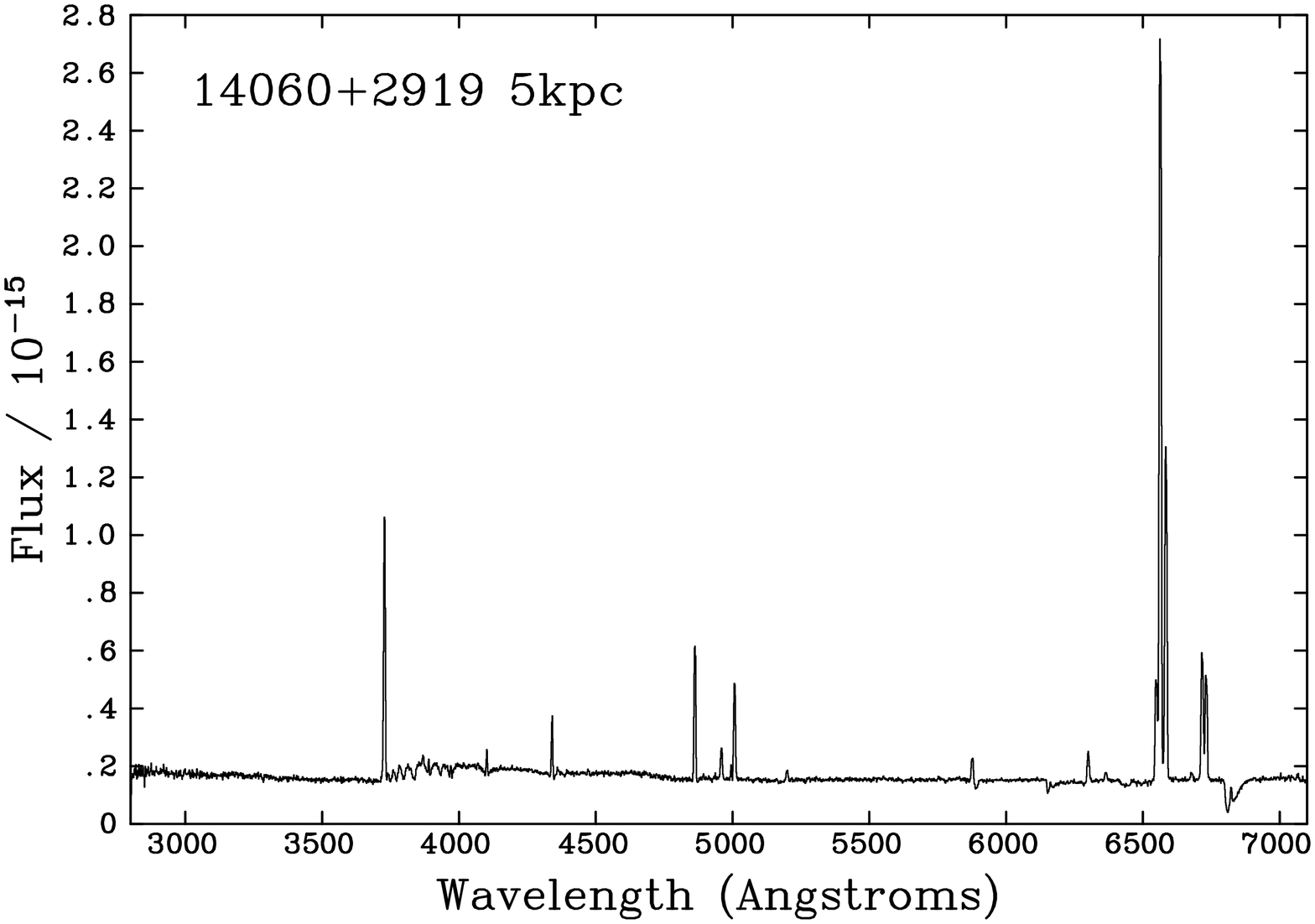,width=5.5cm,angle=-90.}\\
\hspace*{0cm}\psfig{file=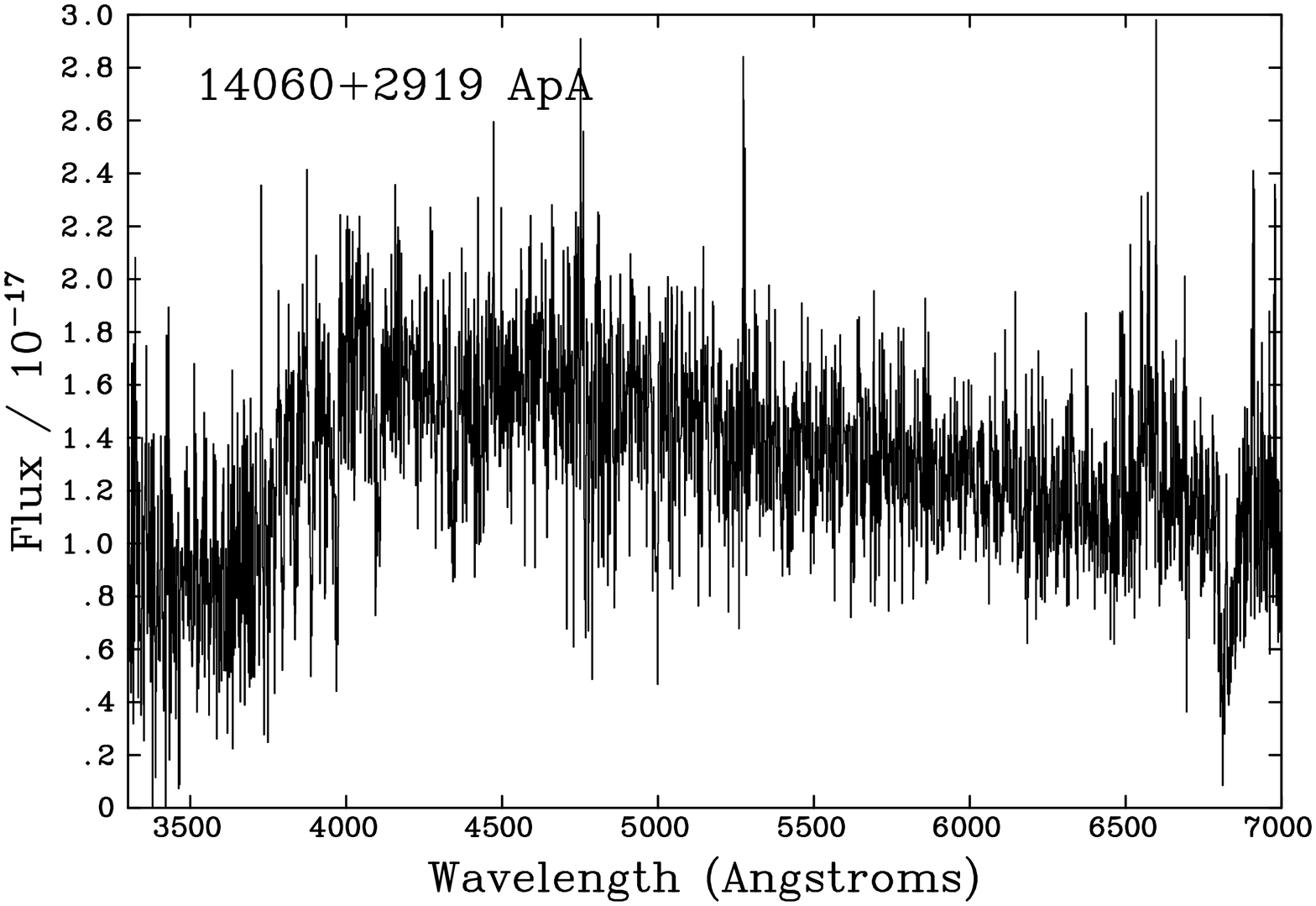,width=5.5cm,angle=-90.}&
\psfig{file=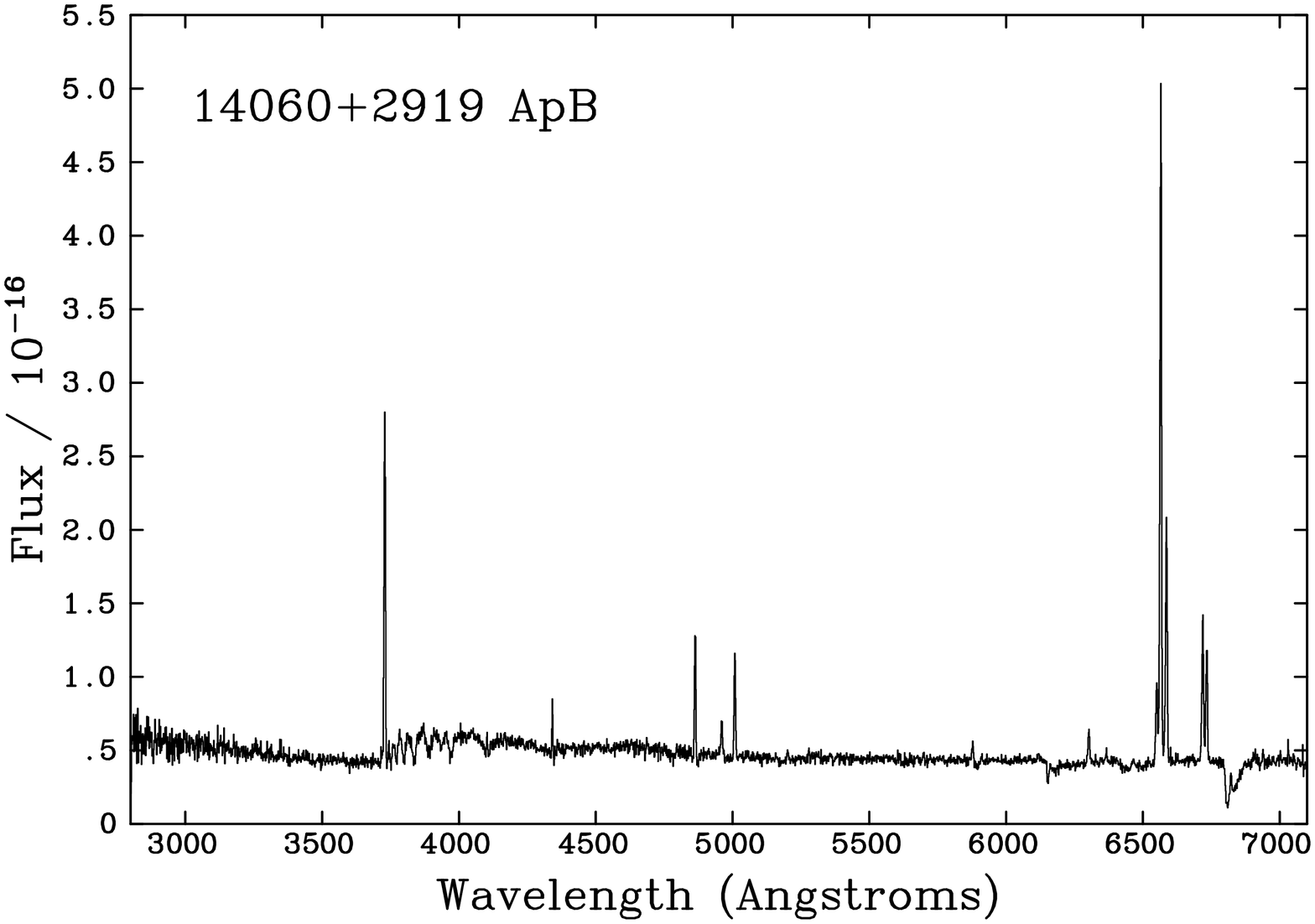,width=5.5cm,angle=-90.}\\
\hspace*{0cm}\psfig{file=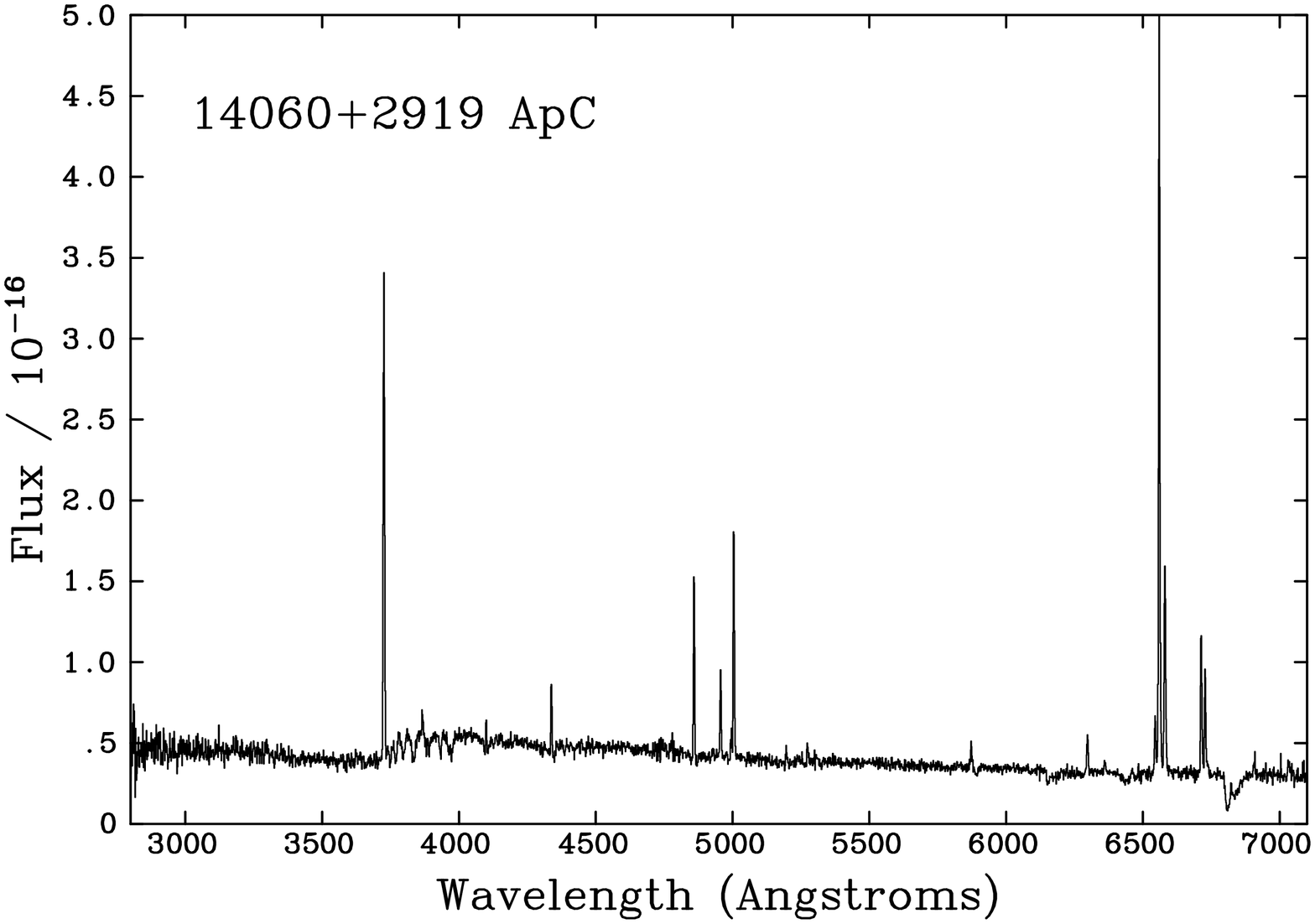,width=5.5cm,angle=-90.}&
\psfig{file=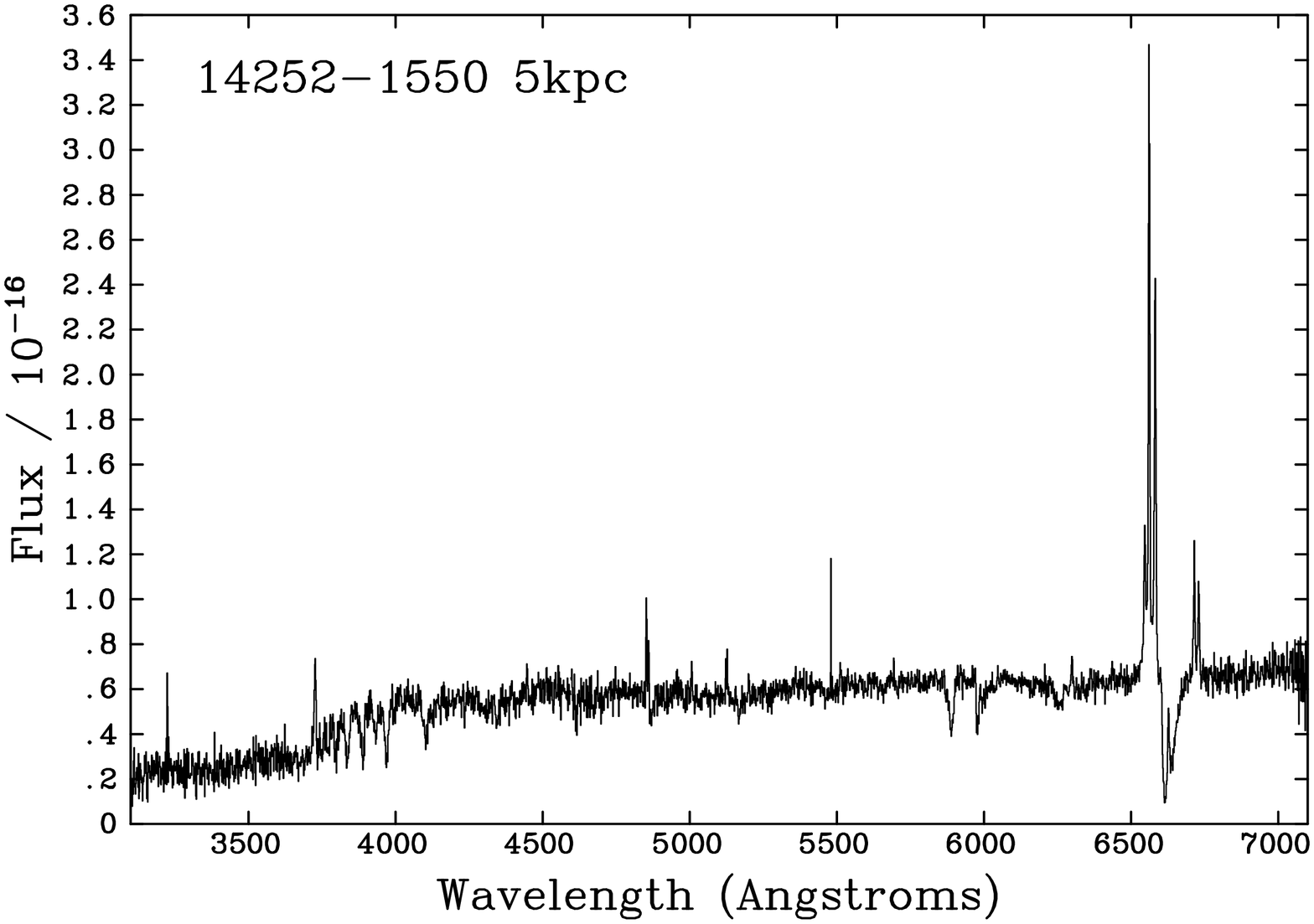,width=5.5cm,angle=-90.}
\end{tabular}
\caption[{\it Continued}]{Continued}
%\label{fig:SED}
\end{minipage}
\end{figure*}
\addtocounter{figure}{-1}
\begin{figure*}
\begin{minipage}{170mm}
\begin{tabular}{cc}
\hspace*{0cm}\psfig{file=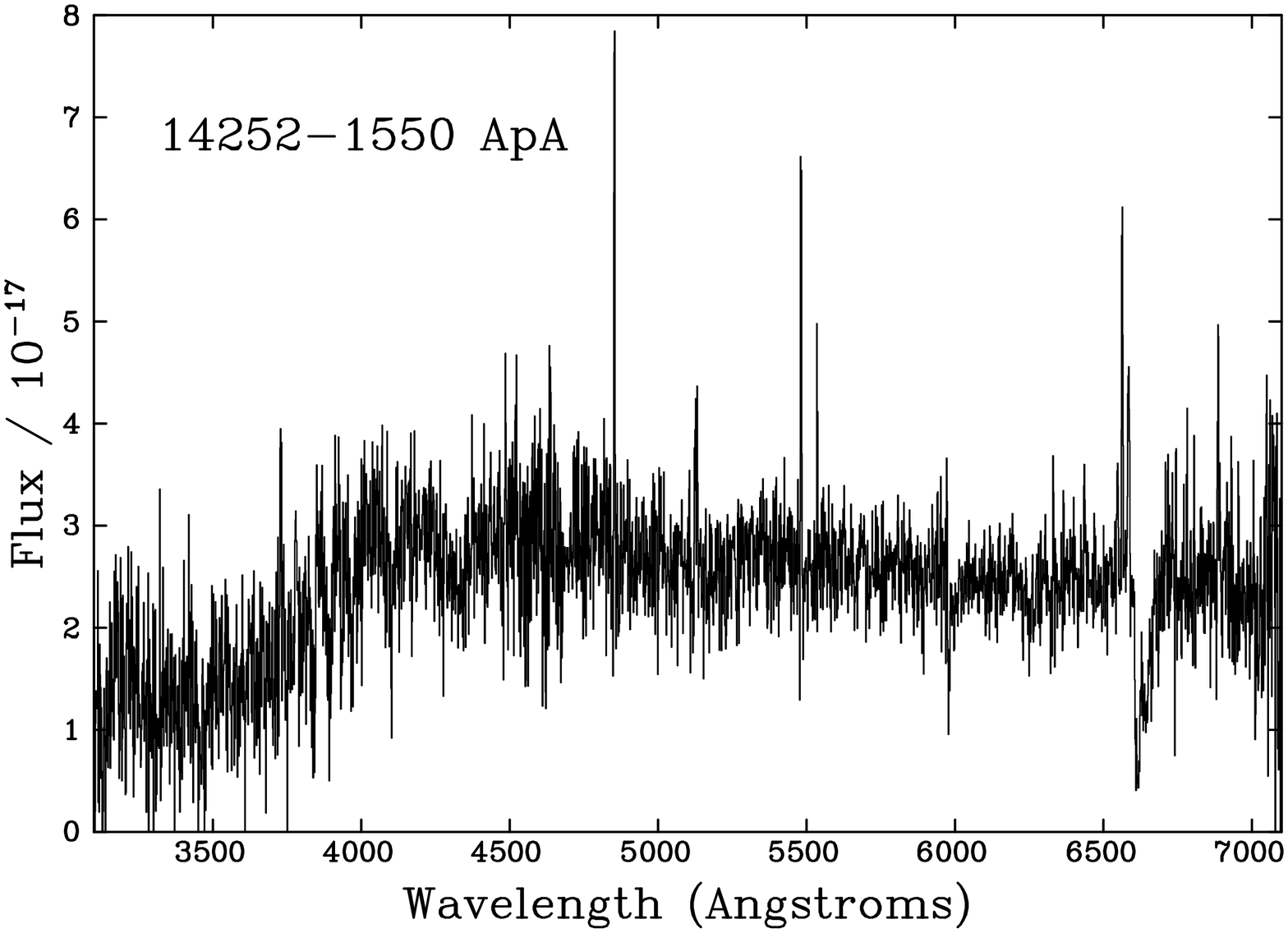,width=5.5cm,angle=-90.}&
\psfig{file=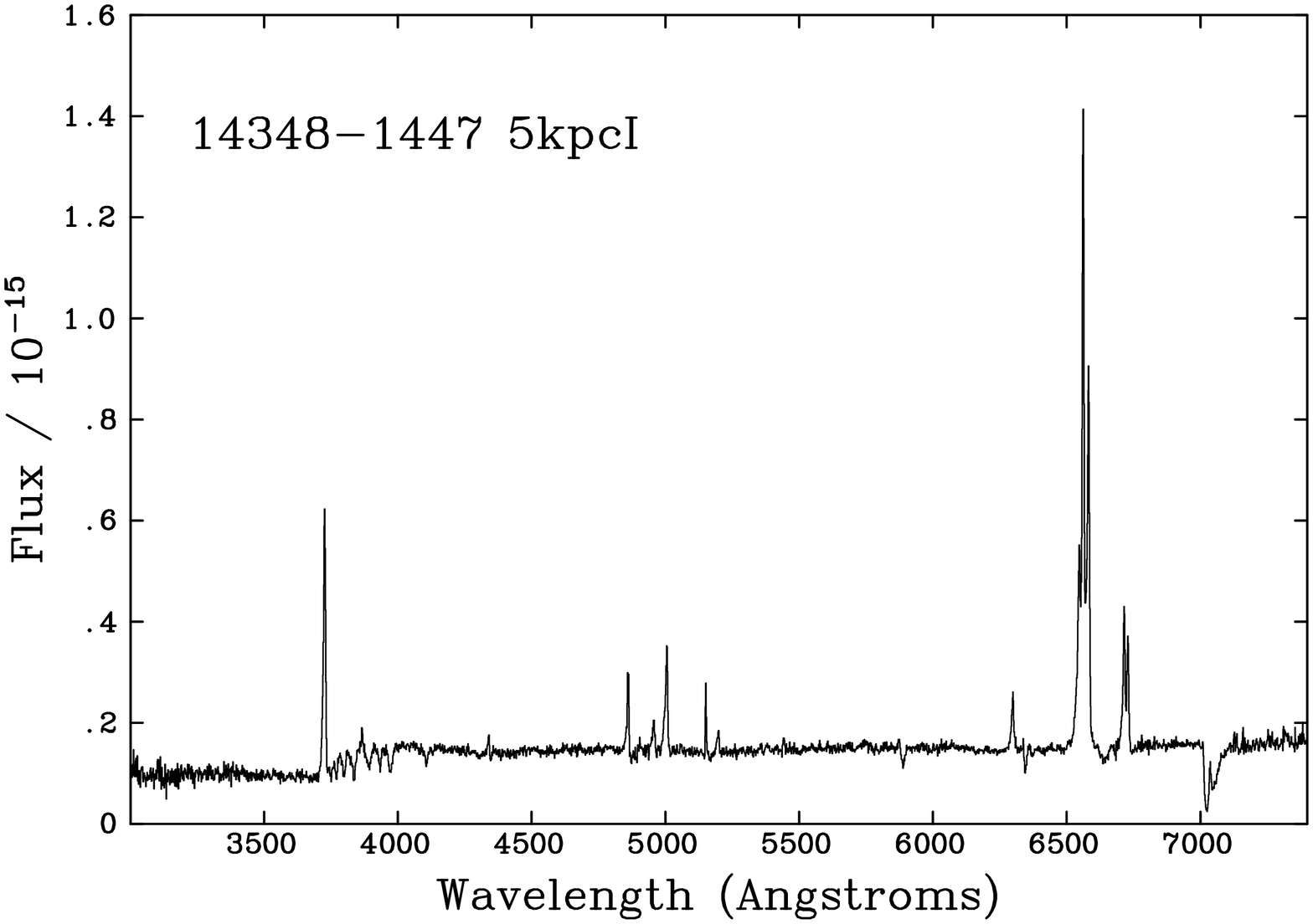,width=5.5cm,angle=-90.}\\
\hspace*{0cm}\psfig{file=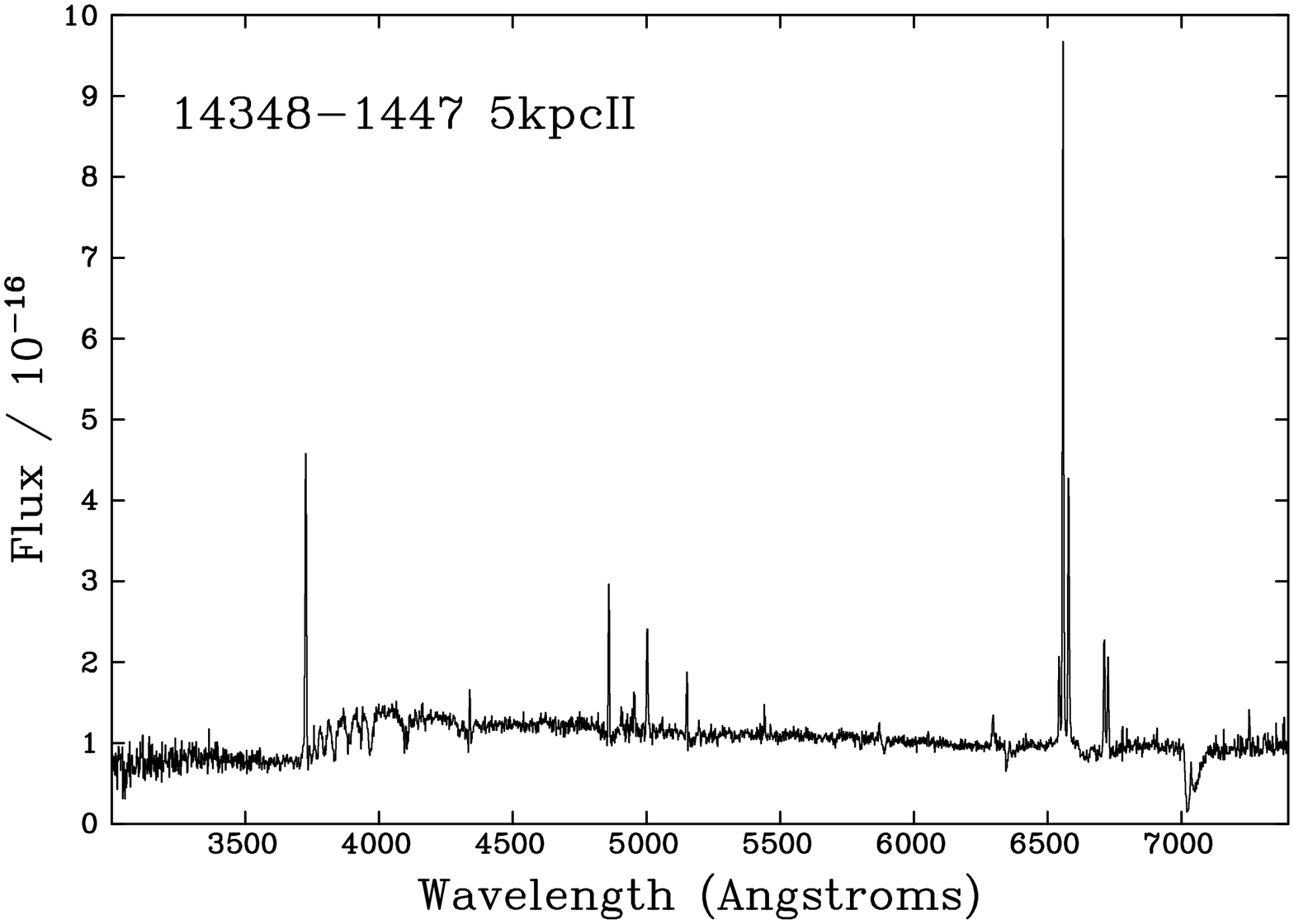,width=5.5cm,angle=-90.}&
\psfig{file=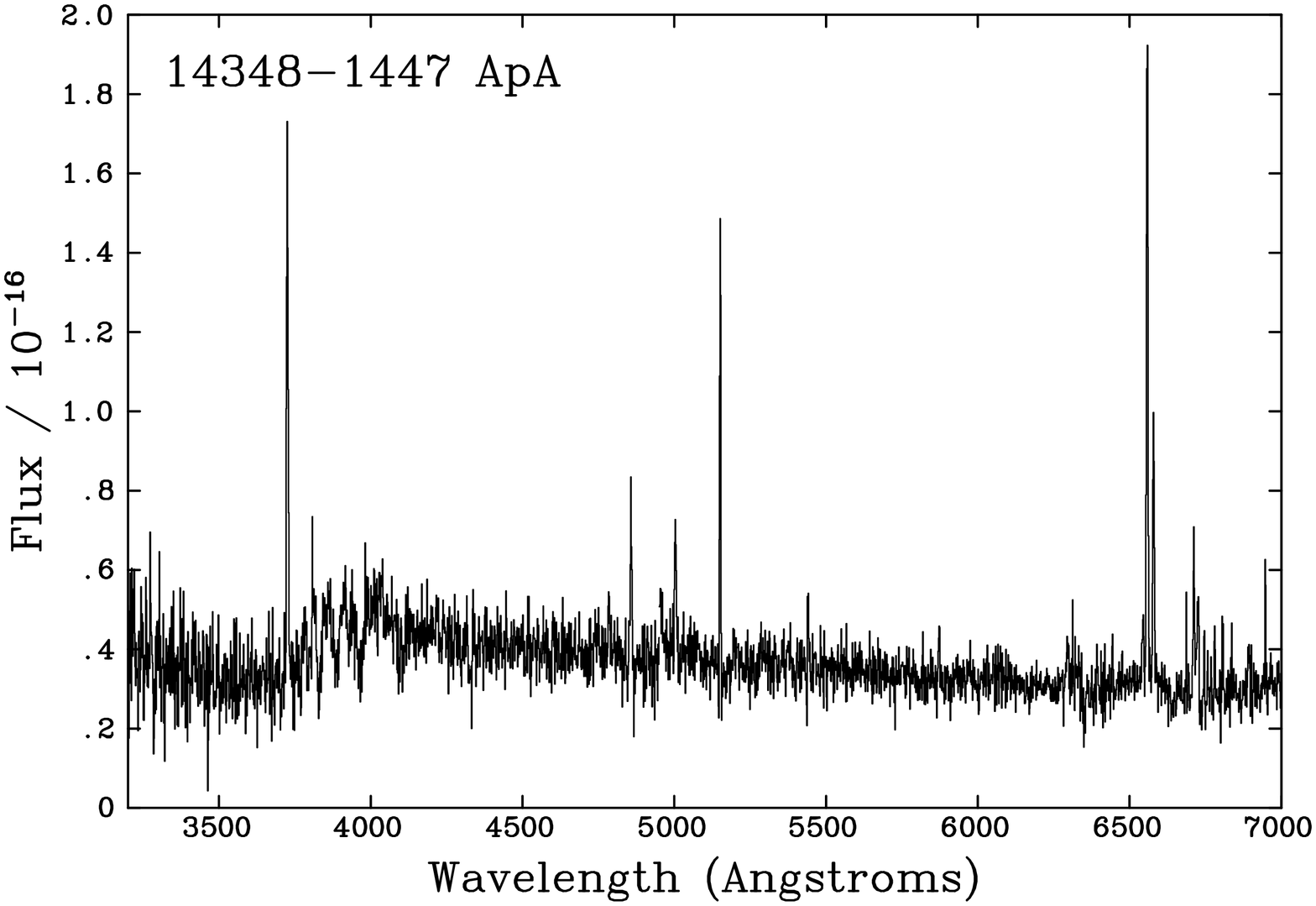,width=5.5cm,angle=-90.}\\
\hspace*{0cm}\psfig{file=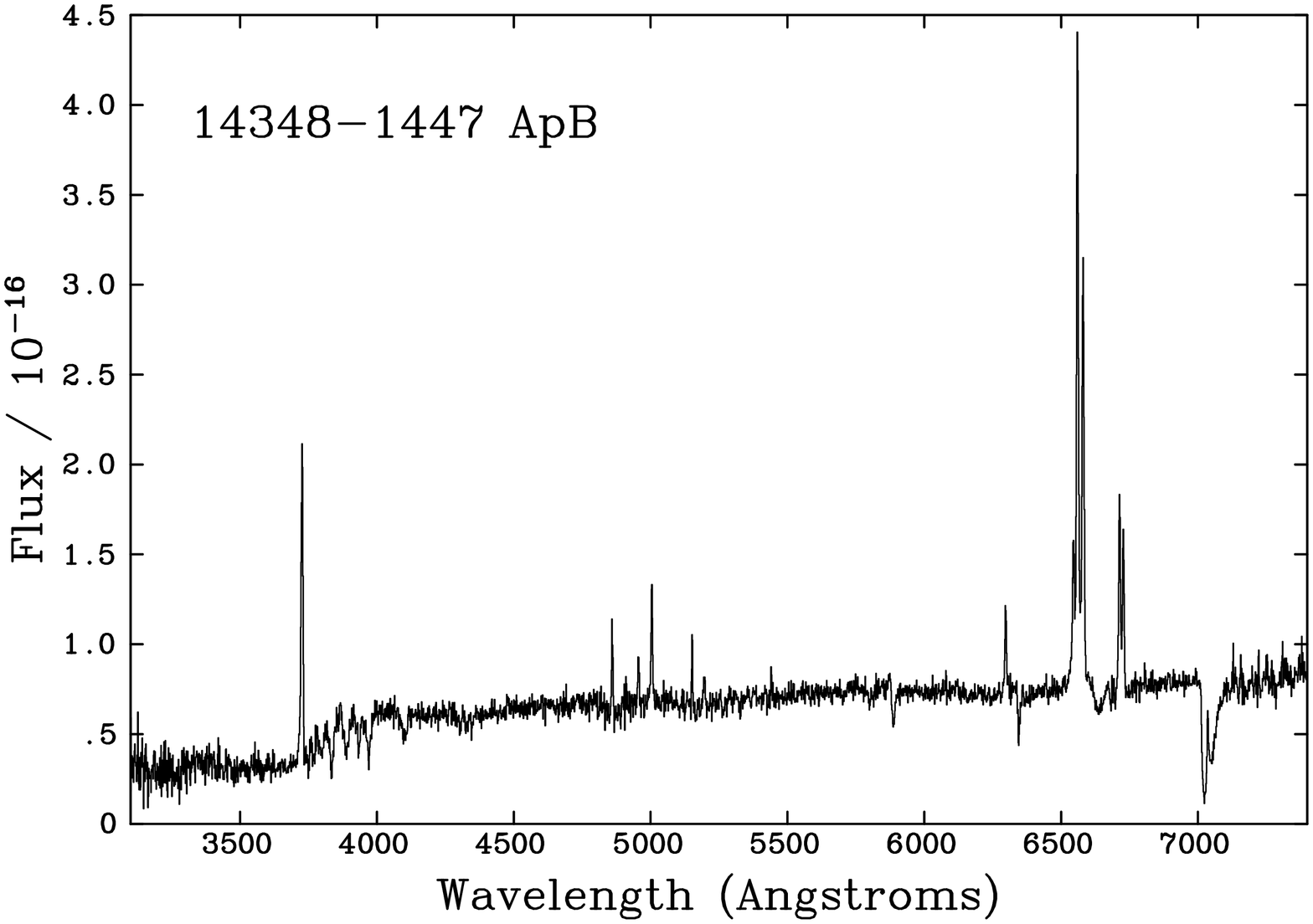,width=5.5cm,angle=-90.}&
\psfig{file=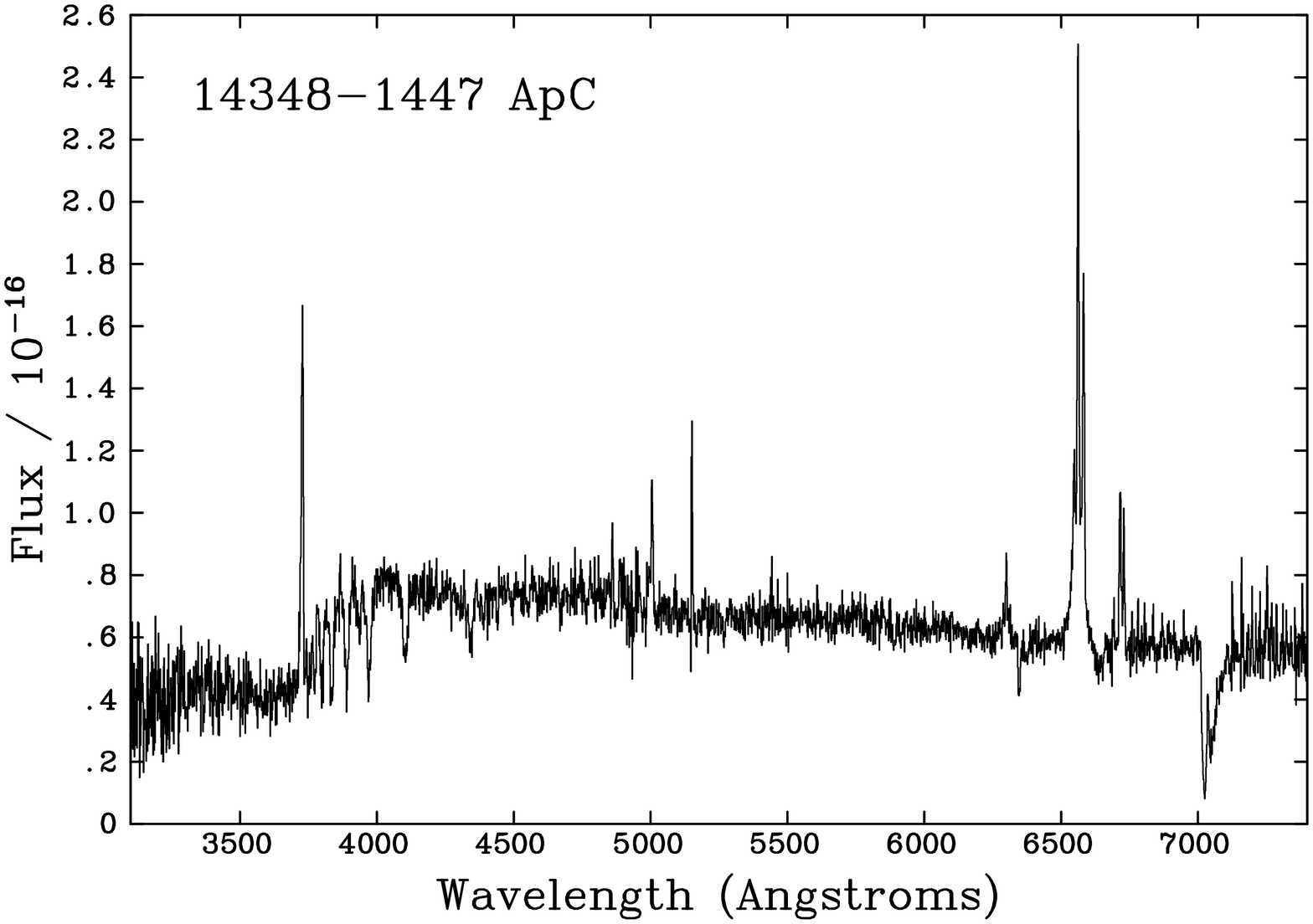,width=5.5cm,angle=-90.}\\
\hspace*{0cm}\psfig{file=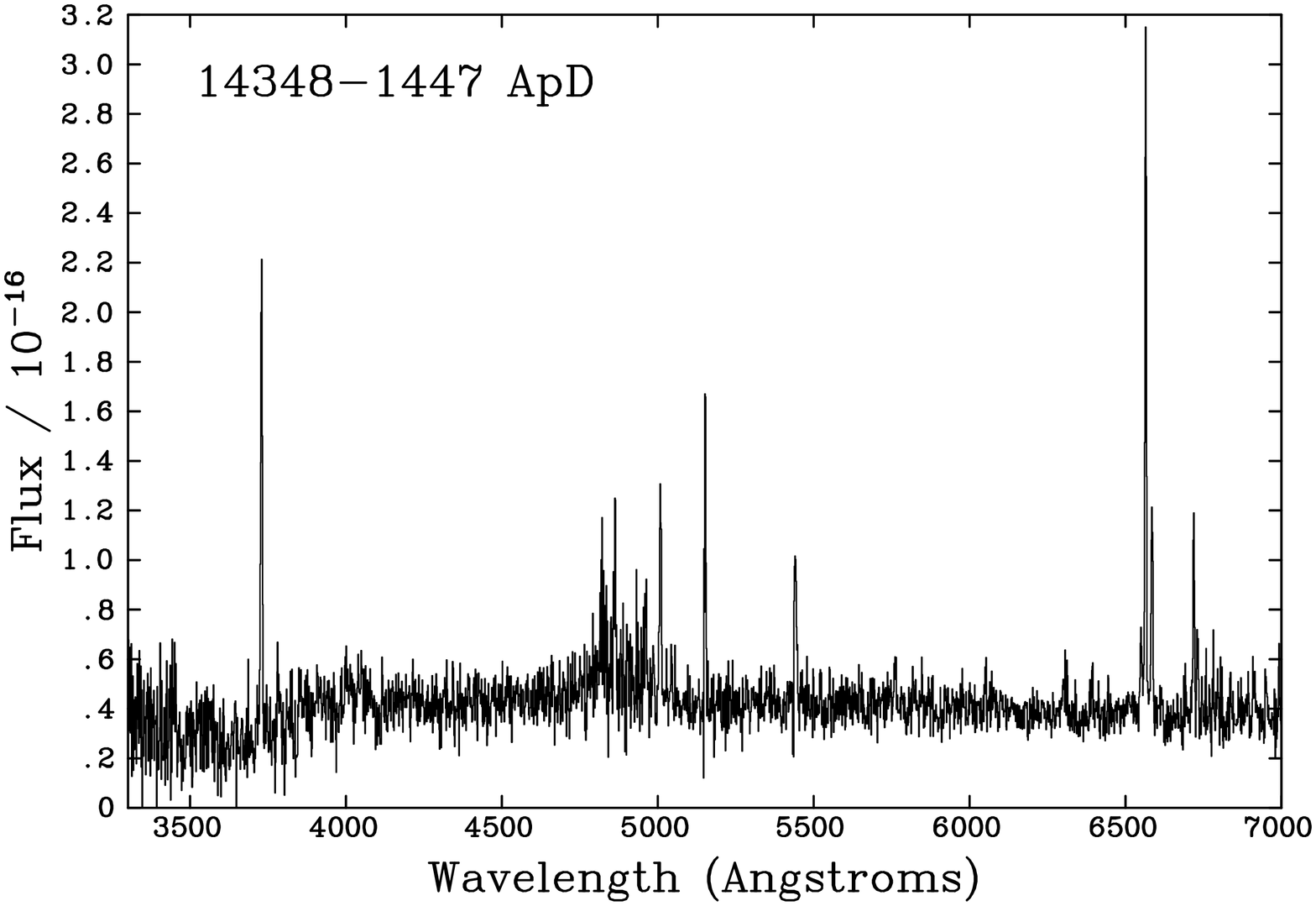,width=5.5cm,angle=-90.}&
\psfig{file=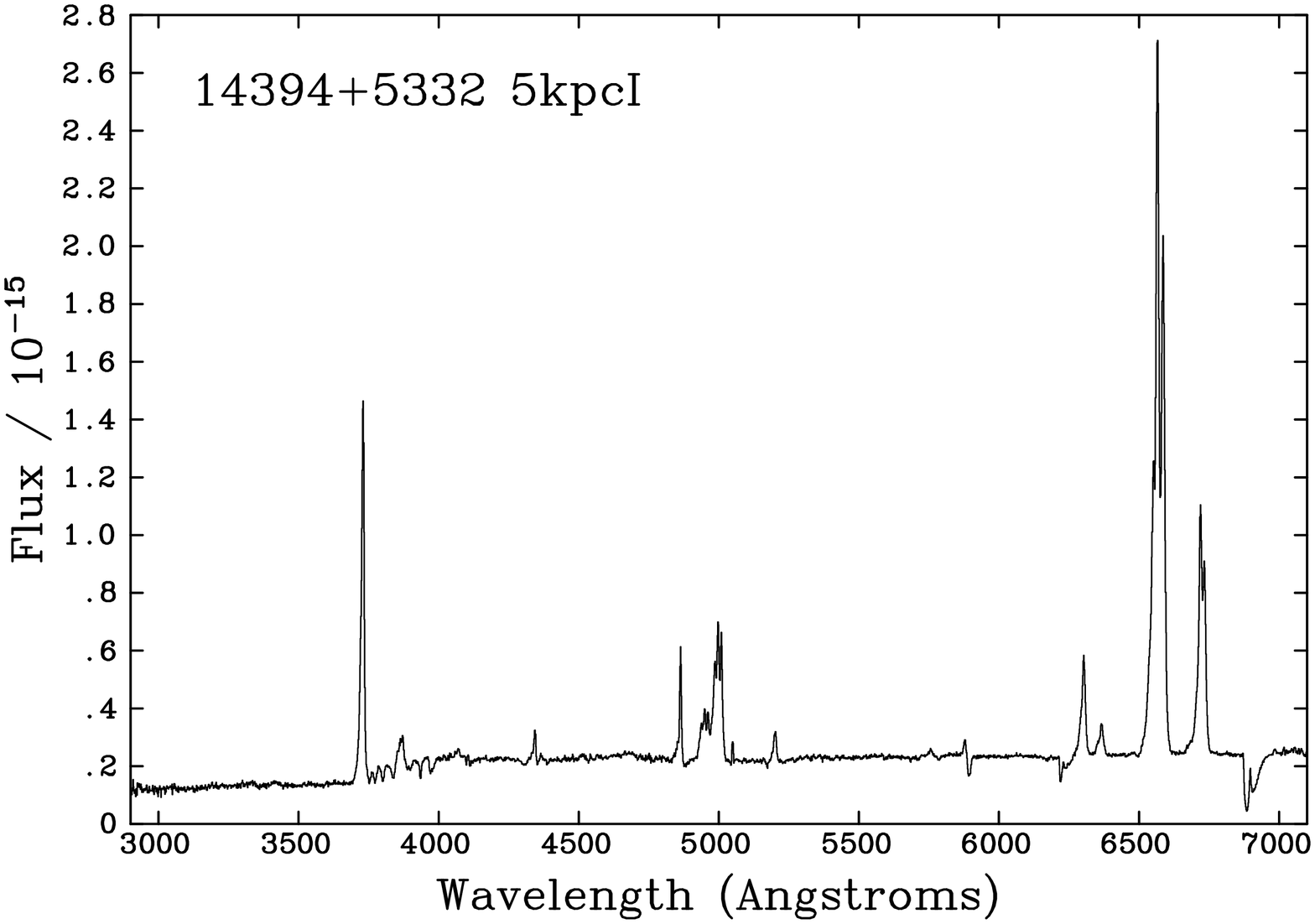,width=5.5cm,angle=-90.}
\end{tabular}
\caption[{\it Continued}]{Continued}
%\label{fig:SED}
\end{minipage}
\end{figure*}
\addtocounter{figure}{-1}
\begin{figure*}
\begin{minipage}{170mm}
\begin{tabular}{cc}
\hspace*{0cm}\psfig{file=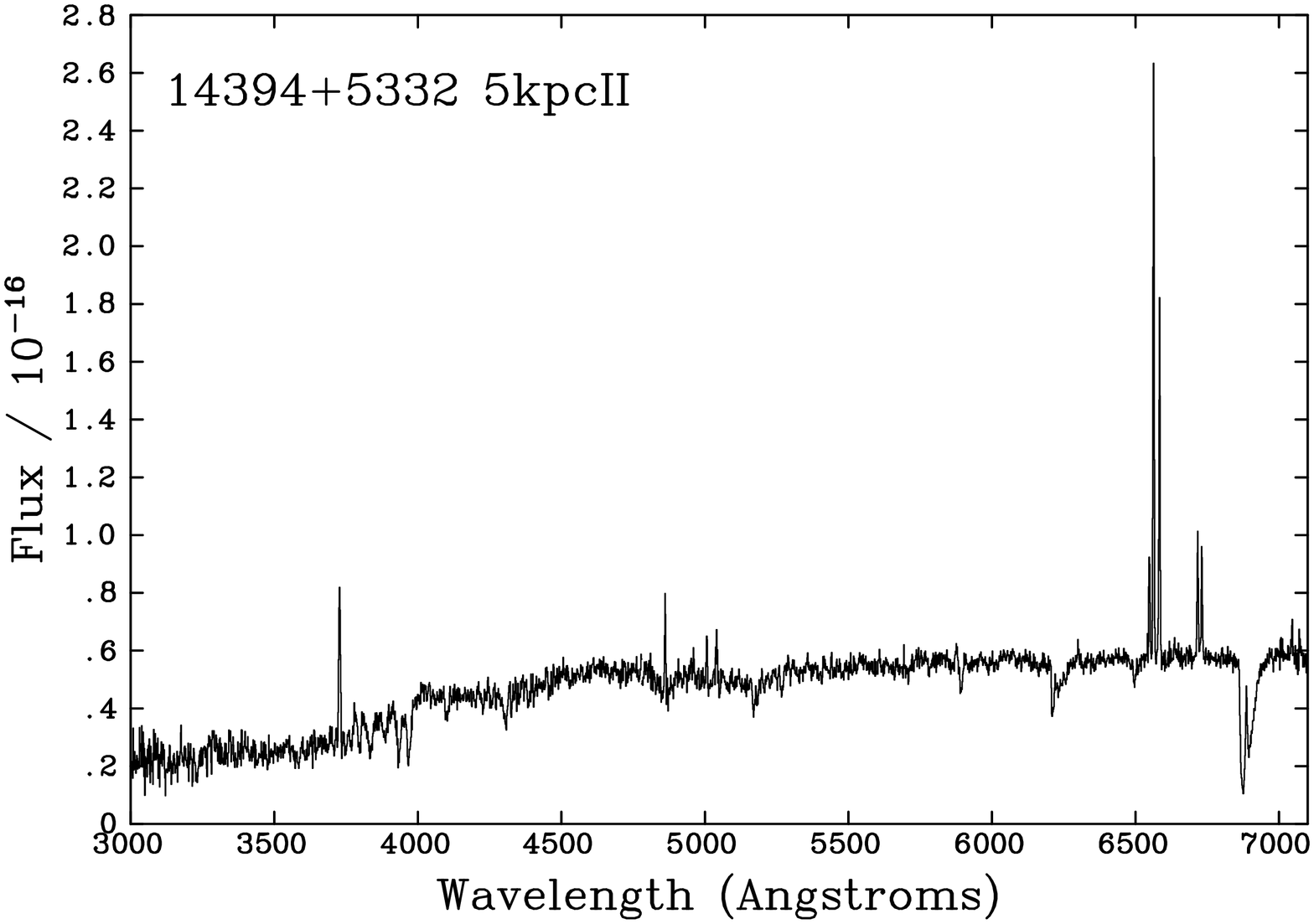,width=5.5cm,angle=-90.}&
\psfig{file=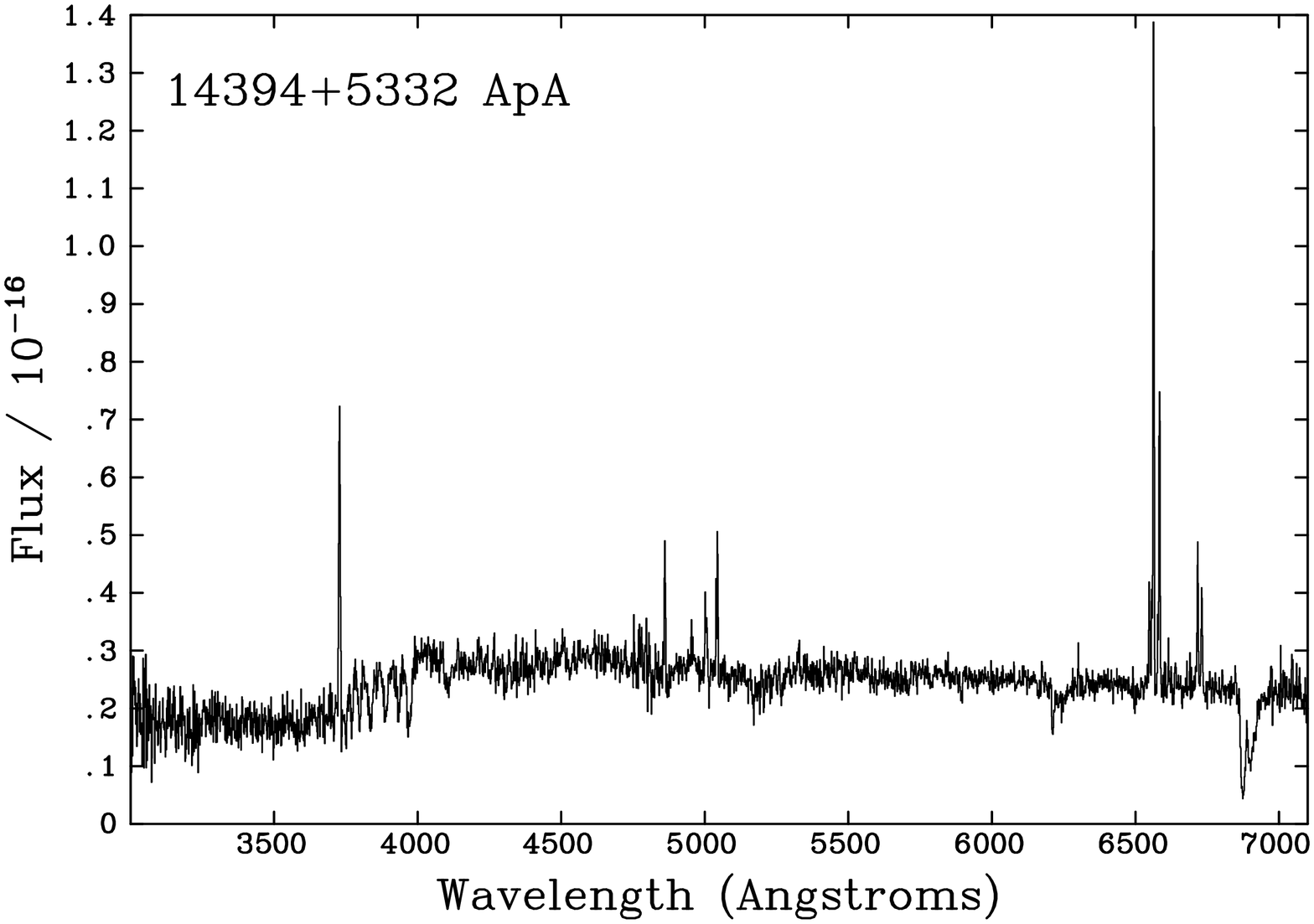,width=5.5cm,angle=-90.}\\
\hspace*{0cm}\psfig{file=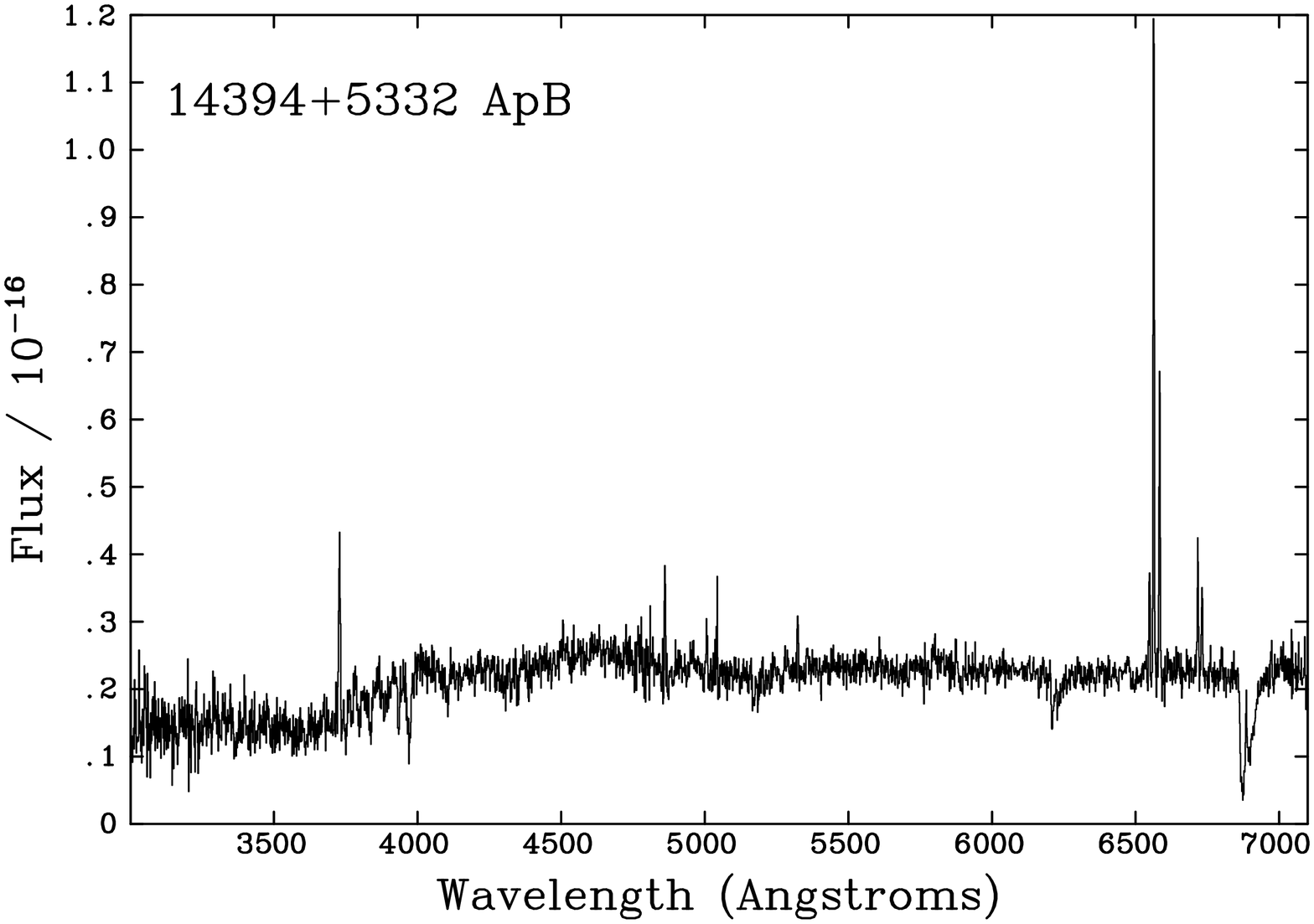,width=5.5cm,angle=-90.}&
\psfig{file=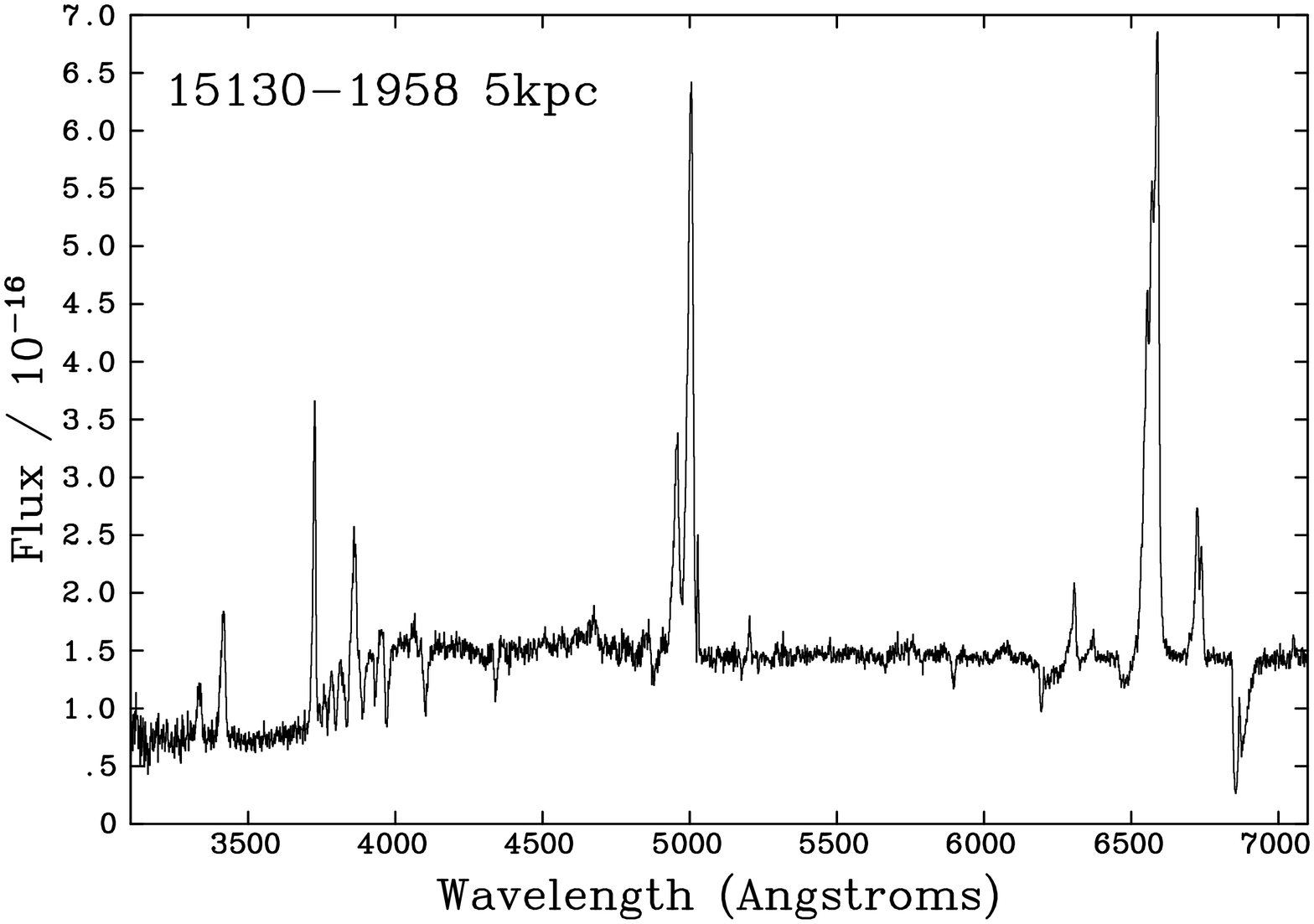,width=5.5cm,angle=-90.}\\
\hspace*{0cm}\psfig{file=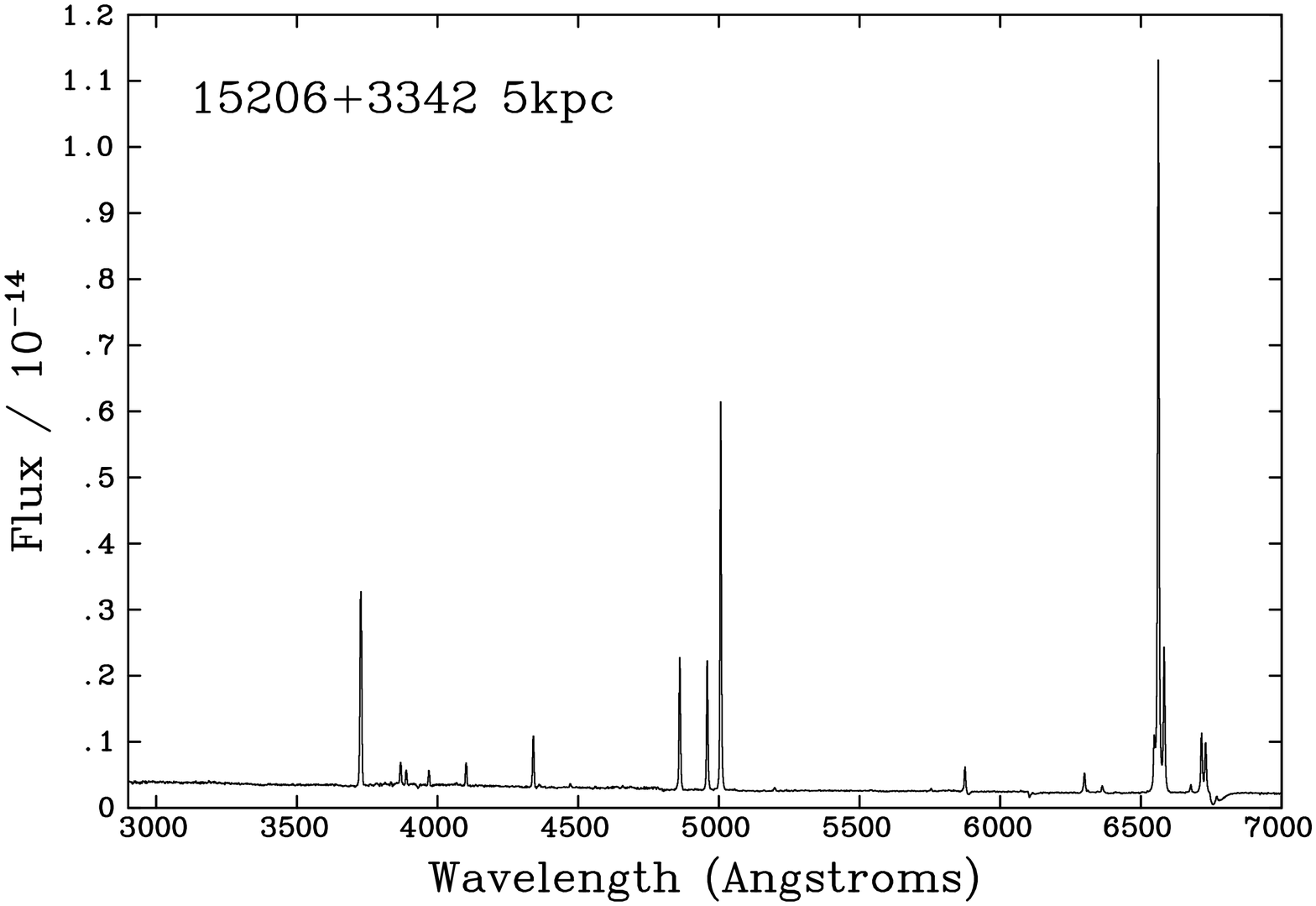,width=7.8cm,angle=0.}&
\psfig{file=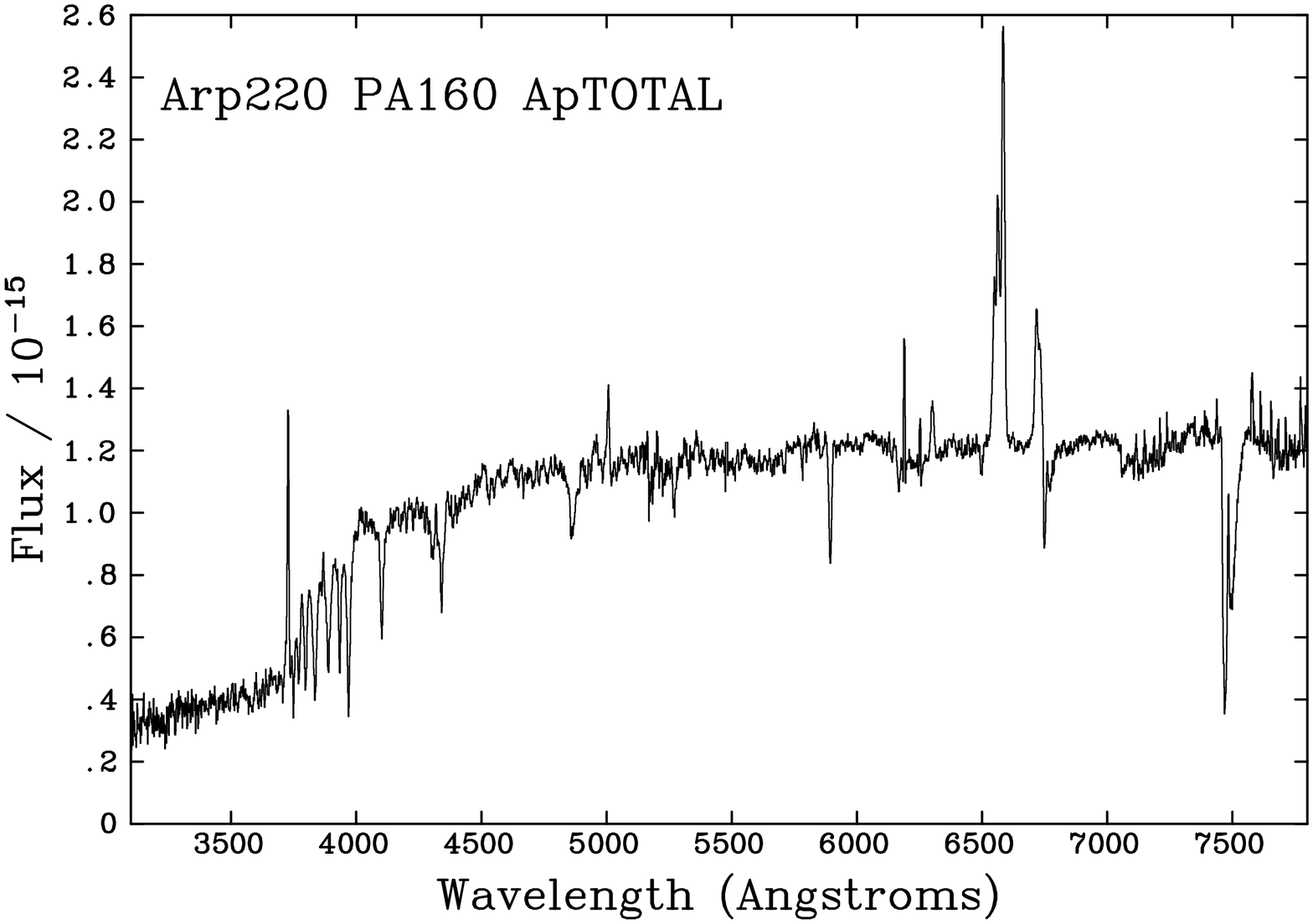,width=7.8cm,angle=0.}\\
\hspace{0cm}\psfig{file=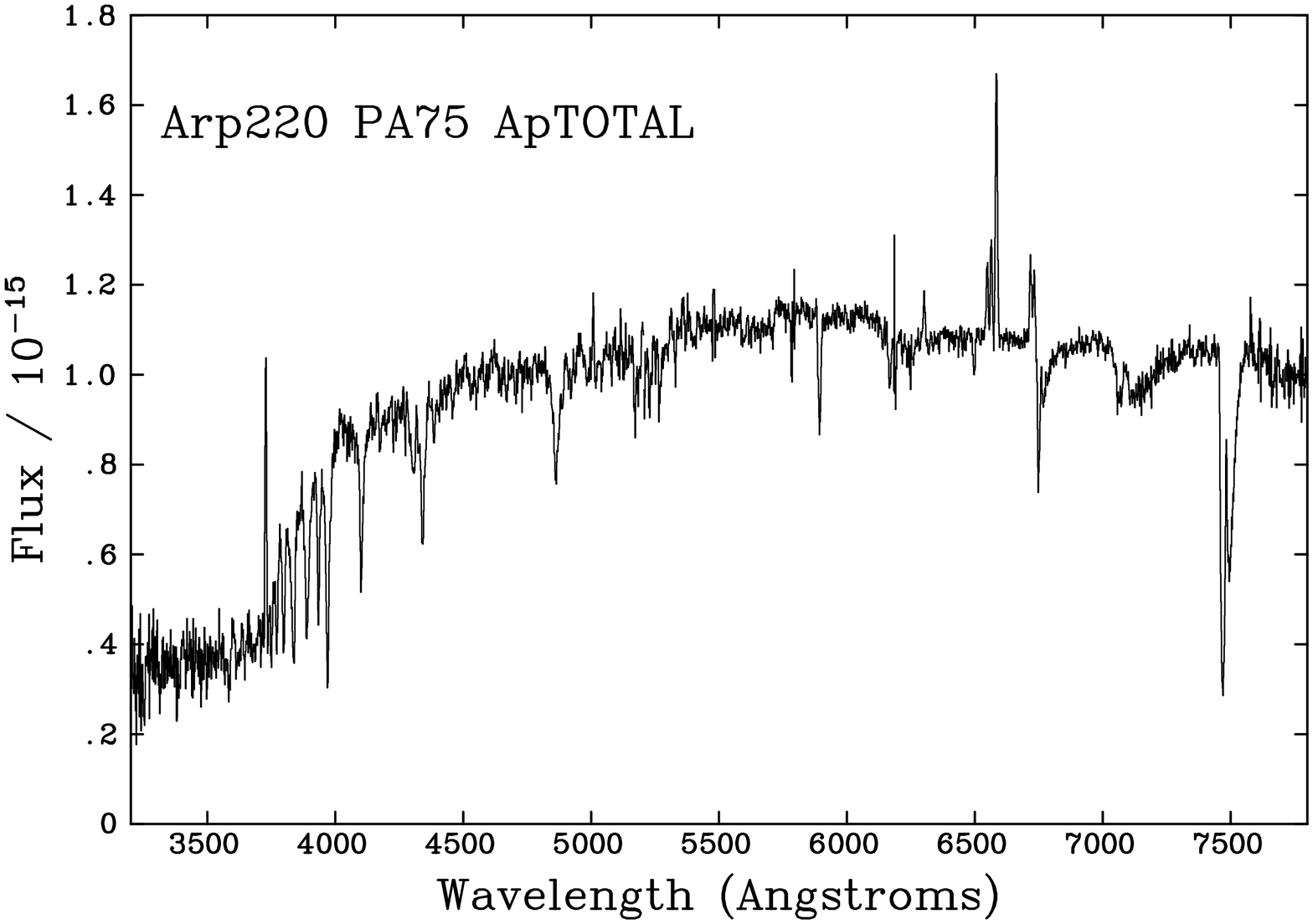,width=7.8cm,angle=0.}&
\psfig{file=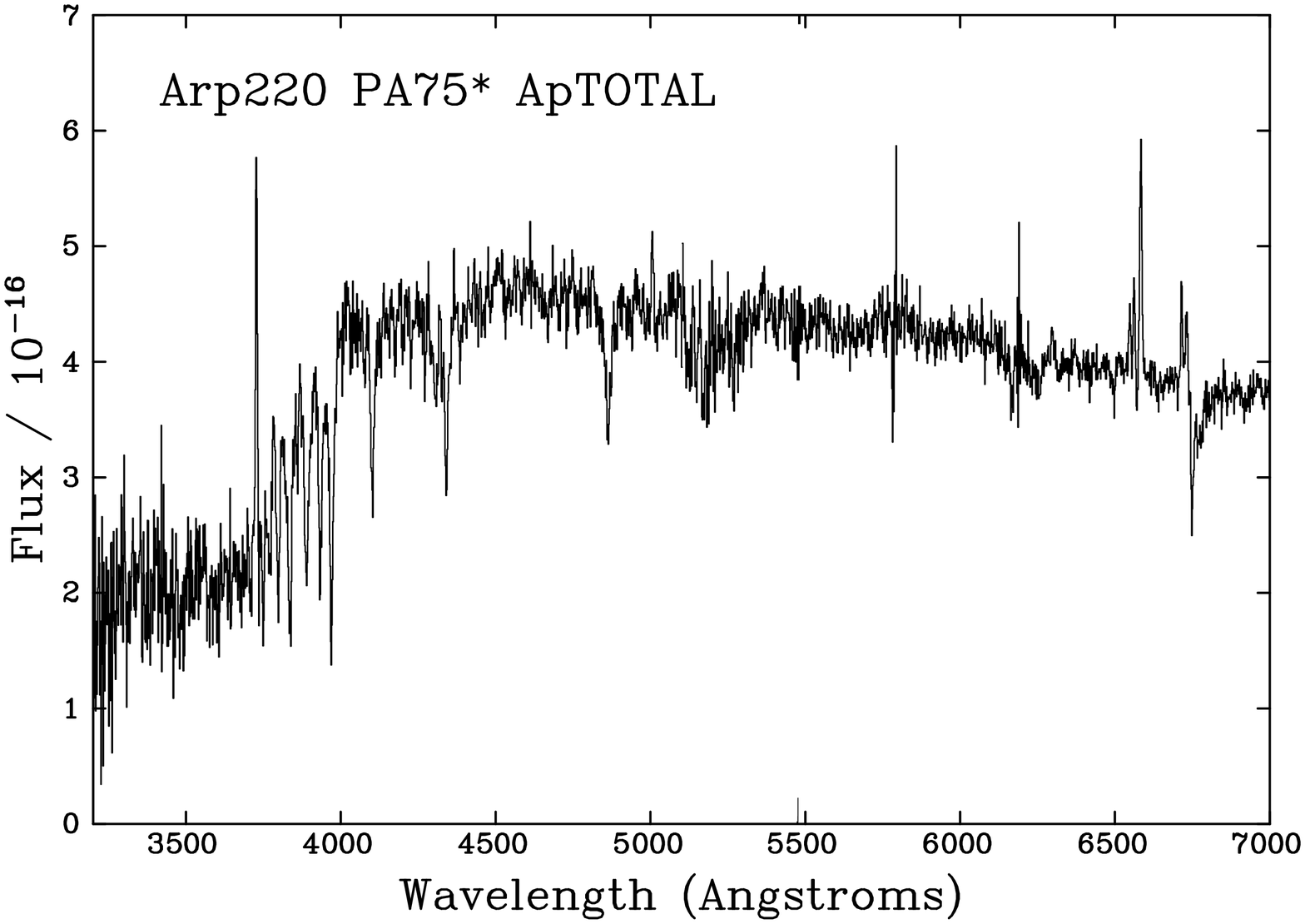,width=7.8cm,angle=0.}\\
\end{tabular}
\caption[{\it Continued}]{Continued}
%\label{fig:SED}
\end{minipage}
\end{figure*}
\addtocounter{figure}{-1}
\begin{figure*}
\begin{minipage}{170mm}
\begin{tabular}{cc}
\hspace*{0cm}\psfig{file=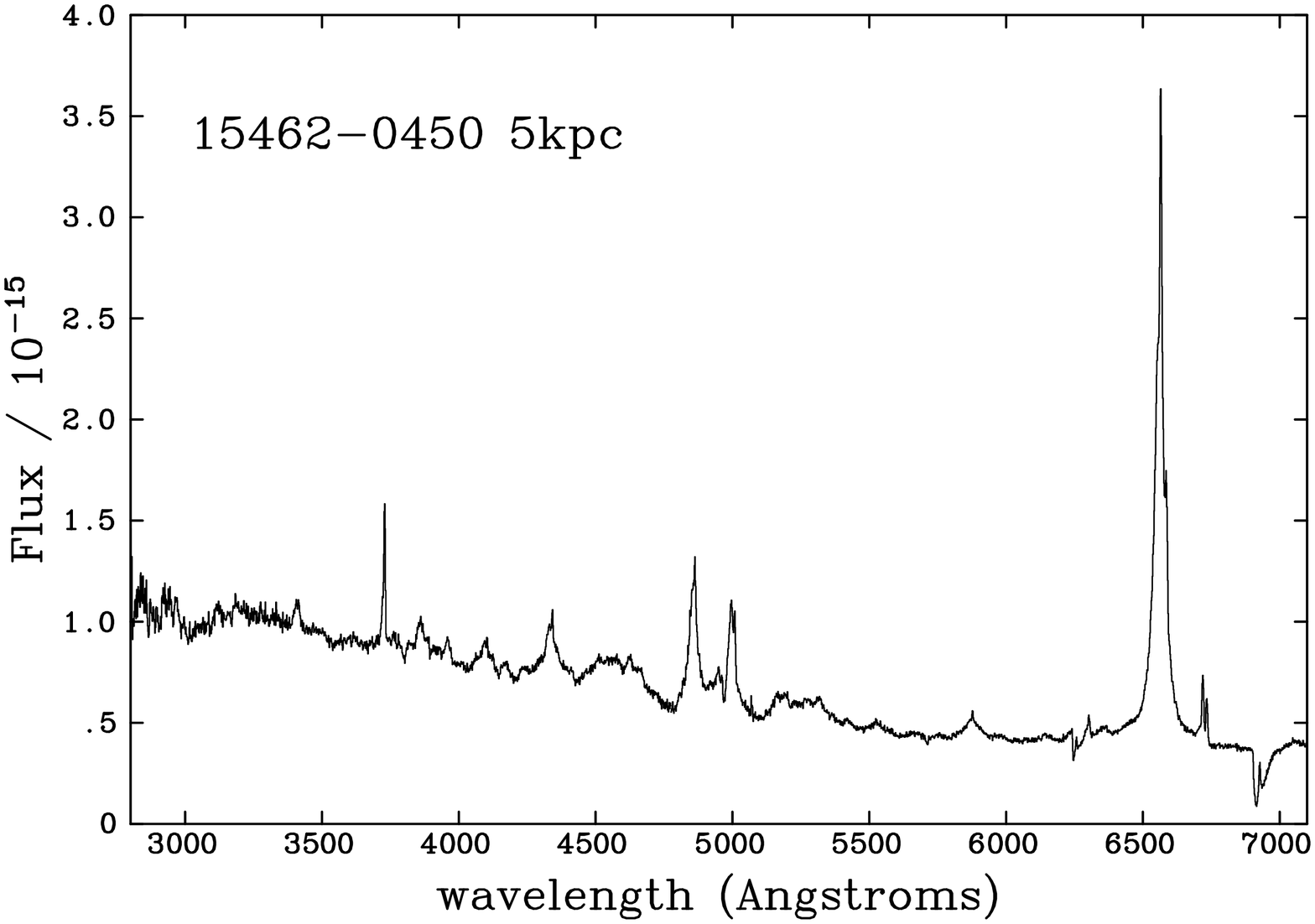,width=5.5cm,angle=-90.}&
\psfig{file=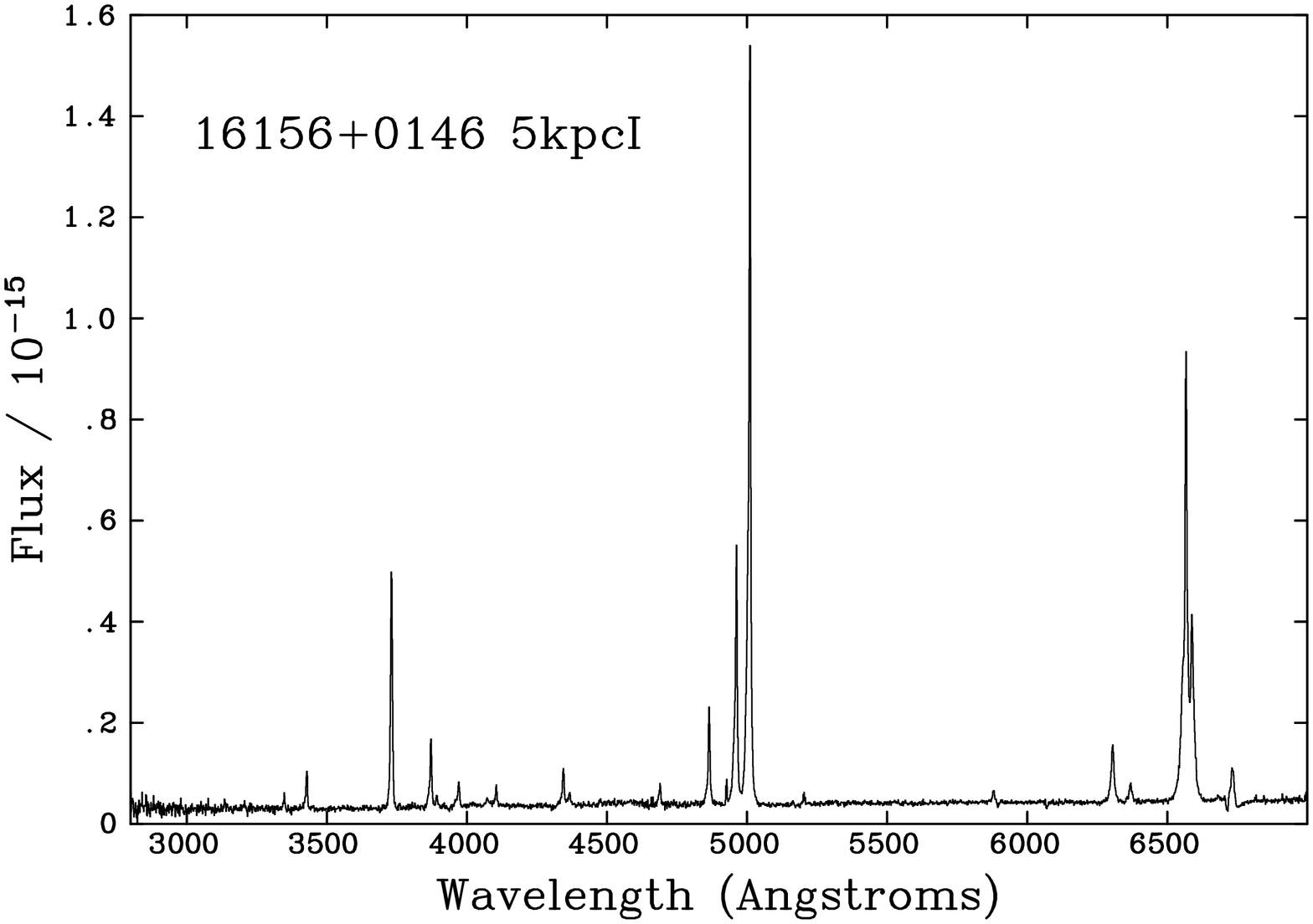,width=5.5cm,angle=-90.}\\
\hspace*{0cm}\psfig{file=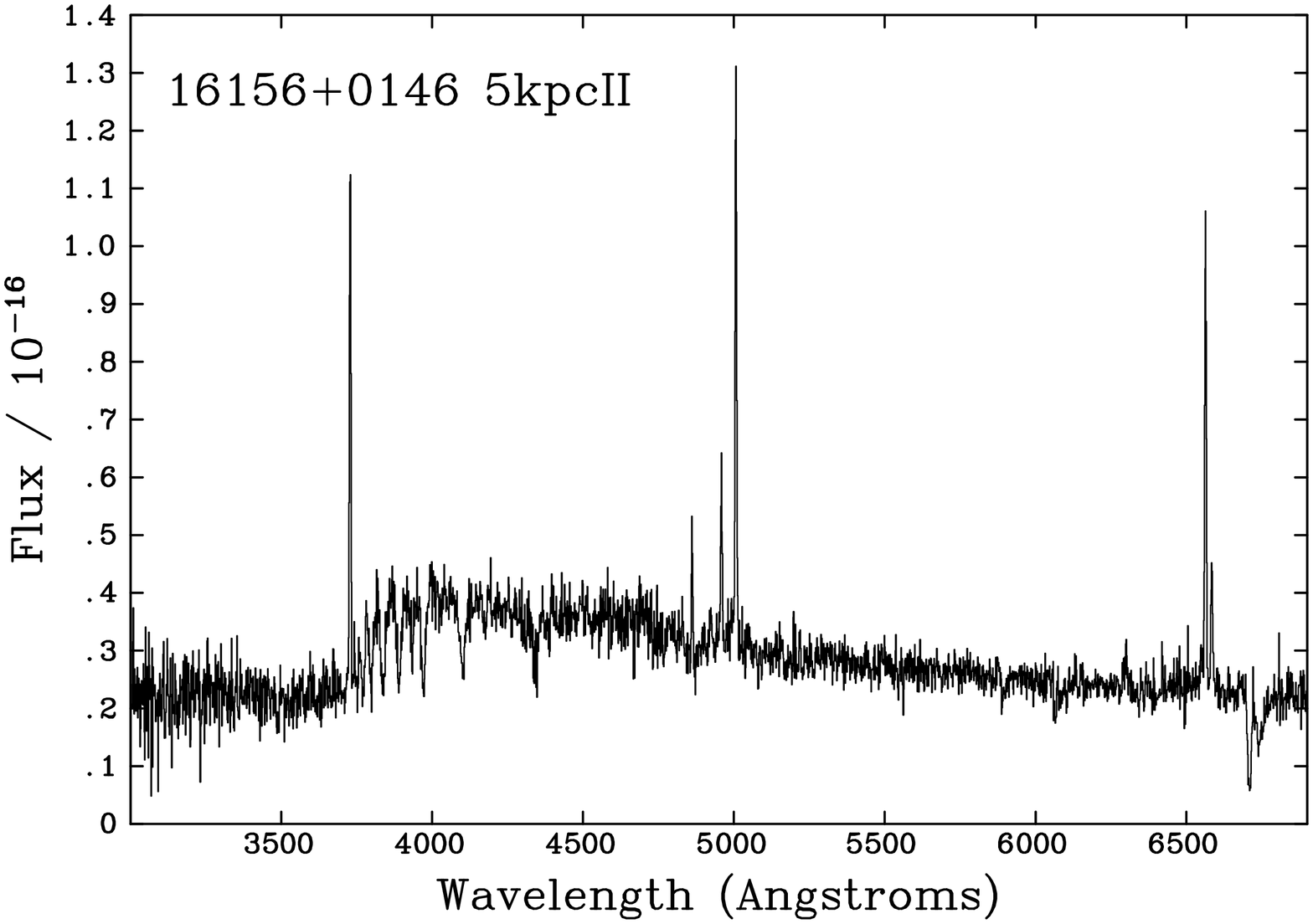,width=5.5cm,angle=-90.}&
\psfig{file=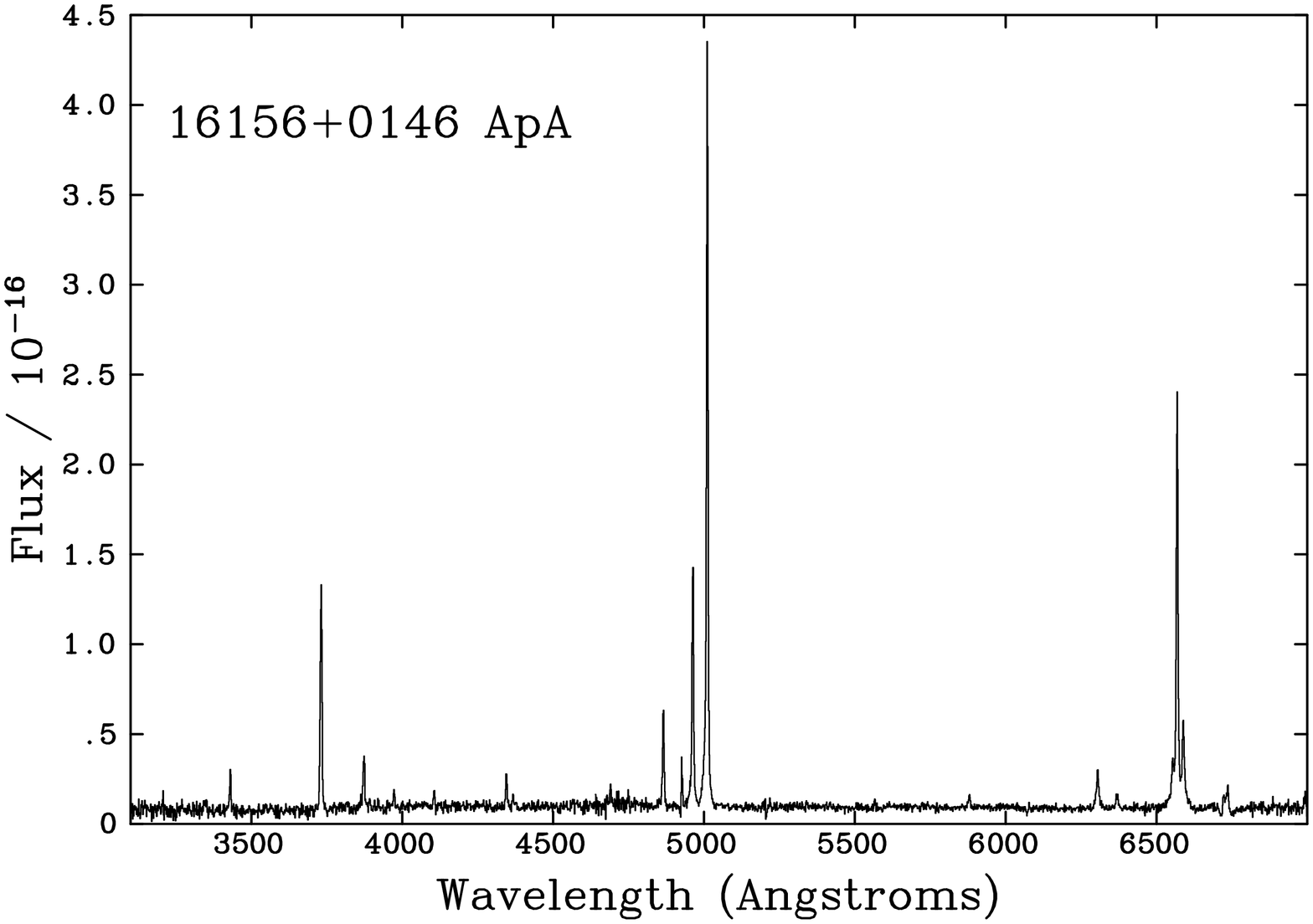,width=5.5cm,angle=-90.}\\
\hspace*{0cm}\psfig{file=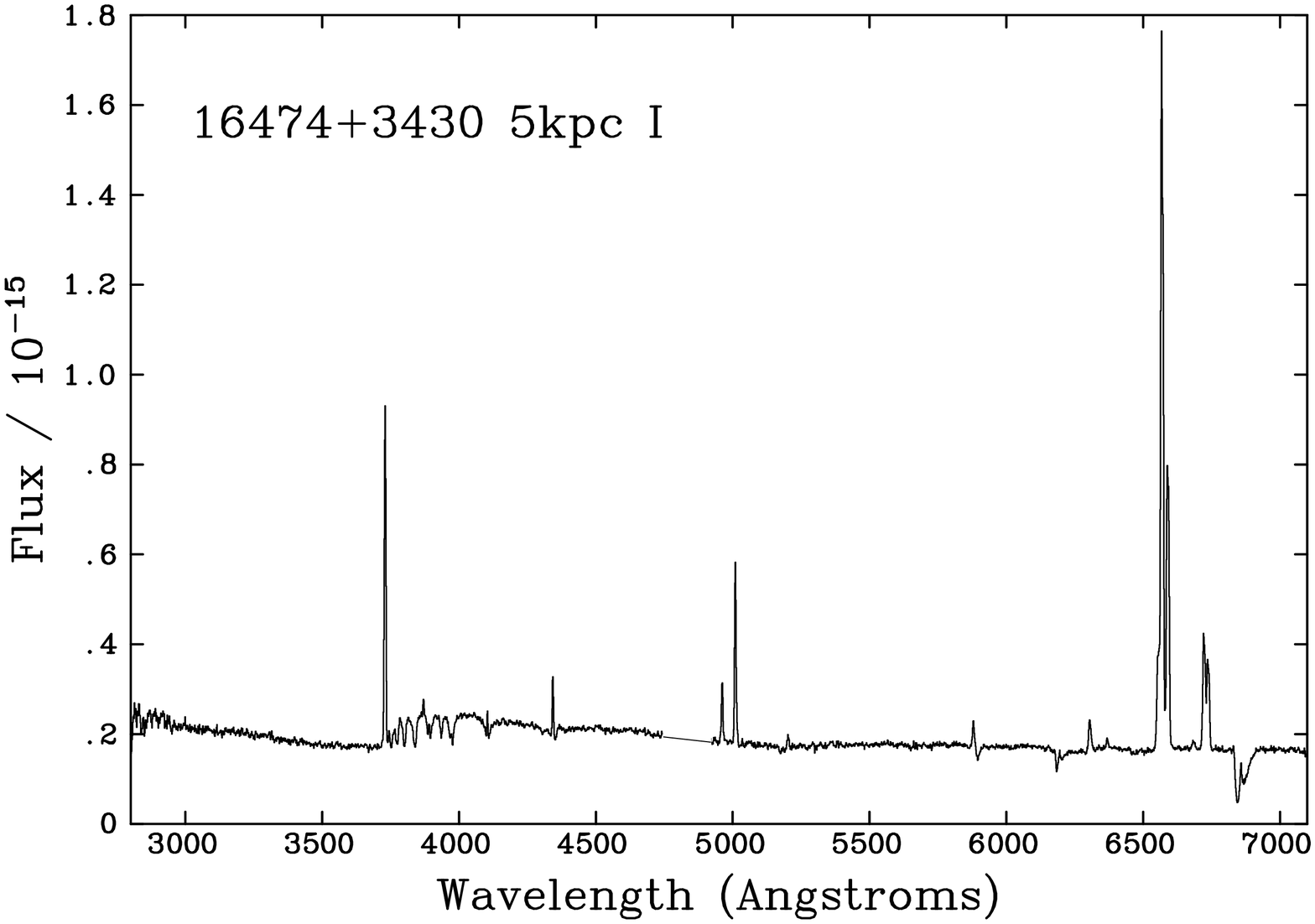,width=7.8cm,angle=0.}&
\psfig{file=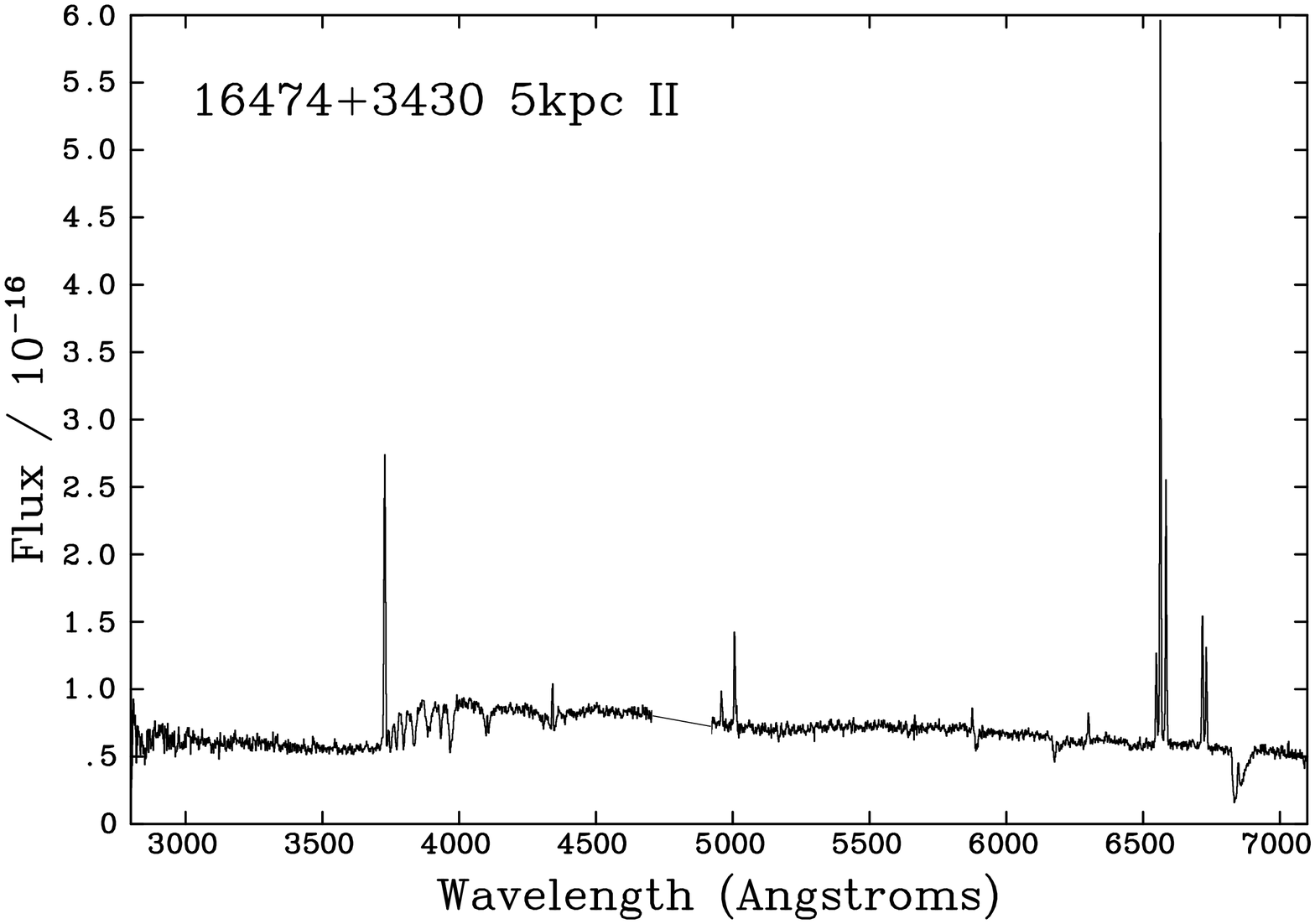,width=7.8cm,angle=0.}\\
\hspace*{0cm}\psfig{file=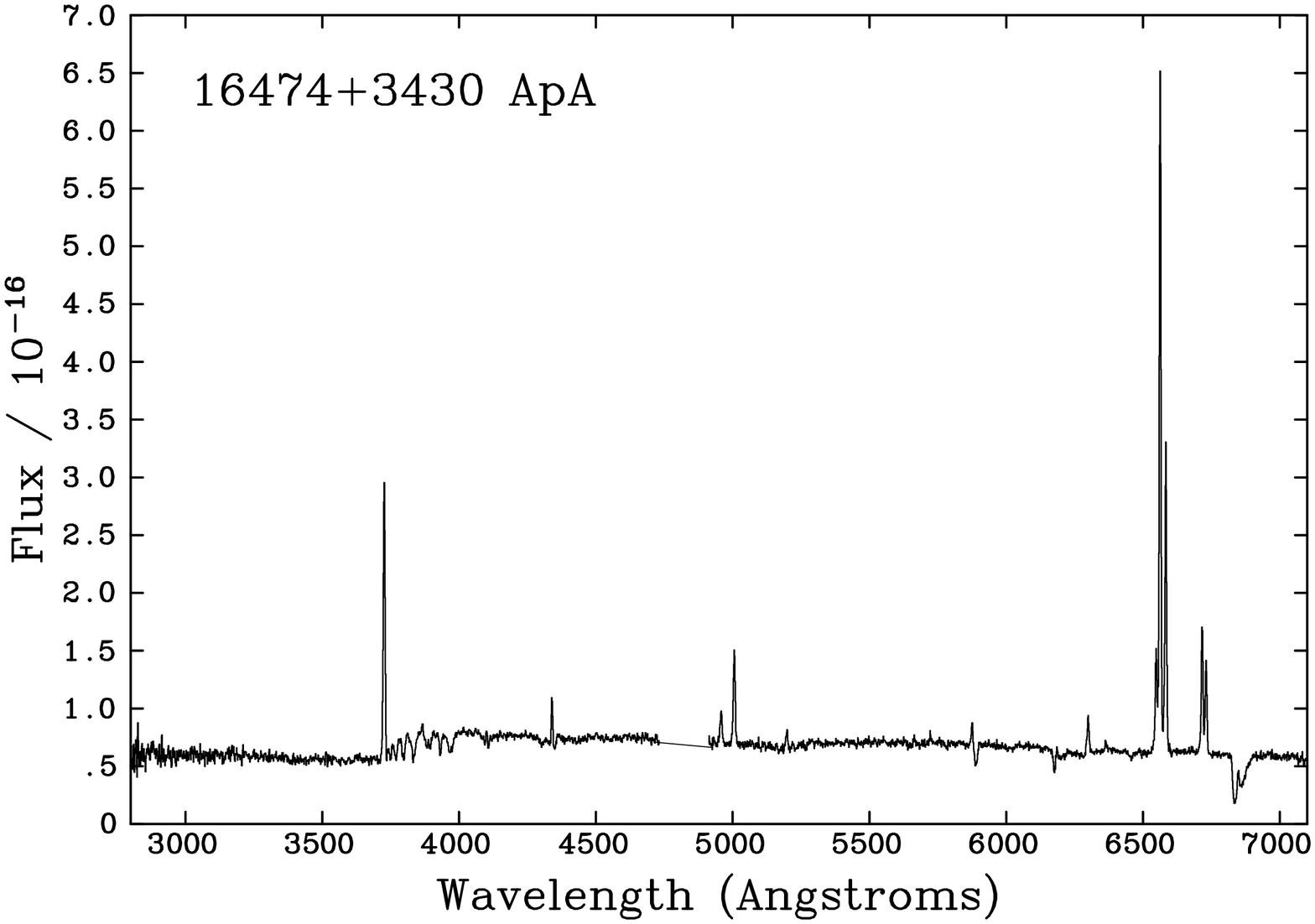,width=7.8cm,angle=0.}&
\psfig{file=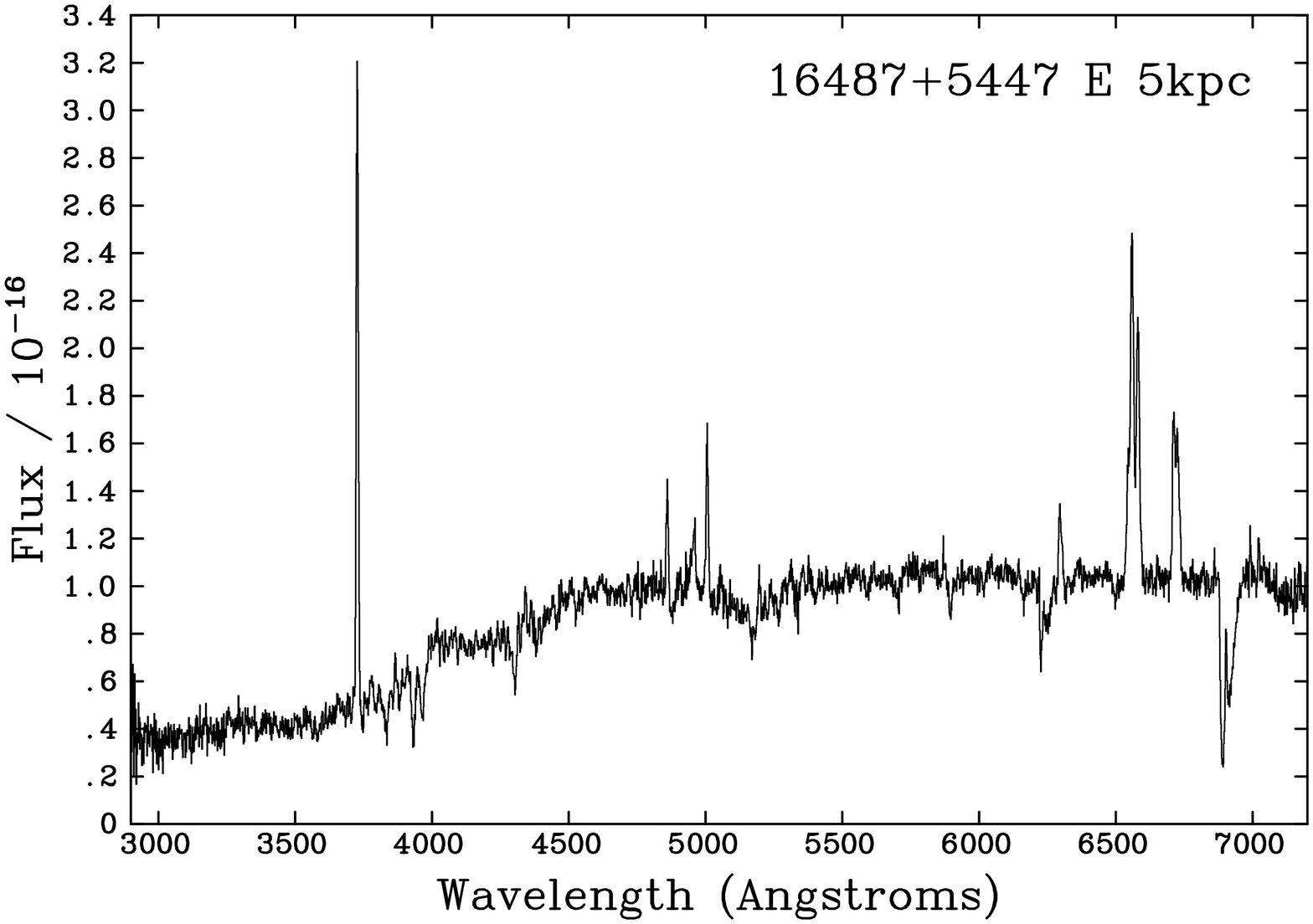,width=7.8cm,angle=0.}\\
\end{tabular}
\caption[{\it Continued}]{Continued}
%\label{fig:SED}
\end{minipage}
\end{figure*}
\addtocounter{figure}{-1}
\begin{figure*}
\begin{minipage}{170mm}
\begin{tabular}{cc}
\hspace*{0cm}\psfig{file=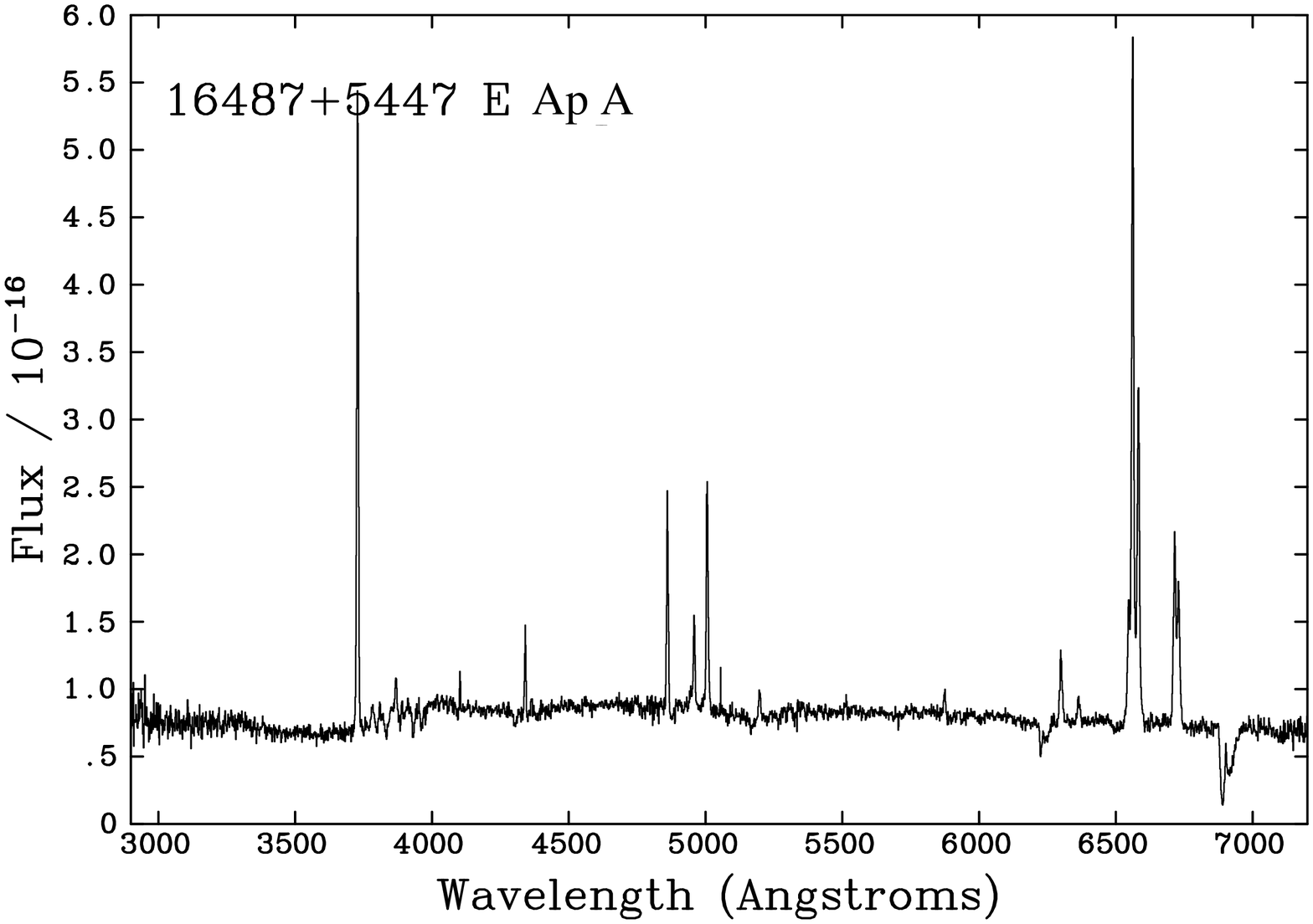,width=7.8cm,angle=0.}&
\psfig{file=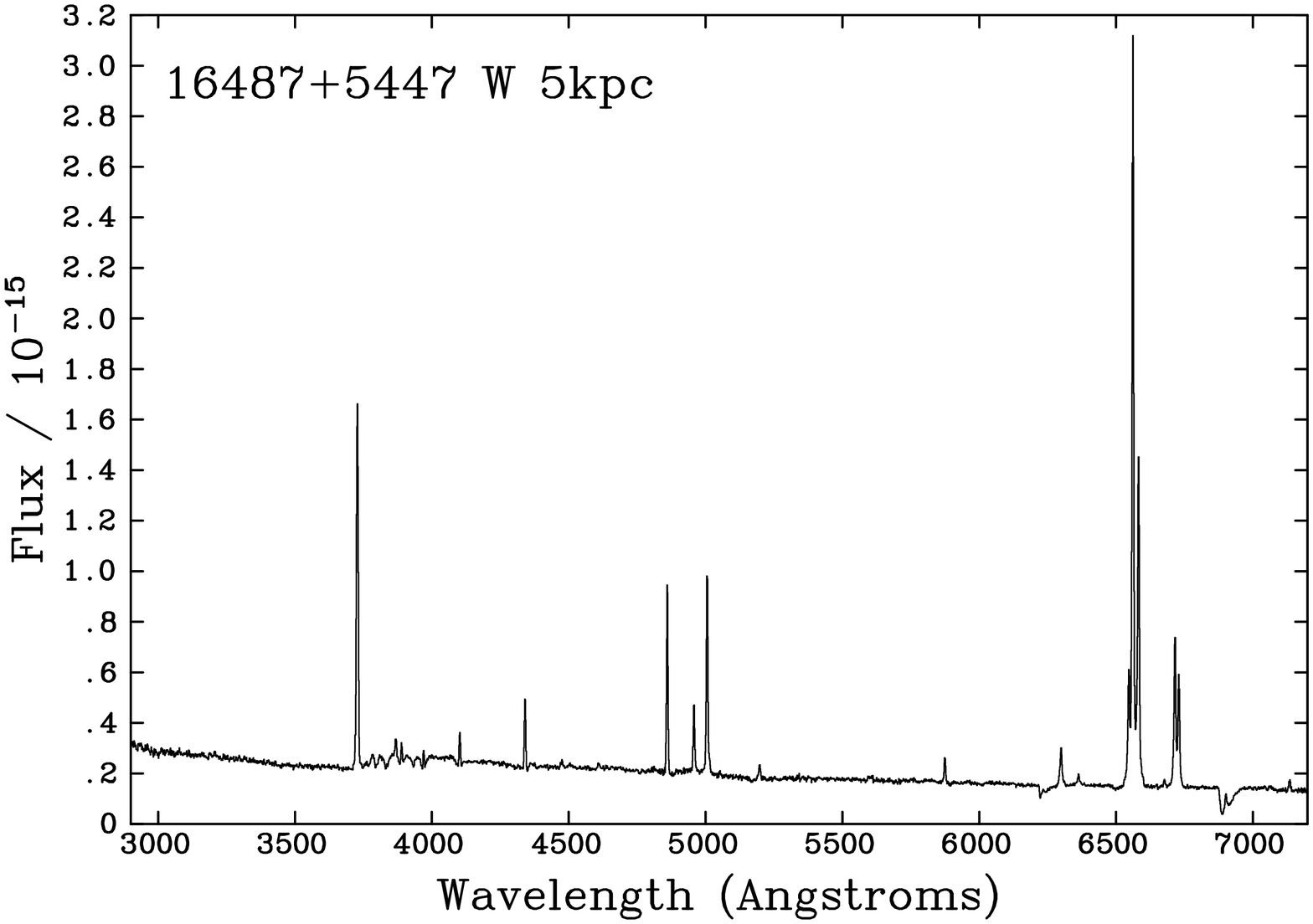,width=7.8cm,angle=0.}\\
\hspace*{0cm}\psfig{file=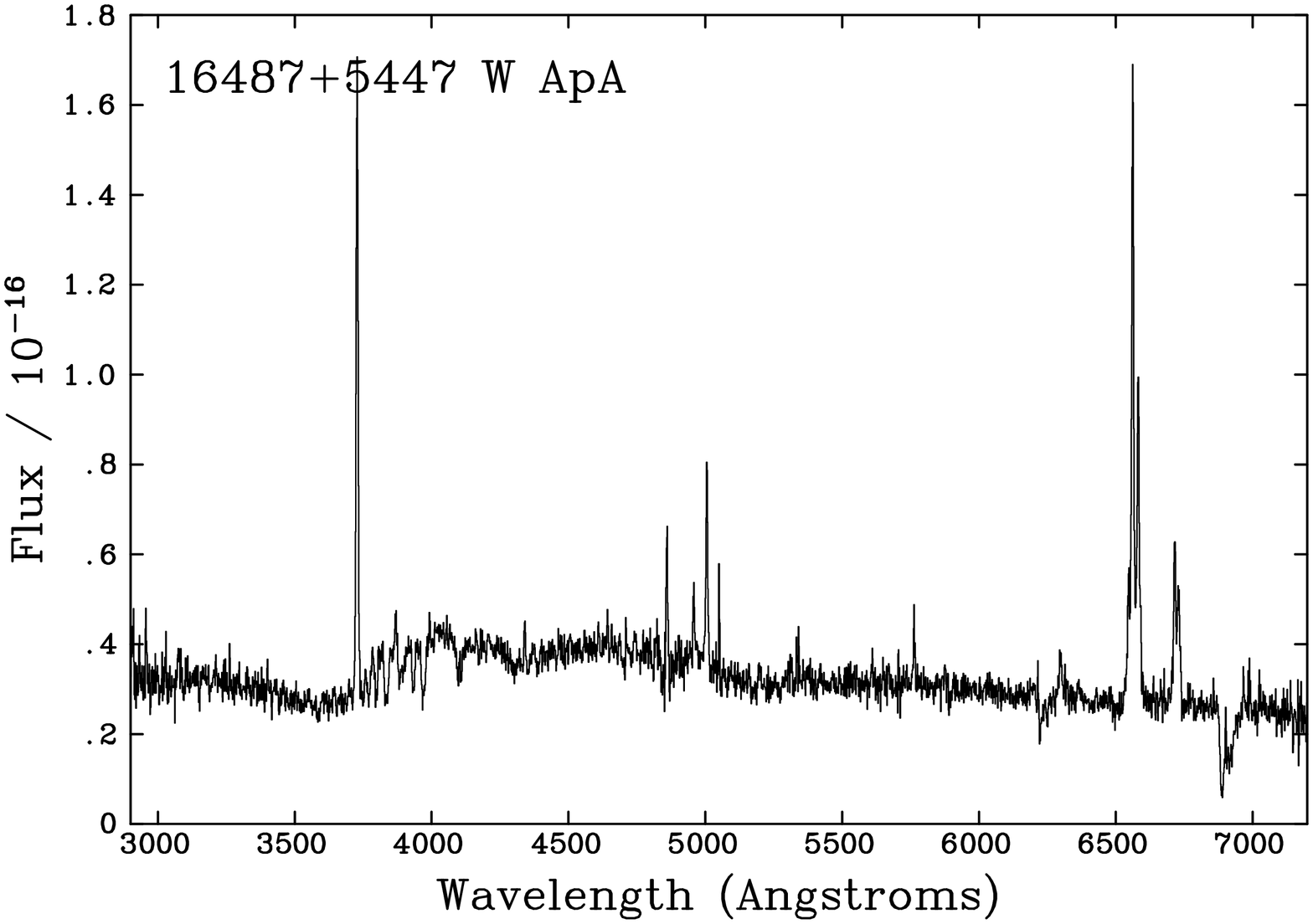,width=5.5cm,angle=-90.}&
\psfig{file=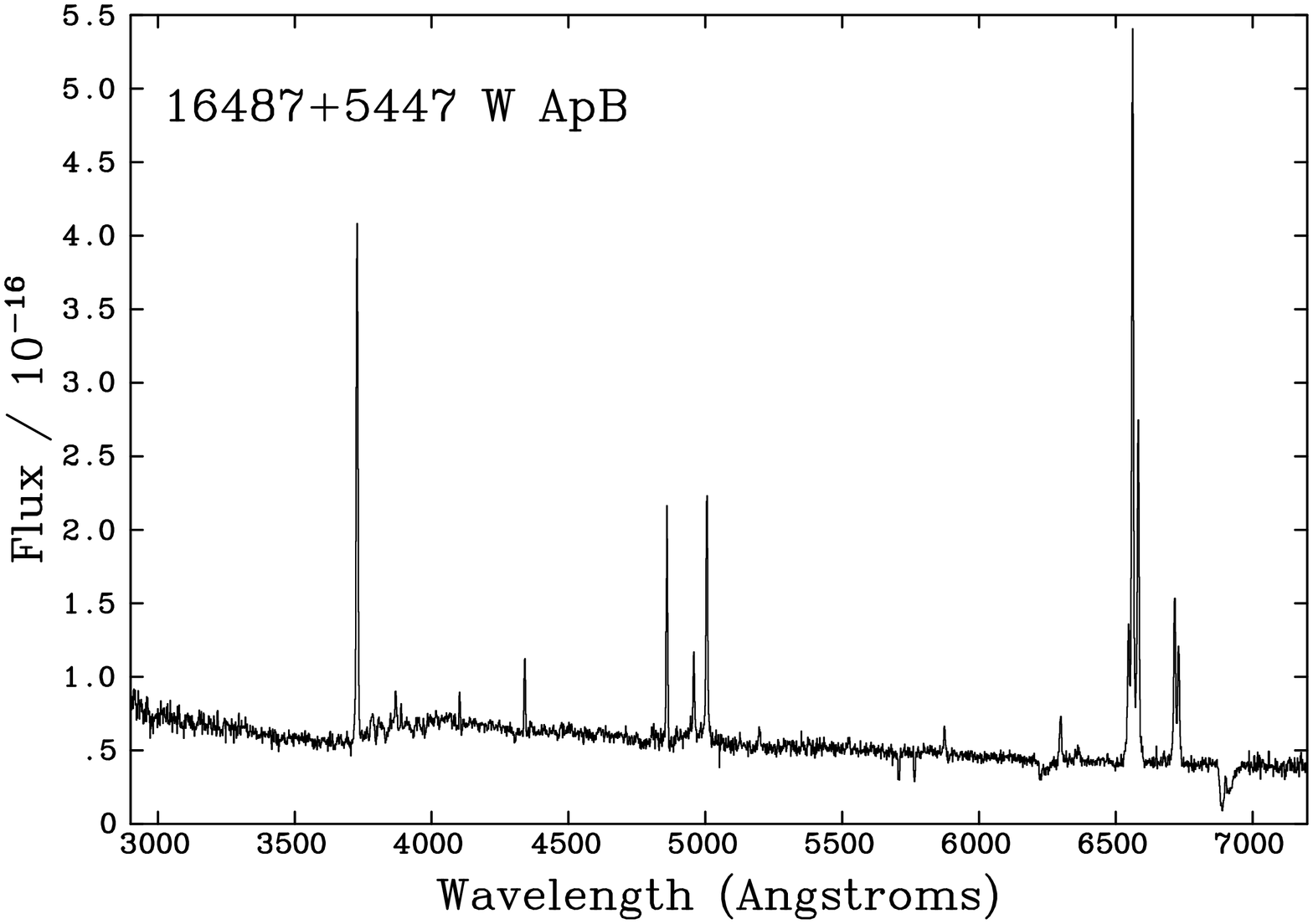,width=5.5cm,angle=-90.}\\
\hspace*{0cm}\psfig{file=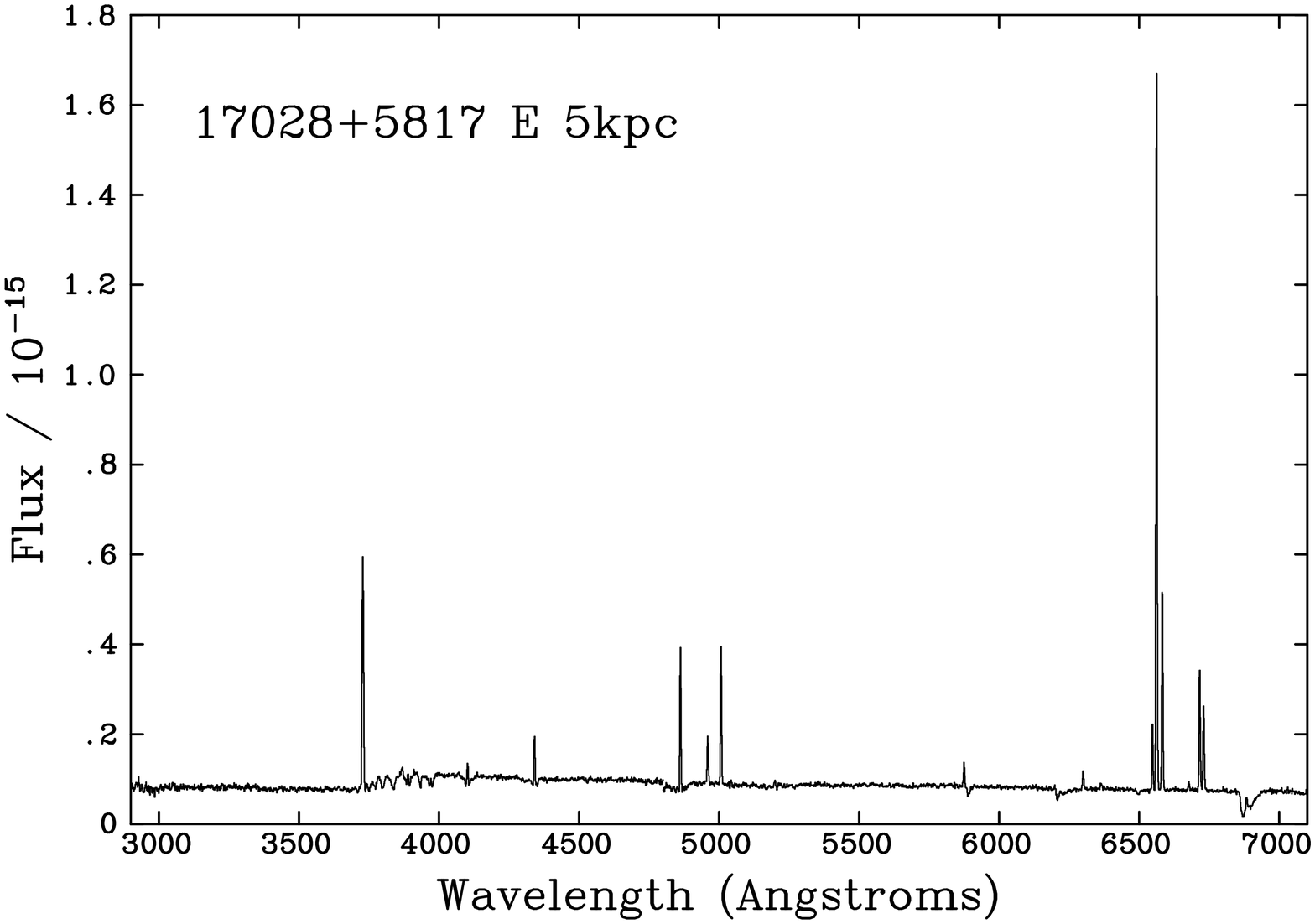,width=7.8cm,angle=0.}&
\psfig{file=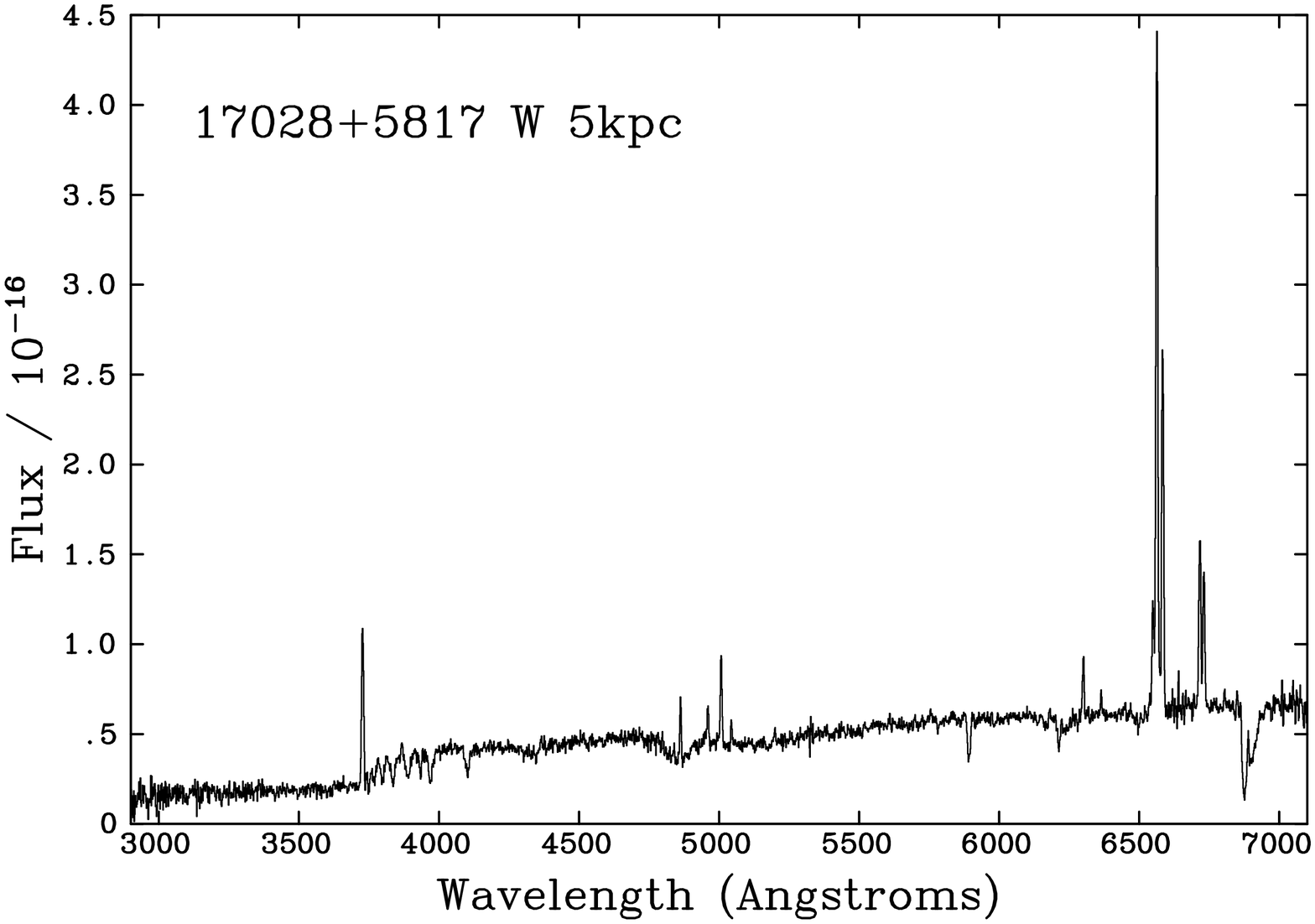,width=7.8cm,angle=0.}\\
\hspace*{0cm}\psfig{file=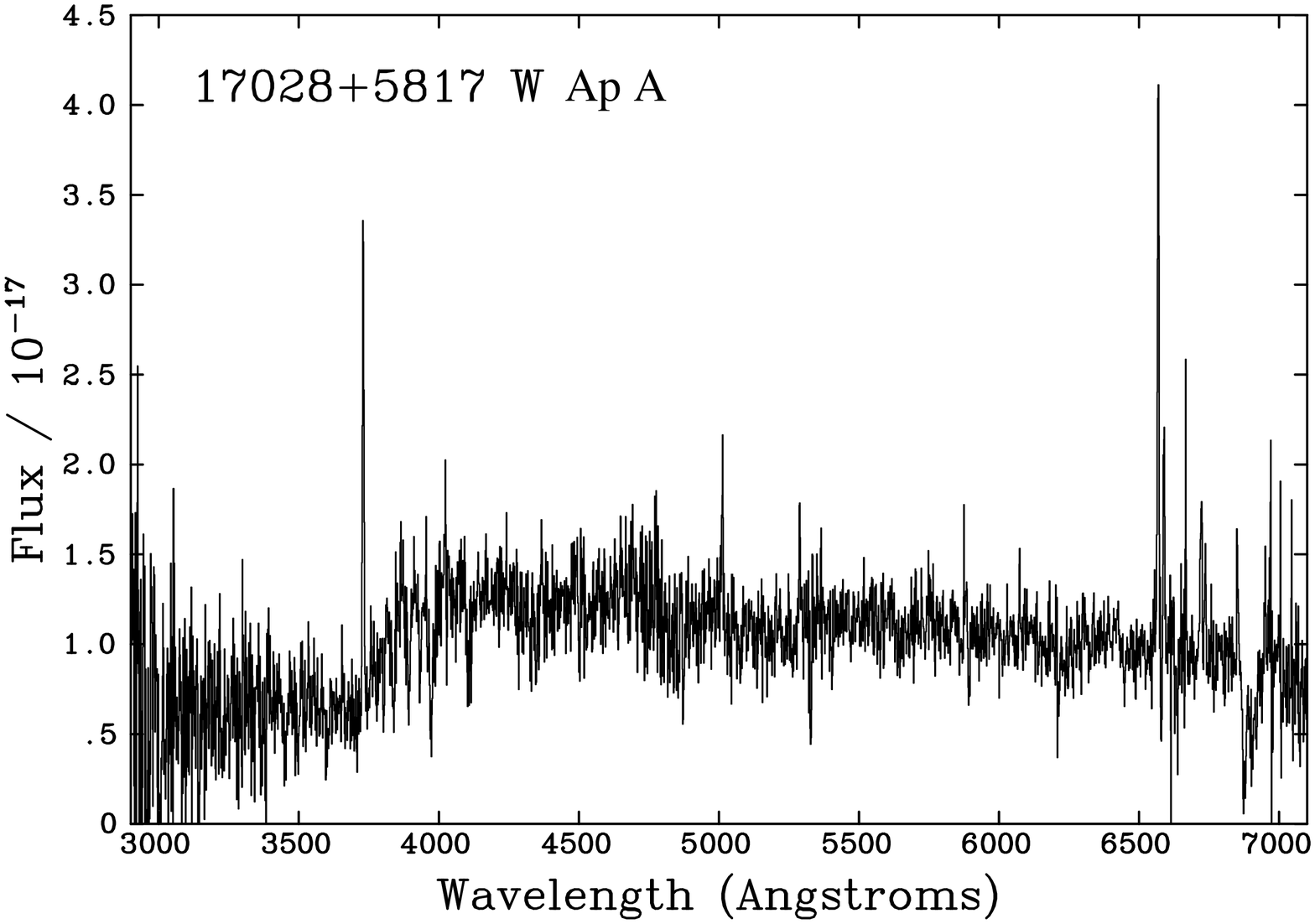,width=7.8cm,angle=0.}&
\psfig{file=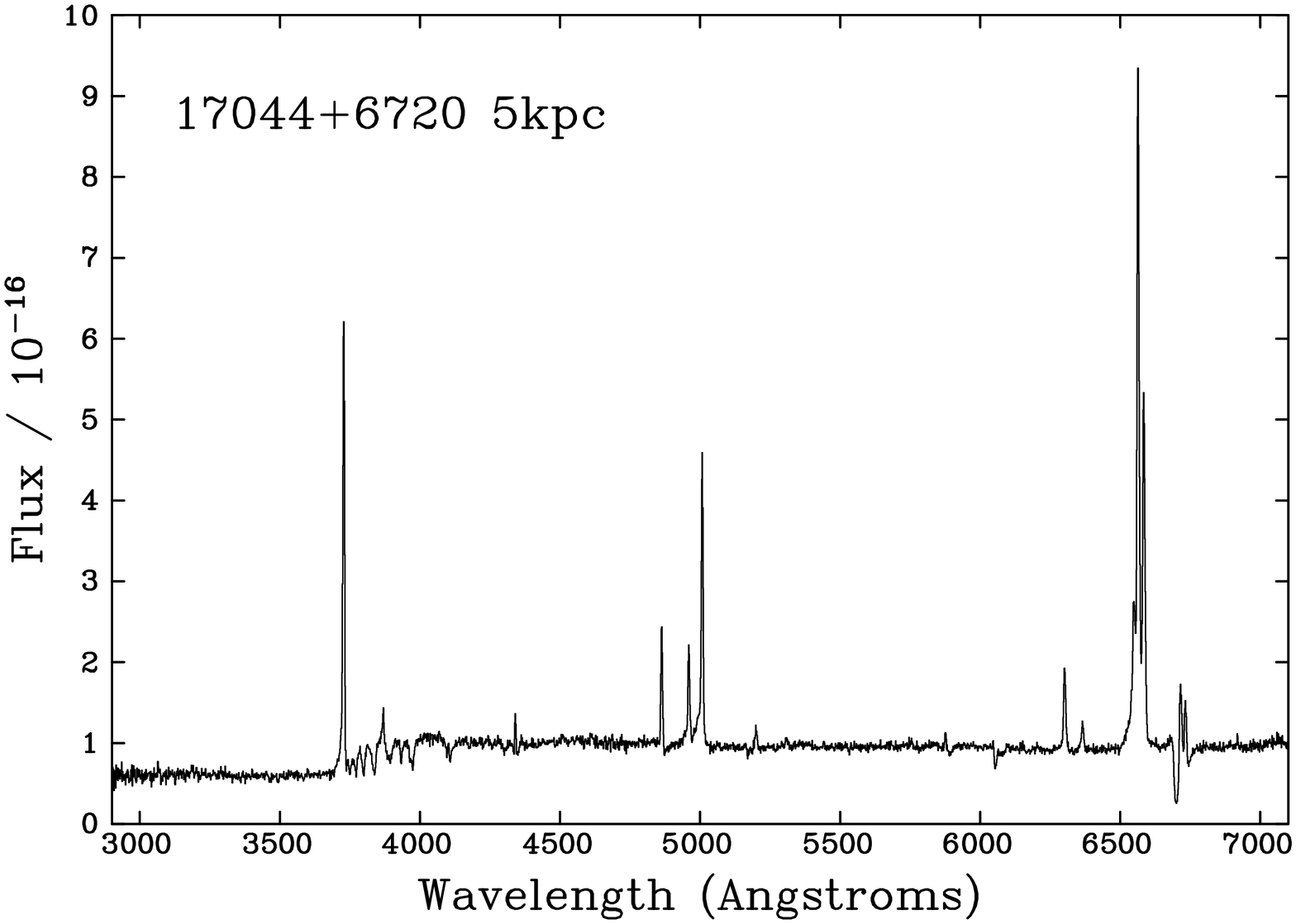,width=7.8cm,angle=0.}\\
\end{tabular}
\caption[{\it Continued}]{Continued}
%\label{fig:SED}
\end{minipage}
\end{figure*}
\addtocounter{figure}{-1}
\begin{figure*}
\begin{minipage}{170mm}
\begin{tabular}{cc}
\hspace*{0cm}\psfig{file=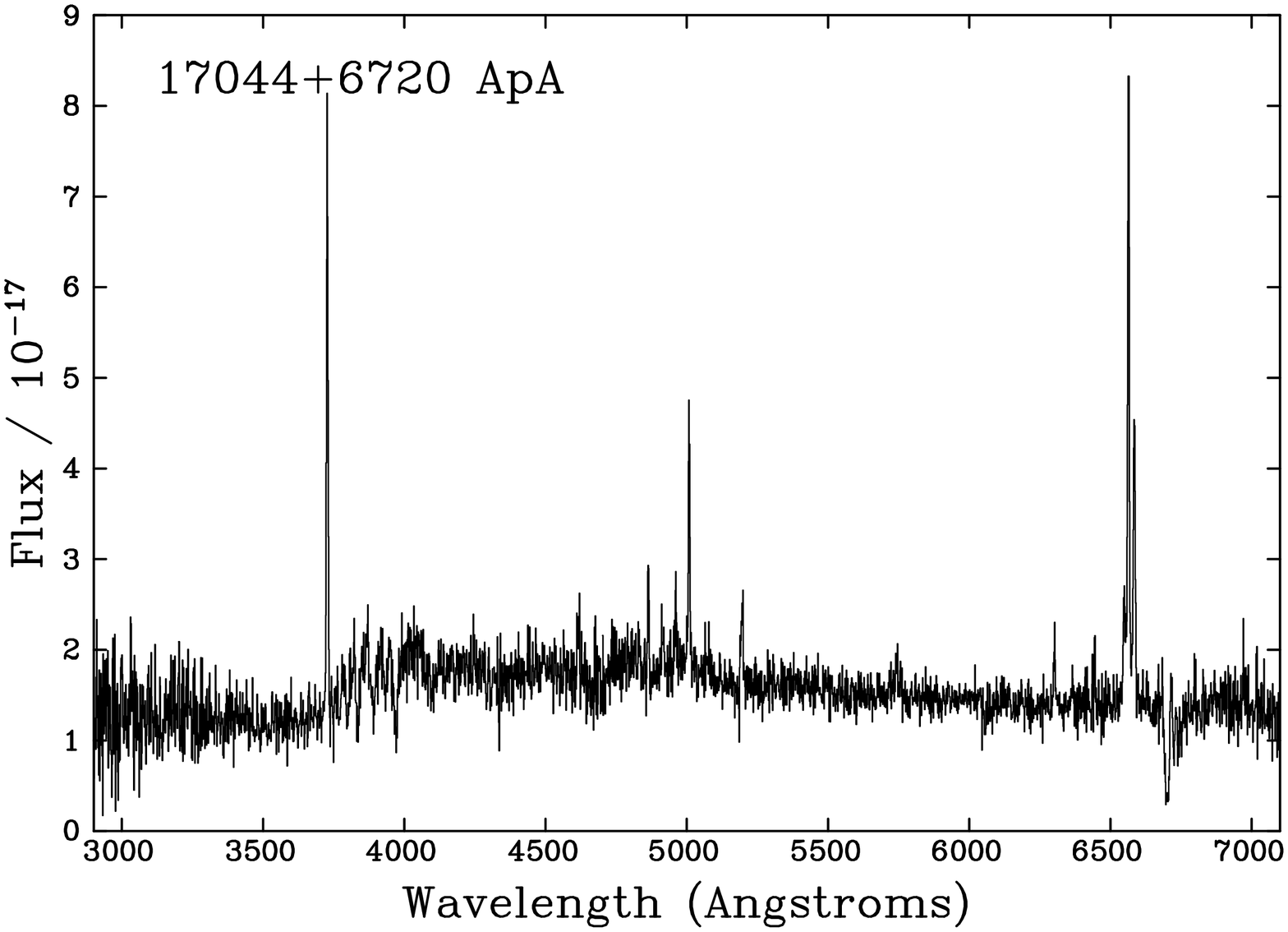,width=5.5cm,angle=-90.}&
\psfig{file=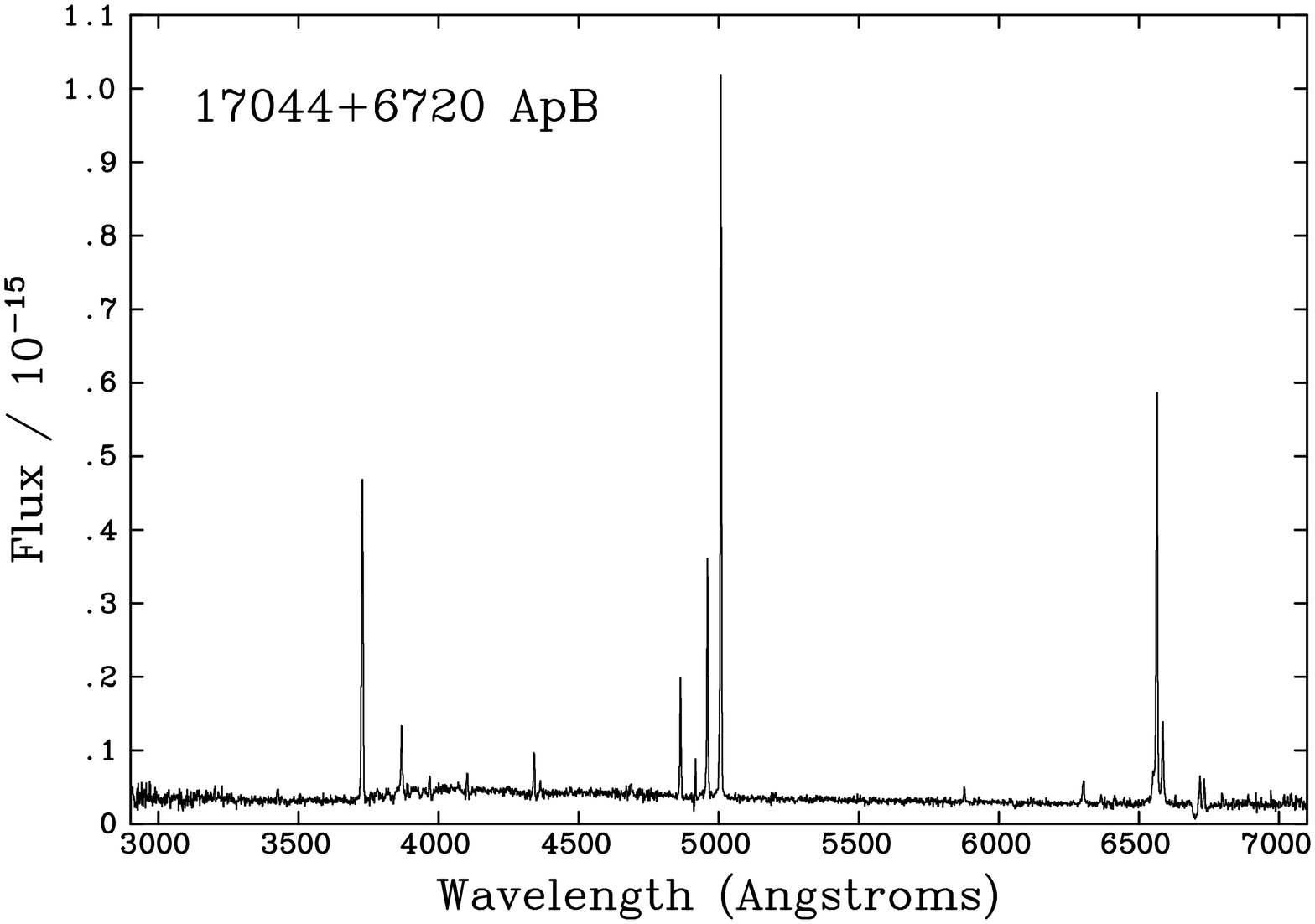,width=5.5cm,angle=-90.}\\
\hspace*{0cm}\psfig{file=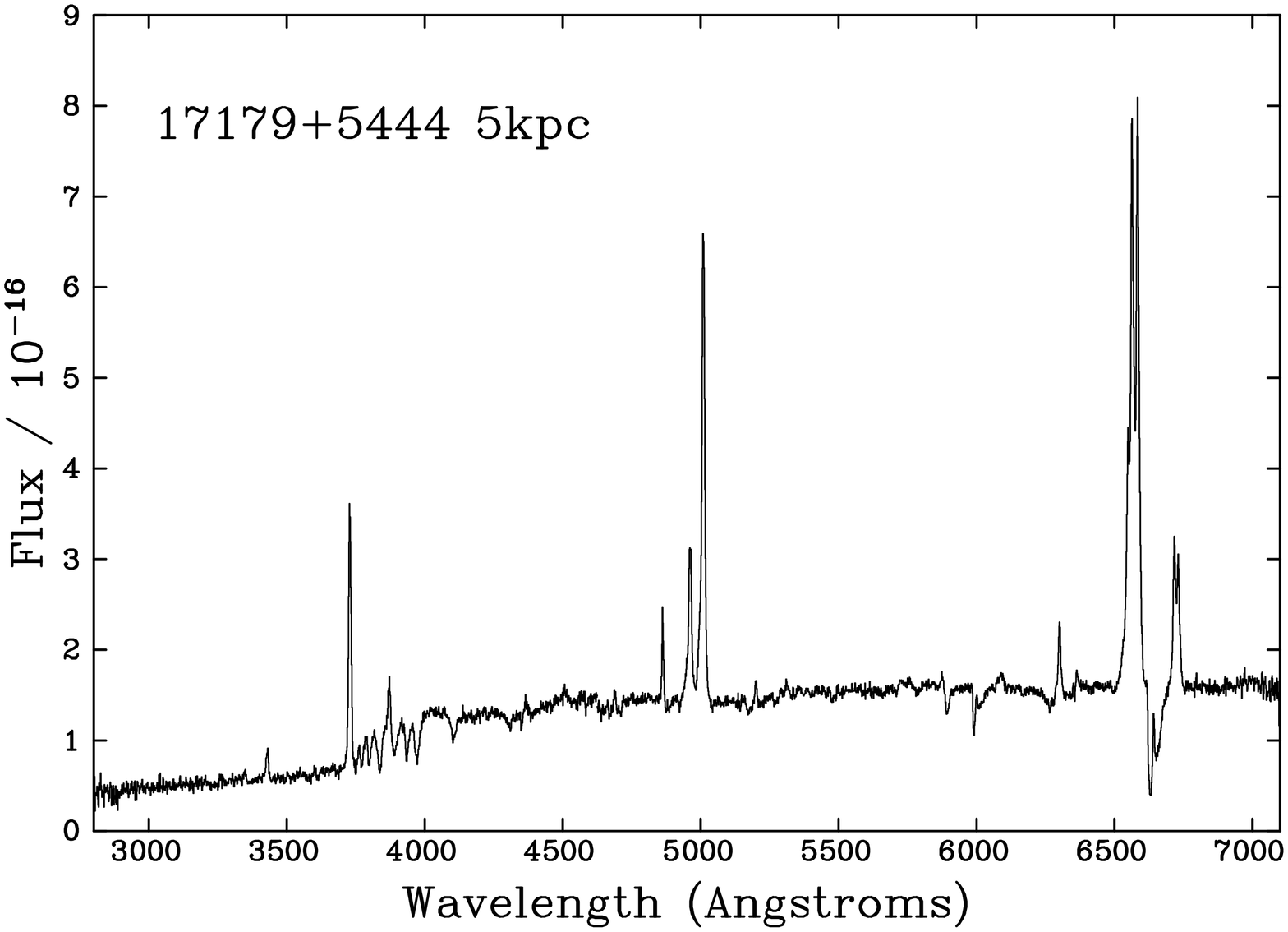,width=5.5cm,angle=-90.}&
\psfig{file=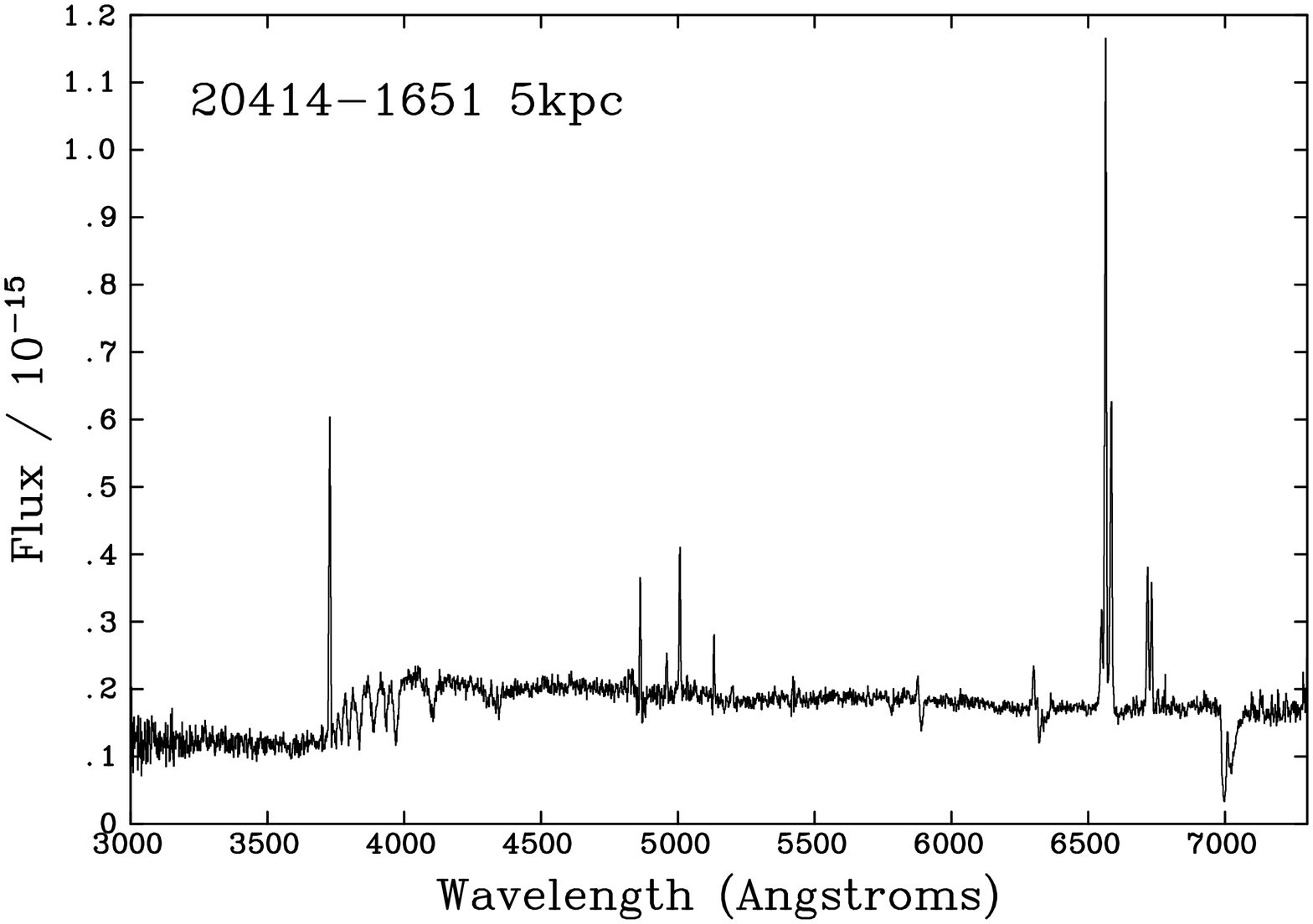,width=5.5cm,angle=-90.}\\
\hspace*{0cm}\psfig{file=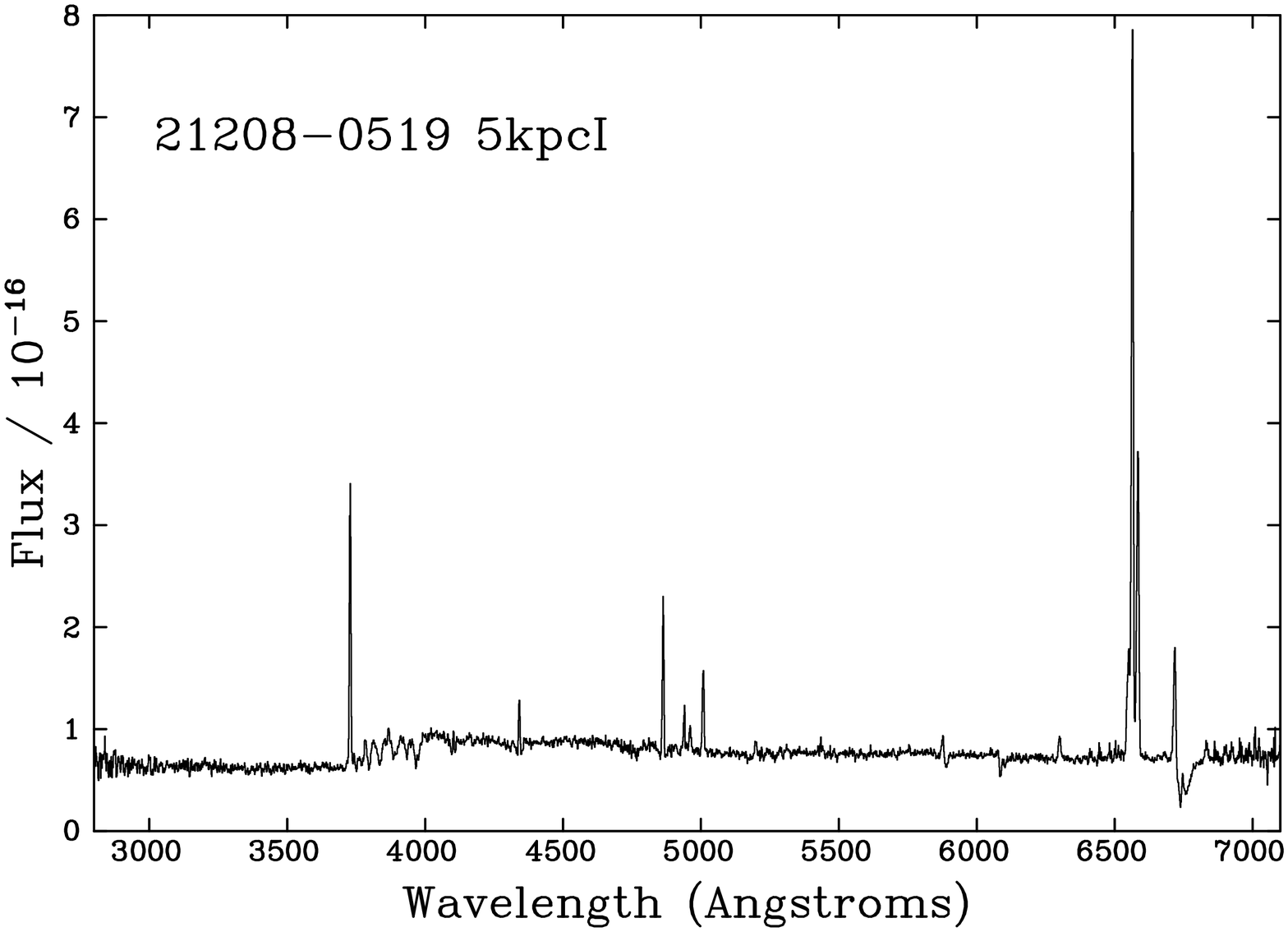,width=7.8cm,angle=0.}&
\psfig{file=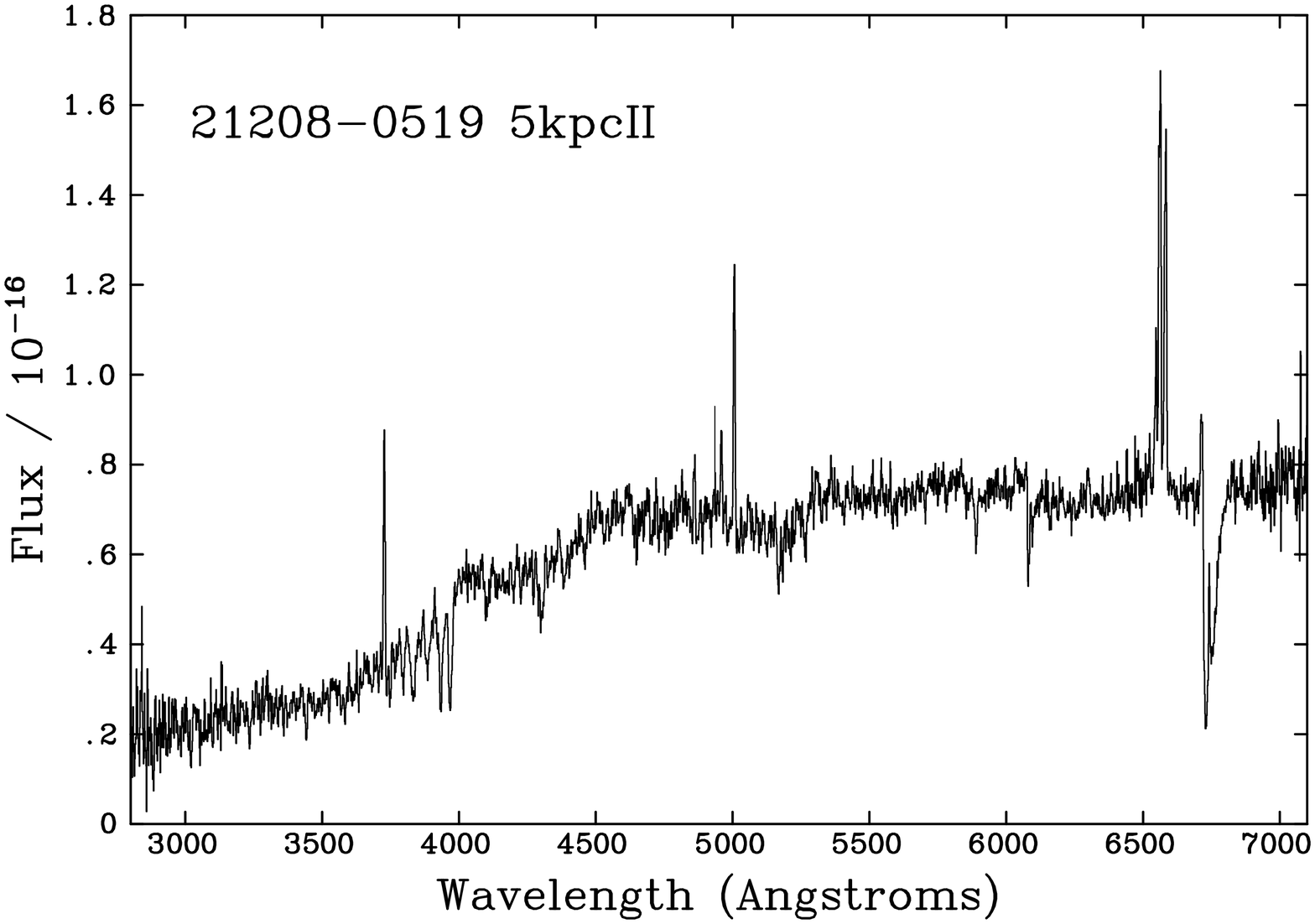,width=7.8cm,angle=0.}\\
\hspace*{0cm}\psfig{file=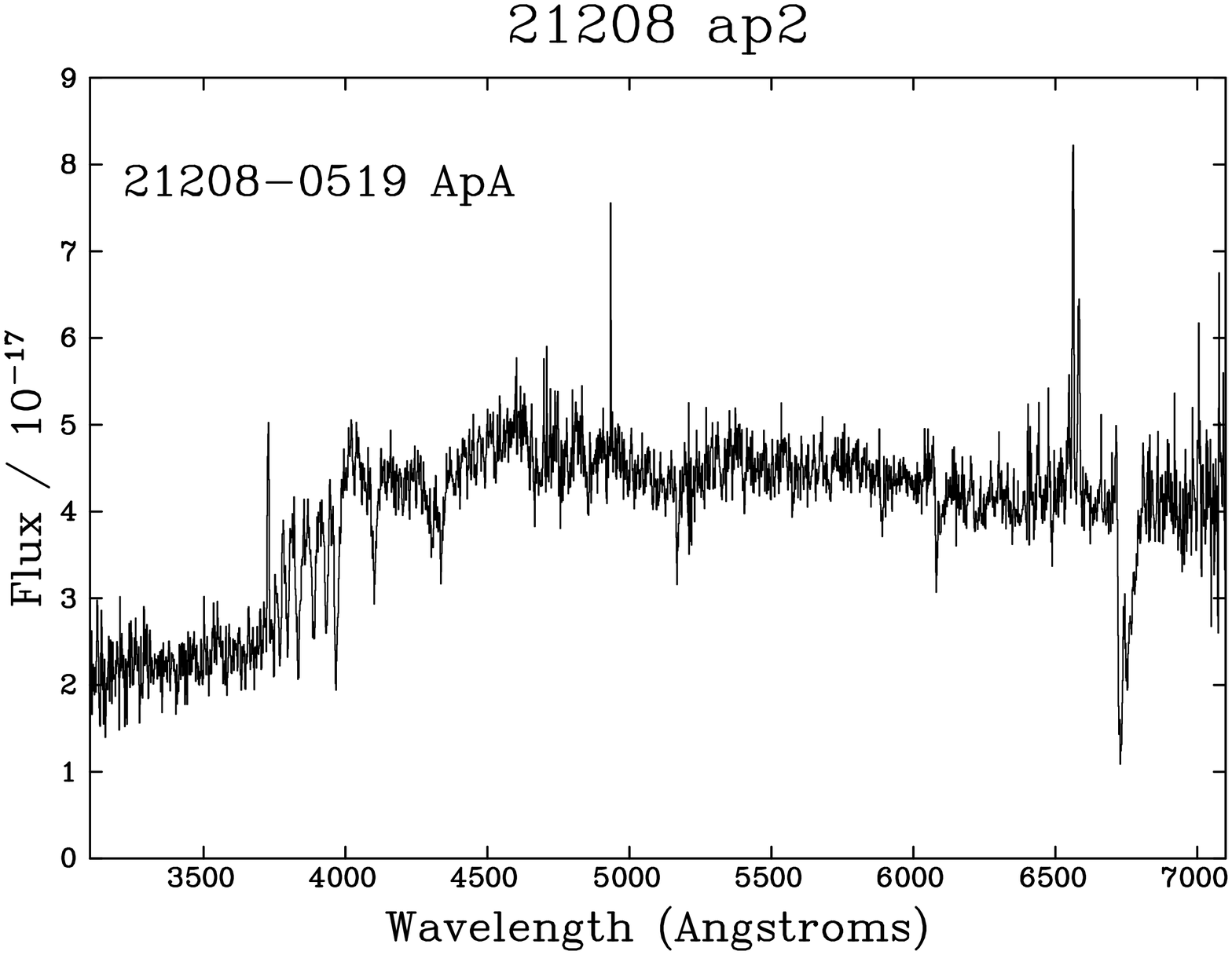,width=7.8cm,angle=0.}&
\psfig{file=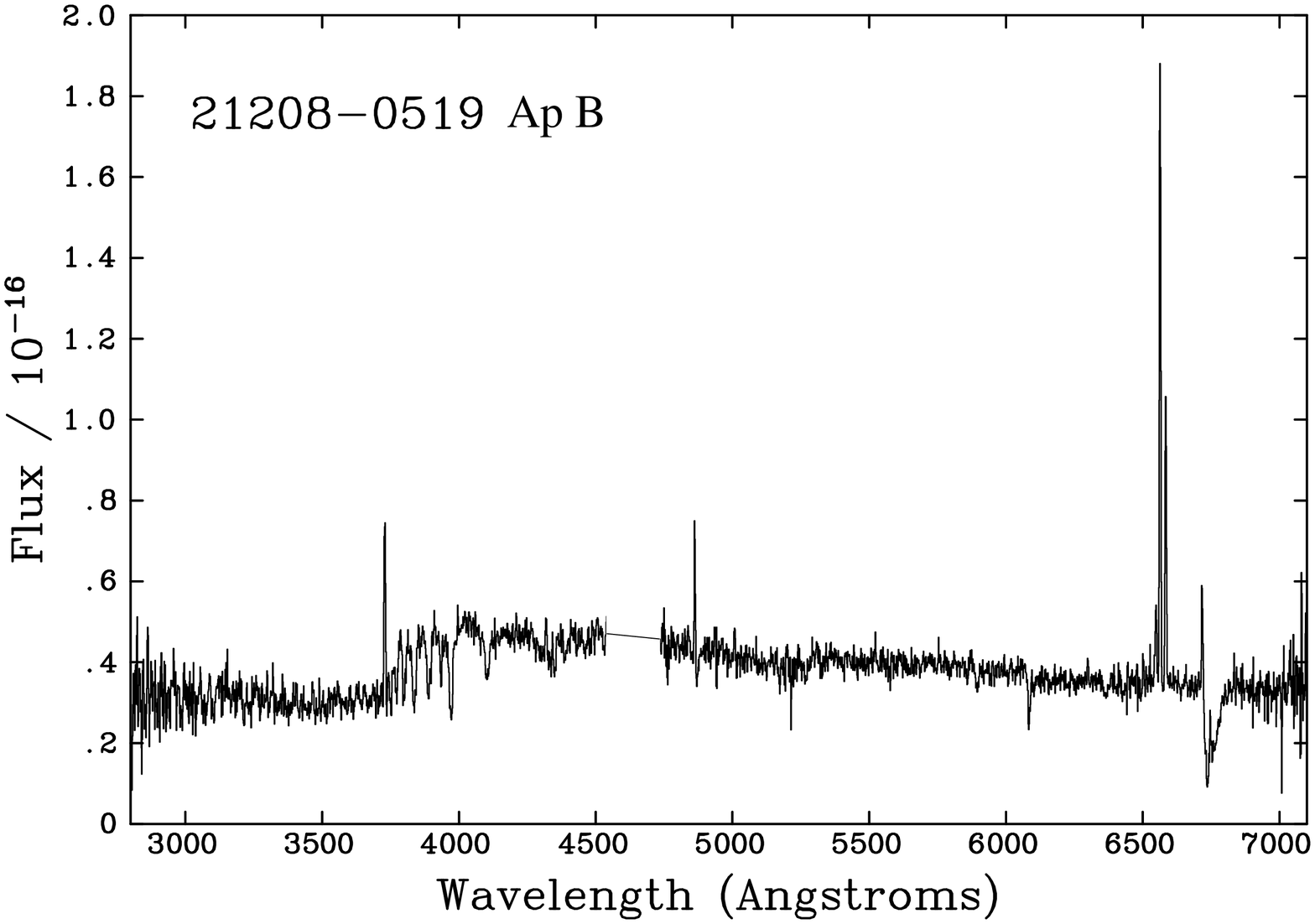,width=7.8cm,angle=0.}\\
\end{tabular}
\caption[{\it Continued}]{Continued}
%\label{fig:SED}
\end{minipage}
\end{figure*}
\addtocounter{figure}{-1}
\begin{figure*}
\begin{minipage}{170mm}
\begin{tabular}{cc}
\hspace*{0cm}\psfig{file=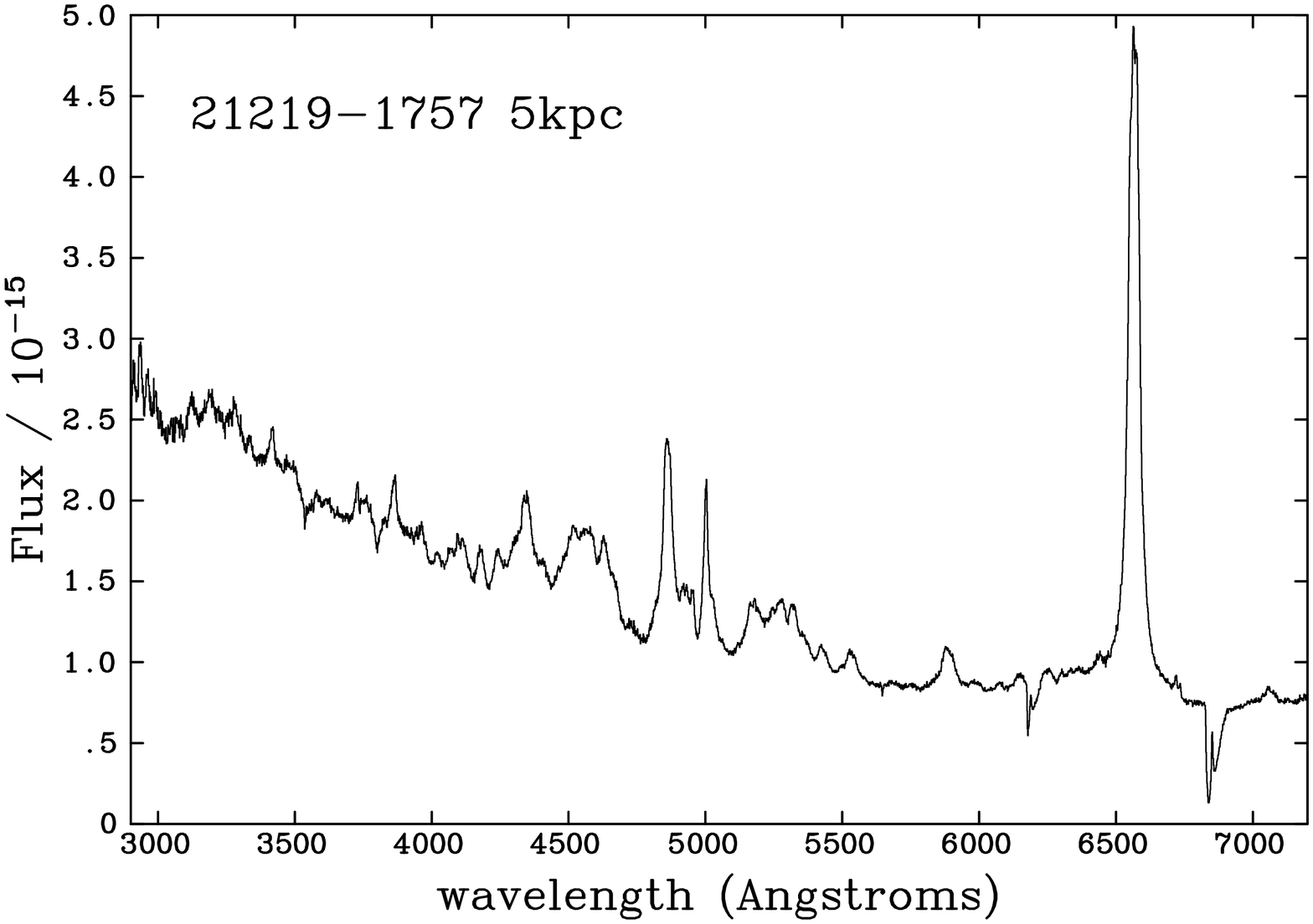,width=7.8cm,angle=0.}&
\psfig{file=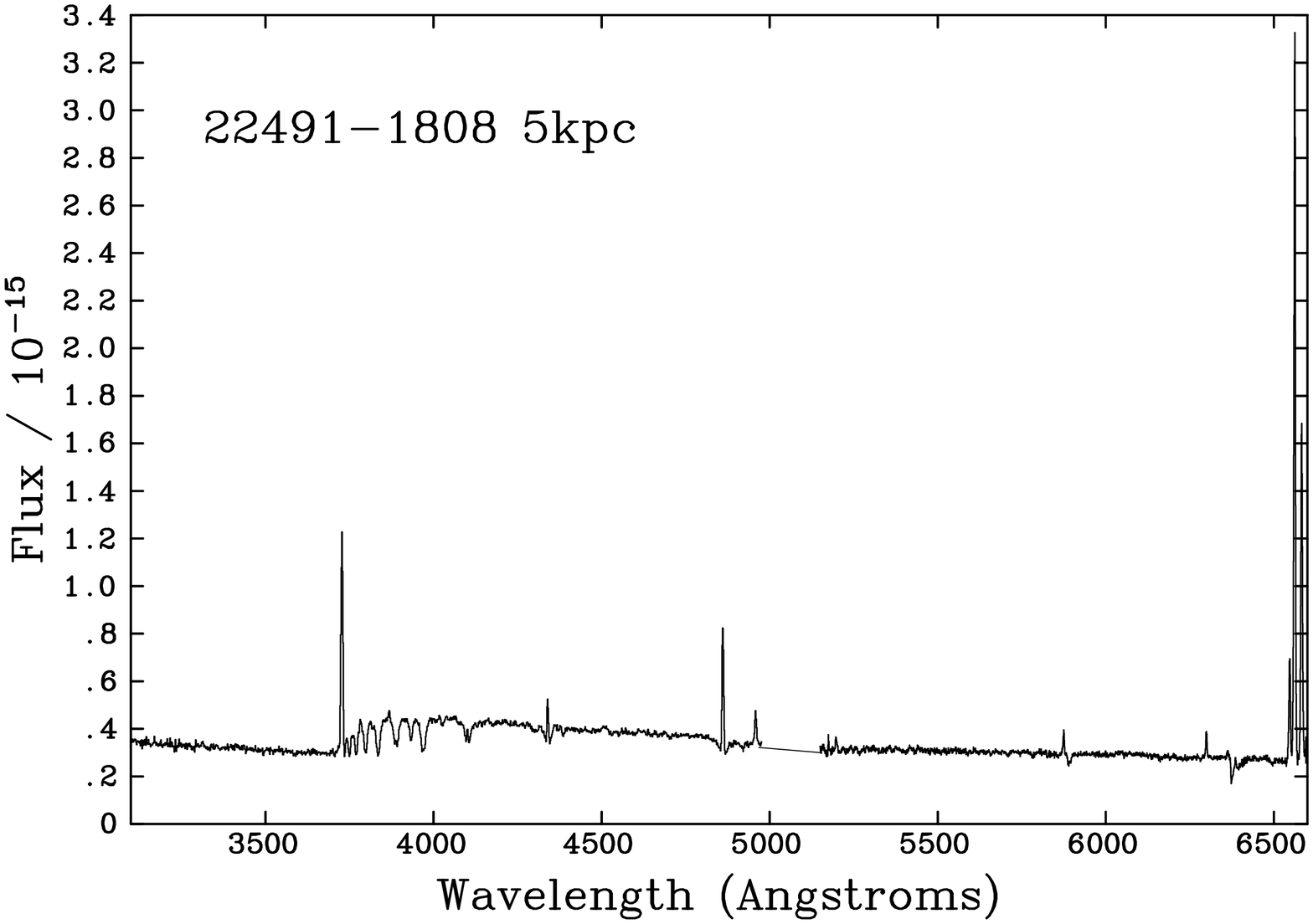,width=7.8cm,angle=-0.}\\
\hspace*{0cm}\psfig{file=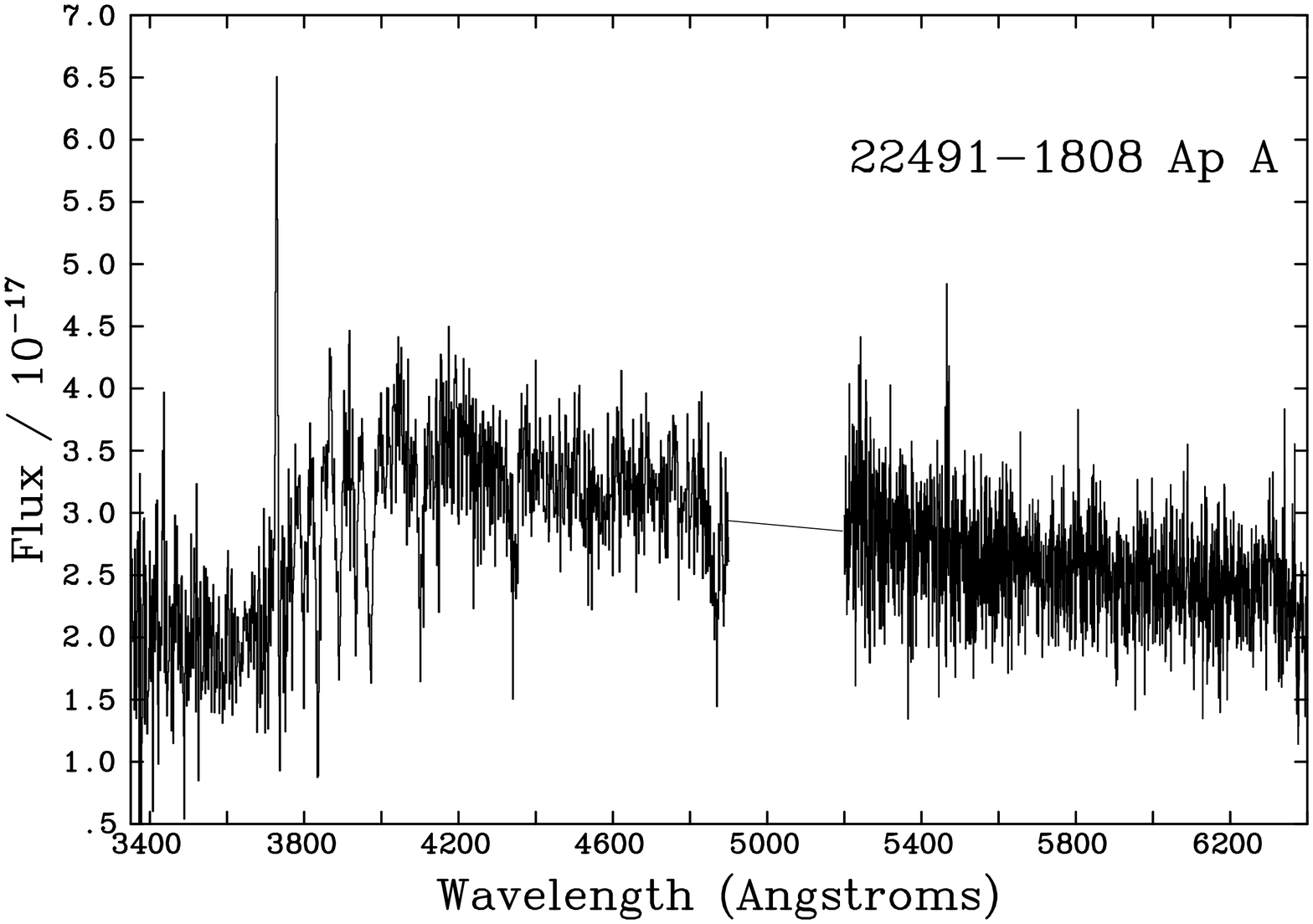,width=7.8cm,angle=-0.}&
\psfig{file=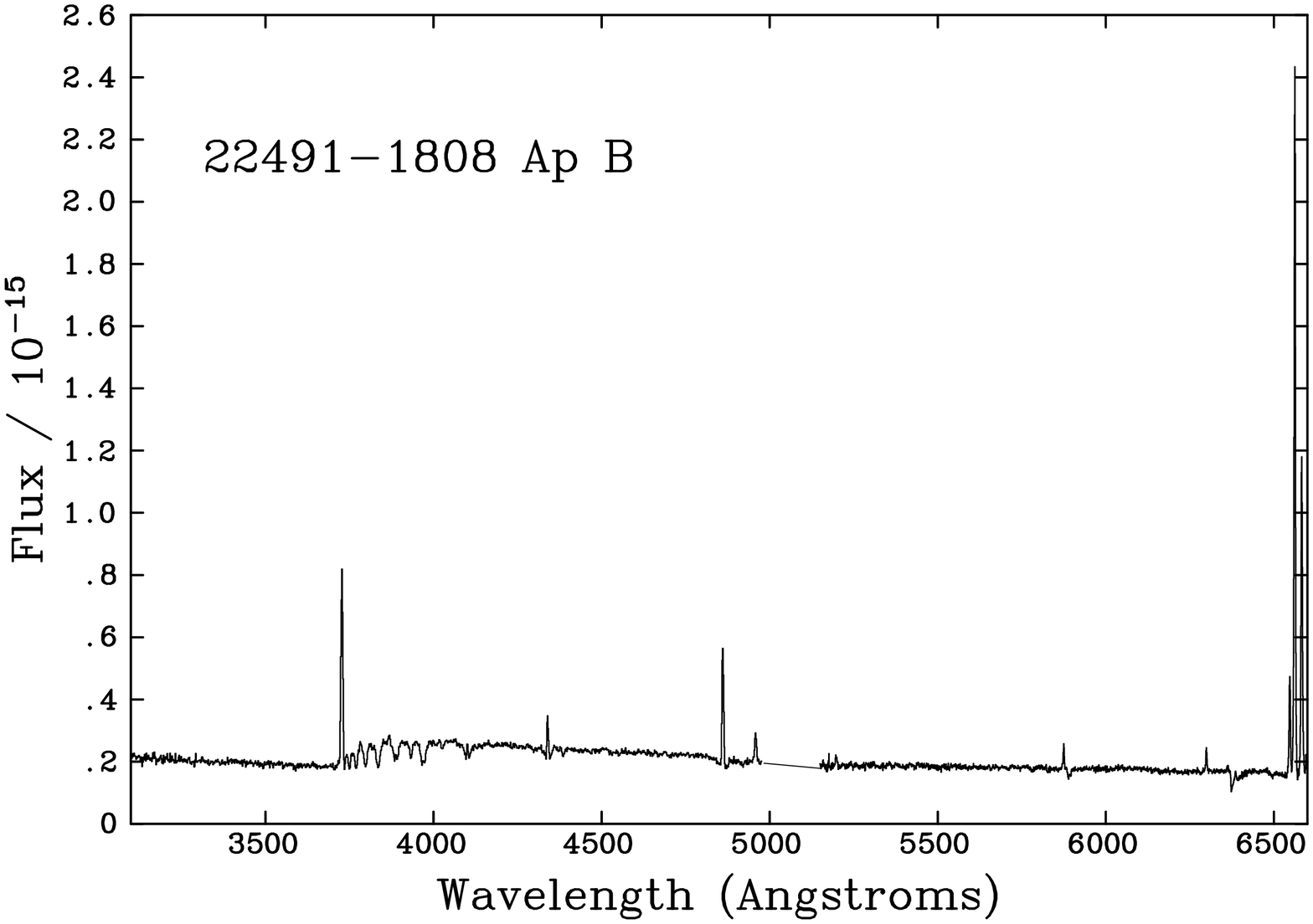,width=7.8cm,angle=-0.}\\
\hspace*{0cm}\psfig{file=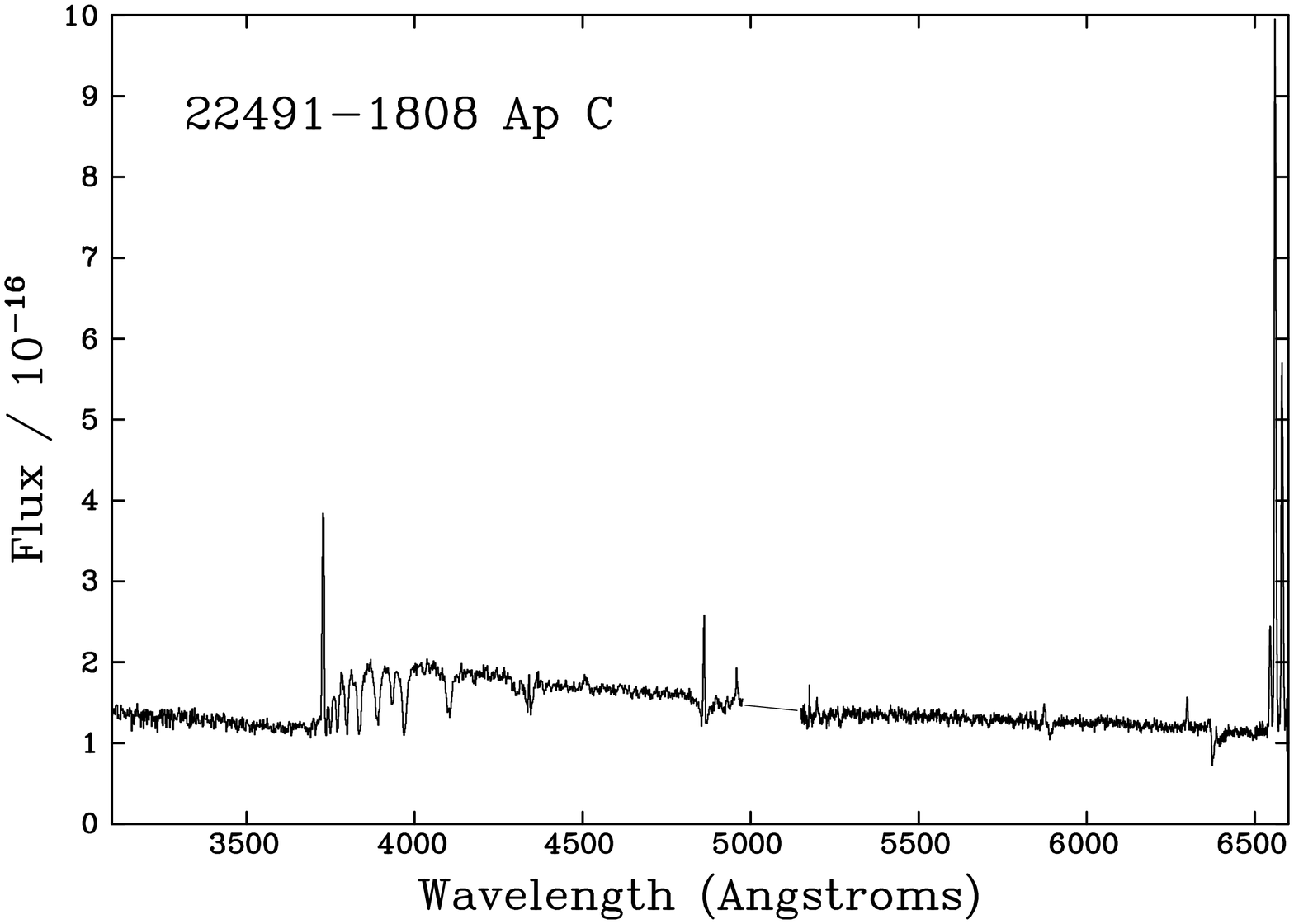,width=7.8cm,angle=-0.}&
\psfig{file=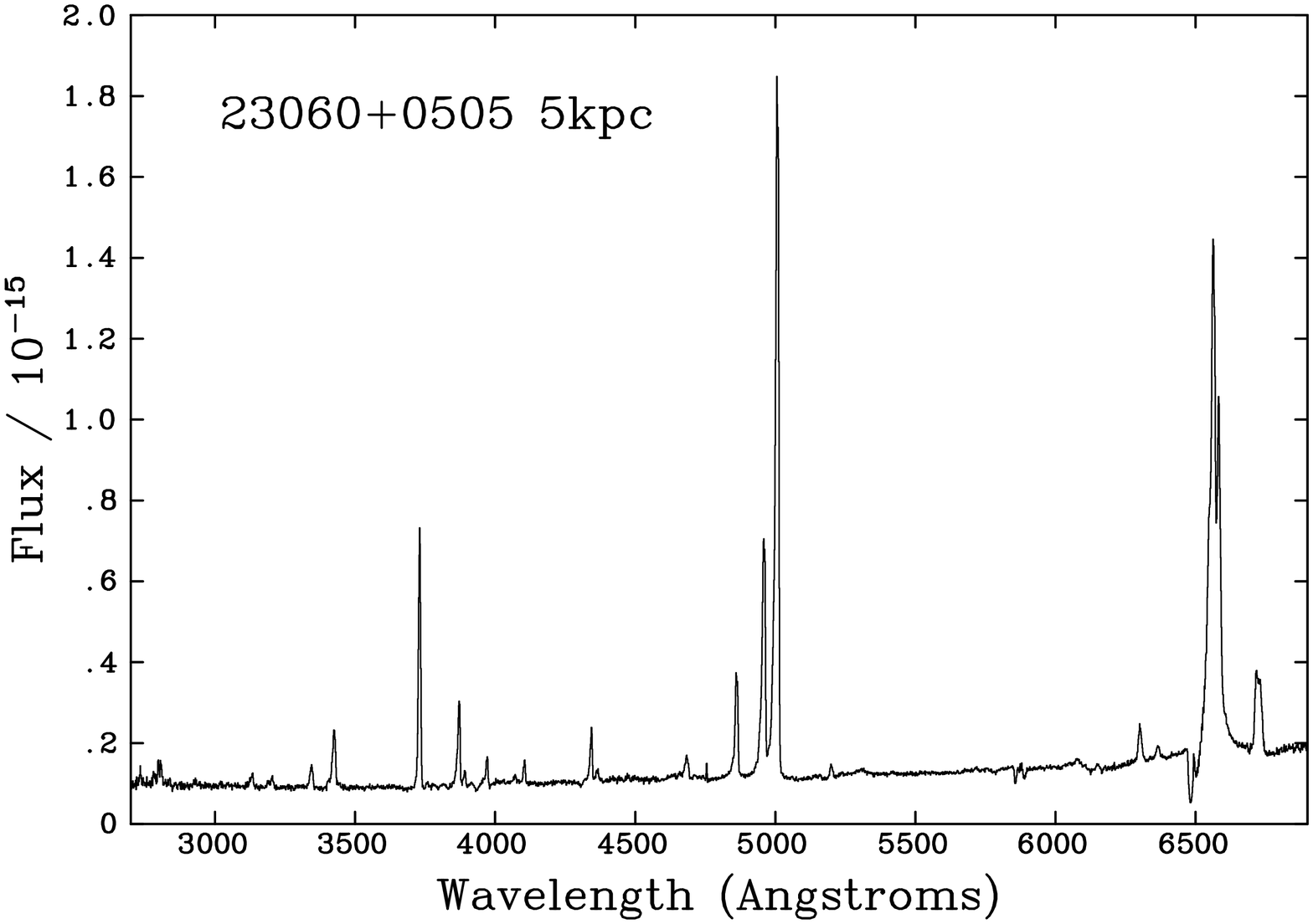,width=7.8cm,angle=0.}\\
\hspace*{0cm}\psfig{file=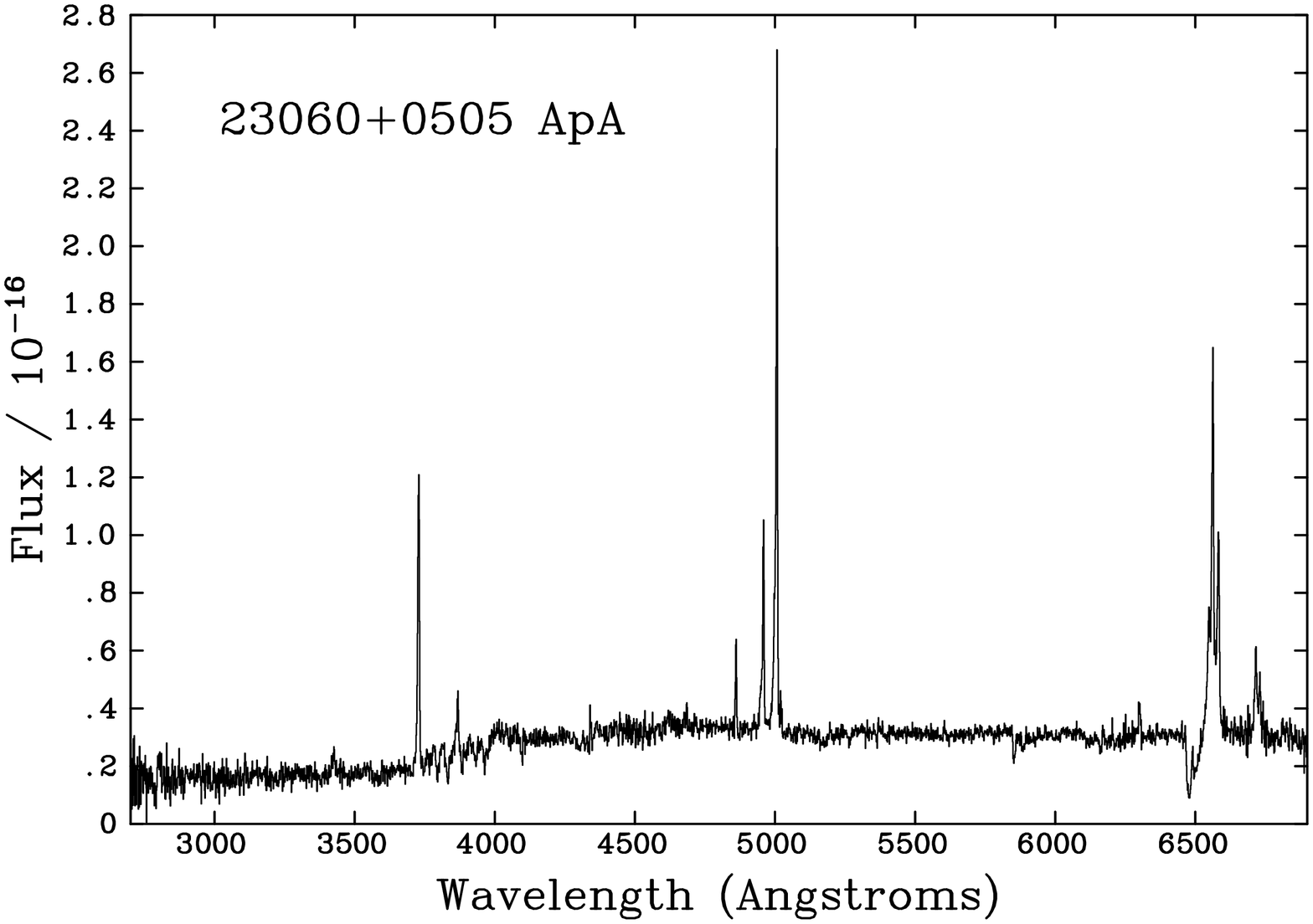,width=5.5cm,angle=-90.}&
\psfig{file=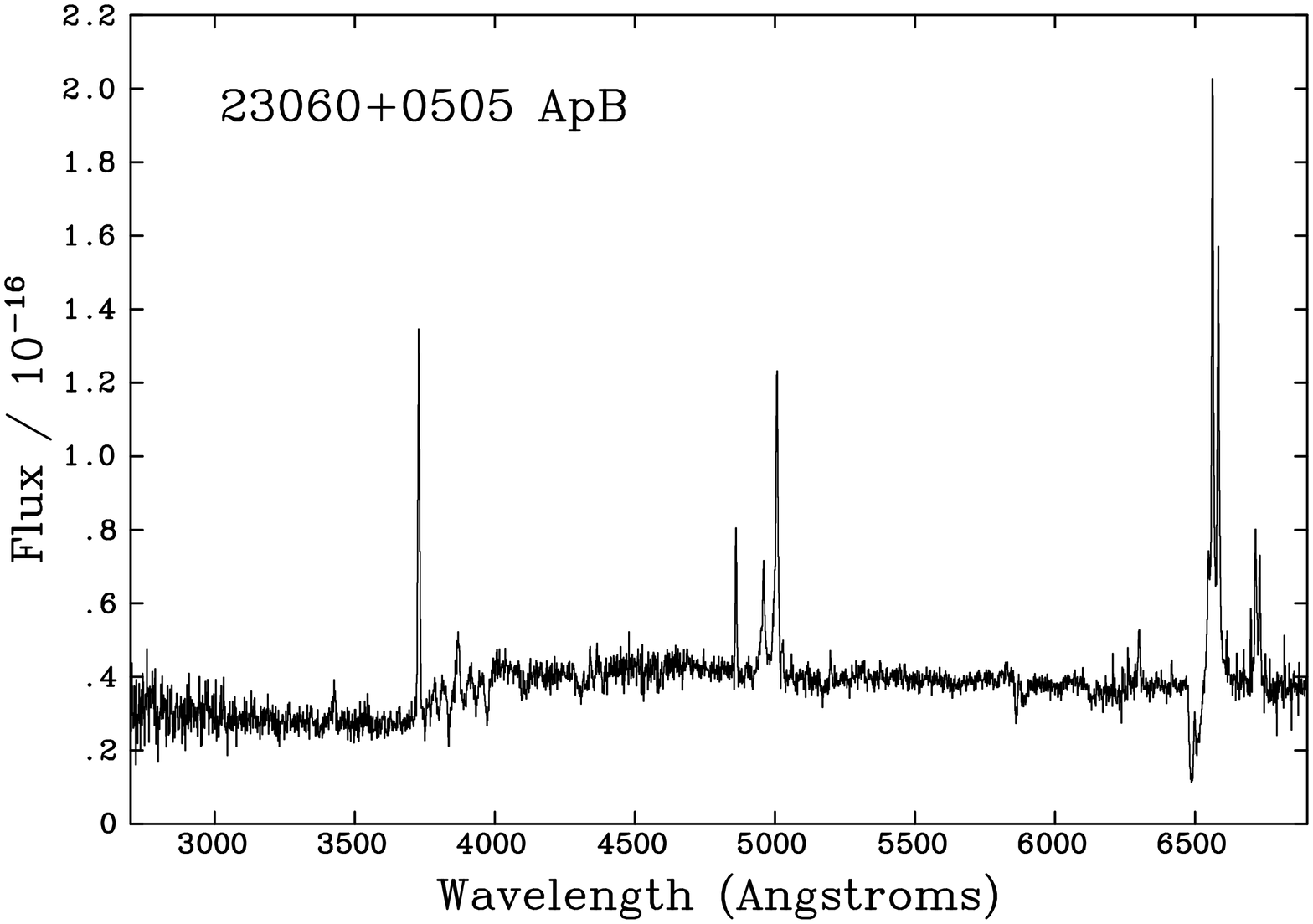,width=5.5cm,angle=-90.}\\
\end{tabular}
\caption[{\it Continued}]{Continued}
%\label{fig:SED}
\end{minipage}
\end{figure*}
\addtocounter{figure}{-1}
\begin{figure*}
\begin{minipage}{170mm}
\begin{tabular}{cc}
\hspace*{0cm}\psfig{file=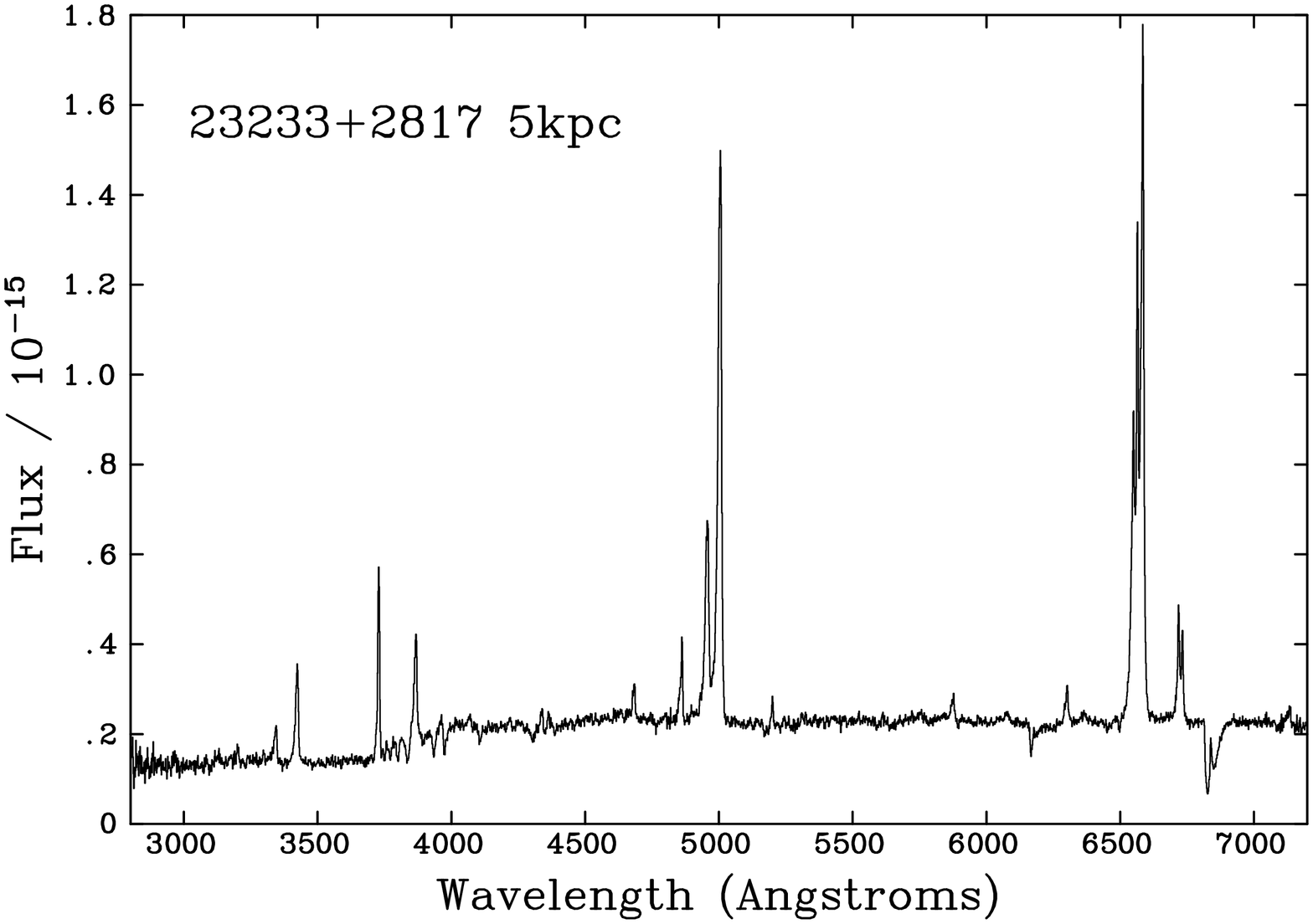,width=5.5cm,angle=-90.}&
\psfig{file=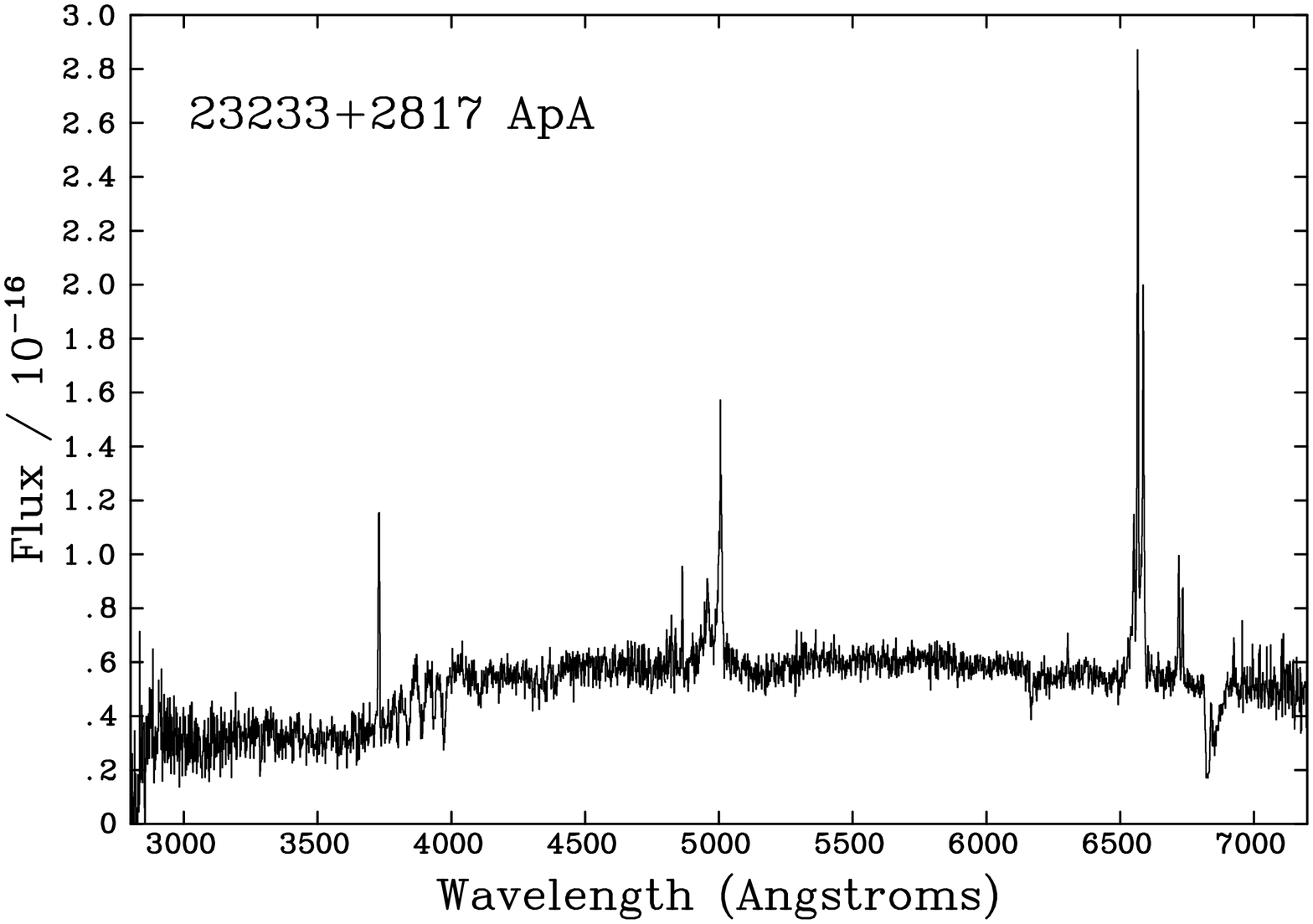,width=5.5cm,angle=-90.}\\
\hspace*{0cm}\psfig{file=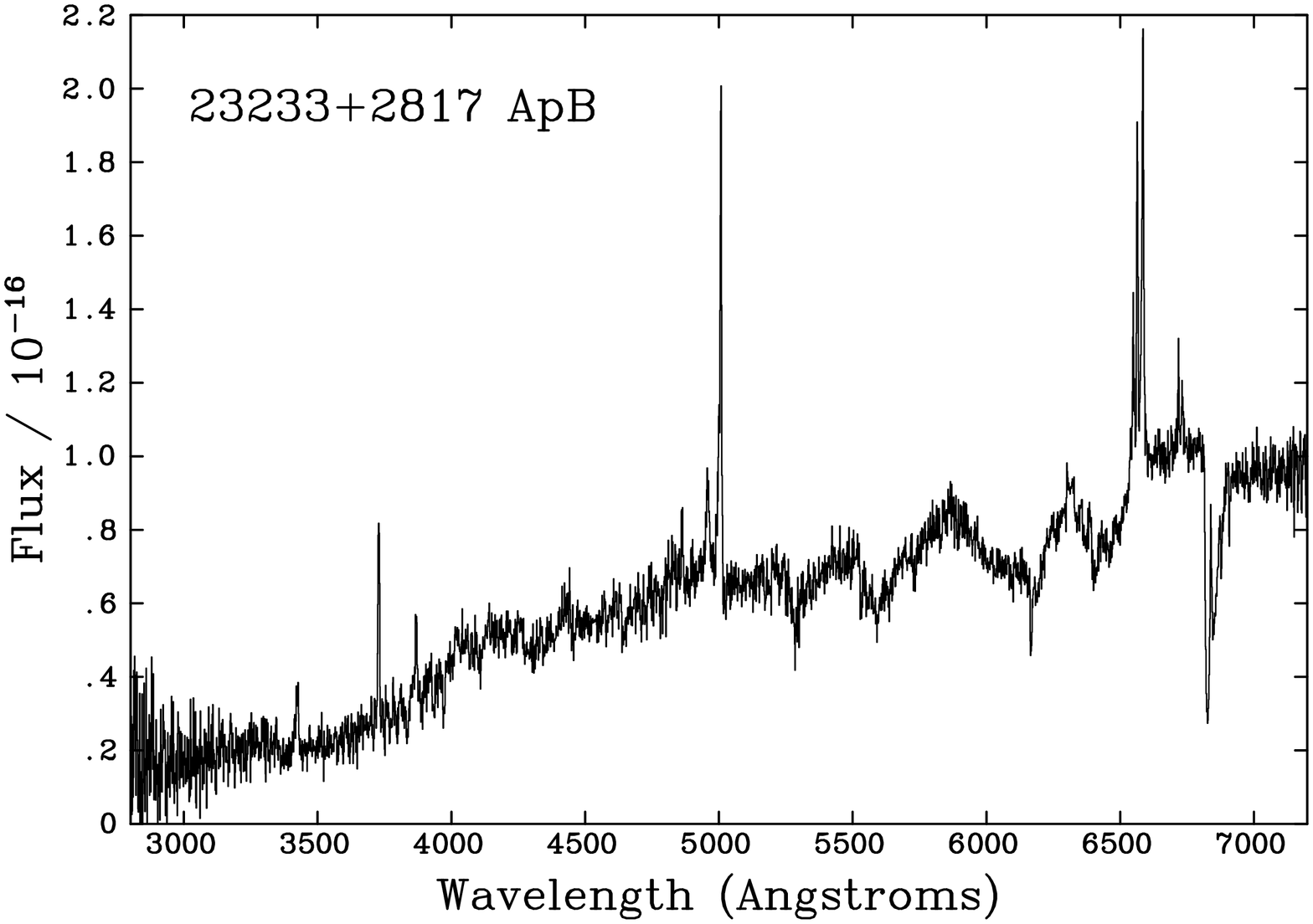,width=5.5cm,angle=-90.}&
\psfig{file=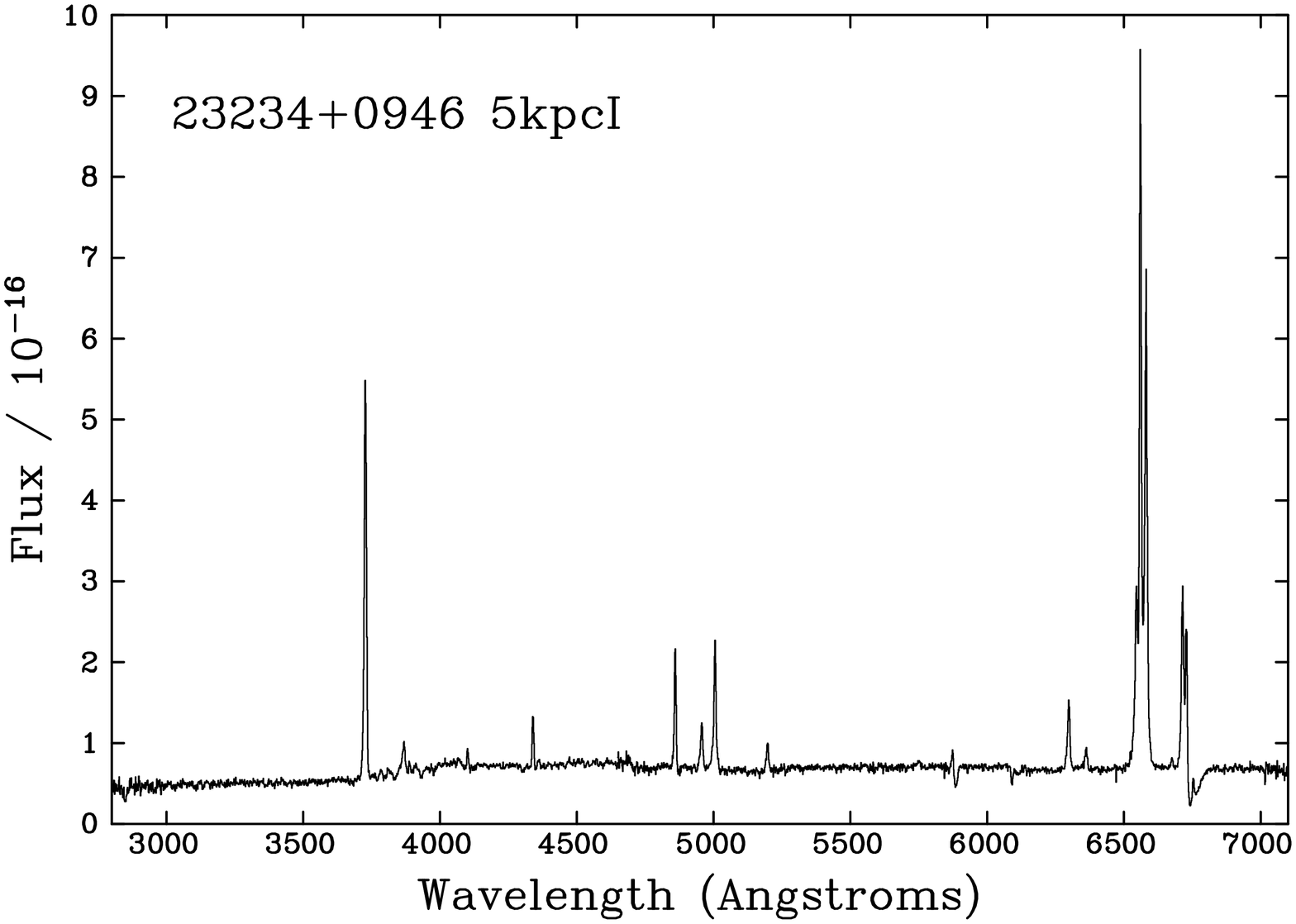,width=5.5cm,angle=-90.}\\
\hspace*{0cm}\psfig{file=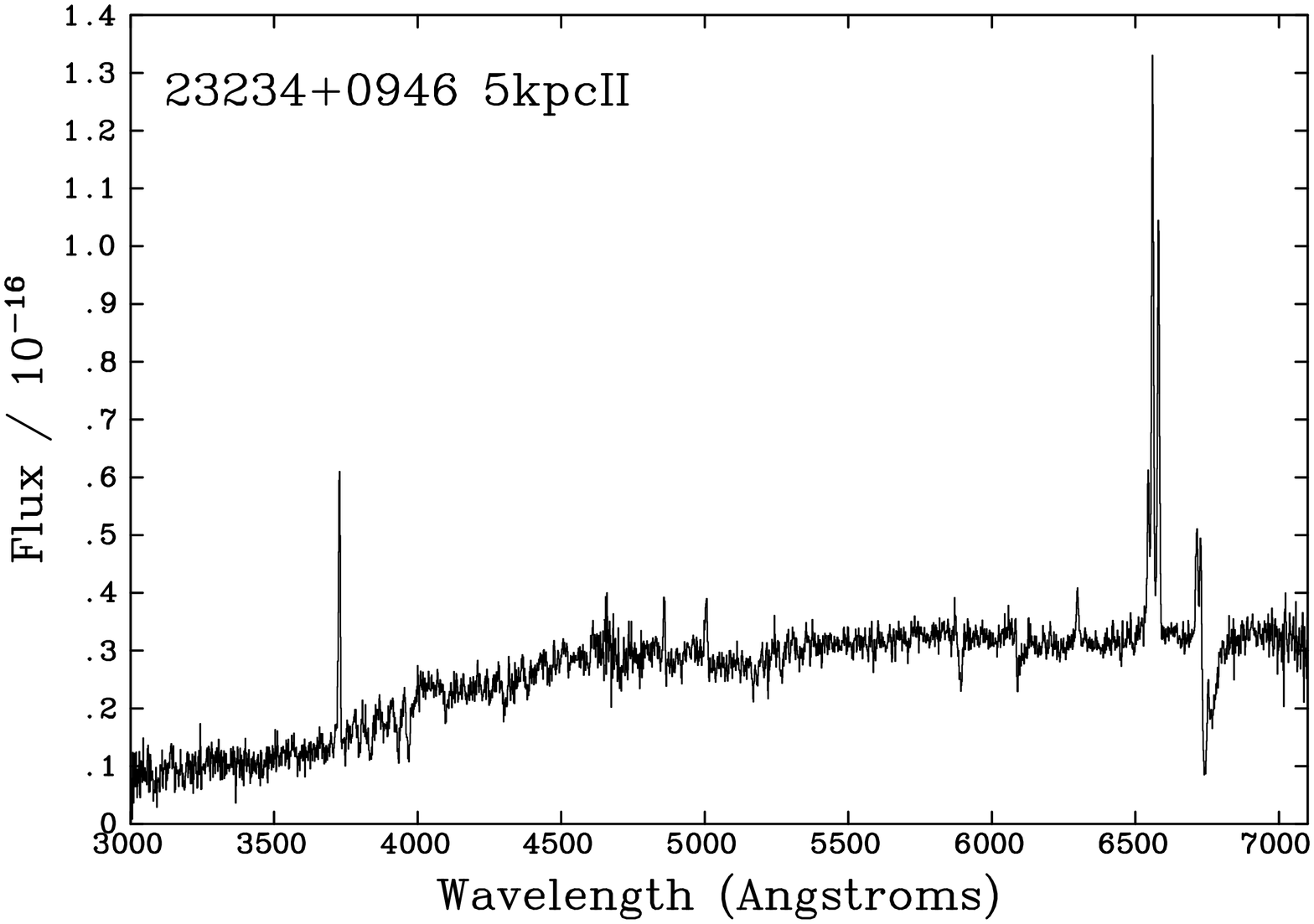,width=5.5cm,angle=-90.}&
\psfig{file=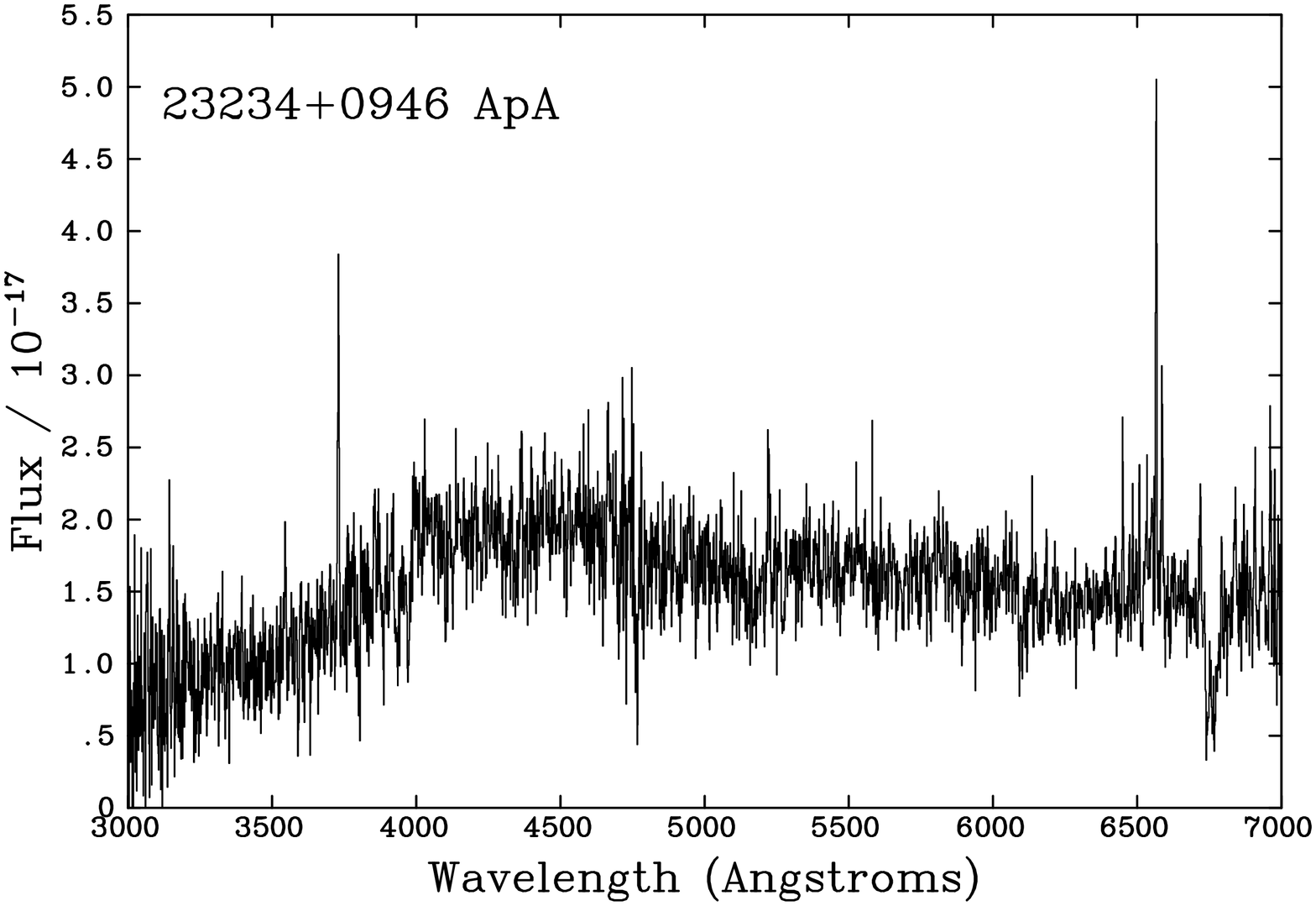,width=5.5cm,angle=-90.}\\
\hspace*{0cm}\psfig{file=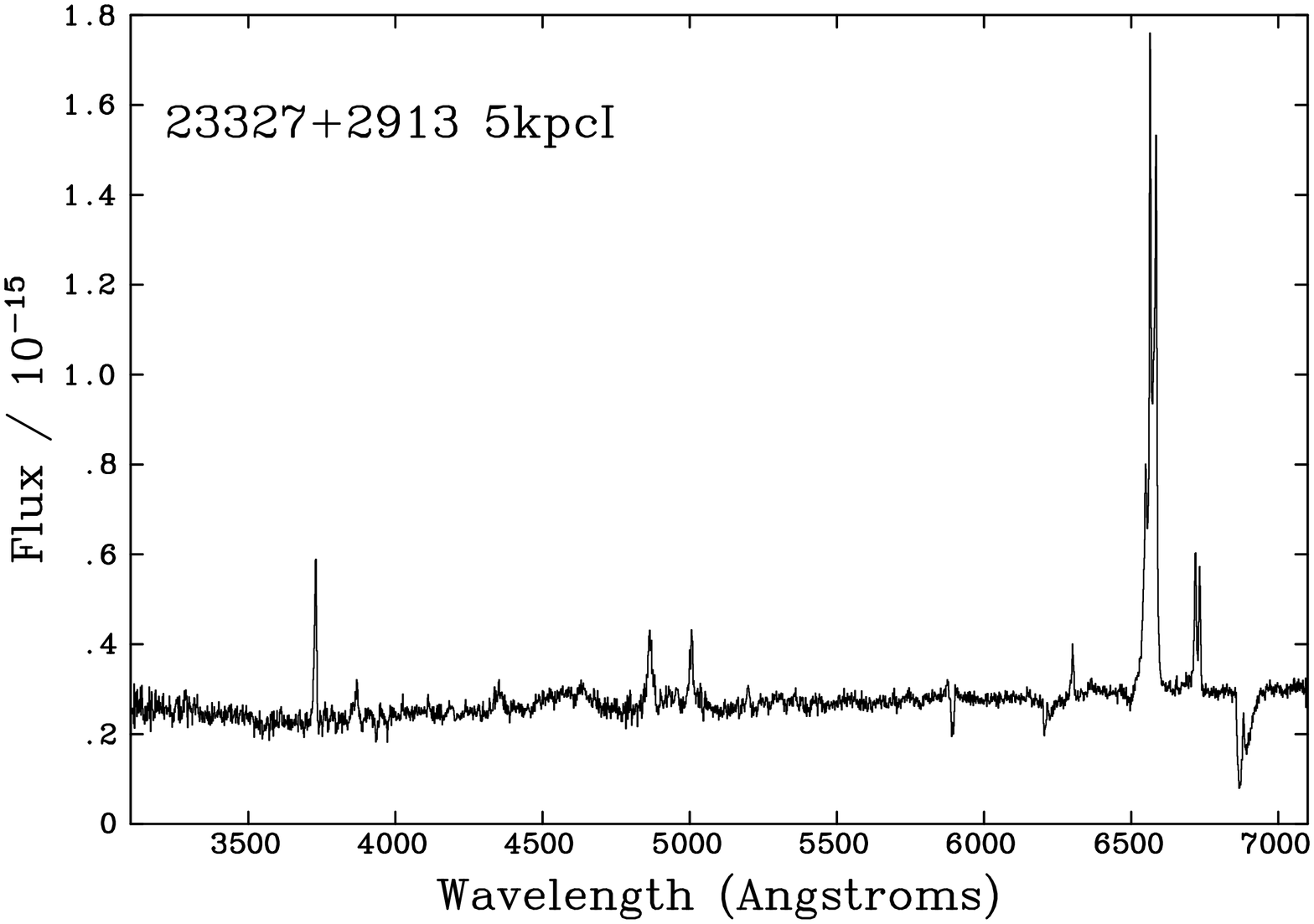,width=5.5cm,angle=-90.}&
\psfig{file=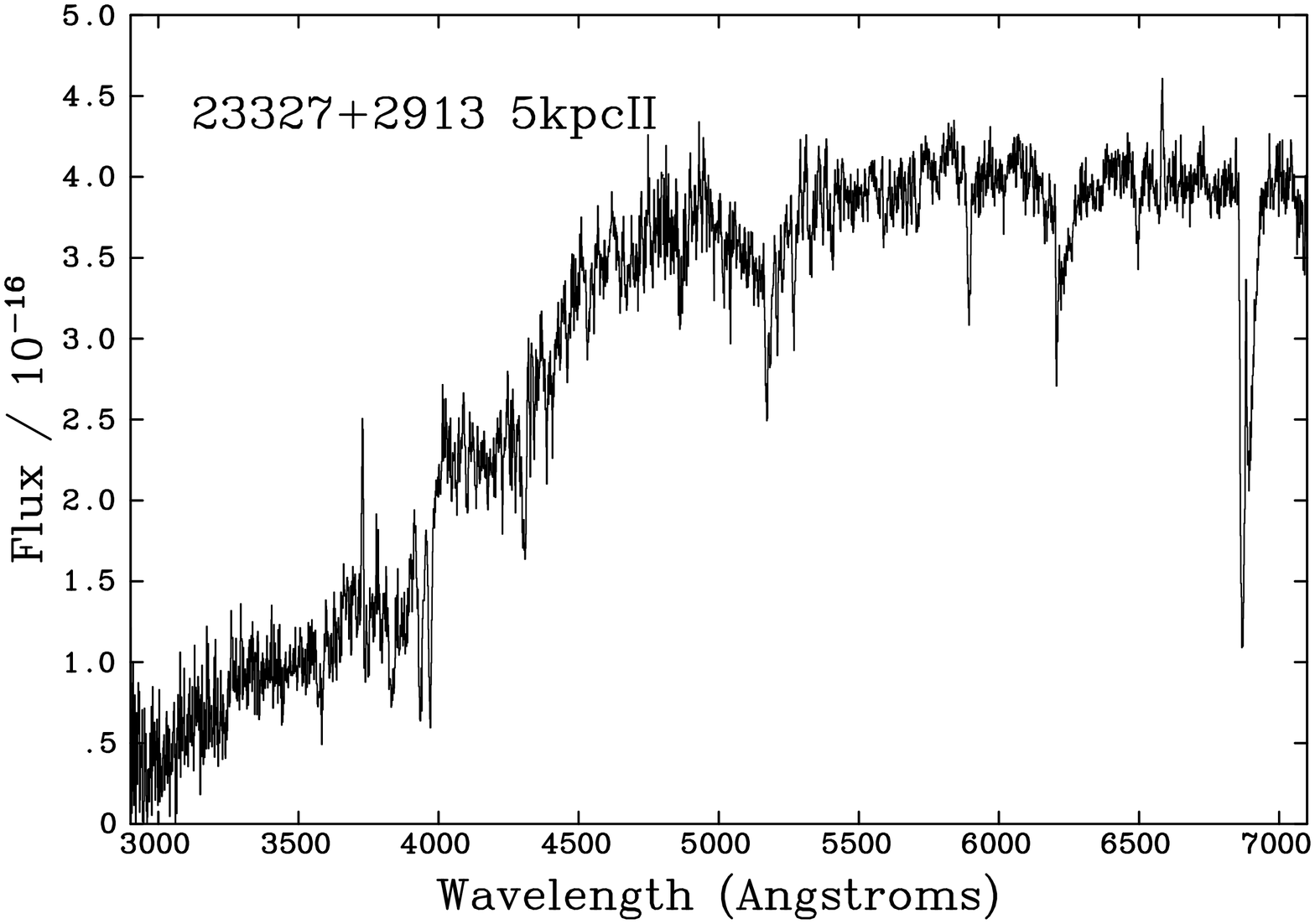,width=5.5cm,angle=-90.}\\
\end{tabular}
\caption[{\it Continued}]{Continued}
%\label{fig:SED}
\end{minipage}
\end{figure*}
\addtocounter{figure}{-1}
%\clearpage
\begin{figure*}
\begin{minipage}{170mm}
\begin{tabular}{cc}
\hspace*{0cm}\psfig{file=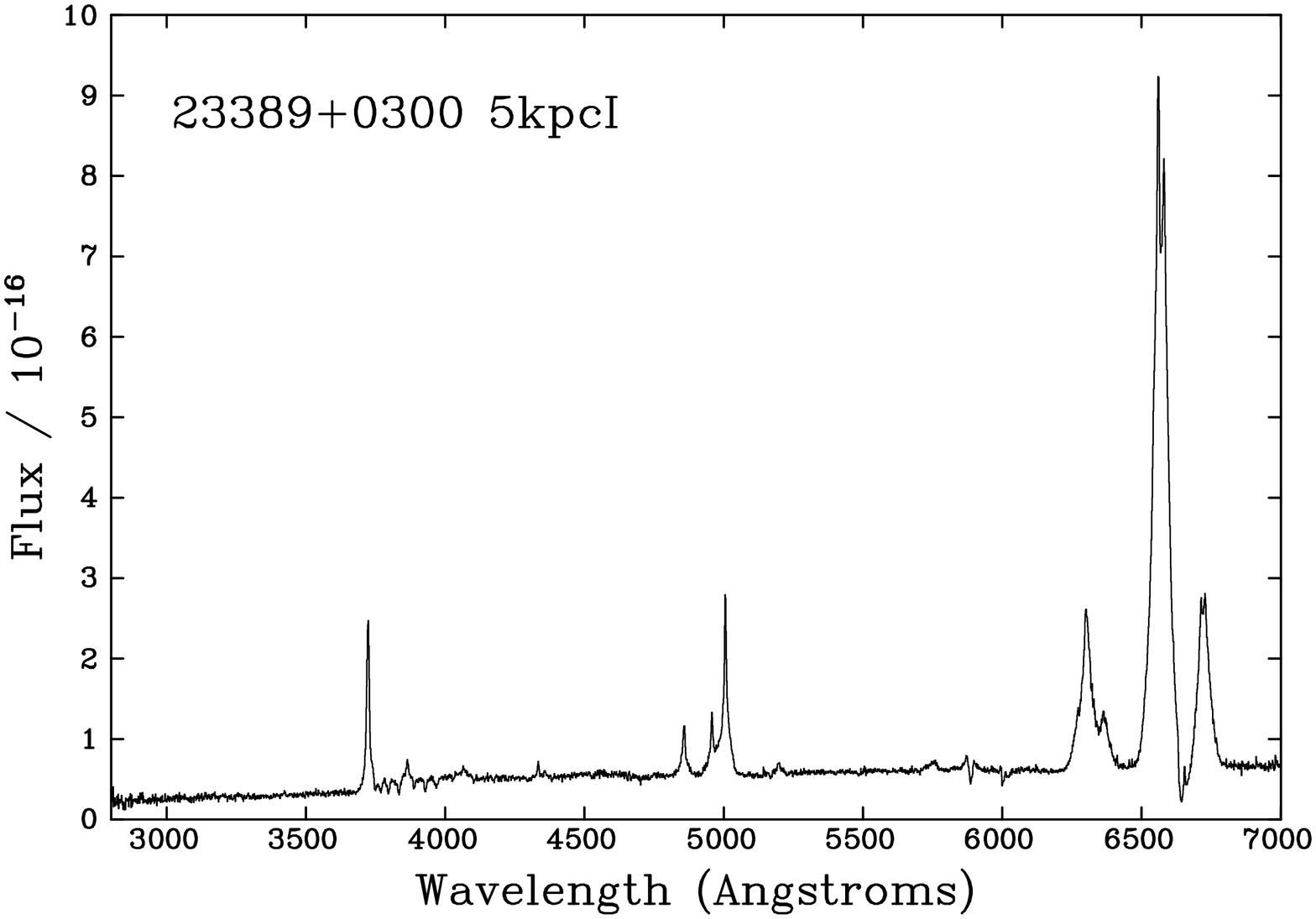,width=5.5cm,angle=-90.}&
\psfig{file=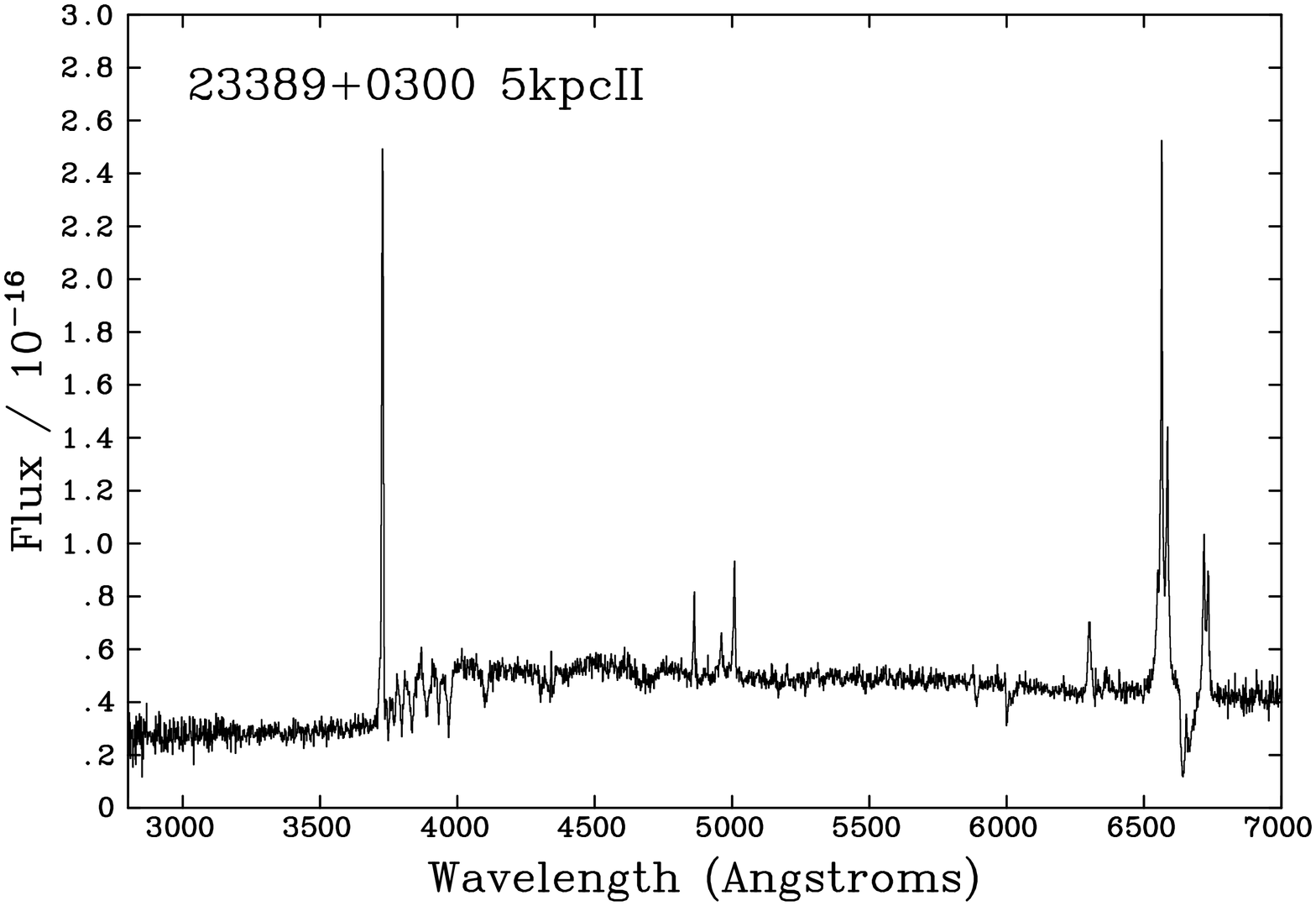,width=5.5cm,angle=-90.}\\
\hspace*{0cm}\psfig{file=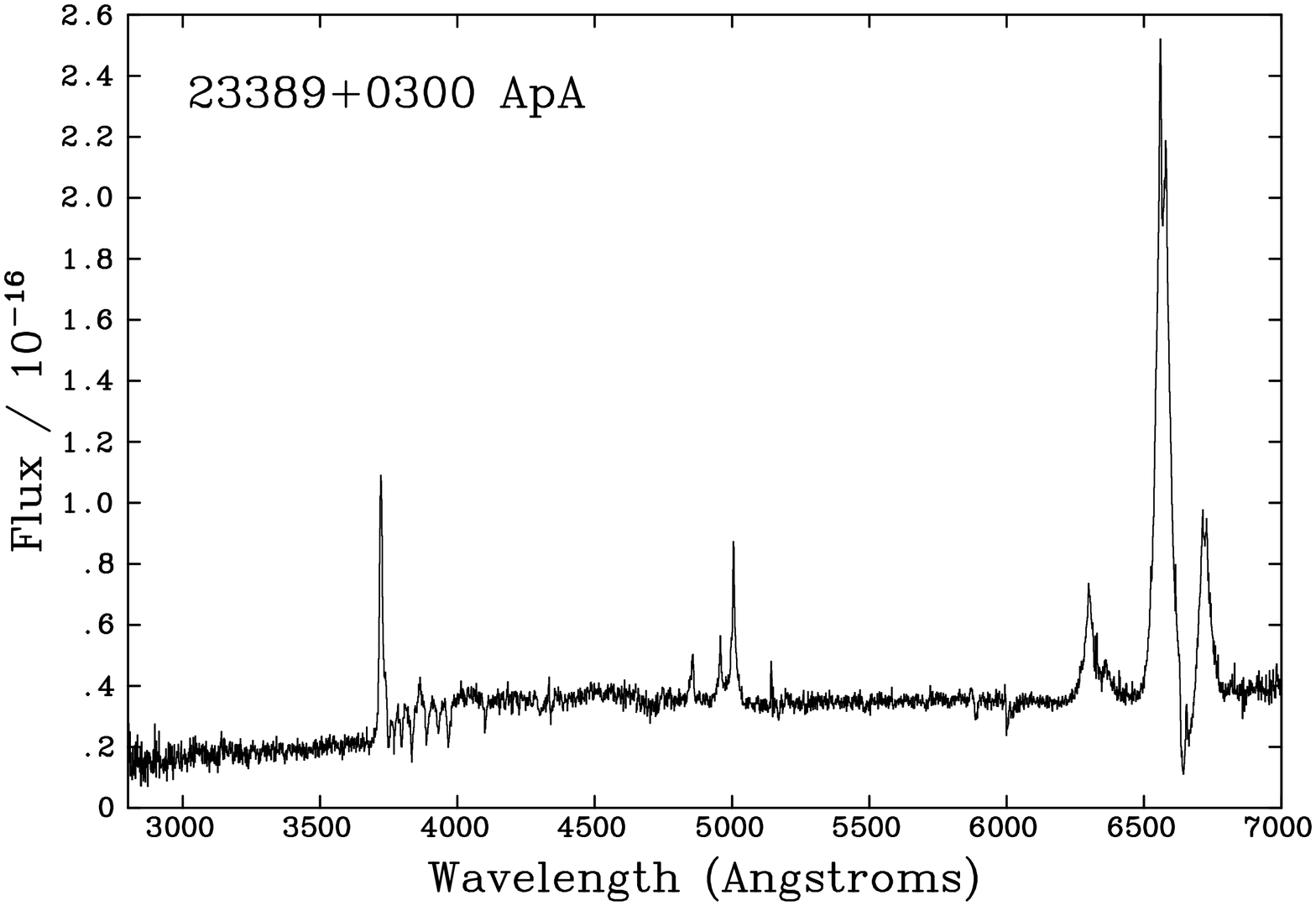,width=5.5cm,angle=-90.}&
\psfig{file=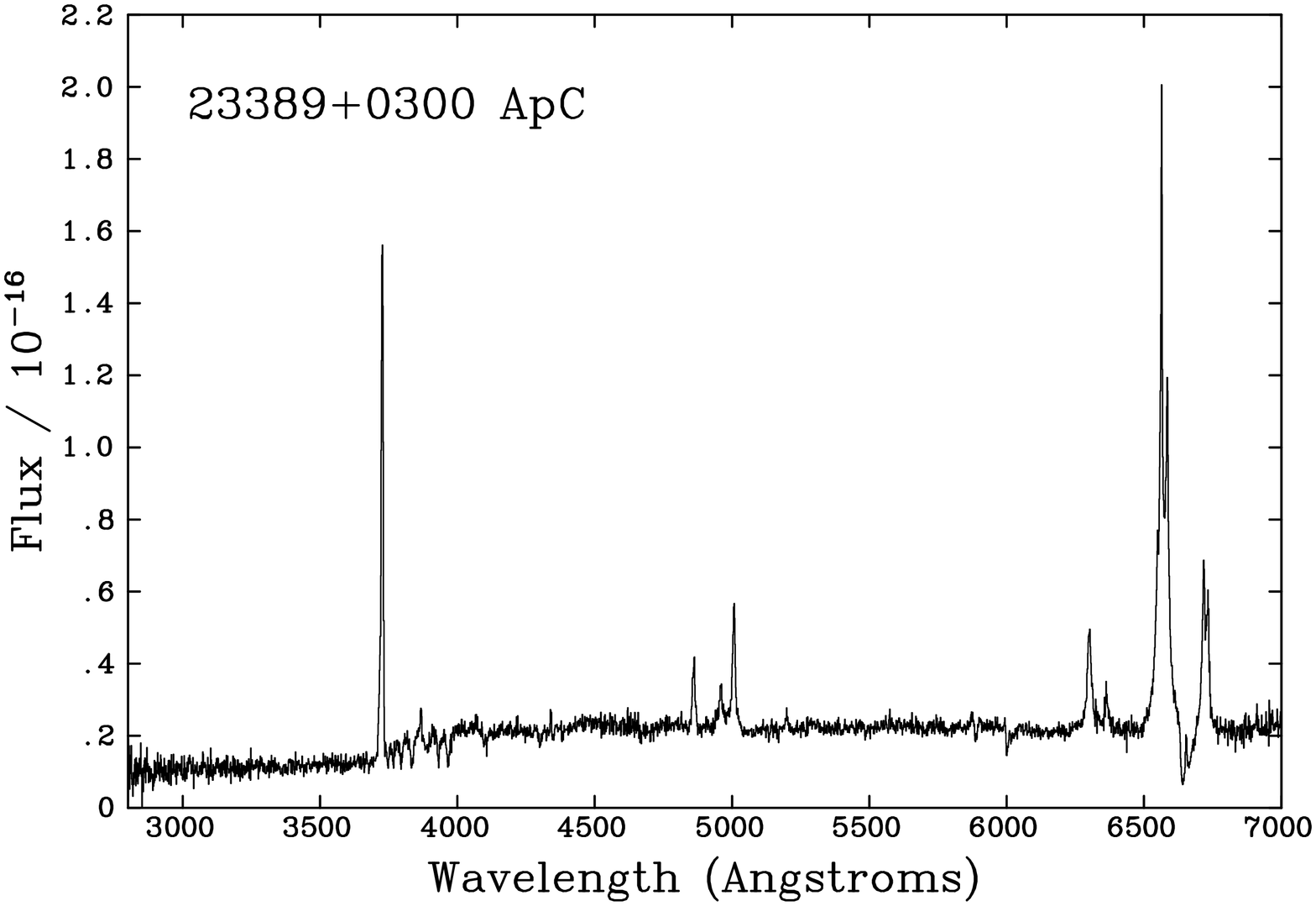,width=5.5cm,angle=-90.}\\
\end{tabular}
\caption[{\it Continued}]{Continued}
%\label{fig:SED}
\end{minipage}
\end{figure*}

\clearpage

\begin{table*}
%\hspace*{3.5cm}
\centering
{\small
%\centering
\begin{tabular}{@{}lccccc@{}}
%\hline  
% &&  Comb I &&&\\
\hline  
Object & Apertures & Detailed  &Age of     &E(B-V)&$\%$YSP\\
Name   &           & features& YSP       &  	& \\ 
IRAS   &           & fitted? &(Gyr)      &  	& \\
 (1)   &  (2)      & (3)     &(4)	 &(5)	&(6)\\
 \hline 
00091-0738 & 5kpc    & N & 0.05 - 0.1 & 0.6 & 75 - 100\\
           & Ap A    & Y & 0.2 - 1.0  & 0.0 -0.1 & 45 - 50\\
           & Ap B    & Y & 0.08 - 0.2 & 0.0 - 0.3 & 60 - 90\\
00188-0856 & 5kpc    & Y & 0.3 - 0.5 & 0.4 & 80 - 100\\
01004-2237 & 5kpc    & Y & 0.006 - 0.03 & 0.2 - 0.3 & 73 - 90\\
12072-0444 & 5kpc    & N & 0.04 - 0.05 & 0.4 - 0.5 & 75 - 100\\
           & Ap A    & Y & 0.1 - 0.3 & 0.0 - 0.1 & 50 - 65\\
           & Ap B    & N & 0.06 - 0.1 & 0.0 - 0.2 & 60 - 95\\
12112+0305 & 5kpc-I  & N & 0.04 - 0.06 & 0.6 & $\sim$100\\ 
           & 5kpc-II & Y & 0.06 - 0.1 & 0.0 - 0.1 & 65 - 70\\
           & Ap A    & N & 0.05 - 0.06 & 0.6 - 0.7 & 80 - 85\\
12540+5708 & 5kpc    & - & - & - &\\
           & Ap A    & Y & 0.005 - 0.007 & 0.1 & $\sim$40\\
           & Ap B    & Y & 0.04 - 0.1 & 0.0 - 0.5 & 45 - 75\\
           & Ap C    & Y & 0.07 - 0.08 & 0.1 & 95\\
           & Ap D    & - & - & - &\\
           & Ap E    & Y & 0.06 - 0.1 & 0.2 - 0.5 & 50 - 70\\
13428+5608 & 5kpc    & N & 0.05 - 0.06 & 0.5 & $\sim$75\\
           & Ap A    & Y & 0.2 - 0.3 & 0.0 & 45 - 55\\
           & Ap B    & Y & 0.08 - 0.2 & 0.0 - 0.1 & 35 - 40\\
           & Ap C    & N & 0.09 - 0.2 & 0.0 - 0.3 & 45 - 60\\
           & Ap D    & N & 0.07 - 0.1 & 0.1 - 0.3 & 50 - 75\\
           & Ap E    & N & 0.05 - 0.06 & 0.5 & $\sim$75\\
           & Ap F    & N & 0.05 - 0.1 & 0.0 - 0.4 & 40 - 55\\
13451+1232(PA160) & 5kpc    & Y & $\leq$ 0.06 & 0.8 - 1.3 &  20 - 35\\
                  & Ap A    & Y & 0.1 - 0.2 & 0.0 & 25\\
                  &         &   & 1.0 - 2.0 & 0.0 & 50 - 100\\
                  & Ap A neb& Y & 0.2 - 1.2 & 0.0 & 20 - 80 \\   
                  & Ap B    & Y & 0.001 - 0.1 & 0.0 - 1.0 & 5 - 30\\
                  & Ap B neb& Y & 0.004 - 0.4 & 0.0 - 1.0 & 10 - 20\\
                  & Ap C    & Y & 0.004 - 0.005 & 0.5 - 0.8  & 40 - 50 \\
                  &         &   & 0.04 - 0.1 & 0.0 - 0.3  & 30 - 50 \\
                  & Ap D    & Y & 0.2 - 0.6 & 0.0 - 0.1 & 30 - 50\\
13451+1232(PA230) & Ap A  & Y & 0.4 - 2.0 & 0.0 - 0.2 & 30 - 90\\ 
                  & Ap B    & Y & $\leq$ 2.0 & 0.0 - 1.0 & 10 - 90\\
13539+2920  & 5kpc-I   & N & 0.07 - 0.2 & 0.0 - 0.3 & 65 - 100\\
            & 5kpc-II  & N & 0.05 - 0.06 & 0.8 & $\sim$ 95\\
14060+2919  & 5kpc     & N & 0.007        & 0.4       & $\sim$ 85\\
            &          &   & 0.03        & 0.3       & $\sim$ 85\\    
            & Ap A     & Y & 0.4 - 0.6 & 0.0 & 85 - 100\\
            & Ap B     & Y & 0.03 - 0.04 & 0.3 & 88 - 100\\
            & Ap C     & Y & 0.04 - 0.06 & 0.0 - 0.2 & 88- 97\\
14348-1447  & 5kpc-I   & N & 0-04 - 0.06 & 0.5 & $\sim$ 95\\
            & 5kpc-II  & Y & 0.06 - 0.08 & 0.2 - 0.3 & 73 - 83\\
            & Ap A     & N & 0.04 - 0.07 & 0.0 - 0.3 & 75 - 100\\
            & Ap B     & N & 0.04 - 0.07 & 0.5 - 0.6 & 70 - 100\\
            & Ap C     & Y & 0.06 - 0.2 & 0.0 - 0.3 & 70 - 75\\
            & Ap D     & Y & 0.04 - 0.1 & 0.0 - 0.4 & 50 - 70\\
14394+5332  & 5kpc-I   & N & 0.05 - 0.07 & 0.5 & 80 - 90\\
            & 5kpc-II  & Y & 0.004 - 0.007 & 0.6 - 0.8 & 44 - 55 \\
            &          &   & 0.04 - 0.06   & 0.4 - 0.5 & 50 - 55 \\        
            & Ap A     & Y & 0.07 - 0.1 & 0.0 - 0.1 & 50 - 55\\
            & Ap B     & N & 0.04 - 0.1 & 0.0 - 0.4 & 45 - 55\\
15130+1958  & 5kpc     & N & 0.07 - 100 & 0.4 - 0.5 & 75 - 90\\
15206+3342  & 5kpc     & Y & 0.01 & 0.3 & 91\\
\hline
\end{tabular}}
\caption[Modelling results for the CS using Comb I]{
Modelling results for the CS using Comb I (12.5 Gyr OSP +
YSP). Col (1): object name. Col (2): aperture label as indicated in Figure
\ref{fig:spatial_cuts}. Col (3): this column indicates whether
or not Comb I fits both the overall shape of the continuum and the
detailed absorption features: Y = yes, N = No. Col (4): range of ages
for the YSP component that adequately fit the data ($\chi^{2}
\leq 1.0$). Col (5): range of $E(B - V)$ values of the YSPs in Col
(4). Col (6): the percentage flux contribution of the YSP component to 
the model in the normalising bin.}
\label{tab:CS_combI}
\end{table*}
\addtocounter{table}{-1}
\begin{table*}
{\small
\centering
\begin{tabular}{@{}llcrrr@{}}
%\hline  
%&& Comb I &&&\\
\hline  
Object & Apertures & Detailed  &Age of     &E(B-V)	&$\%$YSP\\
Name   &           & features& YSP       &  	& \\ 
IRAS   &           & fitted? &(Gyr)      &  	&\\
 (1)   & (2)       & (3)     &(4)	 &(5)	&(6)\\
\hline 
15327+2340(PA160) & 5kpc     & N & 0.07 - 0.1 & 0.8 & $\sim$ 70\\
                  & Ap A      & Y & 0.4 - 0.6  &0.0 - 0.1 &60 - 85\\
                  & Ap B      & Y & 0.4 - 0.6  &0.0       &50 - 70\\
                  & Ap C      & Y & 0.4 - 0.6  &0.4 - 0.5 &50 - 90\\
                  & Ap D      & N & 0.05 - 0.1 &0.9       &$\sim$90\\
                  & Ap E      & N & 0.08 - 0.1 &0.9       &$\sim$75\\
                  & Ap F      & Y & 0.1 - 0.2  &0.4 - 0.5 &$\sim$60\\
                  & Ap G      & Y & 0.2 - 0.6  &0.0 - 0.1 &50 - 70\\
                  & Ap$_{\rm TOTAL}$ & N & 0.09 - 0.2 &0.6       &$\sim$70\\
15327+2340(PA75)  & Ap A     & Y & 0.2 - 0.6  &0.1 - 0.4 &50 - 95\\
                  & Ap B     & Y & 0.09 - 0.2 &0.5       &55 - 70\\
                  & Ap C     & Y & 0.2 - 0.3  &0.5 - 0.6 &$\sim$65\\
                  & Ap D     & Y & 0.2 - 0.4  &0.3 - 0.4 &55 - 75\\
                  & Ap E     & Y & 0.2 - 0.3  &0.2 - 0.3 &$\sim$60\\
                  & Ap F     & Y & 0.08 - 0.6 &0.1 - 0.5 &45 - 70\\
                  & Ap G     & Y & 0.05 - 0.6 &0.1 - 0.5 &45 - 80\\
                  & Ap$_{\rm TOTAL}$ & Y & 0.2 - 0.3  &0.3 - 0.4  & 50 - 60\\
15327+2340(PA75*) & Ap A & Y & 0.2 - 0.6 & 0.0 - 0.1 &50 - 95\\
                  & Ap B & Y & 0.4 - 0.6 & 0.0 - 0.1 &60 - 90\\ 
                  & Ap C & Y & 0.4 - 0.6 & 0.0 - 0.1 &70 - 100\\
                  & Ap D & Y & 0.4 - 0.6 & 0.0 - 0.1 &60 - 90\\
                  & Ap E & Y & 0.4 - 0.8 & 0.0 - 0.1 &50 - 100\\
                  & Ap F & Y & 0.4 - 1.6 & 0.0 - 0.1 &50 - 100\\
                  & Ap$_{\rm TOTAL}$ & Y & 0.2 - 0.6 & 0.0 - 0.1 &45 - 85\\
%15462-0450 & 5kpc      & - & - & - & -\\
16474+3430 & 5kpc-I    & N & 0.007 & 0.4 & 85\\
           & 5kpc-II   & Y & 0.07 - 0.08 & 0.1 & 70\\
           & Ap A      & Y & 0.04 & 0.3 & 82\\
16487+5447NE & 5kpc    & Y & 0.004 - 0.006 & 0.5 - 0.7 & 25 - 45 \\
             & Ap A    & Y & 0.004 - 0.007 & 0.3 - 0.5 & 55 - 70\\
16487+5447NW & 5kpc    & Y & 0.005 - 0.007 & 0.2 & 75\\
             & Ap A    & N & 0.05 - 0.09 & 0.0 - 0.2 & 55 - 75\\
             & Ap B    & Y & 0.004 - 0.006 & 0.2 - 0.3 & 70 - 74\\ 
17028+5817NE & 5kpc    & N & 0.04 - 0.08 & 0.0 - 0.3 & 70 - 90\\
17028+5817NW & 5kpc    & N & 0.05 - 0.07 & 0.7 & 80 - 95\\
             & Ap A    & Y & 0.2 - 0.6 & 0.0 & 70 - 100\\
20414-1651   & 5kpc    & N & 0.07 - 0.1 & 0.1 - 0.4 & 65 - 90\\
21208-0519   & 5kpc-I  & N & 0.004 - 0.005 & 0.6  & 80 - 90 \\
             &         &   &  0.04 - 0.05  &  0.3 & 90 - 100\\
             & 5kpc-II & Y & 0.09 - 0.1 & 0.2 - 0.3 & 30 - 35\\
             & Ap A    & Y & 0.1 - 0.2 & 0.0 - 0.2 & $\sim$ 50\\
             & Ap B    & Y & 0.1 & 0.0 & 65\\
%21219-1757   & 5kpc    & - & - & - & \\
22491-1808   & 5kpc    & N & 0.04 & 0.3 & 100 \\
             & Ap A    & N & 0.05 - 0.1 & 0.0 - 0.3 & 65 - 100\\ 
             & Ap B    & N & 0.04 & 0.2 & 87\\
             & Ap C    & N & 0.05 - 0.07& 0.1 - 0.2 & 80 - 90\\
23233+2817   & 5kpc    & Y & 0.04 - 0.05 & 0.3 - 0.4 & 50 - 65\\
             & Ap A    & - & - & - & - \\
             & Ap B    & Y & 0.06 - 0.1 & 0.1 - 0.3  & 50 - 60\\
23234+0946   & 5kpc-I  & N & 0.006 & 0.6 & 88 \\
             &         &   & 0.04  & 0.4 & 80 \\
             & 5kpc-II & Y & 0.07 - 0.3 & 0.2 - 0.4 & 40 - 50\\
             & Ap A    & N & 0.08 - 0.2 & 0.0 - 0.3 & 60 - 75\\
23327+2913   & 5kpc-I  & Y & 0.009 & 0.3 - 0.4 & 51 - 66\\
             & 5kpc-II & Y & $\leq$ 2 & $\leq$ 2.0 & $\leq$ 50 \\
             & Ap A    & Y & 0.009 & 0.3 - 0.4 & 51 - 66\\
             & Ap B    & Y & $\leq$ 2 & $\leq$ 2.0 & $\leq$ 50 \\
\hline
\end{tabular}
\caption[{\it Continued}]{{\it Continued}}}
\end{table*}
\begin{table*}
{\small
\centering
\begin{tabular}{@{}llcrrr@{}}
%\hline  
%&& Comb I &&&\\
\hline  
Object & Apertures & Detailed  &Age of     &E(B-V)	&$\%$YSP\\
Name   &           & features& YSP       &  	& \\ 
IRAS   &           & fitted? &(Gyr)      &  	& \\
 (1)   &  (2)      & (3)     &(4)	 &(5)	&(6)\\	 
 \hline 
08572+3915 & 5kpc   & N & 0.06 - 0.1 & 0.0 - 0.3 & 65 - 90\\
           & Ap A   & Y & 0.1 - 0.3 & 0.0 - 0.1 & 45 - 55\\
           & Ap B   & N & 0.07 - 0.1 & 0.0 - 0.2 & 60 - 85\\
           & Ap C   & Y & 0.07 - 0.1 & 0.2 - 0.3 & 90 - 100\\
10190+1322 & 5kpc-I & N & 0.05 - 0.08 & 0.3 - 0.5 & 65 - 90\\
           & 5kpc-II& N & 0.2 & 0.5 & 85\\
           & Ap A   & Y & 0.4 - 0.7 & 0.0 - 0.2 & 60 - 95\\
           & Ap B   & N & 0.04 - 0.06 & 0.5 - 0.6 & 80 - 95\\
10494+4424 & 5kpc   & N & 0.07 - 0.1 & 0.7 & 95 - 100\\
           & Ap A   & Y & 0.4 & 0.0 - 0.1 & 72 - 87\\
           & Ap B   & N & 0.04 - 0.1 & 0.0 - 0.4 & 75 - 100\\
13305-1739 & 5kpc   & N & 0.05 - 0.1 & 0.0 - 0.4 & 65 - 90\\
14252-1550 & 5kpc   & N & 0.05 - 0.1 & 0.5 - 0.6 & 80 - 85\\
           & Ap A   & Y & 0.2 - 0.7 & 0.0 - 0.2 & 50 - 100\\
16156+0146 & 5kpc-I & Y & 0.01 & 0.5 & 75\\
           & 5kpc-II& N & 0.08 - 0.1 & 0.1 - 0.3 & 75 - 98\\
           & Ap A   & Y & 0.004 - 0.007 & 0.3 - 0.5 & 40 - 80 \\
           &        &   & 0.03 - 0.04    & 0.1 - 0.3 & 50 - 75\\ 
17044+6720 & 5kpc   & Y & 0.05 - 0.08 & 0.4 & 80 - 85\\
           & Ap A   & Y & 0.05 - 0.1 & 0.0 - 0.3 & 70 - 90\\
           & Ap B   & N & 0.05 - 0.1 & 0.0 - 0.2 & 77 - 95\\
17179+5444 & 5kpc   & N & 0.07 - 0.1 & 0.4 - 0.6 & 50 - 85\\
23060+0505 & 5kpc   & N &  - & - & -\\
           & Ap A   & Y & 0.05 - 0.1 & 0.1 - 0.4 & 43 - 65\\
           & Ap B   & Y & 0.04 - 0.06 & 0.2 - 0.3 & 65\\
23389+0303 & 5kpc I & N & 0.004 & 0.9 & $\sim$100 \\
           &        &   & 0.04  & 0.6 &  90\\  
           & 5kpc II& N & 0.07 - 0.1 & 0.2 - 0.4 & 70 - 85\\
           & Ap A   & N & 0.06  & 0.5 & 85\\
           & Ap B   & N & 0.005 - 0.006 & 0.8 - 0.9 & 85\\
           & Ap C   & N & 0.05 - 0.08 & 0.4 & 63 -76\\  
           & Ap D   & N & 0.1 & 0.2 & 72\\
\hline
\end{tabular}
\caption[Modelling results for the 10 ``extra'' ULIRGs included in the
ES using Comb I]{ Same as Table \ref{tab:CS_combI} but for the 10
additional ULIRGs included in the ES.}}
\label{tab:ES_combI}
\end{table*}
\begin{table*}
%\hspace*{1.0cm}
\centering
{\small
%\centering
\begin{tabular}{@{}lllcrrr@{}}
%\hline   
%&& Comb II &&&&\\ 
\hline
Object Name & Aperture &Age of YSP &E(B - V)&$\%$YSP &$\%$p.l &$\%$OSP \\
IRAS        &          &(Gyr)      &        &        &        &        \\
 (1)        & (2)      & (3)       &(4)     &(5)     &(6)     &(7)     \\
\hline
00091-0738 & 5kpc     & 0.3 - 0.6 & 0.0 - 0.2 & 48 - 60 & 20 - 35 & 5 - 30\\
           & Ap A     & 0.4 - 1.0 & 0.0 - 0.3 & 50 - 100 & 6 - 12 & $\leq$ 25\\ 
           & Ap B     & 0.2 - 0.6 & 0.0 - 0.2 & 66 - 75 & 10 - 25 & $\leq$ 24\\
00188-0856 & 5kpc     & 0.3 - 0.6 & 0.3 - 0.4 & 75 - 90 & 8 - 10 & $\leq$ 16\\
01004-2237 & 5kpc     & 0.03 - 0.1 & 0.0 - 0.5 & 30 - 50 & $\sim$40 & 10 - 30\\
12072-0444 & 5kpc$^1$ & - & - & - & - & -\\
           & Ap A     & 0.2 - 0.6 & 0.0 - 0.2 & 40 - 65 & 7 - 30 & $\leq$ 50\\
           & Ap B     & 0.5 - 0.6 & 0.0 & 50 - 60 & 40 & $\leq$ 10\\ 
12112+0305 & 5kpc I   & 0.4 - 0.5 & 0.0 - 0.1 & 50 - 60 & $\sim$40 & $\leq$ 10\\  
           & 5kpc II  & 0.2 - 0.5 & 0.0 - 0.2 & 50 - 60 & 30 - 40 & $\leq$ 15\\ 
           & Ap A     & 0.2 - 0.8 & 0.0 - 0.1 & 40 - 70 & 20 - 40 & $\leq$ 45\\
12540+5708 & 5kpc     & - & - & - & - & -\\
           & Ap A     &0.4 - 1.4 & 0.1 - 0.5 & 40 -75 & 25 - 40 & $\leq$ 35\\
           & Ap B     & 0.2 & 0.0 & 40 & 30 & 30 \\
           & Ap C     & 0.09 - 0.2 & 0.0 - 0.1 & 70 -88 & 10 - 30 & 0\\
           & Ap D     & - & - & - & - & -\\
           & Ap E     & 0.1 - 0.2 & 0.0 - 0.1 & 30 & 40 & 30 \\
13428+5608 & 5kpc     & 0.5 - 0.8 & 0.0 - 0.1 & 45 - 60 & $\sim$40 & $\leq$ 15\\ 
           & Ap A     & 0.4 - 0.8 & 0.0 - 0.3 & 50 - 85 & 10 - 15 & $\leq$ 35\\
           & Ap B     & 0.2 - 0.7 & 0.0 - 0.2 & 33 - 64 & 6 - 15 & 30 - 65 \\
           & Ap C     & 0.4 - 0.7 & 0.0 - 0.2 & 50 - 90 & 6 - 9 & $\leq$ 45\\ 
           & Ap D     & 0.5 - 0.8 & 0.0 - 0.3 & 50 - 80 & 18 - 23 & $\leq$ 30\\  
           & Ap E     & 0.5 - 0.8 & 0.0 - 0.1 & 40- 60 & $\sim$40 & $\leq$ 20\\   
           & Ap F     & 0.5 - 0.7 & 0.0 - 0.4 & 45 - 80 & 15 - 30 & $\leq$ 30\\   
13451+1232(PA160) &5kpc$^3$ & - & - & - & - & -\\   
                  &Ap A     & 0.2 - 0.8 & 0.0 - 0.2 & 20 - 35 & $\sim$ 5 & 60 - 75\\  
                  &Ap B$^3$ & - & - & - & - & -\\   
                  &Ap C     & 0.4 - 0.8 & 0.0 - 0.3 & 35 - 75 & 15- 25 & 5 - 45\\   
                  &Ap D     & 0.4 - 1.2 & 0.0 - 0.5 & 30 - 85 & 10 - 16 & $\leq$ 45\\    
13451+1232(PA230) &Ap A     & 0.5 - 1.4 & 0.0 - 0.5 & 40 - 90 & 3 - 10 & $\leq$ 50\\  
                  &Ap B     & 0.4 - 1.0 & 0.0 - 0.2 & 25 - 50 & 5 -10 & 45 - 65 \\ 
13539+2920 & 5kpc-I   & 0.5 - 0.6 & 0.0 - 0.2 & 65 - 80 & 20 - 30 & 0\\    
           & 5kpc-II  & 0.5 - 0.8 & 0.2 - 0.5 & 50 - 70 & 35 - 40 & $\leq$ 15\\  
14060+2919 & 5kpc     & 0.004 & 0.5 & 93 & 6 & 0\\   
           & Ap A     & 0.3 - 0.8 & 0.0 - 0.4 & 60 - 90 & 10 - 15 & $\leq$ 25\\ 
           & Ap B     & 0.04- 0.06 & 0.0 - 0.4 & 60 - 85 & 10 - 25 & $\leq$ 10\\  
           & Ap C     & 0.2 - 0.3 & 0.0 & 50 & 40 & 11\\  
14348-1447 & 5kpc-I$^1$ &  - & - & - & - & -\\    
           & 5kpc-II  & 0.5 & 0.0 & 62 & 38 & 0\\ 
           & Ap A     & 0.1 - 0.3 & 0.0 - 0.1 & 40 - 50 & 40 & 10 - 20\\
           & Ap B     & 0.2 - 0.6 & 0.0 - 0.1 & 30 - 40 & 30 - 40 & 12 - 40\\ 
           & Ap C     & 0.09 - 0.4 & 0.0 - 0.4 & 45 - 80 & 5- 35 & $\leq$ 35\\   
           & Ap D     & 0.05 - 0.4 & 0.0 - 0.5 & 40 - 75 & 5 - 30 & $\leq$ 50\\  
14394+5332 & 5kpc-I$^1$ & - & - & - & - & -\\  
           & 5kpc-II  & 0.7 - 2.0 & 0.0 - 0.2 & 40 - 70 & 30 - 35 & $\leq$ 30\\ 
           & Ap A       & 0.3 - 0.8 & 0.0 - 0.2 & 42 - 60 & 30 - 35 & $\leq$ 25\\ 
           & Ap B       & 0.4 - 1.0 & 0.0 - 0.3 & 40 - 60 & 23 - 35 & $\leq$ 30\\ 
15130+1958 & 5kpc       & 0.2 - 0.5 & 0.0 - 0.2 & 55 - 70 & $\sim$ 30 &$\leq$ 15\\  
15206+3342 & 5kpc$^2$   &  - & - & - & - & -\\  
\hline
\end{tabular}}
\caption[Modelling results for the CS using Comb II]{Modelling
results for the CS using Comb II (OSP + YSP + power-law). Col (1):
object name. Col (2): aperture label as indicated in Figure
\ref{fig:spatial_cuts}. Col (3): range of ages for the YSP components that
adequately fit the data (both the continuum and the detailed features).
Col (4): range of $E(B - V)$ values of the YSPs in Col (3). Col (5): the
percentage flux contribution of the YSP component to the model, in the 
normalising bin. Cols (6) and (7): same as Col (5) for the power-law and OSP
respectively. \newline $^1$ No adequate fits (fitting both the
continuum and the detailed absorption features) are found using Comb
II. \newline $^2$ No adequate fits are found with a non negligible
contribution of the p.l.~component (i.e., different than those of Comb
I) and therefore, no results are presented in the
table. \newline$^{3}$ When a p.l. is included, adequate fits are found
for the entire range of YSP combinations used and, therefore, no results
are presented in the table. }
\label{tab:CS_combII}
\end{table*}
\addtocounter{table}{-1}
\begin{table*}
%\hspace*{1.0cm}
\centering
{\small
%\centering
\begin{tabular}{@{}lllcrrr@{}}
%\hline   
%&& Comb II &&&&\\ 
\hline
Object Name & Aperture &Age of YSP &E(B - V)&$\%$YSP &$\%$p.l &$\%$OSP \\
IRAS        &          &(Gyr)      &        &        &        &        \\
 (1)        & (2)      &  (3)      &(4)     &(5)     &(6)     &(7)     \\
\hline
15327+2340(PA160) & 5kpc & 0.4 - 0.7 & 0.4 - 0.5 & 64 - 85 & 15 - 18 & $\leq$20\\  
                  & Ap A  & 0.5 - 0.7 & 0.0 - 0.1 & 75 - 98 & $\sim$2 & $\leq$20\\
                  & Ap B  & 0.5 - 0.7 & 0.1 - 0.2 & 85 - 95 & 3 - 5   & $\leq$20\\
                  & Ap C  & 0.4 - 0.7 & 0.3 - 0.5 & 70 - 98 & 4 - 5   & $\leq$20\\
                  & Ap D  & 0.4 - 0.6 & 0.5 - 0.8 & 62 - 75 & 27 - 35 & $\leq$8\\
                  & Ap E  & 0.4 - 0.6 & 0.6 - 0.8 & 75 - 85 & 10 - 16 & $\leq$15\\
                  & Ap F  & 0.1 - 0.7 & 0.2 - 0.4 & 53 - 85 & 4 - 17  & $\leq$43\\
                  & Ap G  & 0.4 - 0.7 & 0.0 - 0.2 & 55 - 90 & 4 - 8   & $\leq$40\\
                  & Ap$_{\rm TOTAL}$ & 0.3 - 0.7 & 0.2 - 0.4 & 48 - 87 & 5 - 15 & $\leq$45\\
15327+2340(PA75)  & Ap A  & 0.3 - 0.7 & 0.1 - 0.3 & 60 - 93 & 3 - 7  & $\leq$37\\
                  & Ap B   & 0.2 - 0.6 & 0.2 - 0.5 & 50 - 75 & 6 - 18 & 10 - 44\\
                  & Ap C   & 0.3 - 0.6 & 0.4 - 0.5 & 65 - 85 & 6 - 12 & 8 - 27\\
                  & Ap D   & 0.3 - 0.7 & 0.2 - 0.4 & 61 - 95 & 3 - 9  & $\leq$37\\
                  & Ap E   & 0.3 - 0.7 & 0.1 - 0.4 & 52 - 85 & 4 - 10 & 9 - 45\\
                  & Ap F   & 0.3 - 0.9 & 0.1 - 0.5 & 30 - 85 & 2 - 13 & 10 - 70\\
                  & Ap G   & 0.2 - 1.0 & 0.0 - 0.3 & 38 - 90 &7 - 16  & $\leq$60\\
                  & Ap$_{\rm TOTAL}$ & 0.2 - 0.6 & 0.2 - 0.4 & 57 - 75 & 4 - 12 & 12 - 40\\
15327+2340(PA75*) & Ap A   & 0.3 - 0.6 & 0.0 - 0.2 & 55 - 90 & 3 - 10 & $\leq$40\\
                  & Ap B    & 0.4 - 0.7 & 0.0 - 0.2 & 59 - 95 & 2 - 5  & $\leq$41\\
                  & Ap C    & 0.4 - 0.6 & 0.0 - 0.2 & 63 - 90 & 4 - 8  & $\leq$33\\
                  & Ap D    & 0.4 - 0.7 & 0.0 - 0.2 & 55 - 90 & 7 - 10 & $\leq$40\\
                  & Ap E    & 0.4 - 1.2 & 0.0 - 0.3 & 40 - 90 & 9 - 11 & $\leq$52\\
                  & Ap F    & 0.5 - 1.6 & 0.0 - 0.3 & 45 - 95 & 2 - 10 & $\leq$48\\
                  & Ap$_{\rm TOTAL}$ & 0.4 - 0.7 & 0.0 - 0.3 & 51 - 93 & 7 - 9  & $\leq$45\\
%15462-0450 & 5kpc        & - & - & - & - & -\\
16474+3430 & 5kpc-I$^1$   & - & - & - & - & -\\
           & 5kpc-II      & 0.2 - 0.4 & 00 - 0.2 & 50 - 65 & 20 - 30 & $\leq$ 25\\
           & Ap A$^2$     & - & - & - & - & -\\
16487+5447NE & 5kpc     & 0.5 - 1.0 & 0.1 - 0.3 & 30 - 50 & 10 - 15 & 40 - 60\\
             & Ap A       &0.04 - 0.1 & 0.0 - 0.3 & 25 - 45 & 15 - 33 & 33 - 40\\
16487+5447NW & 5kpc$^1$   & 0.04 & 0.2 & 60 & 30 & 10\\
             & Ap A       & 0.2 - 0.4  & 0.0 - 0.1 & 48 - 55 & 16 - 20 & 25 - 30 \\
             & Ap B       & 0.04 - 0.3 & 0.0 - 0.2 & 50 - 66 & 10 - 35 & 20 - 40 \\
17028+5817NE & 5kpc$^2$   & - & - & - & - & -\\
17028+5817NW & 5kpc       & 0.3 - 0.6  & 0.1 - 0.4 & 50 - 60 & 25 - 35 & 10 - 25 \\
             & Ap A       & 0.3 - 0.6  & 0.0	   & 64 - 75 & 16 - 20 & $\leq$ 15\\
20414-1651 & 5kpc         & 0.3 - 0.6  & 0.0 - 0.2 & 50 - 65 & 20 - 40 & $\leq$ 15\\
21208-0519 & 5kpc-I       & 0.1 - 0.2  & 0.0 - 0.2 & 45 - 50 & 30 - 40 & 10 - 25 \\
           & 5kpc-II      & 0.2 - 1.2  & 0.0 - 0.5 & 25 - 85 & 10 - 22 & $\leq$ 65\\
           & Ap A         & 0.2 - 0.6  & 0.0 - 0.2 & 46 - 70 & 5 - 20  & 10 - 50\\
           & Ap B         & 0.2 - 0.6  & 0.0 - 0.1 & 47 - 55 & 25 - 35 & 10 - 22 \\
%21219-1757 & 5kpc         & - & - & - & - & -\\  
22491-1808 & 5kpc$^1$     & - & - & - & - &\\ 
           & Ap A         & 0.4 - 0.6  & 0.0 - 0.2 & 67 - 85 & 15 - 30 & 0 \\
           & Ap B$^1$     & - & - & - & - & - \\
           & Ap C$^1$     & - & - & - & - & - \\	    
23233+2817 & 5kpc     & 0.05 - 0.5 & 0.0 - 0.3 & 25 - 45 & 15 - 40 & 20 - 40\\  
           & Ap A     & - & - & - & - & -\\ 
           & Ap B     & 0.07 - 0.4 & 0.0 - 0.3 & 35 - 44 & 6 - 20  & 40 - 50\\   
23234+0946 & 5kpc-I   & 0.04 - 0.2 & 0.0 - 0.4 & 20 - 63 & 17 - 40 & 20 - 40 \\  
           & 5kpc-II  & 0.1 - 1.0  & 0.0 - 0.4 & 35 - 80 & 5 - 20  & $\leq$ 60 \\ 
           & Ap A     & 0.5 - 0.8  & 0.0	   & 65 - 85 & 10 - 20 & $\leq$ 15\\
23327+2913 & 5kpc-I   & 0.001 - 0.003 & 0.5 & 40 - 45  & 25 - 35 & 25\\
           &          & 0.01          & 0.1 & 30       & 40      & 30\\        
           & 5kpc-II$^3$  & - & - & - & - & -\\   
           & Ap A     & 0.001 - 0.003 & 0.5 & 40 - 45  & 25 - 35 & 25\\
           &          & 0.01          & 0.1 & 30       & 40      & 30\\
           & Ap B$^3$ & - & - & - & - & - \\ 
\hline			     
\end{tabular}}
\caption[{\it Continued}]{{\it Continued}}
\end{table*}
\begin{table*}
\centering
{\small
%\centering
\begin{tabular}{@{}lllcrrr@{}}
%\hline   
%&& Comb II &&&&\\ 
\hline
Object Name & Aperture &Age of YSP &E(B - V)&$\%$YSP &$\%$p.l &$\%$OSP \\
IRAS        &          &(Gyr)      &        &        &        &        \\
 (1)        &   (2)    &  (3)      &(4)     &(5)     &(6)     &(7)     \\
\hline
08572+3915 & 5kpc       & 0.2 - 0.6 & 0.0 - 0.2 & 60 - 70 & 30 - 36 & $\leq$ 25\\  
           & Ap A       & 0.2 - 0.6 & 0.0 - 0.3 & 50 - 85 & 5 - 20  & $\leq$ 45\\  
           & Ap B       & 0.4 & 0.0 - 0.1 & 65 - 75 & 25 - 30 & 0\\   
           & Ap C       & 0.09 - 0.4 & 0.0 - 0.3 & 73 - 95 & 6 - 35 & $\leq$ 12\\  
10190+1322 & 5kpc-I$^1$ & - & - & - & - & - \\   
           & 5kpc-II$^1$& - & - & - & - & - \\ 
           & Ap A       & 0.4 - 0.7 & 0.0 - 0.2 & 70 - 95 & $\sim$ 5 & $\leq$ 25\\ 
           & Ap B$^1$   & - & - & - & - & - \\  
10494+4424 & 5kpc$^1$   & - & - & - & - & - \\  
           & Ap A$^2$   & - & - & - & - & - \\ 
           & Ap B       & 0.07 - 0.2 & 0.0 - 0.2 & 65 - 85 & 10 - 40 & $\leq$ 10\\  
13305-1739 & 5kpc       & 0.3 - 0.5 & 0.0 - 0.2 & 50 - 60 & 40 & $\leq$ 10\\  
14252-1550 & 5kpc       & 0.2 - 0.6 & 0.0 - 0.2 & 40 - 66 & 15 - 35 & $\leq$ 35\\  
           & Ap A       & 0.2 - 0.8 & 0.0 - 0.2 & 60 - 85 & 10 - 20 & $\leq$ 45\\ 
16156+0146 & 5kpc-I     & $\leq$ 0.03 & .2 - 0.7 & 25 - 55 & 30 - 40 & 14 - 40\\ 
           & 5kpc-II    & 0.4 - 0.5 & 0.0 - 0.1 & 70 - 80 & 20 - 26 & 0\\   
           & Ap A       & 0.05 - 1.0 & 0.2 - 0.5 & 15 - 66 & 30 - 40 & $\leq$ 30\\  
17044+6720 & 5kpc       & 0.2 - 0.4 & 0.0 - 0.1 & $\sim$ 50 & 36 - 40 & 10 - 15 \\  
           & Ap A       & 0.2 - 0.7 & 0.0 - 0.2 & 50 - 62 & 35 - 40 & $\leq$ 15\\ 
           & Ap B       & 0.4 - 0.5 & 0.0 & 55 & 40 & 5 \\   
17179+5444 & 5kpc       & 0.2 - 0.6 & 0.1 - 0.3 & 40 - 62 & 22 - 35 & 6 - 50\\  
23060+0505 & 5kpc       & 0.01 & 0.4 & 72 & 6 & 22 \\  
           & Ap A       & 0.07 - 0.6 & 0.0 - 0.3 & 35 - 50 & 6 - 30 & 20 - 35 \\ 
           & Ap B       & 0.05 - 0.6 & 0.0 - 0.4 & 40 - 60 & 8 - 35 & 5 - 30\\   
23389+0303 & 5kpc-I$^1$ & - & -& - & - & -\\  
           & 5kpc-II    & 0.2 - 0.7 & 0.0 & 53 - 70 & 12 - 30 & $\leq$ 35\\      
           & Ap A       & 0.4 - 0.6 & 0.0 - 0.1 & $\sim$ 60 & $\sim$ 40 & 0\\ 
           & Ap B$^1$   & - & -& - & - & -\\ 
           & Ap C       & 0.5 - 0.8 & 0.0 - 0.1 & 50 - 70 & $\sim$ 30 & $\leq$20\\    
           & Ap D       & 0.2 - 0.5 & 0.0 - 0.1 & 55 - 65 & 10 - 30 & 2 - 30 \\   
\hline			  
\end{tabular}}		  
\caption[Modelling results for the 10 ``extra'' ULIRGs included in the
ES using Comb II]{Same as Table \ref{tab:CS_combII}, but for
the 10 additional ULIRGs included in the ES.}
\label{tab:ES_combII}	  
\end{table*}
\begin{table*}
%\hspace*{1.0cm}
\centering
{\small
%\centering
\begin{tabular}{@{}lllcrrr@{}}
%\hline
%&&& Comb III &&&\\ 
\hline    
Object      &Aperture &Age of      &E(B - V)    & Age of      &E(B -V)    &\%VYSP \\
Name        &         & IYSP       & of         & VYSP        & of        &\\
IRAS        &         &(Gyr)       & IYSP       &(Myr)        &VYSP       & \\ 
 (1)        &   (2)   &(3)         &(4)         &(5)          &(6)        &(7)\\
\hline
00091-0738 & 5kpc     & 0.5 - 1.0 & 0.0 - 0.4 & 6 - 50  & 0.5 - 1.0 & 30 - 65 \\   
           & Ap A     & 0.7 - 2.0 & 0.0 - 0.2 & $\leq$ 100 & 0.0 - 1.5 & $\leq$ 50\\ 
           & Ap B     & 0.3 - 2.0 & 0.0 - 0.4 & $\leq$ 100 & 0.2 - 1.0 & 10 - 90\\	
00188-0856 & 5kpc     & 0.5 - 0.7 & 0.0 - 0.4 & $\leq$ 100 & 0.2 - 1.6 & $\leq$ 40\\   
01004-2237 & 5kpc     & 0.5 - 0.7 & 0.2 - 0.4 & 9 - 20 & 0.1 - 0.3 & 60 - 75\\ 
12072-0444 & 5kpc     & 0.7 - 2.0 & 0.0 - 0.4 & 7 - 20 & 0.4 - 0.6 & 45 - 65 \\   
           & Ap A     & 0.5 - 2.0 & 0.0 - 0.4 & $\leq$ 100  & 0.0 - 1.2 & 20 - 65 \\  
           & Ap B     & 1.0 - 2.0 & 0.0 & 40 - 100 & 0.0 - 0.3 & 40 - 60\\    
12112+0305 & 5kpc-I   & 0.5 - 1.0 & 0.0 - 0.4 & 7 - 40 & 0.5 - 0.8 & 50 - 90 \\     
           & 5kpc-II  & 0.3 - 2.0 & 0.0 - 0.4 & 4 - 60 & 0.0 - 0.5 & 35 - 90\\    
           & Ap A     & 0.5 - 2.0 & 0.0 - 0.4 & $\leq$ 50 & 0.6 - 1.0 & 35 - 80\\     
12540+5708 & 5kpc     & - & - & - & - & - \\   
           & Ap A     & 2 & 0.0 & 5 - 10 & 0.0 - 0.1 & 30 - 35\\
           & Ap B     & 0.5 - 2.0 & 0.0 - 0.4 & 10 - 100 & 0.1 - 0.5 & 30 - 65\\ 
           & Ap C     & 0.3 - 2.0 & 0.0 - 0.4 & 50 - 80 & 0.0 - 0.1 & 70 - 96\\  
           & Ap D     & - & - & - & - & -\\  
           & Ap E     & 0.7 - 2.0 & 0.0 - 0.4 & 6 - 70 & 0.5 - 1.0 & 40 - 75\\ 
13428+5608 & 5kpc     & 0.7 - 2.0 & 0.0 - 0.4 & 6 - 50 & 0.4 - 0.8 & 20 - 50 \\ 
           & Ap A     & 0.7 - 2.0 & 0.0 - 0.2 & $\leq$ 100 & 0.0 - 0.6 & 5 - 50\\
           & Ap B     & 2.0 & 0.0 & 50 - 100 & 0.4 - 0.5 & 25 - 33\\
           & Ap C     & 0.7 - 2.0 & 0.0 - 0.2 & $\leq$ 100 & 0.0 - 0.8 & 10 - 50\\ 
           & Ap D     & 0.7 - 2.0 & 0.0 - 0.4 & 5 - 60 & 0.3 - 0.8 & 25 - 60 \\  
           & Ap E     & 0.7 - 2.0 & 0.0 - 0.4 & 6 - 50 & 0.4 - 0.8 & 20 - 50 \\   
           & Ap F     & 0.7 - 2.0 & 0.0 - 0.4 & 5 - 60 & 0.3 - 0.6 & 25 - 55\\   
13451+1232(PA160) & 5kpc  & 2.0 & 0.0 - 0.2 & $\leq$ 100 & 0.0 - 2.0 & $\leq$ 10\\   
                  & Ap A  & 2.0 & 0.0 - 0.1 & $\leq$ 100 & 0.0 - 2.0 & $\leq$ 10 \\  
                  & Ap B  & 2.0 & 0.1 - 0.2 & $\leq$ 100 & 0.0 - 1.4 & 5 - 20 \\   
                  & Ap C  & 1.0 - 2.0 & 0.1 - 0.2 & $\leq$ 100 & 0.1 - 0.8 & 20 - 35 \\  
                  & Ap D  & 1.0 - 2.0 & 0.0 - 0.2 & $\leq$ 100 & 0.0 - 0.9 & 5 - 15\\	
13451+1232(PA230) & Ap A  & 2.0 & 0.0 - 0.1 & $\leq$ 100 & 0.0 - 2.0 & $\leq$ 12\\  
                  & Ap B  & 2.0 & 0.1 & $\leq$ 100 & 0.0 - 2.0 & $\leq$ 11\\
13539+2920 & 5kpc-I   & 0.5 - 2.0 & 0.0 - 0.2 & 10 - 100 & 0.2 - 0.5 & 15 - 60\\	 
           & 5kpc-II  & 0.7 - 1.0 & 0.0 - 0.4 & $\leq$ 50 & 0.8 - 1.5 & 40 - 70\\  
14060+2919 & 5kpc     & 0.5 - 0.7& 0.4 & 7 & 0.4 - 0.5 & 60 - 100\\
           & Ap A     & 0.5 - 2.0 & 0.0 - 0.2 & $\leq$ 100 & 0.0 - 0.9 & 25 - 55\\
           & Ap B     & 0.3 - 0.7 & 0.0 - 0.4 & 7 - 30 & 0.1 - 0.4 & 45 - 90\\
           & Ap C     & 0.3 - 0.7 & 0.0 - 0.4 & 4 - 7 & 0.2 - 0.4 & 45 - 82\\
14348-1447 & 5kpc-I   & 0.3 - 0.7 & 0.0 - 0.6 & 5 - 20 & 0.4 - 0.7 & 50 - 80\\
           & 5kpc-II  & 0.3 - 2.0 & 0.0 - 0.4 & 7 - 8   & 0.4 - 0.6 & 40 \\
           &          &           &           & 50 - 80 & 0.0 - 0.3 & 50 - 80\\ 
           & Ap A     & 0.3 - 2.0 & 0.0 - 0.4 & 5 - 60 & 0.0 - 0.5 & 45 - 100\\
           & Ap B     & 0.5 - 1.0 & 0.0 - 0.4 & $\leq$ 30 & 0.5 - 1.2 & 35 - 60\\
           & Ap C     & 0.3 - 2.0 & 0.0 - 0.4 & $\leq$ 100 & 0.0 - 0.9 & 14 - 80\\
           & Ap D     & 0.5 - 0.7 & 0.2 - 0.4 & 4 - 50 & 0.0 - 0.6 & 20 - 80\\
14394+5332 & 5kpc-I   & 0.5 - 0.7 & 0.2 & 6 - 7 & 0.7 & 75\\  
           & 5kpc-II  & 2.0 & 0.0 - 0.2 & 6 - 100 & 0.0 - 0.8 & 25 -40 \\ 
           & A        & 0.5 - 2.0 & 0.0 - 0.4 & 4 - 100 & 0.0 - 0.5 & 30 - 75 \\ 
           & B        & 0.7 - 2.0 & 0.0 - 0.2 & 7 - 40 & 0.3 - 0.5 & 30 - 55 \\ 
15130+1958 & 5kpc     & 0.3 - 1.0 & 0.0 - 0.4 & $\leq$ 100 & 0.0 - 1.3 & 20 - 70 \\  
15206+3342 & 5kpc     & 0.5 - 0.7 & 0.5 - 0.6 & 1 - 3  & 0.4 - 0.5 & 70 - 80 \\
           &          & 0.3 - 2.0 & 0.0 - 0.4 & 10     &  0.3      & 91\\       
\hline    
\end{tabular}}
\caption[Modelling results for the CS using Comb III]{
Modelling results for the CS using Comb III (IYSP + VYSP). Col (1):
object name.  Col (2): aperture label as indicated in Figure
\ref{fig:spatial_cuts}. Col (3): range of ages for the IYSP component that
adequately fit the data. Col (4): range of $E(B - V)$ values for the
IYSPs in Col (3). Col (5): range of ages of the VYSP that adequately
fit the data. Col(6): range of $E(B - V)$ values for the VYSPs in Col
(5). Col (7): flux contributions of the VYSP components to the model, in
the normalising bin.}
\label{tab:CS_combIII}
\end{table*}
\addtocounter{table}{-1}
\begin{table*}
%\hspace*{1.0cm}
\centering
{\small
%\centering
\begin{tabular}{@{}lllcrrr@{}}
%\hline
%&&& Comb III &&&\\ 
\hline    
Object      &Aperture &Age of      &E(B - V)    & Age of      &E(B -V)    &\%VYSP \\
Name        &         & IYSP       & of         & VYSP        & of        &\\
IRAS        &         &(Gyr)       & IYSP       &(Myr)        &VYSP       & \\ 
 (1)        &   (2)   &(3)         &(4)         &(5)          &(6)        &(7)\\
\hline
15327+2340(PA160) & 5kpc  & 0.5 - 0.7 & 0.4       & $\leq$ 60 & 0.8 - 1.3 & 16 - 45\\  
                  & Ap A  & 0.6 - 0.7 & 0.0 - 0.2 & $\leq$ 100 & 0.0 - 0.2 & $<$ 5\\
                  & Ap B  & 0.5 - 0.7 & 0.0 - 0.2 &  $\leq$ 100 & 0.0 - 0.3 & $<$ 20 \\
                  & Ap C  & 0.5 - 0.7 & 0.3 - 0.5 &  $\leq$ 100  & 0.0 - 1.0 & 5 - 15 \\
                  & Ap D  & 0.5 - 0.9 & 0.0 - 1.0 &  $\leq$ 80 & 0.9 - 1.8 & 30 - 63\\
                  & Ap E  & 0.5 - 0.9 & 0.2 - 0.6 &  $\leq$ 60 & 0.9 - 1.6 & 22 - 47\\
                  & Ap F  & 0.5 - 0.7 & 0.2 - 0.6 & $\leq$ 80  & 0.0 - 1.5 &  5 - 36\\
                  & Ap G  & 0.5 - 0.7 & 0.0 - 0.3 &  $\leq$ 70 & 0.0 - 0.7 & 5 - 20\\
                  & Ap$_{\rm TOTAL}$ & 0.5 - 0.7 & 0.2 - 0.6 & $\leq$ 100 & 0.0 - 1.5 &  5 - 35 \\
15327+2340(PA75)  & Ap A  & 0.5 - 0.7 & 0.2 - 0.5 & $\leq$ 100  & 0.0 - 2.0 &    $<$ 20\\
                  & Ap B  & 0.5 - 0.7 & 0.2 - 0.6 & $\leq$ 100  & 0.0 - 1.0 & 5 - 25 \\
                  & Ap C  & 0.5 - 0.7 & 0.2 - 0.6 & $\leq$ 50 & 0.0 - 1.2&    5 - 30 \\
                  & Ap D  & 0.5 - 0.7 & 0.2 - 0.6 & $\leq$ 60 & 0.0 - 1.5&    5 - 25\\
                  & Ap E  & 0.5 - 0.7 & 0.2 - 0.5 & $\leq$ 50 & 0.0 - 2.0&    $<$ 30\\
                  & Ap F  & 0.6 - 0.9 & 0.1 - 0.4 & $\leq$ 100  & 0.0 - 1.2& $<$ 25\\
                  & Ap G  & 0.7 - 0.9 & 0.0 - 0.3 & $\leq$ 100  & 0.0 - 0.8& $<$ 25\\
                  & Ap$_{\rm TOTAL}$ & 0.5 - 0.7 & 0.2 - 0.5 & $\leq$ 100 & 0.0 - 1.2& 5 - 25\\
15327+2340(PA75*) & Ap A     & 0.5 - 0.7 & 0.0 - 0.3 & $\leq$ 60 & 0.0 - 0.6 & 10 - 35\\
                  & Ap B     & 0.5 - 0.7 & 0.0 - 0.3 & $\leq$ 100  & 0.0 - 1.5  &   5 - 25\\
                  & Ap C     & 0.5 - 0.7 & 0.0 - 0.3 & $\leq$ 100  & 0.0 - 1.5  &   $<$ 40\\
                  & Ap D     & 0.5 - 0.8 & 0.0 - 0.3 & $\leq$ 60 & 0.0 - 1.5 &   $<$ 40\\
                  & Ap E     & 0.6 - 0.9 & 0.0 - 0.3 & $\leq$ 60 & 0.0 - 1.0& $<$ 25 \\
                  & Ap F     & 0.6 - 0.9 & 0.0 - 0.3 & $\leq$ 100  & 0.0 - 2.0& $<$ 15\\
                  & Ap$_{\rm TOTAL}$ & 0.5 - 0.8 & 0.0 - 0.3 & $\leq$ 60 & 0.0 - 1.5&   5 - 30\\ 
%15462-0450 & 5kpc & - & - & - & - & -\\   
16474+3430 & 5kpc-I  & 0.5 - 0.7 & 0.0 - 0.4 & 7 - 8 & 0.3 - 0.4 & 65 - 80\\   
           & 5kpc-II & 0.5 - 2.0 & 0.0 - 0.2 & 6 - 8   & 0.2 - 0.4  & 30 - 50  \\ 
           &         &           &           & 50 - 90 & 0.0 - 0.2  & 50 \\
           & Ap A & 0.5 - 2.0 & 0.0 - 0.4 & 6 - 8   & 0.3 - 0.5  & 40 - 75\\
           &      &           &           & 30 - 40 & 0.3 - 0.5  & 40 - 75\\  
16487+5447NE & 5kpc    & 2.0 & 0.1 & $\leq$ 80 & 0.0 - 0.8 & 10 - 20 \\
             & Ap A    & 2.0 & 0.0 - 0.2 & 5 - 8   & 0.2 - 0.3 & 45 - 55 \\
             &         &     &           & 30 - 40 & 0.0 - 0.2 & 50 - 60\\
16487+5447NW & 5kpc    & 0.5 - 0.7 & 0.0 - 0.4 & 4 - 20 & 0.1 - 0.3 & 55 - 85\\
             & Ap A    & 0.5 - 2.0 & 0.0 - 0.4 & 4 - 60 & 0.0 - 0.4 & 15 - 65\\
             & Ap B    & 0.7 - 2.0 & 0.0 - 0.4 & 4 - 8 & 0.2 - 0.3 & 65 - 80\\
17028+5817NE & 5kpc    & 0.5 - 1.0 & 0.0 - 0.4 & 4 - 8 & 0.3 - 0.5 & 50 - 65\\
17028+5817NW & 5kpc    & 0.5 - 1.0 & 0.0 - 0.2 & $\leq$ 60 & 0.8 - 1.5 & 40 - 70\\
             & Ap A    & 0.5 - 2.0 & 0.0 - 0.4 & $\leq$ 100 & 0.0 - 0.9 & 30 - 85\\
20414-1651   & 5kpc    & 0.5 - 2.0 & 0.0 - 0.4 & 4 - 100 & 0.4 - 0.8 & 30 - 65\\
21208-0519   & 5kpc I  & 0.5 - 2.0 & 0.0 - 0.4 & 5 - 50 & 0.3 - 0.5 & 50 - 82\\
             & 5kpc II & - & - & - & - & -\\
             & Ap A    & 2.0 & 0.0 & 70 - 100 & 0.3 - 0.4 & 40 - 50 \\
             & Ap B    & 0.5 - 2.0 & 0.0 - 0.4 & 4 - 8     & 0.0 - 0.6    & 35 - 70\\
             &         &           &           &  30 - 100 & 0.0 - 0.6    & 35 - 70\\
%21219-1757   &  5kpc    & - & - & - & - & -\\
22491-1808   &  5kpc & 0.3 - 0.5 & 0.0 - 0.4 & 4 - 8  & 0.2 - 0.4 & 45 - 75\\
             &  Ap A & 0.5 - 2.0 & 0.0 - 0.4 & $\leq$100&0.0 - 0.8& 25 - 57\\
             &  Ap B & 0.3 - 0.7 & 0.0 - 0.4 & 4 - 7  & 0.3 - 0.4 & 60 - 90\\
             &  Ap C & 0.3       & 0.2 - 0.4 & 5 - 6  & 0.2 - 0.4 & 44 - 60\\
23233+2817& 5kpc     & - & - & - & - & -\\
          & Ap A     & - & - & - & - & -\\
          & Ap B     & - & - & - & - & -\\
23234+0946& 5kpc I   & 0.5 - 2.0 & 0.0 - 0.4 & 6 - 30 & 0.4 - 0.6 & 60 - 82\\
          & 5kpc II  & 1.0 - 2.0 & 0.0 - 0.2 & $\leq$ 100  & 0.5 - 1.5 & 17 - 37  \\
          & Ap A     & 0.7 & 0.0 & $\leq$ 20 & 0.5 - 0.8 & 15 - 25\\
23327+2913& 5kpc I   & - & - & - & - & -\\
          & 5kpc II  & - & - & - & - & - \\
          & Ap A     & - & - & - & - & - \\
          & Ap B     & - & - & - & - & - \\
\hline 
\end{tabular}}
\caption[{\it Continued}]{{\it Continued}}
\end{table*}
\begin{table*}
%\hspace*{1.0cm}
\centering
{\small
%\centering
\begin{tabular}{@{}lllcrrr@{}}
%\hline
%&&& Comb III &&&\\ 
\hline    
Object      &Aperture &Age of      &E(B - V)    & Age of      &E(B -V) of &\%VYSP \\
Name        &         & IYSP       & of         & VYSP        & of        &\\
IRAS        &         &(Gyr)       & IYSP       &(Myr)        &VYSP       & \\ 
 (1)        &   (2)   &(3)         &(4)         &(5)          &(6)        &(7)\\
\hline 
08572+3915& 5kpc   & 0.5 - 2.0 & 0.0 - 0.4 & 6 - 70 & 0.1 - 0.6 & 30 - 90\\  
          & Ap A   & 0.5 - 2.0 & 0.0 - 0.4 & $\leq$ 100 & 0.0 - 1.0 & 10 - 50\\  
          & Ap B   & 0.3 - 2.0 & 0.0 - 0.2 & $\leq$ 100 & 0.0 - 0.5 & 10 - 60\\   
          & Ap C   & 0.3 - 2.0 & 0.0 - 0.4 & $\leq$ 100 & 0.0 - 0.7 & 25 - 100\\  
10190+1322& 5kpc I & 0.7 - 2.0 & 0.0 - 0.4 & $\leq$ 60 & 0.7 - 1.1 & 30 - 65\\   
          & 5kpc II& 0.7 & 0.2 - 0.4 & 50 - 100 & 0.8 - 1.1 & 30 - 55 \\ 
          & Ap A   & 0.5 - 0.7 & 0.0 - 0.2 & $\leq$ 100 & 0.0 - 2.0 & $\leq$ 25\\ 
          & Ap B   & 0.5 - 1.0 & 0.0 - 0.4 & 8 - 30 & 0.8 - 0.9 & 52 - 66\\  
10494+4424& 5kpc   & 0.5 & 0.0 - 0.2 & 5 - 10  & 1.1 - 1.3 & 30 - 45 \\  
          &        &     &           & 40 - 60 & 0.8 - 0.9 & 50 - 70\\
          & Ap A   & - & - & - & - & -\\ 
          & Ap B   & 0.3 - 0.7 & 0.0 - 0.4 & 6 - 60 & 0.0 - 0.6 & 45 - 75 \\  
13305-1739& 5kpc   & 0.5 - 2.0 & 0.0 - 0.4 & 4 - 7 & 0.4 - 0.7 & 50 - 80\\  
14252-1550& 5kpc   & 0.5 - 2.0 & 0.0 - 0.4 & $\leq$ 90 & 0.5 - 1.3 & 35 - 80\\  
          & Ap A   & 0.5 - 1.0 & 0.0 - 0.4 & $\leq$ 100 & 0.0 - 2.0 & 5 - 80\\ 
16156+0146& 5kpc I & 0.7 - 2.0 & 0.0 - 0.4 & 9 - 10 & 0.5 & 60 - 80\\ 
          & 5kpc II& 0.3 - 2.0 & 0.0 - 0.4 & $\leq$ 100 & 0.0 - 0.6 & 20 - 70\\   
          & Ap A   & 0.5 - 2.0 & 0.0 - 0.4 & $\leq$ 30 & 0.0 - 0.7 & 15 - 81\\ 
17044+6720& 5kpc   & 0.3 - 1.0 & 0.0 - 0.4 & 7 - 40 & 0.4 - 0.6 & 40 - 87\\  
          & Ap A   & 0.3 - 2.0 & 0.0 - 0.4 & $\leq$ 100 & 0.0 - 0.8 & 30 - 85\\ 
          & Ap B   & 0.3 - 2.0 & 0.0 - 0.4 & 4 - 100 & 0.1 - 0.4 & 40 - 85\\   
17179+5444& 5kpc   & 0.7 - 2.0 & 0.0 - 0.4 & 4 - 100 & 0.3 - 1.1 & 25 - 60\\ 
23060+0505& 5kpc   & - & - & - & - & - \\  
          & Ap A   & 1.0 - 2.0 & 0.0 & 6 - 7   &  0.7       & 50 - 60\\
          &        &           &     & 40 - 60 &  0.4 - 0.5 & 50 - 60 \\
          & Ap B   & 0.7 - 2.0 & 0.0 - 0.4 & 10 - 50 & 0.0 - 0.5 & 30 - 75\\   
23389+0303& 5kpc I & 0.7 - 1.0 & 0.0 - 0.4 & 6 - 8 & 0-6 0.8 & 50 - 72\\  
          & 5kpc II& 0.7 - 2.0 & 0.0 - 0.2 & $\leq$ 100 & 0.2 - 0.9 & 25 - 60\\	      
          & Ap A   & 0.5 & 0.0 & 10 - 20 & 0.7  & $\sim$ 40\\ 
          & Ap B   & 0.7 - 2.0 & 0.0 - 0.4 & 7 & 0.8 & 75\\ 
          & Ap C   & 0.7 - 1.0 & 0.0 - 0.4 & 5 - 60 & 0.2 - 0.8 & 35 - 60\\    
          & Ap D   & 0.7 - 2.0 & 0.0 - 0.2 & 4 - 6     & 0.6 - 0.7 & 40 - 50 \\  
          &        &           &           &  40 - 100 & 0.2 - 0.4 & 45 - 70 \\    
\hline 
\end{tabular}}
\caption[Modelling results for the 10 ``extra'' ULIRGs included in the
ES using Comb III]{Same as Table \ref{tab:CS_combIII}, but for
the 10 additional ULIRGs included in the ES.}
\label{tab:ES_combIII}
\end{table*}

\clearpage

\begin{center}
    {\bf APPENDIX}\\
    {\bf NOTES ON INDIVIDUAL SOURCES}
  \end{center}
 
  {\bf IRAS 00091-0738:} optical and near-IR images reveal the
  presence of several bright knots showing the location of young
  ($\lsim$ 7 Myr) stellar populations \citep{Surace00a} in this double
  nucleus system (NS = 2.2 kpc). The double nucleus structure is not
  resolved in our spectra and a single 5 kpc aperture was used during
  the analysis. Our modelling results reveal the presence of a
  significant contribution from VYSPs in the nuclear region, which is
  consistent with the results obtained for the extended
  apertures. IRAS 00091-0738 is classified as an HII-galaxy at optical
  wavelengths. Therefore, it is possible to use the nebular emission
  lines in order to provide a further constraint on the ages of the
  VYSP component. The VYSP age obtained using the H$_{\alpha}$
  equivalent width and the \cite{Leitherer99} models (EW(H$\alpha$) =
  28~\AA, t$_{VYSP}$ $\sim$ 7 Myr) is consistent with the lower end of
  the range of ages found from the modelling.

  {\bf IRAS 00188-0856:} this compact, single nucleus system is
  classified as a LINER at optical wavelengths. On the other hand,
  \cite{Imanishi07} and \cite{Valdes05} reported evidence of the
  presence of an AGN in the nucleus of the galaxy based on their near-
  and mid-IR studies of the object. However, the AGN does not dominate
  the IR emission, which is likely to be dominated by starburst
  activity \citep{Lutz99,Valdes05}. The modelling results suggest the
  presence of a dominant IYSP with a relatively small percentage
  contribution of any VYSP. Note that, due to the small contribution
  of the VYSP to the optical light from the galaxy, both the age and
  reddening of that component are relatively unconstrained.

  {\bf IRAS 01004-2237:} the optical spectrum of this single nucleus
  system shows a mixture of HII and Seyfert 2 features
  (\citetalias{Veilleux99a}), with clear evidence of a buried AGN at
  IR wavelegths \citep{Imanishi07}. The modelling results clearly
  suggest the presence of a dominant VYSP with relatively low
  reddening. The VYSP ages obtained from the modelling (t$_{\rm VYSP}$
  = 9 -- 20 Myr) differs than those obtained from the H$\alpha$
  equivalent width (EW(H$\alpha$) = 58~\AA, t$_{\rm VYSP} \sim$ 6.5
  Myr ). However, given the uncertainties attached to the synthetic
  templates for very young ages ($\leq$ 10 Myr), we consider that
  $\sim$ 3 Myr is not a major disagreement between the two results.

  {\bf IRAS 08572+3915:} this is double nucleus system consisting of
  two spirals galaxies in the process of merging, rotating in
  different planes \citep{Arribas00}. The nuclear, 5 kpc aperture was
  centred in the northern nucleus (the secondary nucleus was not
  covered by the slit), thought to be responsible for the far-IR
  emission from the source
  \citep{Condon91,Soifer00,Evans02,Nagar03}. The {\it HST}-WFPC2 image
  taken by \cite{Surace98b} reveals the presence of several young
  (t$_{knots}$ = 5 -- 10 Myr), massive knots located in both the tails
  and the nuclear regions. The optical spectrum of IRAS 08572+3915 is
  a composite of HII and LINER-like features. Although no evidence for
  the presence of an AGN is found in the optical and near-IR
  (\citetalias{Veilleux99a,Veilleux99b}), IRAS 08572+3915 shows signs
  of AGN activity at mid-IR wavelengths
  \citep{Imanishi00,Imanishi06,Armus07}. Moderately reddened VYSPs are
  found in the norhern nucleus, which contribute significantly to the
  optical light. The reddenings and ages of the stellar components in
  the extended apertures are consistent with those of the nuclear
  aperture.

  {\bf IRAS 10190+1322:} The north-east galaxy of this double nuleus
  system is brighter at near-IR and radio wavelengths
  \citep{Murphy01,Dasyra06b,Nagar03}, while the south-western nucleus
  appears the brightest in the optical. Recently, \cite{Rupke02}
  classified the north-east nucleus as a LINER and the south-western
  source as an HII galaxy in their high-resolution spectroscopic study
  of ULIRGs. No evidence for AGN activity is found at near- and mid-IR
  wavelengths (Veilleux et al., 1999b, Imanishi et al., 2007), which
  is consistent with the idea of starburst activity powering the IR
  luminosity of the source (Farrah et al., 2003). The results obtained
  from the modelling suggest that the VYSPs are concentrated in the
  two nuclei of the galaxy, and in the region in between, coinciding
  with large concentrations of gas and dust.  In this case, only the
  extracted spectrum from aperture 5kpc-I shows clear HII-like region
  features. The VYSP age derived using the H$\alpha$ equivalent width
  (EW(H$\alpha$) = 47~\AA, t$_{\rm VYSP} \sim$ 6.5 Myr) for that
  region of the galaxy is consistent with the lower end of the range
  of ages found for the VYSPs from modelling the optical continuum.

  {\bf IRAS 10494+4424:} the modelling results obtained using near-IR
  and optical photometric points clearly suggest a starburst as an
  ionizing source for this single nucleus system \citep{Farrah03},
  which is classified as a LINER in the optical
  \citep{Kim98a}. Modelling the optical spectra, we find that the
  VYSPs located in the nuclear and the northern region of the galaxy
  make a significant contribution to the optical light. However,
  adequate fits for Ap A (sampling a region towards the south of the
  galaxy) are only obtained using Comb I, with the optical light
  dominated by a 400 Myr IYSP with little or no reddening, plus a
  small contribution from a 12.5 Gyr OSP. In terms of reddening, our
  modelling results clearly provide evidence for higher reddenings
  towards the nuclear regions of the galaxy.

  {\bf IRAS 12072-0444:} this ULIRG appears as a single nucleus system
  in the optical and near-IR images of \cite{Kim02}. However a second
  nucleus becomes apparent in the {\it HST}-NICMOS H-band images
  \citep{Veilleux06}. This second nucleus was first detected in the
  WFPC2 images of \cite{Surace98b}, but was classified as a super star
  cluster. More recently, \cite{Dasyra06a} confirmed spectroscopically
  the presence of a second genuine nucleus, with a NS = 2.8 kpc. The
  double nucleus structure is unresolved in our spectroscopic
  data. The detection of strong [Si~{\small IV}] emission is evidence
  for the presence of a buried AGN in this source, classified as a Sy2
  galaxy in the optical \citep{Veilleux95,Kim98a}. Moderately reddened
  VYSPs which contribute significantly to the optical light from the
  galaxy are located in the nuclear region. Comparing the nuclear with
  the extended apertures, the modelling results suggest the presence
  of younger, more highly reddened VYSPs in the nuclear regions of the
  galaxy than in the extended apertures.

  {\bf IRAS 12112+0305:} this double nucleus system is classifies as a
  LINER in the optical, shows no evidence for powerful AGN activity
  (\citealt{Genzel98}; Veilleux et al., 1999b), and is classified as a
  starburst at all other wavelengths studied
  \citep{Lutz99,Franceschini03,Risaliti06,Armus07}. The modelling
  results show the presence of dominant, relatively high reddened
  VYSPs located in the north-eastern nucleus (5 kpc-I), which is the
  brightest at optical wavelengths. For this object, the ages and
  percentage contributions to the flux for both nuclei and the region
  in between are consistent. On the other hand, our results suggest
  increasing concentrations of obscuring dust towards the northern
  source.

  {\bf IRAS 12540+5708:} also known as Mrk231, IRAS 12540+5708 is the
  most luminous ULIRG in our sample (log(L$_{\rm IR}$) =~12.54). This
  object is classified as a Sy1 galaxy in the optical, and considered
  an AGN-powered ULIRG based on mid- and near-IR spectroscopy
  \citep{Lutz99,Rigopoulou99}. However, the fraction of the bolometric
  luminosity produced by the AGN remains controversial. Despite its
  warm mid- to far-IR colours, \cite{Davies04} found that star
  formation within the inner 300pc is responsible for 25\% -- 40\% of
  the entire luminosity of the galaxy, and that starburst activity
  seems to power most of the far-IR luminosity
  \citep{Downes98,Farrah03}. On the other hand, ages from a few Myr in
  the inner parts of the galaxy, up to ages of order of Gyr in the
  extended regions, have been found for the stellar populations in the
  galaxy \citep{Surace98b,Canalizo00a,Davies04}

\begin{figure}
\hspace{1.5cm}\psfig{file=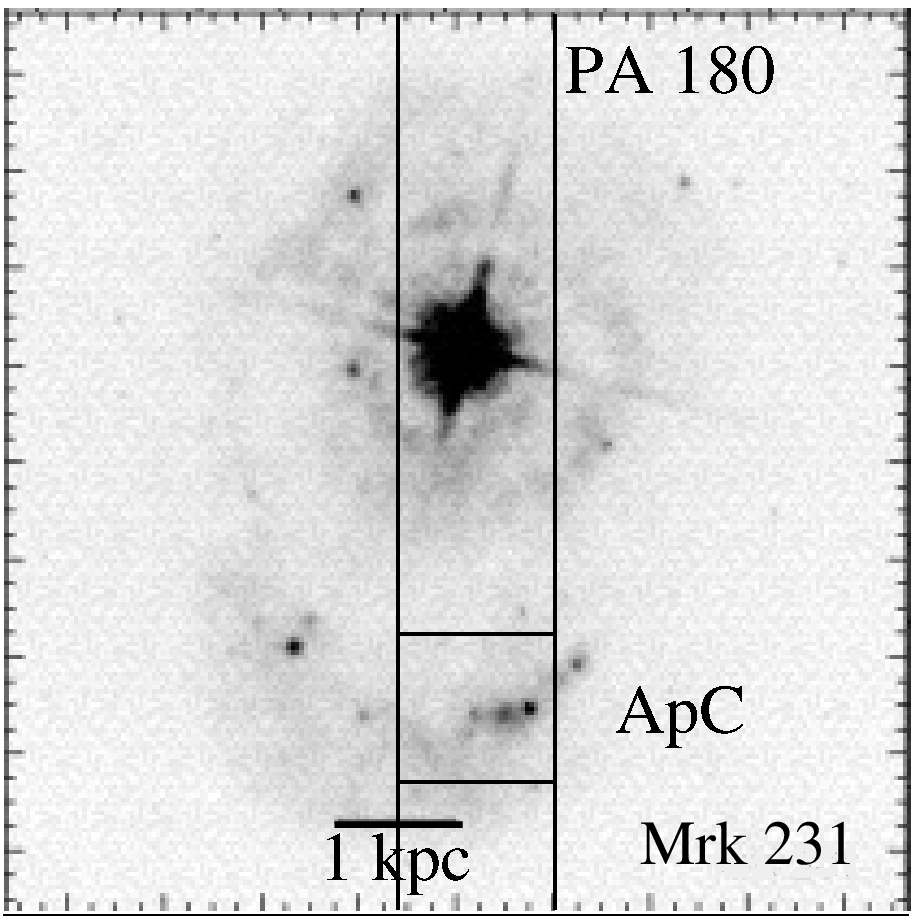,width=5.0cm,angle=0.0}
\caption[{\it HST} WFPC2 image of Mrk231 showing the the bright
condensation to the South of the galaxy ] {Mrk 231: the slit position
and the location of Ap C are shown in the figure superimposed on the
\cite{Surace98b} F439W WFPC2 {\it HST} image. The major ticks are
arcseconds.}
\label{fig:Mrk231_WFPC2_ApC}
\end{figure}

\begin{figure}
\hspace{0.5cm}\psfig{file=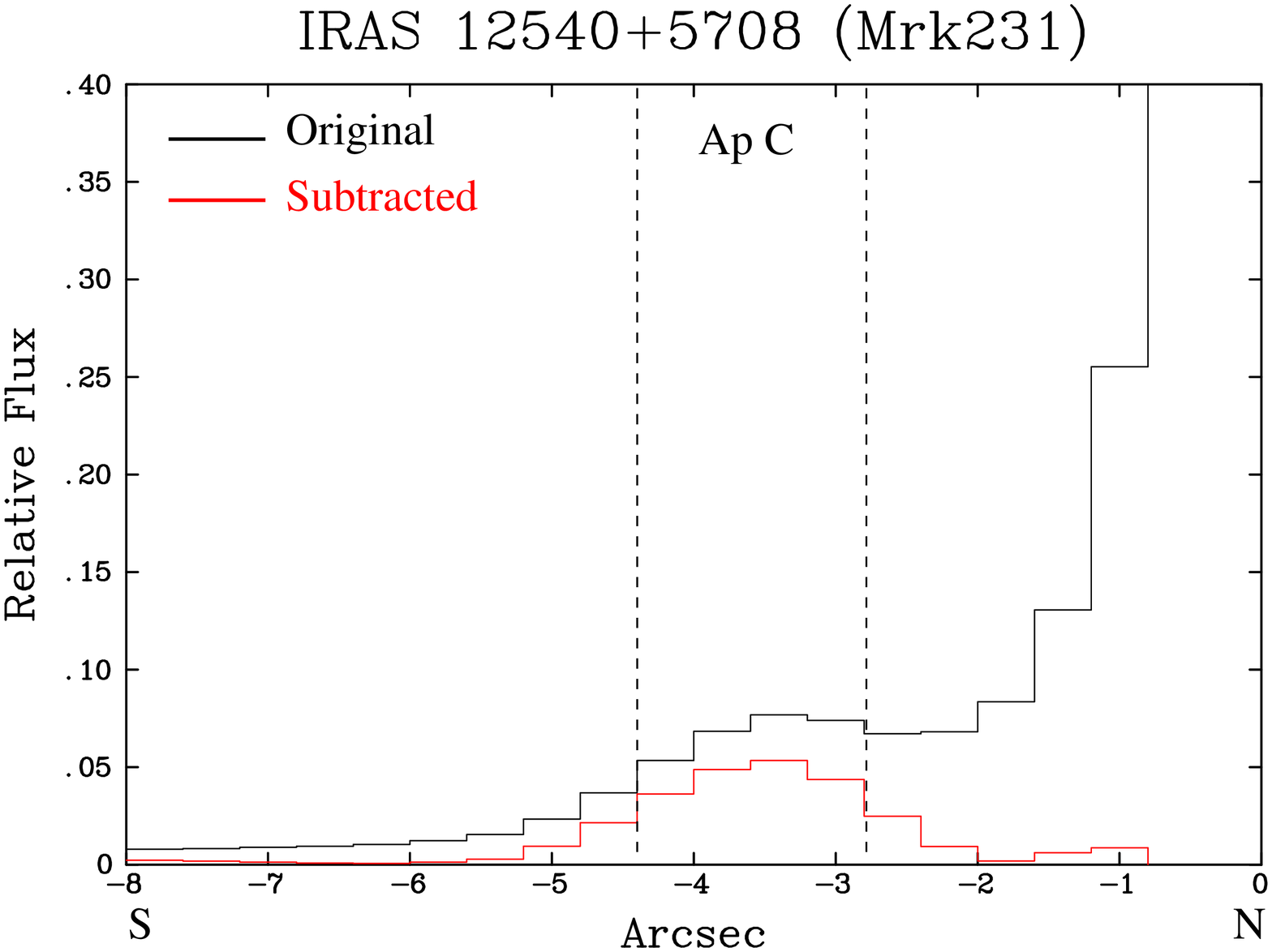,width=7.0cm,angle=0.0}
\caption[Mrk 231: spatial cuts of the region sampling the bright
condensation to the South of the galaxy]{Mrk 231: spatial cuts (4400
-- 4600~\AA) extracted from the subtracted and original 2-D frames
showing the region including the bright condensation to the South of
the galaxy. It is clear from the figure that the AGN contamination has
been almost entirely removed after applying the technique described in
the text. The location of Ap C is also shown in the figure.}
\label{fig:Bright_blob_spatial_cut}
\end{figure}

  Due to ghost images affecting many of the apertures in the red arm,
  it was only possible to model the extracted spectra from the blue
  arm, with the exception of Ap C. In addition, the powerful AGN
  emission prevents any attempt at modelling the stellar populations
  for those apertures sampling the nuclear regions (5 kpc and Ap
  D). Therefore, only the extracted spectra from apertures A, B, C and
  E were used for the analysis of the stellar populations presented
  here. Note that Ap C samples the bright condensation \cite[the
  ``horseshoe'',][]{Hamilton87} $\sim$ 3 kpc to the south of the
  galaxy, as seen in the {\it HST}-WFPC2 images taken by
  \cite{Surace98b}. Figure \ref{fig:Mrk231_WFPC2_ApC} shows the
  position of the slit and the extent of Ap C, superimposed on the
  \cite{Surace98b} {\it HST}-WFPC2 F439W image. In order to minimize
  the potential AGN contamination (including the seeing disk of the
  bright AGN and any dust-scattered AGN light), the spatial axes of
  the blue and red 2-D frame were reversed. The reversed and the
  original frames were registered and then subtracted. A spatial cut
  of the subtracted 2-D frame is shown in Figure
  \ref{fig:Bright_blob_spatial_cut}, as well as the spatial cut of the
  original frame and the extent of Ap C. It is clear from the figure
  that any AGN contamination has been removed, and therefore, we were
  able to model the full spectral range for this aperture.

\begin{figure}
\hspace{0.2cm}\psfig{file=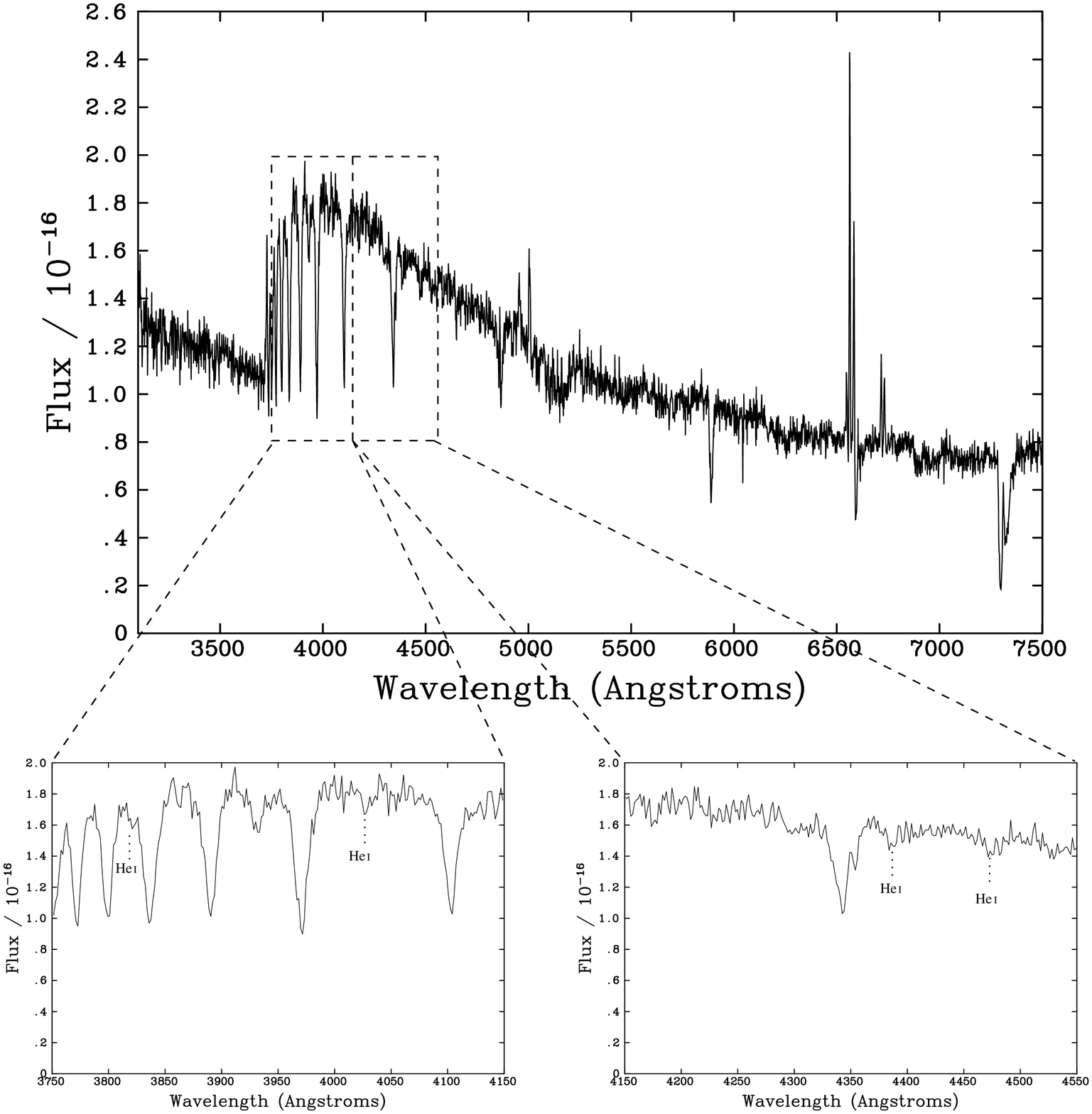,width=7.5cm,angle=0.0}
\caption[Mrk 231: extracted spectrum from Ap C showing the He~{\small
 I} absorption lines] {Extracted spectrum from Ap C for Mrk 231. The
 figure shows the detection of the He~{\small I} absorption lines at
 3820~\AA, 4026~\AA, 4388~\AA~and 4471~\AA. These lines have a maximum
 strength for ages in the range of 20 -- 50 Myr, and are not produced
 by stellar populations older than 100 Myr. }
\label{fig:Mrk231_HeI_B}
\end{figure}

  The modelling results for Ap C reveal the presence of a VYSP with a
  low reddening ($E(B - V)$ = 0.0 -- 0.1) and ages in the range 50
  $\leq t_{\rm VYSP} \leq$ 80 Myr, that accounts for almost the entire
  optical light within the aperture (P$_{\rm VYSP}$ = 70 -- 96\%). It
  is worth noting that Comb I, Comb II and Comb III provide consistent
  results for this aperture. In addition, the extracted spectrum shows
  the clear presence of He~{\small I} absorption features at various
  wavelengths, which are indicated in Figure
  \ref{fig:Mrk231_HeI_B}. In the models, the He~{\small I} lines are
  strongest for single stellar populations (SSP) with ages in the
  range 20 -- 50 Myr, when the stellar population is dominated by B
  stars \citep{Gonzalez05}, and are not observed in stellar
  populations older than 100 Myr (the lifetime of B stars). Therefore,
  the presence of such features helps to provide a further constraints
  on the age of the VYSP component within the aperture. The detection
  of the He~{\small I} absorption lines is consistent with the lower
  end of the range of ages of the VYSP component deduced from
  modelling the continuum.


Previous spectroscopic studies already revealed the presence of YSPs
in the host galaxy of Mrk 231 \citep{Hamilton87, Canalizo00b}.
\cite{Canalizo00b} carried out a similar study to that presented here,
modelling the optical spectra extracted from a series of apertures
sampling various regions of the galaxy. In order to fit the extracted
spectra, they used a combination of a 10 Gyr OSP plus unreddened YSPs
with varying ages. This is similar to the Comb I used here, although
reddening effects are only fully considered in this study. Their
``regions'' a, b, c and k are similar to our Ap C, Ap B, Ap A and Ap
E, respectively, shown in Figure \ref{fig:spatial_cuts}. They found
YSPs with ages of 5 Myr, 140$^{+80}_{-70}$ Myr and 42$^{+22}_{-17}$
Myr for region c, b and a, respectively, which is consistent with the
modelling results found here using Comb I. On the other hand, they
found no evidence for significant YSPs in region k, similar to the Ap
E used here. However, we find important YSPs (YSP\% = 50 -- 70\%)
within that aperture, with ages and reddening in the range of t$_{\rm
YSP}$ = 60 -- 100 Myr and $E(B - V)$ = 0.2 -- 0.5. This difference is
likely to be due to the fact that the rest wavelength range of the
spectra used in their study is 3800 -- 7000~\AA. Therefore, they do
not sample the near-UV, which is a crucial region of the spectrum for
constraining the properties of the YSPs. In addition, it is notable
that only unreddened synthetic templates are used in the
\cite{Canalizo00b} study, whilst reddened YSPs are required to fit the
data in the case of Ap E.

  {\bf IRAS 13305-1739:} the optical spectrum of this very compact
ULIRG is that of a Sy2 galaxy. In addition, \citetalias{Veilleux99b}
found broad Pa$\alpha$ and [Si~{\small IV}] emission at near-IR
wavelengths -- clear evidence of the presence of a quasar/AGN -- in
the nucleus of the galaxy. A single nuclear aperture was used for the
analysis of the stellar populations in this object. We find a
moderately reddened, dominant VYSP component with a remarkably narrow
range of ages (4 $\leq t_{\rm VYSP} \leq$ 7 Myr).

  {\bf IRAS 13428+5608:} also known as Mrk 273, this galaxy shows an
  impressive tidal tail extended over 30 kpc towards the south of the
  galaxy. Seen as a single-nucleus object in UV/optical images
  \citep{Surace00a,Kim02}, a double nucleus structure emerges at
  near-IR \citep{Armus90,Majewski93,Knapen97,Scoville00} and radio
  \citep{Ulvestad84,Condon91,Cole99} wavelengths, with a nuclear
  separation of $\sim$ 700 pc. The double nucleus structure is unresolved
  in our spectroscopic data. Classified as a Sy2 at optical
  wavelengths, IRAS 13428+5608 is a composite object powered by both
  starburst and AGN activity \citep{Armus07}. However, the dominant
  power source remains controversial.

  In order to perform a detailed study of the stellar populations
  detected at optical wavelengths, a total of 7 apertures were
  extracted, sampling the nuclear region and the large tidal tail. In
  the case of the nuclear 5 kpc aperture, we find an ``old'' IYSPs
  (t$_{\rm IYSP}$ = 0.7 -- 2.0 Gyr) with low reddening which dominates
  the optical emission, with a smaller contribution from a VYSP
  component with higher reddening. Previous optical spectroscopy
  already revealed the presence of a mix of stellar populations in the
  nuclear region of Mrk273 \citep{Gonzalez01}. In a broad-brush sense,
  the results presented here are consistent with those of
  \cite{Gonzalez01}. The extended apertures sample the prodigious
  tidal tail to the south of the galaxy and the north of the nuclear
  region. Overall, the results show a remarkable uniformity, with a
  dominant ``old'' IYSP present at all locations in the galaxy, out to
  a radius of 30 kpc. It is also notable that a VYSP component is
  required in order to fit the data for all the apertures, even for
  those sampling the extended tail. No evidence suggesting the
  presence of important age or reddening variations across the body of
  the galaxy is found in the case of Mrk 273.

  {\bf IRAS 13451+1232:} also known as PKS1345+12, this is the only
  ULIRG in our sample also classified as a radio galaxy. The activity
  is concentrated in the western nucleus, which is the only nucleus
  covered by the two slit positions used for this object. This nucleus
  is classified as a Sy2 in the optical (\citetalias{Veilleux99a}),
  and a compact source at radio wavelengths
  \citep{Evans99}. Significant young stellar populations have already
  been detected in PKS1345+12 at optical and near-IR wavelengths
  \citep{Surace98b,Tadhunter05}, which may have been formed as a
  consequence of the merger.

  A detailed analysis of the stellar populations within the galaxy,
  using the so-called Comb I, was already presented for this object in
  \cite{Rodriguez-Zaurin07}. We emphasize that prior to modelling the
  data for both the \cite{Rodriguez-Zaurin07} study and this work, we
  subtracted the best model for the nebular continuum emission from
  the detailed emission line modelling work of Holt et al. (2003), in
  the case of the 5 kpc aperture (i.e. the ``NUC'' aperture in
  Rodr\'iguez Zaur\'in et al., 2007). Such emission is also likely to
  be important in the cases of Ap A and Ap B in PKS 1345+12. For these
  two apertures, both the results from the modelling using the nebular
  corrected and uncorrected spectra are presented in Table
  \ref{tab:CS_combI}

  In this study, we use an updated version of the code that allows a
  more detailed analysis of the different, important regions of the
  optical spectra and, therefore, it is possible to refine the results
  presented in \cite{Rodriguez-Zaurin07}. In addition, the extracted
  spectra are also modelled using Comb II and III. In general terms,
  the results found here are consistents with those of
  \cite{Rodriguez-Zaurin07}, suggesting that a VYSP may not be as
  important in the case of PKS 1345+12 as for other ULIRGs. In
  addition, we find that adequate fits are also obtained for all the
  apertures with an ``old'' IYSPs (t$_{\rm IYSP}$ = 1.0 -- 2.0 Gyr)
  that dominate the flux at all locations sampled by the
  apertures. Due to the small percentage contribution of the VYSP
  component, the age and reddening of this component are relatively
  unconstrained, which explains the wide range of ages and reddening
  presented in Table \ref{tab:CS_combIII}. Finally, note that this is
  the most massive galaxy in our sample, in terms of its stellar
  population.

  {\bf IRAS 13539+2920:} two 5kpc apertures, sampling the two nuclei
  of this double nucleus system, were extracted for this object. For
  the western nucleus (5kpc-II), which is the brightest at optical
  wavelengths, we find highly reddened VYSPs that contribute
  significantly or dominate the optical emission. For the eastern
  nucleus (5kpc-I), we find ages consistent with those of the western
  source, but with lower reddening values. On the other hand, IRAS
  13539+2920 is spectroscopically classifed as an HII-galaxy in the
  optical. However, modelling the emission lines of the extracted
  spectra for aperture 5kpc-II, sampling the nucleus to the west of
  the galaxy, we find [OIII]5007/H$\beta$ and [NII]6583/H$\alpha$
  emission-line ratios consistent with a combination of an HII-galaxy
  and a LINER. Moreover, the 1D-spectrum of that nucleus shows clear
  signs of line splitting, indicating the presence of ionized gas
  outflows. On the other hand, no evidence of HII-like region features
  is found in the case of the eastern source, sampled by aperture
  5kpc-I (it has a LINER spectrum). Therefore, no attempt to estimate
  the age of the VYSP using the H$\alpha$ equivalent width was made
  for this galaxy.

  {\bf IRAS 14060+2919:} we find a moderately reddened, 7 Myr VYSP
  component that dominates the optical emission in the nuclear region
  of this single nucleus system. In general terms, the results for the
  extended apertures are consistent with those in the nuclear region,
  except for perhaps a slightly smaller VYSP contribution in Ap A
  (although Ap A is not so well constrained because it is noisier). In
  addition, IRAS 14060+2919 is classifed as an HII galaxy at optical
  wavelengths. The H$\alpha$ equivalent widths measured for Ap B, Ap C
  and 5kpc ( 77 \AA~$\leq EW(H\alpha) \leq$ 177 \AA) reveal the
  presence of VYSPs younger than $\sim$ 6 Myr, which is consistent
  with the continuum modelling results.

  {\bf IRAS 14252-1550:} classified as a LINER in the optical, no
  signatures of buried quasar activity are found in the 5-35 $\mu$m
  mid-IR spectrum of this double nucleus system
  \citep{Imanishi07}. The modelling results are consistent with a wide
  range of VYSPs ages, reddenings and percentage contributions. No
  clear evidence suggesting the presence of important age or reddening
  variations is found.

  {\bf IRAS 14348-1447:} This is a double nucleus system, in which the
  ULIRG activity is likely associated with the southern nucleus, which
  dominates the near-IR luminosity. Several bright knots are seen in
  the in the U'-, B- and I-band images \citep{Surace00a,Surace00b},
  some of the knots with estimated ages younger than 5-7 Myr
  \citep{Surace00a}. The optical spectrum of IRAS 14348-1447 is that
  of a LINER \citep{Veilleux95,Kim98a}, and is classified as as
  starburst-powered ULIRG at near-IR, mid-IR, and X-ray wavelengths
  \citep{Veilleux97,Genzel98,Lutz99,Franceschini03,Farrah03,Risaliti06}. We
  find significant, or even dominant VYSPs, with a wide range of
  reddening values, located in the nuclei of the system. Note that in
  the case of the north-eastern nucleus (5kpc-II) the model solutions
  for the VYSPs divide into two groups: a younger group of 7 -- 8 Myr,
  and a second, older, group with ages of 50 -- 80 Myr. In this case,
  we cannot distinguish between the two groups using the detailed
  fits. However, modelling the emission lines of the extracted spectra
  for aperture 5kpc-II we find [OIII]5007/H$\beta$ and
  [NII]6583/H$\alpha$ emission-line ratios consistent with an
  HII-galaxy. The VYSP age obtained using the H$_{\alpha}$ equivalent
  width (EW(H$\alpha$) = 51~\AA, t$_{VYSP}$ $\sim$ 6.5 Myr) is
  consistent with the first group of adequate fits, including the
  younger YSP component. Finally, no evidence is found for significant
  variations in the YSP age and reddening across the body of the
  galaxy.

  {\bf IRAS 14394-1447:} this IRAS source is a multiple ($>$ 2
  nucleus) system, with the two main components separated by 54 kpc --
  the widest nuclear separation in the ES. The eastern nucleus is
  classified as a Sy2 and identified with the source responsible for
  the far-IR luminosity (Veilleux et al., 1999a, Rupke et al.,
  2005). Dominant, highly reddened VYSPs are found for that nucleus
  (5kpc-I). On the other hand, the modelling results are consistent
  with a wide ranges of age and reddening for the VYSP component in
  the case of the 5 kpc aperture extracted for the western source
  (5kpc-II). Moreover, adequate fits are only obtained for that
  aperture using a 2 Gyr IYSP which, in fact, dominates the flux
  contribution in the normalising bin. These results clearly
  demonstrate large differences between the stellar populations in the
  two nuclei, perhaps indicating the presence of a more powerful
  starburst activity in the eastern component. In addition, two
  further apertures were extracted to the east and west of the western
  source. We find no clear evidence for the presence of large age
  and/or reddening variations across the body of that source.

\begin{figure}
\hspace{-0.5cm}\psfig{file=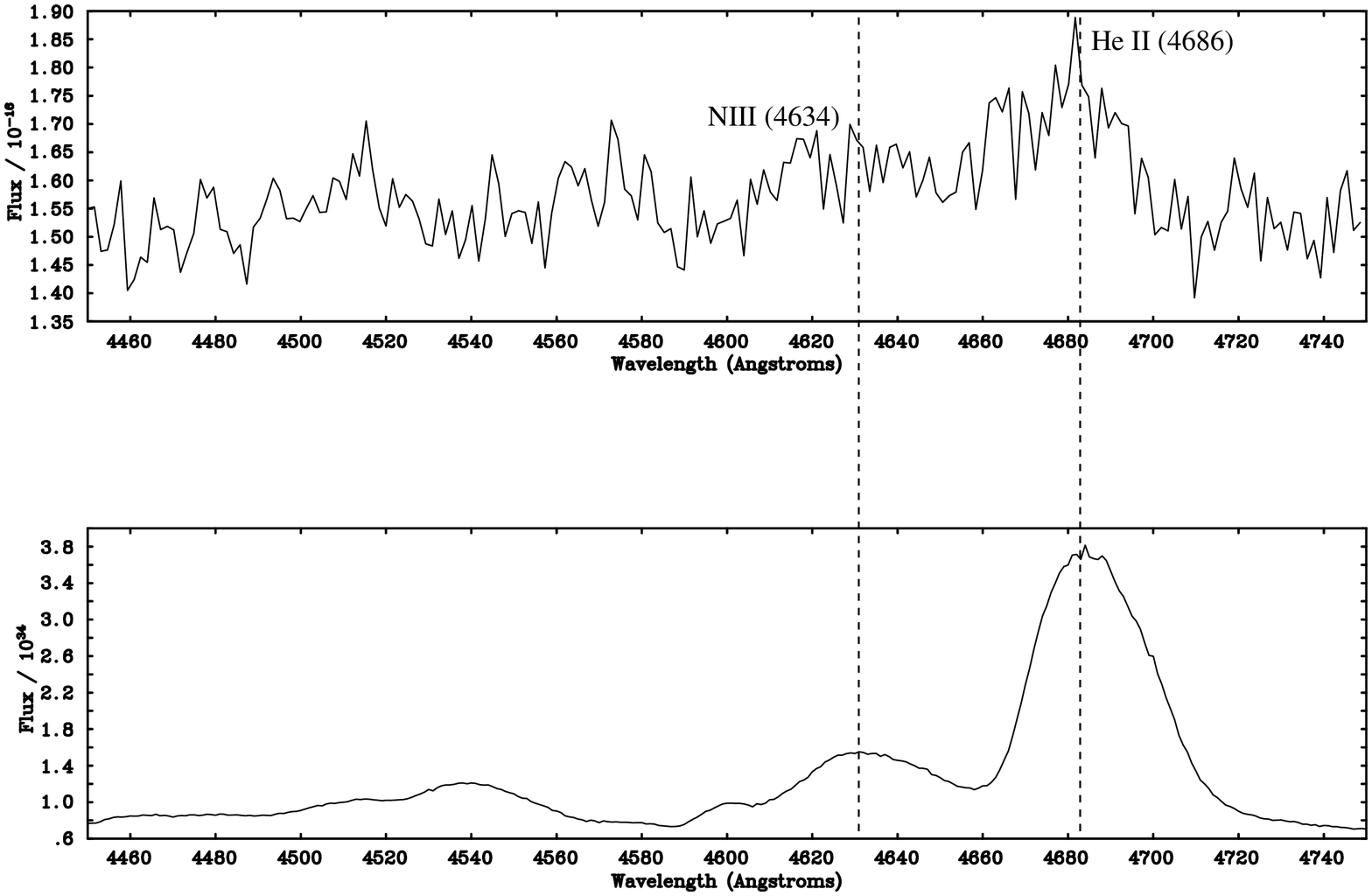,width=9.3cm,angle=0.0}
\caption[WR features in the ULIRG IRAS 15130-1958]{Comparison between
the spectrum of IRAS 15130-1958 (top) and a mid-type WN5-6 star spectrum
in the Milky Way (bottom). The figure shows the detection of the 
NIII4634 and the HeII4686 features, revealing the presence of WR 
stars in the galaxy.}
\label{fig:15130_WR}
\end{figure}

  {\bf IRAS 15130-1958:} the modelling results are consistent with
  VYSPs with a wide range of ages, reddenings and percentage
  contributions for the single 5kpc aperture extracted for this single
  nucleus system, classified as a Sy2 in the optical. In addition,
  Wolf-Rayet (WR) features are detected for this galaxy. The detection
  of such features indicates ongoing star formation activity.  Figure
  \ref{fig:15130_WR} shows the wavelength range 4450 -- 4750 \AA~for
  the spectrum of the galaxy and a mid-type WN5-6 star spectrum in the
  Milky Way (Paul Crowther, private communication). The aim of this
  figure is to compare the morphology of the two spectra in the
  relevant wavelength range. The figure shows the detection of the
  NIII4634 and the HeII4686 features in our spectrum, consistent with
  the presence of WR stars. The detection of such features indicates
  ongoing star formation activity in the case of IRAS
  15130-1958. Among all the galaxies in our sample, this is the only
  case with a clear detection of WR features. Assuming that the VYSP
  component is associated with the WR features, the detection of these
  features provides a further constraint on the VYSP ages, i.e. the
  VYSP ages must be $\lsim$ 10 Myr. This is consistent with the lower
  end of the range of VYSP ages obtained from modelling the continuum.

  {\bf IRAS 15206-3342:} bright knots are observed in the innermost
  region of this single nucleus system, and also spread over even
  larger scales, spanning a wide range of ages \citep[from a few Myr
  up to 2 Gyr,][]{Surace98b}. This object is spectroscopically
  classified as an HII-like galaxy in the optical (Kim et al., 1998;
  Veilleux et al., 1999a). Moreover, no evidence for the presence of a
  buried AGN/quasar is found at IR and radio wavelengths
  \citep{Veilleux97,Nagar03,Imanishi07}. The large nuclear H$\alpha$
  equivalent width and luminosity are also consistent with a
  starburst-dominated ionizing source \citep{Arribas02}. We find
  adequate fits for two groups of combinations of stellar
  populations. The VYSPs in the first group are younger than 3 Myr,
  while such component has a age of 10 Myr in the case of the second
  group. However, the large equivalent width of the H$\alpha$ emission
  line (EW(H$\alpha$) = 315\AA), suggest VYSP ages of t$_{\rm VYSP}
  \lsim$ 5 Myr. This result favours the first group of adequate fits,
  including the younger VYSP component, which is also consistent with
  the upper limit of 20 Myr found by \cite{Farrah05} on the basis of
  their UV/optical spectroscopic study of the stellar populations in
  the galaxy.

  {\bf IRAS 15327+2340:} also known as Arp 220, this double nucleus
  system is the closest ULIRG in our sample (z = 0.018). The double
  nucleus structure is unresoved in our spectroscopic data. A
  biography and a detailed analysis of modelling results are presented
  for this object in \cite{Rodriguez-Zaurin08}. For completeness, the
  modelling results of Comb I, Comb II and Comb III are shown in Table
  \ref{tab:CS_combI}, Table \ref{tab:CS_combII} and Table
  \ref{tab:CS_combIII}, respectively. The extraction apertures and the
  extracted spectra are also shown in Figures \ref{fig:spatial_cuts}
  and \ref{fig:all1dspectra}.

  {\bf IRAS 15462-0450:} due to the powerful AGN emission, it was not
  possible to model the stellar populations in this compact ULIRG,
  which is classified as a Sy1 galaxy at optical wavelengths. The
  extraction aperture and the extracted 1D-spectrum are shown in
  Figures \ref{fig:spatial_cuts} and \ref{fig:all1dspectra}.

  {\bf IRAS 16156+0146:} three apertures were extracted for this
  double nucleus system, sampling the two nuclei and the region in
  between. The optical spectrum is that of a Sy2 galaxy, with the AGN
  activity associated with the north-western source. Moderately
  reddened, dominant VYSPs of age 9 -- 10 Myr are found for that
  source (5 kpc-I). On the other hand, in the case of the
  south-eastern nucleus (5 kpc-II), the modelling results are
  consistent with wide ranges of age, reddening and percentage
  contribution for both the VYSPs and the IYSPs. It is notable that
  the minimum percentage contribution of the VYSPs found for the
  aperture 5kpc-I (60\%), is significantly higher than in the case of
  Ap A and 5kpc-II (20\% and 15\%, respectively). These results
  suggest higher concentrations of VYSPs towards the north-west of the
  system, although we cannot rule out the idea of significant VYSPs at
  all locations.

  {\bf IRAS 16474+3430:} three apertures were used for the analysis of
  the stellar populations in this double nucleus system, centered on
  the nuclei and the region in between. The modelling results found
  for the southern nucleus (5kpc-I), the brightest at optical
  wavelengths, reveal the presence of moderately reddened VYSPs
  (t$_{\rm VYSP}$ = 7 -- 8 Myr) that dominate the flux contribution at
  optical wavelengths. For the other two apertures, the VYSPs are
  divided into two groups: a younger group with ages in the range 6 -
  8 Myr, and a second, older, group with ages in the range 50 -- 90
  Myr. However, IRAS 16474+3430 is classified as an HII-galaxy,
  exhibiting HII-like region features at all locations covered by the
  slit. The values of the H$\alpha$ equivalent widths ( 101 \AA, 49
  \AA~and 69 \AA~for 5kpc-I, 5kpc-II and Ap A, respectively) are only
  attained for stellar populations of $\sim$ 6.5 Myr. This is
  consistent with modelling results found for the southern nucleus
  (5kpc-I), and the younger group of VYSPs in the cases of 5kpc-II and
  Ap A.

  {\bf IRAS 16487+5447:} this galaxy is classified as a LINER galaxy
\citep{Kim98a} in the optical, and shows no evidence for a buried AGN
at mid-IR wavelengths \citep{Imanishi07}, consistent with a starburst
as the power source \citep{Lutz99}. Two slit positions were used
during the obervations of this double nucleus system. For the eastern
nucleus, the modelling results are consistent with the combination of
a dominant ``old'' IYSP plus a VYSP with a wide range of ages and
reddenings. It is notable that adequate fits are also obtained for
this nucleus using Comb I, with a dominant 12.5 Gyr OSP. On the other
hand, for the western source, we find VYSPs with relatively low
reddening, which represent a large fraction of the optical emission
from the galaxy. In addition, modelling the emission lines of the 5 kpc
aperture for this source we find [OIII]5007/H$\beta$ and
[NII]6583/H$\alpha$ emission-line ratios consistent with an
HII-galaxy. The VYSP age obtained using the H$_{\alpha}$ equivalent
width (EW(H$\alpha$) = 28~\AA, t$_{VYSP}$ $\sim$ 7 Myr) is consistent
with the lower end of the range of ages found from the modelling.
These results clearly demonstrate large differences between the
stellar populations in the two nuclei. Finally, no clear evidence is
found for the presence of age or reddening trends in the case of the
western source.

  {\bf IRAS 17028+5817:} again, two slit PAs were used during the
  observations of this system, sampling the two, widely separated (25
  kpc) galaxies. We find that VYSPs make a large contribution, or
  even dominate, the flux at optical wavelengths in the two
  nuclei. However, in the case of the eastern galaxy, classified as an
  HII-galaxy, we find moderately reddened VYSPs that span a narrow
  range of ages. On the other hand, the modelling results are
  consistent with a wider range of ages and significantly higher
  reddening values for the VYSPs located in the western galaxy, which
  is classifed as a LINER in the optical. The modelling results
  obtained for the eastern source are consistent with the age of
  $\sim$6 Myr estimated using the H$\alpha$ equivalent width
  (EW(H$\alpha$) = 102 \AA).  In addition, we extracted a third,
  extended aperture (labelled as Ap A in Figure
  \ref{fig:spatial_cuts}), sampling the region towards the south of
  that source. We do not find clear evidence suggesting the presence
  of large age or reddening variations.

  {\bf IRAS 17044+6720:} classified as a LINER in the optical, this
  single nucleus system shows strong evidence for the presence of
  obscured AGN activity at mid-IR wavelengths \citep{Imanishi07}. The
  modelling results found in the nuclear region of the galaxy are
  consistent with VYSPs of relatively low reddening, that make a large
  contribution to the flux at optical wavelengths. Comparing these
  results with those found for the extended apertures, we do not find
  clear evidence for the presence of significant age and/or reddening
  variations across the body of the galaxy. Note that Ap B (see Figure
  \ref{fig:all1dspectra}) has line ratios typical of an HII region
  ([NII]6583/H$\alpha$ = 0.24 and [OIII]5007/H$\beta$ = 3.0). The
  estimated H$\alpha$ equivalent width is EW(H$\alpha$) = 113
  \AA. Such high equivalent width is only attained for stellar
  populations younger than $\sim$6 Myr, which is consistent with the
  lower end of the age range obtained for Ap B from the continuum
  modelling.

  {\bf IRAS 17179+5444:} a single 5 kpc aperture was extracted for the
  analysis of the stellar populations in this single nucleus
  object. The optical spectrum of the nucleus is that of a Sy2
  galaxy. The modelling results are consistent with a wide range of
  VYSPs ages, reddenings and percentage contributions.

  {\bf IRAS 20414-1651:} this galaxy is classified as a single nucleus
  system in the new {\it HST} H-band imaging survey carried out by
  \cite{Veilleux06}. Optically classified as an HII galaxy
  \citep{Kim98a}, IRAS 20414-1651 shows no evidence for a buried AGN at
  near- and mid-IR wavelengths \citep{Imanishi07}. \cite{Farrah03}
  modelled optical and IR photometric data and found that almost the
  entire IR emission in this source is powered by the starburst. Our
  modelling results reveal the presence of an important contribution
  to the flux from relatively highly reddened VYSPs. The age of
  $\sim$6 Myr estimated using the H$\alpha$ equivalent width
  (EW(H$\alpha$) = 43 \AA) is consistent with the lower end of the age
  range obtained from the modelling.

  {\bf IRAS 21208-0519:} this double nucleus system is classified as
  an HII-galaxy in the optical. However, the northern nucleus, the
  brightest at optical and near-IR wavelengths, is the only region of
  the galaxy clearly showing HII-like features in our spectra. The
  modelling results found for that nucleus (5kpc-I) are consistent the
  presence of moderately reddened, dominant VYSPs. The H$\alpha$
  equivalent width found for that region of the galaxy (EW(H$\alpha$)
  = 45~\AA) indicates VYSPs of age $\sim$6.6 Myr. This is consistent
  with the lower range of ages found from the modelling for this
  component. However the modelling results found for southern nucleus
  (5kpc-II), reveal the presence of a dominant 12.5 Gyr OSP plus an
  ``old'' VYSP (90 -- 100 Myr) with low reddening, that makes a small
  contribution to the flux in the optical. Intriguingly, an important
  contribution from a VYSP component is again required in order to
  model the extracted spectrum from Ap B, centred in the bright blob
  to the south of the galaxy \citep[see][their Figure
  1]{Kim02}. Finally, the modelling results for Ap A, sampling the
  region between the nuclei, are consistent with a mix of stellar
  populations from the two nuclei.

  {\bf IRAS 21219-1757:} this object is classified as a Sy1 galaxy
  at optical wavelengths. Therefore, due to the powerful AGN emission,
  it was not possible to model the stellar populations of this compact
  ULIRG. The extraction aperture and the extracted 1D-spectrum are
  shown in Figures \ref{fig:spatial_cuts} and \ref{fig:all1dspectra}.

  {\bf IRAS 22491-1808:} several knots and condensations located in
  both the tidal features and the nuclear region are visible in the
  high-resolution BIHK$'$-band images presented in
  \cite{Surace00a}. The estimated ages cover a wide range from 5 -- 7
  Myr up to a few hundred Myr. With the optical spectrum of an HII
  galaxy \citep{Veilleux95}, IRAS 22491-1808 shows no signs of
  obscured nuclear activity at IR wavelengths (Genzel et al., 1998;
  Veilleux et al., 1999b; Imanishi et al., 2007) and it is therefore
  classified as a starburst-powered ULIRG \citep{Lutz99,Armus07}. Due
  to the small nuclear separation, and the fact that the slit is not
  perfectly aligned with the double nucleus structure, the 5 kpc
  aperture includes emission from both nuclei. Therefore, we decided
  to use another two apertures to sample the eastern (Ap B) and
  western (Ap C) nuclei individually. 

\begin{figure}
\centering
\hspace{0.0cm}
\psfig{figure=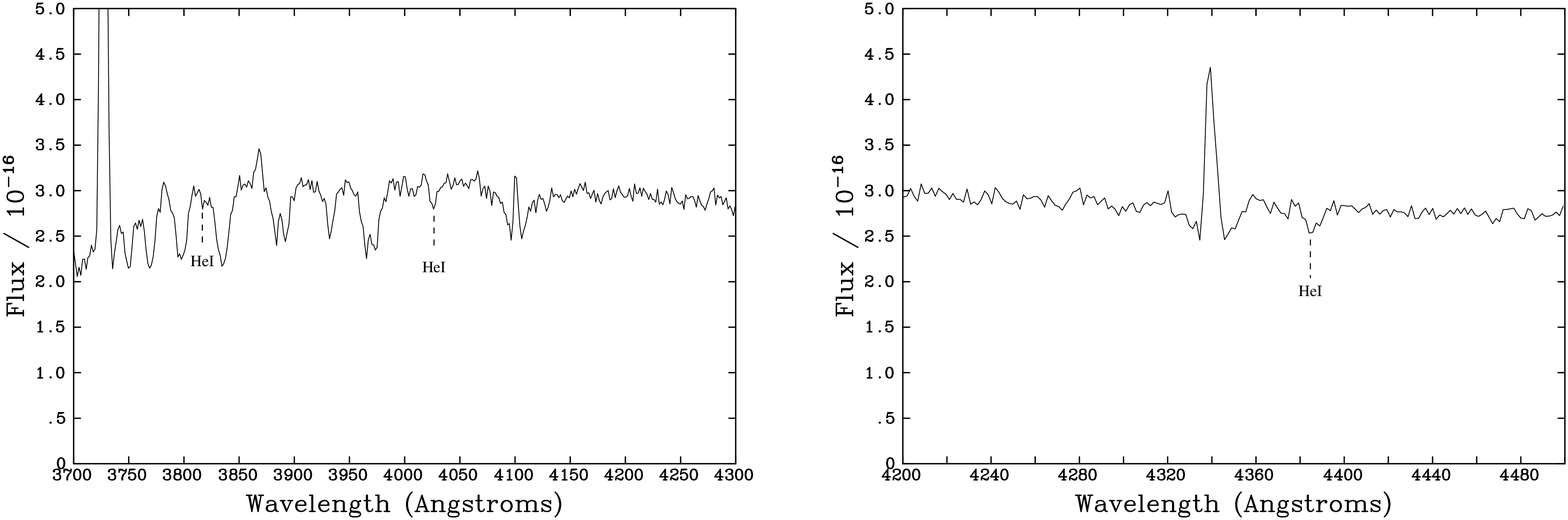,width=8.0cm,angle=0.}\\
\hspace{0.0cm}\psfig{figure=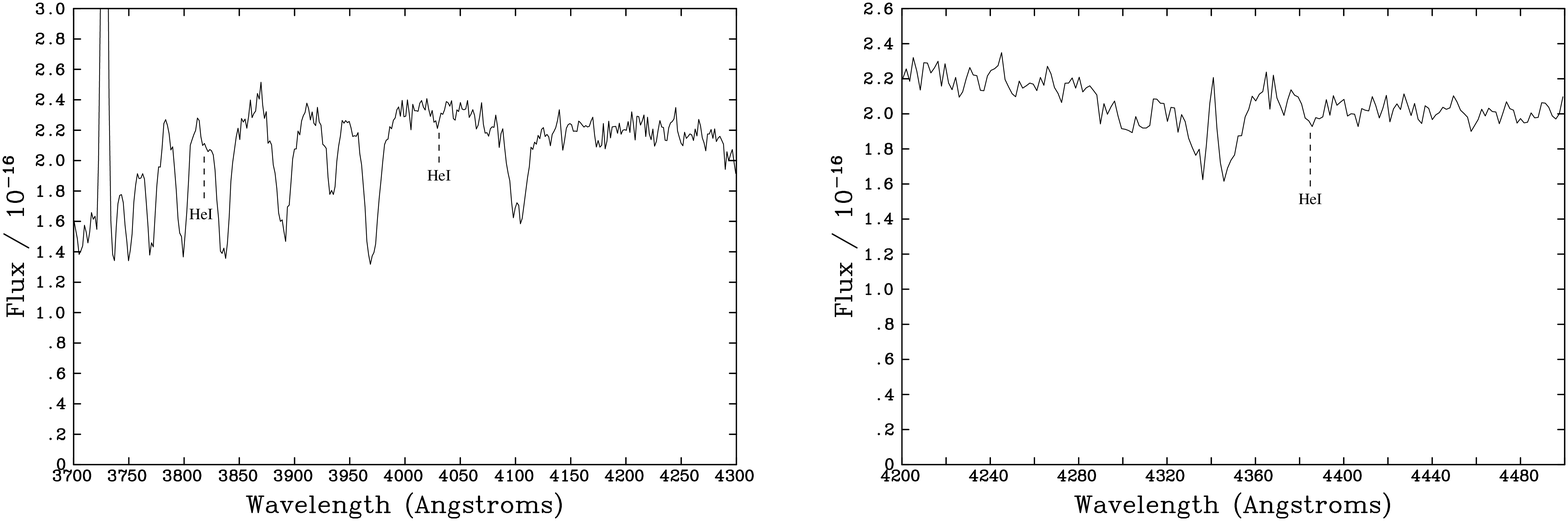,width=8.0cm,angle=0.}
\caption[IRAS 22491-1808: 1-D spectrum showing the location of 
the He~{\small I} absorption features ]{Extracted spectra from Ap B
 (upper panel) and Ap C (lower panel) for IRAS 22491-1808. The figure
 shows the detection of the He~{\small I} absorption lines at
 3820~\AA, 4026~\AA~and 4388~\AA. These lines are not observed in
 stellar populations older than 100 Myr.}
\label{fig:22491_HeI_B}
\end{figure}

  The modelling results found for both nuclei are consistent with the
  presence of moderatelly reddened VYSPs that make an important
  contribution, or even dominate, the flux at optical
  wavelengths. Unsurprisingly, the modelling results for the 5 kpc
  aperture are consistent with the mix of stellar population in the
  two nuclei. The VYSP age determined from the H$\alpha$ equivalent
  widths ( 64 \AA~and 46 \AA~for Ap B and C, respectively,
  corresponding to VYSPs of age $\sim$6.5 Myr), is consistent with the
  results obtained from the modelling. Furthermore, the extracted
  spectra for Ap B and C (and also 5kpc) show the clear presence of
  He~{\small I} absorption features at various wavelengths, which are
  indicated in Figure \ref{fig:22491_HeI_B}. These lines are not
  observed in stellar populations older than 100 Myr and, therefore,
  provide further evidence for the presence of significant VYSPs in
  the two nuclei of the galaxy. Finally, Ap A was extracted to
  sample the tidal tail towards the north-west of the galaxy. Overall
  we do not find clear evidence for the presence of significant
  age/reddening gradients across the body of the galaxy.

  {\bf IRAS 23060+0505:} this single nucleus system is classified as a
  Sy2 in the optical, and shows clear signatures of AGN activity at
  mid-IR wavelengths \citep{Genzel98,Lutz99}. For the nuclear region
  adequate fits are only obtained using Comb II, including a 10 Myr
  VYSP, a ``red'' power law component (p.l. index $\alpha$ = 6.1) and
  a 12.5 Gyr OSP making contributions to the optical light of 72\%,
  6\% and 22\% respectively. It is worth remembering that these are
  percentage contributions in the normalising bin, which in the case
  of IRAS 23060+0505 is located at a wavelength range of 4450 -
  4550~\AA. The contribution of the power-law component becomes more
  important with increasing wavelength, and is as high as 43\% at
  6800~\AA. This is consistent with the results of \cite{Veilleux97}
  who found a strong broad Pa$_{\alpha}$ line at near-IR wavelength,
  providing evidence for a significant contribution from a highly
  obscured AGN. For the other two apertures, Ap A and B, adequate fits
  are obtained using all combinations (i.e. the YSP properties are
  relatively unconstrained).

  {\bf IRAS 23233+2817:} \cite{Kim02} described this object as: ``the
  only object in the 1 Jy sample with no obvious signs of
  interaction''. Our Ap A samples a region to the south of the
  nucleus, including the bright knot detected in the \cite{Kim02}
  K-band image. These authors described the latter as a bright HII
  regions on the basis of their luminosities and colours. However we
  identified this knot as an M star in the galaxy based in our
  spectroscopic data (see Figure \ref{fig:all1dspectra}). In the case
  of the 5 kpc aperture, adequate fits were obtained using Comb I and
  Comb II, but not Comb III. Since the galaxy is classified as a Sy2,
  we concentrate on the results of Comb II. We find YSPs in
  combination with a 12.5 Gyr OSP, and power-law that makes a
  significant contribution to the flux at optical wavelengths. These
  results suggest that the AGN component is important in the case of
  IRAS 23233+2817.

  {\bf IRAS 23234+0946:} this double nucleus system is classified as a
LINER on the basis of its optical spectrum. The western source is the
brightest at optical wavelengths, and shows no sign of AGN activity in
the mid-IR \citep{Imanishi07}. The modelling results found for this
nucleus (5 kpc-I) are consistent with the presence of VYSPs that
dominate the flux in the optical. Comparing these results with those
of the other two apertures, it is notable that the percentage
contribution of the VYSP component in the western nucleus (60 -- 82\%)
is significantly higher than in the eastern nucleus and the extended
structure to the east of the galaxy (17 -- 37\%). Intriguingly, while
high reddening values are found for the VYSP component ($E(B - V)$ =
0.5 -- 1.5) to the east of the galaxy (5kpc-II and Ap A), lower
reddening values ($E(B - V)$ = 0.4 -- 0.6) are found for the western
source.

  {\bf IRAS 23327+2913:} initially, two 5 kpc apertures were extracted
  sampling the two nuclei of this double nucleus system. In order to
  look for possible variations of the stellar populations within the
  galaxy when sampling larger regions, two additional, wider apertures
  were extracted, also centered on each nucleus. The extracted spectra
  for these apertures are identical to those of the 5 kpc apertures
  and therefore are not shown in Figure \ref{fig:all1dspectra}. In the
  case of the southern source (5kpc-I) an OSP is always required in
  order to model the extracted spectrum. The modelling results of Comb
  I are consistent with a mix of a 12.5 Gyr OSP plus a moderately
  reddened VYSP of 9 Myr that dominates the flux at optical
  wavelengths. On the other hand, this object is classified as a LINER
  at optical wavelengths and shows no evidence for a buried AGN
  \citep{Imanishi07}. Therefore, it is likely that the power-law
  component used in Comb II accounts for a VYSP in this case, and it
  is possible that we are detecting the contribution of three
  different stellar components to the optical light: a 12.5 Gyr OSP
  plus important contribution from a highly reddened VYSP younger than
  $\sim$ 5 Myr (represented by a power-law with power-law index
  $\alpha$ = 1.54) and a VYSP of age 10 Myr and lower reddening. In
  the case of the northern galaxy (5kpc-II), it is possible to model
  the extracted spectrum using a single OSP of age 12.5 Gyr, with no
  need for any additional stellar component. Again, an OSP is always
  required to model the data. This demonstrates large differences
  between the stellar populations in the two nuclei.

  {\bf IRAS 23389+0300:} the northern source of this double nucleus
  system is the brightest galaxy in the radio, known to have a
  milli-arcsecond radio core and extended radio emission
  \citep{Nagar03}. Two 5 kpc apertures were extracted for this Sy2
  galaxy, centred on the two nuclei. In order to perform a more
  detailed analysis of the stellar populations, an additional set of
  four smaller apertures was extracted sampling various regions of
  the galaxy. The extracted spectra from Ap B and D are identical to
  those of 5kpc-I and 5kpc-II, and hence, they are not shown in Figure
  \ref{fig:all1dspectra}. The modelling results suggest the presence
  of younger VYSPs towards the north of the system, although we cannot
  rule out the idea of VYSPs of similar age at all locations sampled
  by the various apertures. In terms of reddening, we find no clear
  evidence for the presence of significant variations across the body
  of the galaxy.

\end{document}